\documentclass[12pt,a4paper]{book}
\usepackage{amssymb}
\usepackage{german}
\usepackage{graphicx}
\DeclareGraphicsExtensions{.eps}
\usepackage{psfrag}
\newcommand{\aeff}{\alpha_{\rm eff}}
\newcommand{\alDW}{\`{a} la Dokshitzer, Webber}
\newcommand{\ao}{\overline{\alpha}_1}
\newcommand{\an}{\overline{\alpha}_0}
\newcommand{\as}{\alpha_s}
\newcommand{\asmz}{\alpha_s(M_Z)}
\newcommand{\Bf}{^\star}
\newcommand{\BF}{{^\star}}
\newcommand{\chin}{\chi^2/{\rm dof}}
\newcommand{\cm}{\,{\rm cm}}
\newcommand{\gev}{\,{\rm GeV}}
\newcommand{\gevq}{\,{\rm GeV}^2}
\newcommand{\grad}{{^\circ}}
\newcommand{\hftwo}{\hspace*{\fill}}
\newcommand{\highq}{\mbox{high $Q^2$}}
\newcommand{\ich}{{\scriptscriptstyle i \in {\rm CH}}}
\newcommand{\ee}{\mbox{$e^+e^-$~}}
\newcommand{\mean}[1]{\left< #1 \right>}
\newcommand{\fmean}{\mean{F}}
\newcommand{\fpert}{\fmean^{\rm pert}}
\newcommand{\fpow}{\fmean^{\rm pow}}
\newcommand{\lmsb}{\Lambda_{5,\overline{\rm MS}}}
\newcommand{\lowq}{\mbox{low $Q^2$}}
\newcommand{\mev}{\,{\rm MeV}}
\newcommand{\mf}{\mu_{\scriptscriptstyle F}}
\newcommand{\mi}{\mu_{\scriptscriptstyle I}}
\newcommand{\mr}{\mu_{\scriptscriptstyle R}}
\newcommand{\anmi}{\overline{\alpha}_0(\mi=2\gev)}
\newcommand{\order}{{\cal O}}
\newcommand{\rbthm}{\rule[-2ex]{0ex}{5ex}}
\newcommand{\rbthr}{\rule[-1.7ex]{0ex}{5ex}}
\newcommand{\rbtrm}{\rule[-2ex]{0ex}{5ex}}
\newcommand{\rbtrr}{\rule[-0.8ex]{0ex}{3.2ex}}
\newcommand{\rmvec}[1]{{\rm \vec{#1}}}
\newcommand{\qi}{``}
\newcommand{\qo}{''}
\newcommand{\eprint}[2]{{hep-#1/}#2.}
\newcommand{\eprintk}[2]{{hep-#1/}#2,}

\newcommand{\ejrnl}[4]{{#1} #2 (#3) #4.}

\newcommand{\jrnl}[4]{{#1} {\bf #2} (#3) #4.}
\newcommand{\jrnls}[4]{{#1} {\bf #2} (#3) #4;}
\newcommand{\CPC}{{\em Comp.\ Phys.\ Comm.}}
\newcommand{\EPJC}{{\em Eur.\ Phys.\ J.\ }{\bf C}}
\newcommand{\IJMPA}{{\em Int.\ J.\ of Mod.\ Phys.\ }{\bf A}}
\newcommand{\JPC}{{\em J.\ Comp.\ Phys.}}
\newcommand{\JHEP}{\bf JHEP}
\newcommand{\JPG}{{\em J.~Phys.\ }{\bf G}}

\newcommand{\NIMA}{{\em Nucl.~Instr.\ and Meth.\ }{\bf A}}
\newcommand{\NPB}{{\em Nucl.~Phys.\ }{\bf B}}
\newcommand{\NPPS}{\em Nucl.~Phys.\ Proc.~Suppl.}
\newcommand{\PLB}{{\em Phys.~Lett.\ }{\bf B}}
\newcommand{\PRL}{\em Phys.\ Rev.\ Lett.}
\newcommand{\PRD}{{\em Phys.~Rev.\ }{\bf D}}
\newcommand{\ZPC}{{\em Z.~Phys.\ }{\bf C}}
\newcommand{\PInst}{I.}                               
\newcommand{\Pdat}{Dezember 1998}                     
\newcommand{\Pnum}{98/44}                             
\newcommand{\Name}{Rabbertz}                          
\newcommand{\Vname}{Klaus}                            
\newcommand{\Ptit}{%
  Power Corrections to\\\bigskip
  Event Shape Variables\\\bigskip
  measured in\\\bigskip
  ep~Deep-Inelastic Scattering\\} 
\newcounter{cut}
\newcommand{\cut}{\refstepcounter{cut}}
\begin{document}
\pagestyle{empty}
%
%
%
\addtolength{\oddsidemargin}{-0.5cm}
\addtolength{\evensidemargin}{+0.5cm}
\addtolength{\topmargin}{-1.2cm}
%
%
\begin{figure}[ht]
  \includegraphics[height=2.2cm]{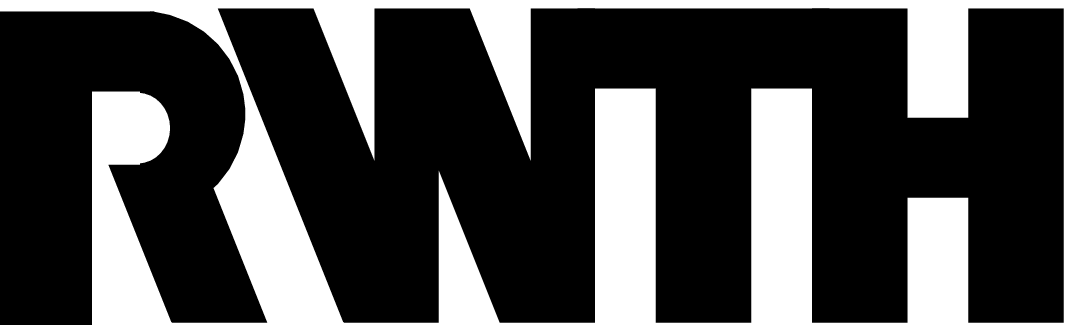}\hfill%
  \raisebox{-0.45cm}{\parbox[b]{3.0cm}{\small\sf%
      \begin{flushleft}
        {RHEINISCH\\     
          WESTF"ALISCHE\\ 
          TECHNISCHE\\   
          HOCHSCHULE\\    
          AACHEN}
      \end{flushleft}
      }}\hfill%
  \raisebox{-0.05cm}{\parbox[b]{4.0cm}{%
      \begin{flushright}
        {\large\sf PITHA \Pnum}\\
        \rule[0.2cm]{4.0cm}{1.0mm}\\[-0.1cm]
        {\large\sf \Pdat }\\
      \end{flushright}
      }}%
  \vspace{3.18cm}%
  \begin{center}
    {\LARGE\sf\Ptit}
    \vspace{2.35cm}
    {\Large\sf\Vname~\Name}\\
  \end{center}%
  \vspace*{5.cm}%
  \begin{center}
    {\normalsize\sf\PInst\ Physikalisches Institut der
      Technischen Hochschule Aachen\\}
  \end{center}%
  \vspace*{-0.35cm}%
  \rule{15.8cm}{0.2mm}
  \begin{center}
    \LARGE\sf
    PHYSIKALISCHE INSTITUTE\\
    RWTH AACHEN\\
    52056 AACHEN, GERMANY
  \end{center}
\end{figure}
%
%
\addtolength{\oddsidemargin}{+0.5cm}
\addtolength{\evensidemargin}{-0.5cm}
\addtolength{\topmargin}{+1.2cm}


\cleardoublepage
\begin{titlepage}
\begin{center}
  {\LARGE \bf
    Power Corrections to\\
    Event Shape Variables\\
    measured in\\
    \boldmath$ep$\unboldmath~Deep-Inelastic Scattering\\[3.0cm]}
  Von der Mathematisch-Naturwissenschaftlichen Fakult"at\\
  der Rheinisch-Westf"alischen Technischen Hochschule Aachen\\
  genehmigte Dissertation zur Erlangung des akademischen Grades\\
  eines Doktors der Naturwissenschaften\\[3.0cm]
  vorgelegt von\\[1.0cm]
  {\large Diplom-Physiker}\\[0.25cm]
  {\large Klaus Rabbertz}\\[1.0cm]
  aus M"onchengladbach\\[3.0cm]
  \begin{tabular}{lcl}
    Berichter & : & Universit"atsprofessor Dr.~Ch.~Berger\\
    & : & Universit"atsprofessor Dr.~S.~Bethke\\
    &&\\
    Tag der m"undlichen Pr"ufung & : & 23.12.1998
  \end{tabular}
\end{center}
\end{titlepage}


\cleardoublepage
\vspace*{9.75cm}
\begin{flushright}
  \large\em\ Meinen Eltern
\end{flushright}


\cleardoublepage
\vspace*{-1.5cm}
\hspace{9.cm}
\parbox[t]{0.3\textwidth}
{\large{\em Ceci n'est pas une pipe.}---Ren\'{e} Magritte}


\vspace*{2.24cm} \centerline{\Huge \bf Abstract} \vspace{1.0cm}
Deep-inelastic $ep$ scattering data, taken with the H1 detector at HERA,
are used to study event shape variables over a large range of
\qi relevant energy\qo\ $Q$ between $7\gev$ and $100\gev$. Previously
published analyses on thrust, jet broadening, jet mass and $C$~parameter
are substantially refined and updated; differential two-jet rates treated
as event shapes are presented for the first time.

The $Q$ dependence of the mean values is fit to second order calculations
of perturbative QCD applying power law corrections proportional to $1/Q^p$
to account for hadronization effects. The concept of these power corrections
is tested by a systematic investigation in terms of a non-perturbative
parameter $\overline{\alpha}_{p-1}$ and the strong coupling constant.
\\[2.0cm]
\vspace{1.0cm}\centerline{\Huge \bf Kurzfassung}
Ereignisformvariablen in der tief"|inelastischen $ep$-Streuung, gewonnen
aus Daten des H1-Detektors bei HERA, werden "uber einen gro"sen Bereich
der "`relevanten Energieskala"' $Q$ von $7\gev$ bis $100\gev$ untersucht.
Bereits publizierte Studien zu {\em thrust}, Jetbreite, Jetmasse und
$C$~Parameter werden erheblich verbessert und aktualisiert; Ergebnisse zu
differentiellen Zweijetraten als Ereignisformen werden erstmals vorgestellt.

Bei der Anpassung der $Q$-Abh"angigkeit der Mittelwerte an Berechnungen der
perturbativen QCD in zweiter Ordnung sollen potenzartige Korrekturterme der
Form $1/Q^p$ Hadronisierungseffekte ber"ucksichtigen. Eine "Uberpr"ufung des
Konzepts solcher Potenzkorrekturen erfolgt im Rahmen der Bestimmung eines
nichtperturbativen Parameters $\overline{\alpha}_{p-1}$ und der
Kopplungskonstanten der starken Wechselwirkung.


\cleardoublepage
\pagestyle{headings}
\pagenumbering{arabic}
\tableofcontents
\chapter{Introduction}
\label{chap:intro}

Curiosity is one of mankind's elementary driving forces ---~the more
so when scientists are concerned. In European history the appearance
of the first \qi professional scientists\qo\ is usually dated back to
antique Greece, where, besides others, philosophers laid the
foundations for a great part of European culture.  Two of them living
around 500 B.C., Leukippos and Demokritos, suggested first that matter
consists of tiny indivisible particles called {\em atoms}. They may be
considered the forefathers of today's elementary particle physicists.

Their reasoning was of a basically theoretical nature.  Turning to
experiment, a \qi quantum leap\qo\ leads us to Antoni van Leeuwenhoek
(1632--1723). Employing microscopes he constructed himself, he was
able to observe \qi creatures that swarm and multiply in a drop of
water\qo\footnote{H.G.~Wells, {\em The War of the Worlds}, 1898.}  for
the first time.

With the advent of the modern microscope {\bf H}adron-{\bf
  E}lektron-{\bf R}ing-{\bf A}nlage HERA at the {\bf D}eutsches {\bf
  E}lektronen-{\bf SY}nchrotron DESY in Hamburg three centuries later,
the resolution power could be increased by about twelve orders of
magnitude. Now it is possible to study \qi gluons that propagate and
split in a droplet of nuclear matter called proton.\qo

The theory that is believed to describe the {\em strong}\/
interaction, responsible for the structure of hadronic (nuclear)
matter, is called {\bf Q}uantum {\bf C}hromo{\bf D}ynamics (QCD) in
analogy to {\bf Q}uantum {\bf E}lectro{\bf D}ynamics (QED) dealing
with the {\em electromagnetic}\/ interaction. Both are part of a more
comprehensive theory usually referred to as the {\em Standard Model}\/
(for introductory textbooks s.\ e.g.~\cite{Berger,HM,Perkins,Quigg}).
It unites three, the {\em weak}, electromagnetic and strong
interaction, out of the four fundamental forces observed in nature
into a common framework. Gravity so far does not fit in.  Comprising
merely twelve elementary particles of spin $1/2$, six quarks and six
leptons, and the spin $1$ exchange quanta of the three forces, i.e.\ 
the $Z^0$ and $W^\pm$ bosons, the photon $\gamma$ and eight gluons
$g$, the Standard Model gives a very good account of a tremendous
amount of data gathered by experiments during the last
decades~\cite{PDG}.

The HERA collider offers the unique possibility to investigate the
structure of the proton and its interacting constituents in a
completely new kinematic domain unreached by fixed-target experiments.
With respect to this analysis, the large accessible range in \qi
relevant energy\qo\ $Q$ is of great advantage.  Observables can that
way be studied and compared to theory in dependence of the available
energy $Q$ in a single experiment.  To that goal, the thesis is
organized as follows:

Chapter two describes the machinery necessary to perform $ep$
scattering experiments at the required high energies with special
attention to the parts relevant for this analysis.  The third chapter
explains the basic kinematics of the scattering process.
Subsequently, the observables to measure are defined in chapter four.
The series of chapters five, six and seven deals in detail with the
selection of data, their comparison with simulations and the
correction for detector imperfections. Fully corrected data are
presented.  The two following chapters eight and nine are dedicated to
the derivation of the necessary theoretical input employing {\bf
  p}erturbative {\bf QCD} (pQCD).  Finally, experimental and
theoretical results are compared and combined to test the power
correction approach to hadronization effects that was initiated by
Dokshitzer and Webber~\cite{pc:DWform1}.  The last chapter presents a
summary and an outlook.


\chapter{HERA and H1}
\label{chap:HERAH1}
\section{The Storage Ring HERA}
\label{sec:HERA}

Scattering experiments are the basic tools to investigate the
elementary building blocks of matter, i.e.\ quarks and leptons, and
the fundamental forces acting between them.  Two rather
straightforward set-ups are \ee~and $p\bar{p}$ colliders. In the last
decades, a lot of valuable information has been accumulated with such
machines~\cite{PDG}.  However, for the detailed study of a strongly
bound, hadronic object like the proton, it is preferable to use a
probe that does not itself interact strongly.  Fixed-target
experiments employ electrons and myons as well as neutrinos to
determine the structure of nucleons.  To enlarge the resolving power,
it is necessary to increase the energy available in the collisions by
accelerating the \qi targets,\qo\ i.e.\ a collider experiment is asked
for.

\begin{figure} 
  \centering \includegraphics[width=15.1cm]{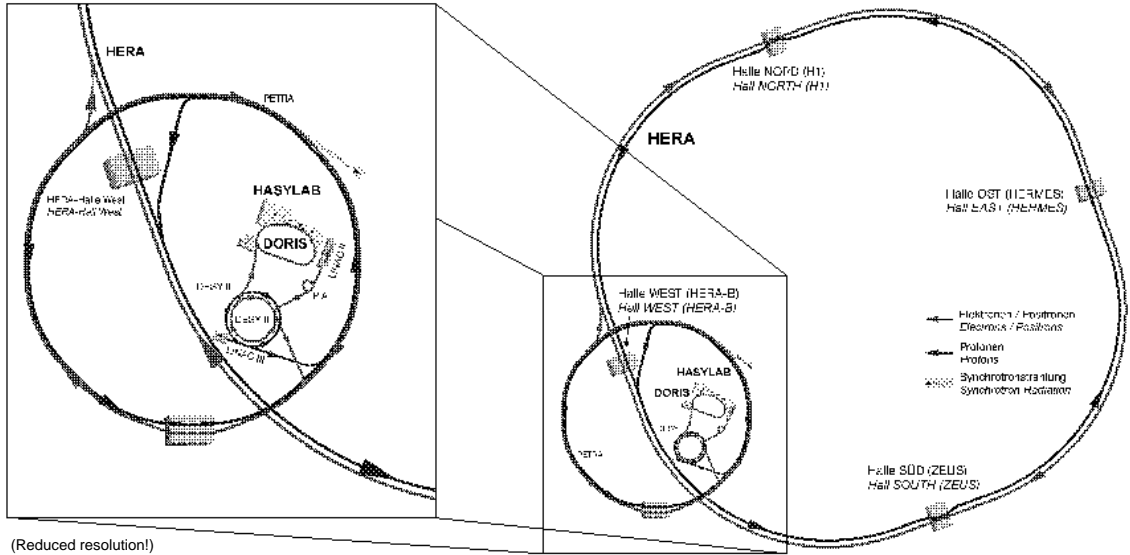}
  \caption{The storage ring HERA at the DESY laboratory in Hamburg, Germany.}
  \label{fig:HERAZOOM}
\end{figure}

The Hadron-Elektron-Ring-Anlage HERA, shown in
fig.~\ref{fig:HERAZOOM}, is situated at DESY in Hamburg, Germany, and
is currently the world's only facility where electrons\footnote{Due to
  a considerably higher lifetime at large currents, the electron beam
  has been replaced by a positron beam in July 1994. Since for the
  purpose of this study it does not make a difference, the term {\em
    electron}\/ will henceforth be used synonymously for positrons,
  too.}  and protons are accelerated in two separate storage rings to
final energies of $27.5\gev$ and $820\gev$ respectively.  The
resulting center of mass energy of about $300\gev$ corresponds to
electron beams of $50\,{\rm TeV}$ for fixed-target experiments.  For a
recent overview of the knowledge gained on the structure of nucleons
consult e.g.~\cite{DIS:EiselePIC98}.

Along the circumference of $6.3\,{\rm km}$ two locations, the north
hall and the south hall, are assigned to the study of $ep$ scattering.
The interaction regions, where during operation every $96\,{\rm ns}$
particle bunches may collide, are surrounded almost hermetically by
complex detectors, H1~\cite{H1:NIMH1} and ZEUS~\cite{ZEUS}.  They are
dedicated to the task of measuring as many collision products as
precisely as possible.  Two other experiments, HERMES and
\mbox{HERA-B}, are situated in the east and west halls and are
committed to spin and $B$ meson physics respectively.

\section{The H1 Detector}
\label{sec:H1}

The basis of this experimental analysis are data collected with the H1
detector, which is located at the northern interaction point of
HERA\@.  Only a brief overview of the total system, shown in
fig.~\ref{fig:H1}, will be given in the next section.  The parts that
provide the main information needed in this study will subsequently be
discussed in more detail. A complete description of the detector and
its performance in the first three years of operation (1992--1994) can
be found in~\cite{H1:NIMH1,H1:NIMTCM}.

During the winter shutdown 1994/95, a major upgrade of the H1 backward
region was undertaken~\cite{H1:upgrade}.  However, none of the
components relevant to this analysis were significantly affected.
Therefore, these improvements will not be covered here, although data
from 1994 up to 1997 are used. Events where the scattered electron was
found in the backward calorimeter are taken from 1994 data only.

\subsection{Overview of the H1 Detector}
\label{sec:H1overview}

The most obvious feature of the H1 detector in fig.~\ref{fig:H1} is
its asymmetric design. Because the momentum of the protons is about
$30$ times higher than that of the electrons, the $ep$ center of mass
system is moving along the direction of the proton beam. As a result,
the density of collision products hitting the equipment in the \qi
forward\qo\ region is very high which is accordingly reflected in the
more massive as well as finer granulated material of that part. The
term \qi forward\qo\ refers to the conventional coordinate system used
in the H1 collaboration where the proton beam direction is defined to
be the $+z$-axis. The $x$-axis ($y$-axis) points from the nominal
interaction point at the center of the HERA ring~(upwards).
Subsequently, the detector components are briefly described proceeding
from the innermost parts outwards:

\begin{figure} 
  \centering \includegraphics[width=14.25cm]{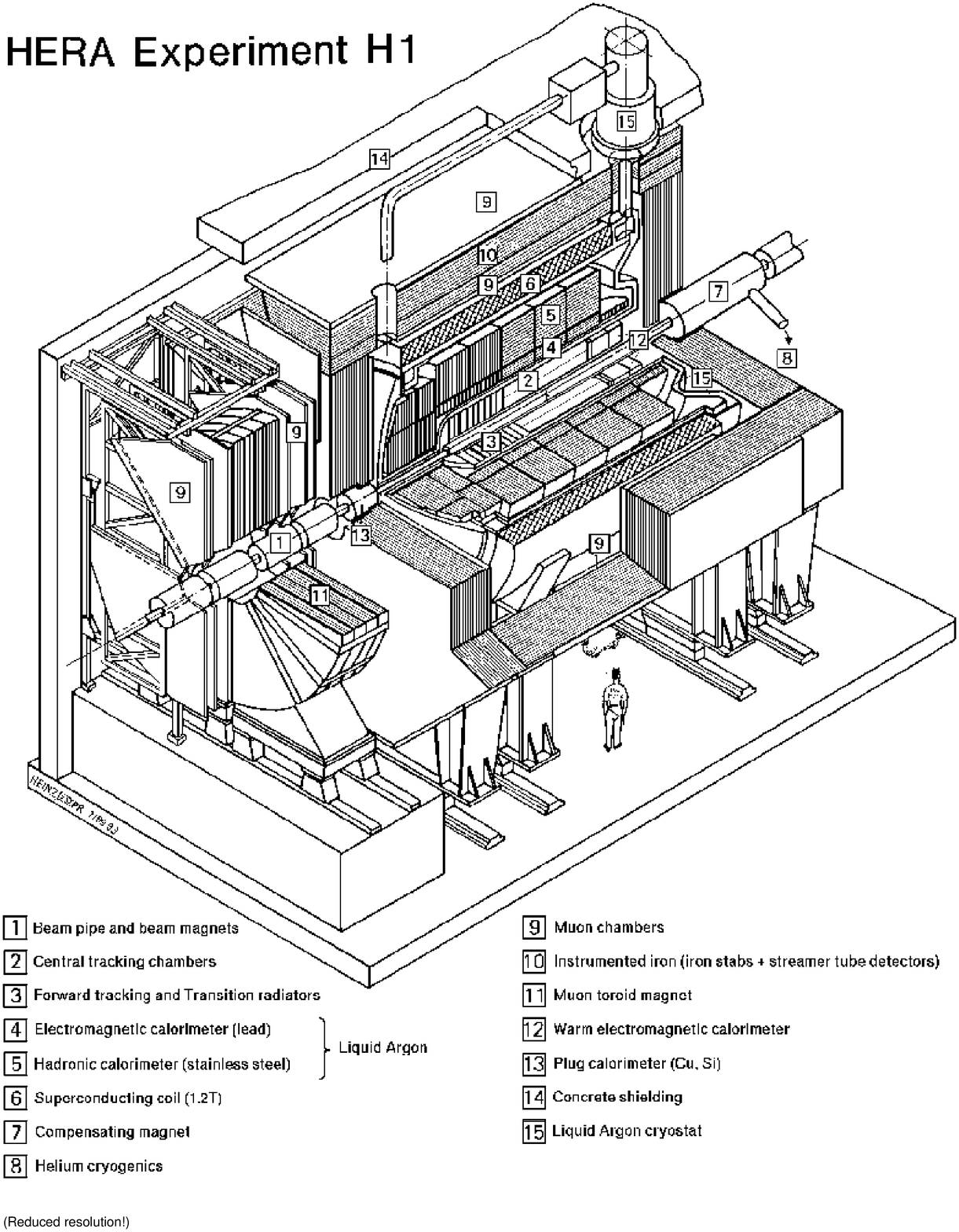}
  \caption{Schematic layout of the H1 detector.}
  \label{fig:H1}
\end{figure}

\begin{figure} 
  \centering \includegraphics[viewport=78 359 513 646]{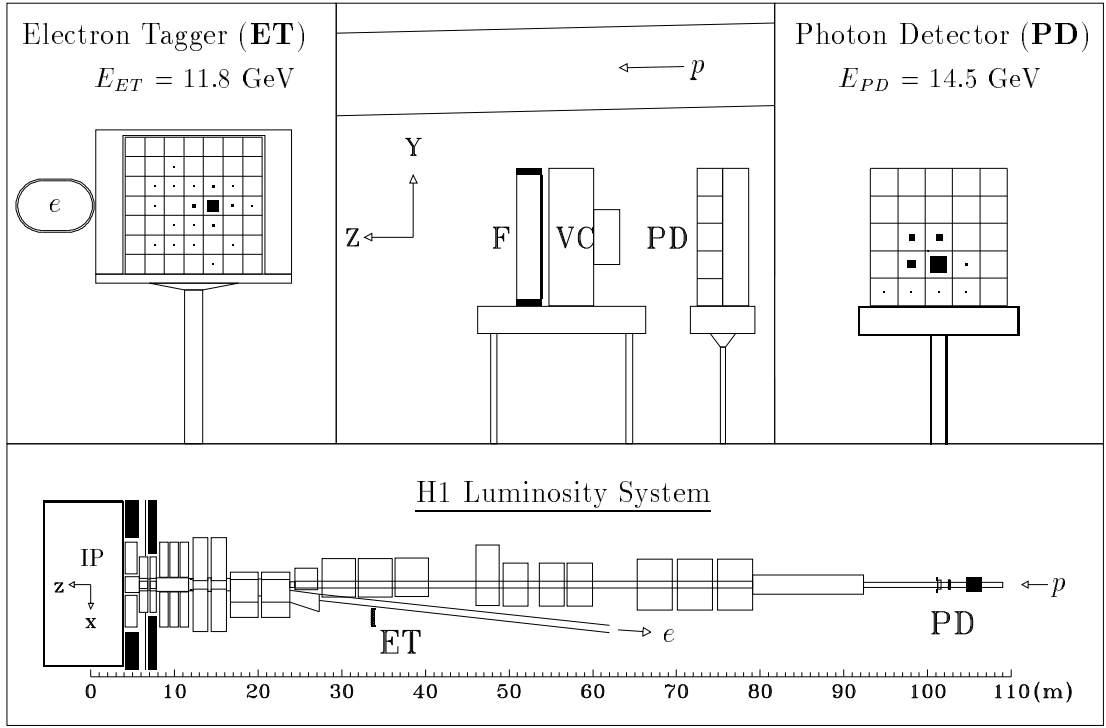}
  \vspace{0.5cm}\\
  \caption[The layout of the H1 luminosity system.]
  {The layout of the H1 luminosity system~\cite{H1:NIMH1}.}
  \label{fig:LUMI}
\end{figure}

\begin{itemize}
\item The tracking system:
  
  Directly surrounding the beam pipe and beam magnets, a tracking
  system is installed to measure the momentum of charged particles.
  It consists of two main parts: the {\bf C}entral {\bf T}racking {\bf
    D}evice (CTD) covering the region around the nominal interaction
  vertex and the {\bf F}orward {\bf T}racking {\bf D}evice (FTD)
  supplementing it in the $+z$-direction.
  
\item The calorimeters:
  
  To complement the momentum measurement and to detect neutral
  particles, the CTD and FTD are enclosed in the forward and central
  region by a sampling calorimeter (LAr) with {\bf L}iquid {\bf Ar}gon
  as active material.  For the innermost absorber stacks, lead has
  been chosen to ensure a good containment and energy determination of
  electromagnetic showers produced by electrons and photons ({\bf
    E}lectromagnetic {\bf CAL}orimeter, ECAL).  The outer part of the
  LAr ({\bf H}adronic {\bf CAL}orimeter, HCAL) predominantly measures
  hadronic showers and is equipped with steel absorber plates, which
  also serve as mechanical support structure.
  
  The remaining holes of the LAr around the beam pipe are closed with
  a silicon-copper calorimeter for polar angles below $4\grad$ (PLUG)
  and a lead-scintillator calorimeter in the backward direction ({\bf
    B}ackward {\bf E}lectro{\bf M}agnetic {\bf C}alorimeter, BEMC).
  During the upgrade in 1994/95 the BEMC was replaced by a
  lead-scintillating fibre {\bf Spa}ghetti {\bf Cal}orimeter
  (SpaCal)~\cite{H1:SpaCal}, which is subdivided into a first
  electromagnetic section and an additional second part to determine
  energy leakage and to improve the containment of hadronic showers.
  For the purpose of this analysis, it was only employed to supplement
  eventual energy deposits in the backward direction not due to the
  scattered electron.
  
\item The superconducting coil:
  
  A cylindrical superconducting coil providing the magnetic field of
  $1.15\,{\rm T}$ for the trackers envelops the calorimeters. Thereby,
  the amount of dead material in front of the calorimeters is reduced
  and the time of flight of myons within the magnetic field is
  increased improving the resolution of their momentum measurement.
  
\item The instrumented iron yoke and myon system:
  
  The {\bf IRON} return yoke (IRON) of the magnet, enclosing almost
  completely all other parts of the detector, is sandwiched with
  streamer tubes for the measurement of myons and energy leakage from
  the inner calorimeters ({\bf T}ail {\bf C}atcher, TC).
  Supplementary chambers inside the IRON further improve the
  evaluation of myon tracks. Myons with high momenta in the forward
  direction are analyzed by a spectrometer consisting of four drift
  chambers in the magnetic field of $1.5\,{\rm T}$ of a toro\"{\i}dal
  coil in front of the IRON.
  
\item The luminosity system:
  
  The luminosity system sketched in fig.~\ref{fig:LUMI} utilizes the
  Bethe-Heitler process $ep \, \rightarrow \, ep\gamma$. The electrons
  and photons are detected in coincidence by a hodoscope of crystal
  Cherenkov counters.  Whereas the electrons from these grazing
  collisions, i.e.\ $\theta_{e'} \approx 180\grad$, are deflected by
  magnets from the beam line to hit the {\bf E}lectron {\bf T}agger
  (ET) at $z = -33.4\,{\rm m}$, this is not possible for the photons,
  of course. They leave the proton beam through a window at $z =
  -92.3\,{\rm m}$ instead, where the beam pipe bends upwards, and
  reach the {\bf P}hoton {\bf D}etector (PD) at $z = -102.9\,{\rm m}$.
  
  The main background is caused by bremsstrahlung from residual gas
  atoms $eA \, \rightarrow \, eA\gamma$. The technique of electron
  pilot bunches that do not have colliding proton counterparts allows
  to correct for it.
\end{itemize}

\pagebreak
\subsection{Details on the Main Components}
\label{sec:H1details}

The basic ingredient for this analysis is the calorimetric information
provided by the LAr and BEMC in the form of electromagnetic and
hadronic clusters reconstructed from the primary energy depositions.
The fragmentation region of the proton remnant, which is related to
\qi soft\qo\ physics, can be excluded more easily after performing a
Lorentz transformation into the Breit frame (s.\ 
section~\ref{sec:Breit}).  This procedure requires a good electron
identification. On the one hand, the electron cluster has to be
separated from the hadronic final state, and on the other hand, it is
necessary for the extraction of the event kinematics which is crucial
in the calculation of the boost to the Breit frame.  Therefore,
additional information of the tracking system, especially the CTD, is
exploited to determine the properties of the scattered electron and
the actual event vertex.

The main attention in this analysis rests on mean values and
normalized distributions.  Hence, the luminosity system needed to
measure absolute cross sections is not of prevalent importance.

\begin{figure}
  \centering
  \begin{minipage}[b]{0.47\textwidth}
    \centering
    \hspace*{-0.75cm}\includegraphics[width=1.25\textwidth]{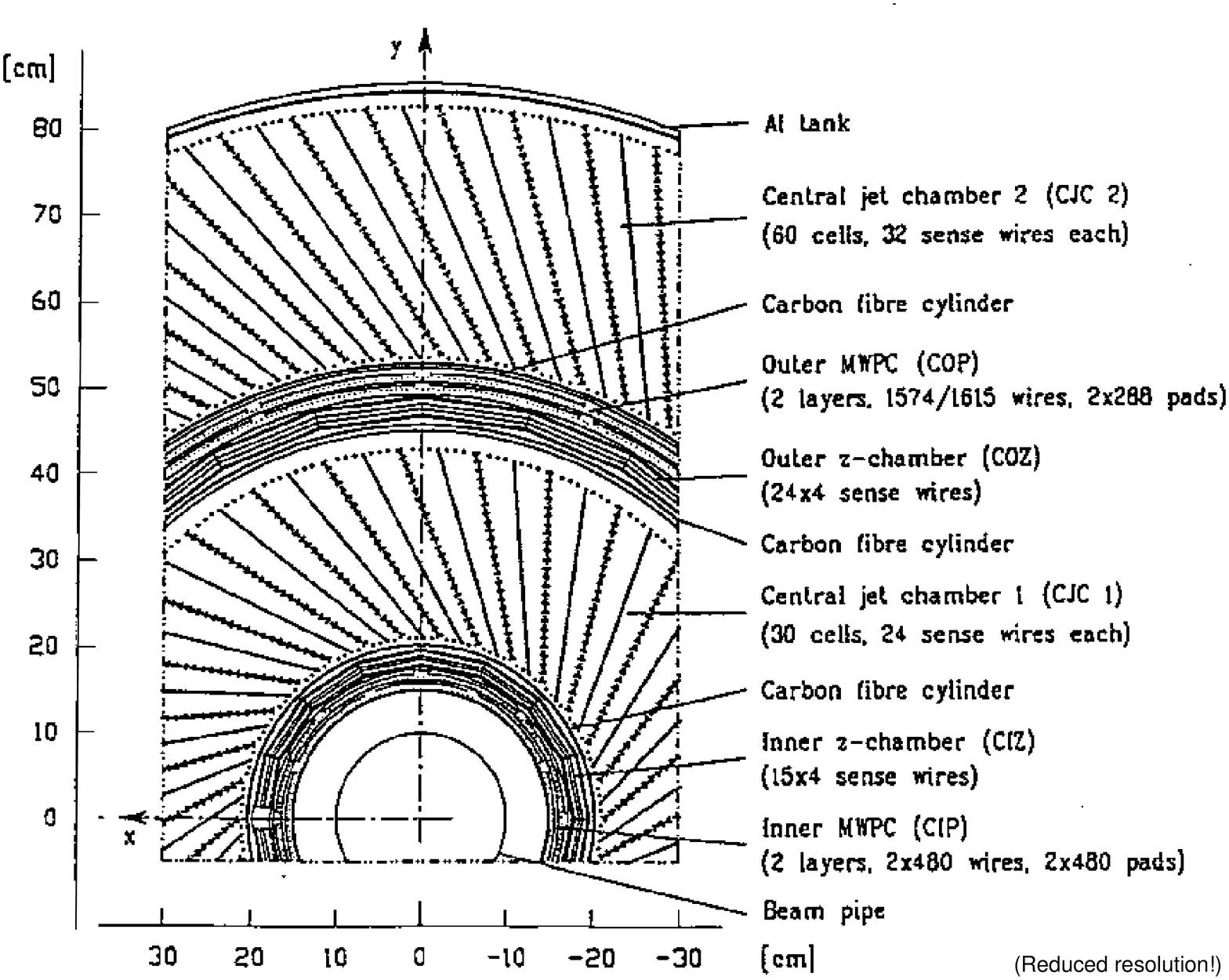}
    \caption[Central tracking system, section perpendicular to the
    beam.]{Central tracking system, section perpendicular to the
      beam~\cite{H1:NIMTCM}.}
    \label{fig:CTD}
  \end{minipage}\hfill
  \begin{minipage}[b]{0.47\textwidth}
    \centering \includegraphics[width=0.85\textwidth, viewport=40 0
    500 500,clip]{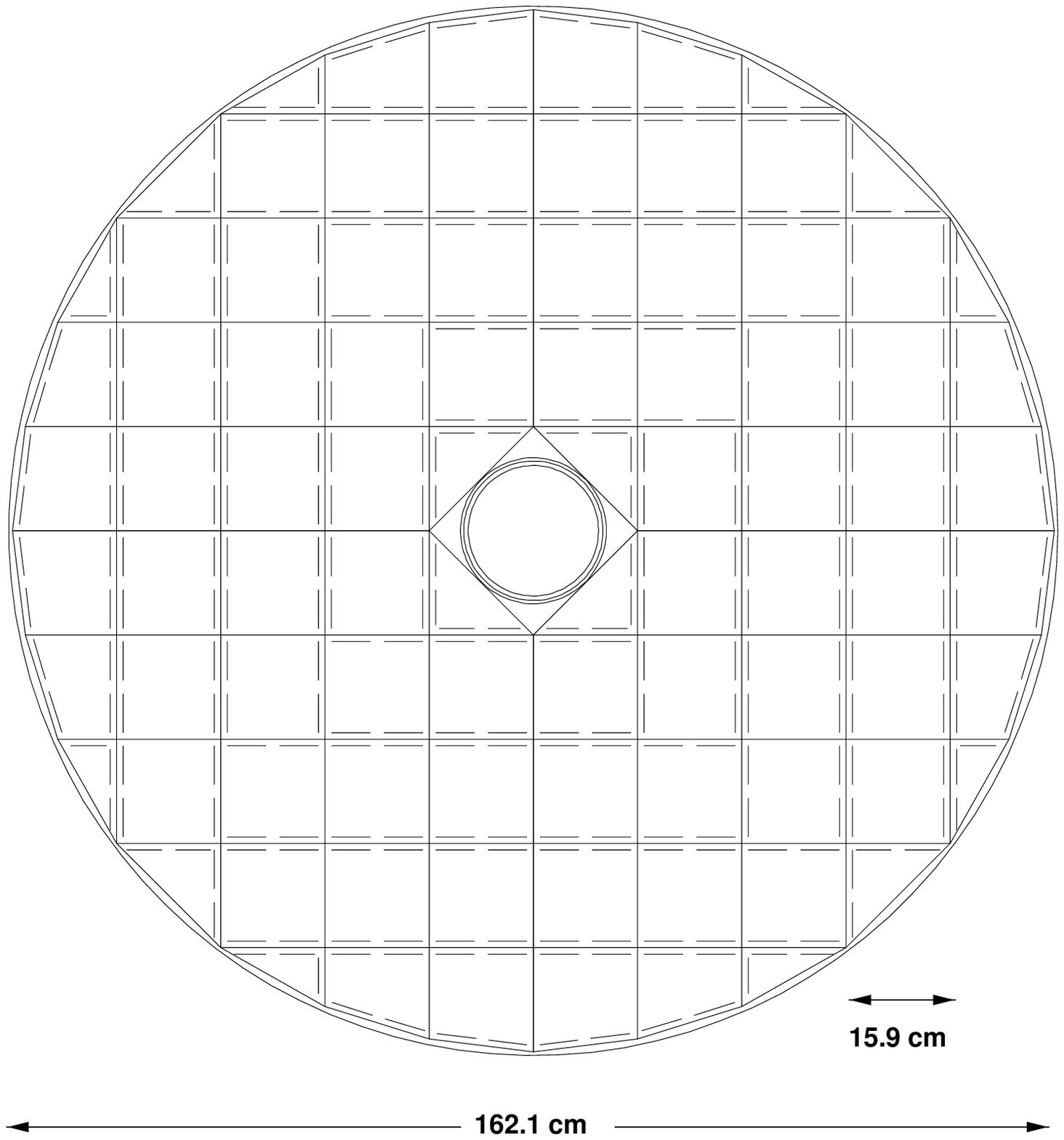}
    \caption[Transverse view of the stack
    segmentation of the BEMC\@.]{Transverse view of the stack
      segmentation of the BEMC~\cite{H1:NIMTCM}.}
    \label{fig:BEMC}
  \end{minipage}\\[0.75cm]
  \begin{minipage}[b]{\textwidth}
    \includegraphics[width=\textwidth,viewport=80 325 485 510]
    {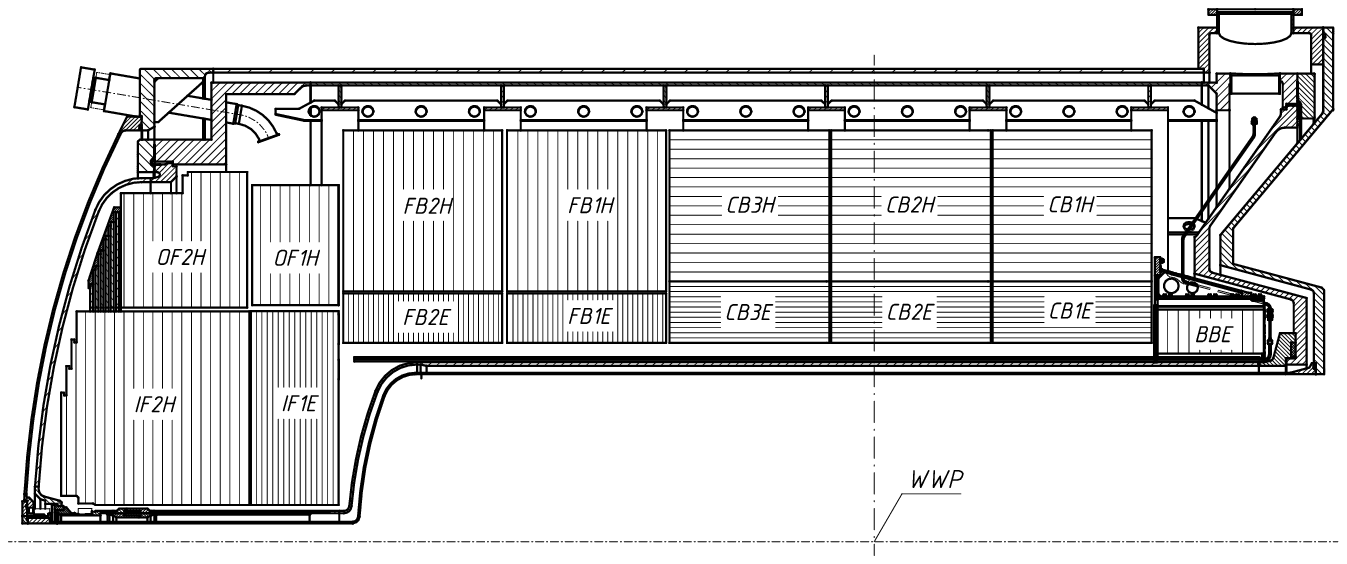} \vspace{-1.4cm}\\\hspace*{0.3cm}
    \includegraphics[viewport=80 30 360 120,width=0.96\textwidth,clip]
    {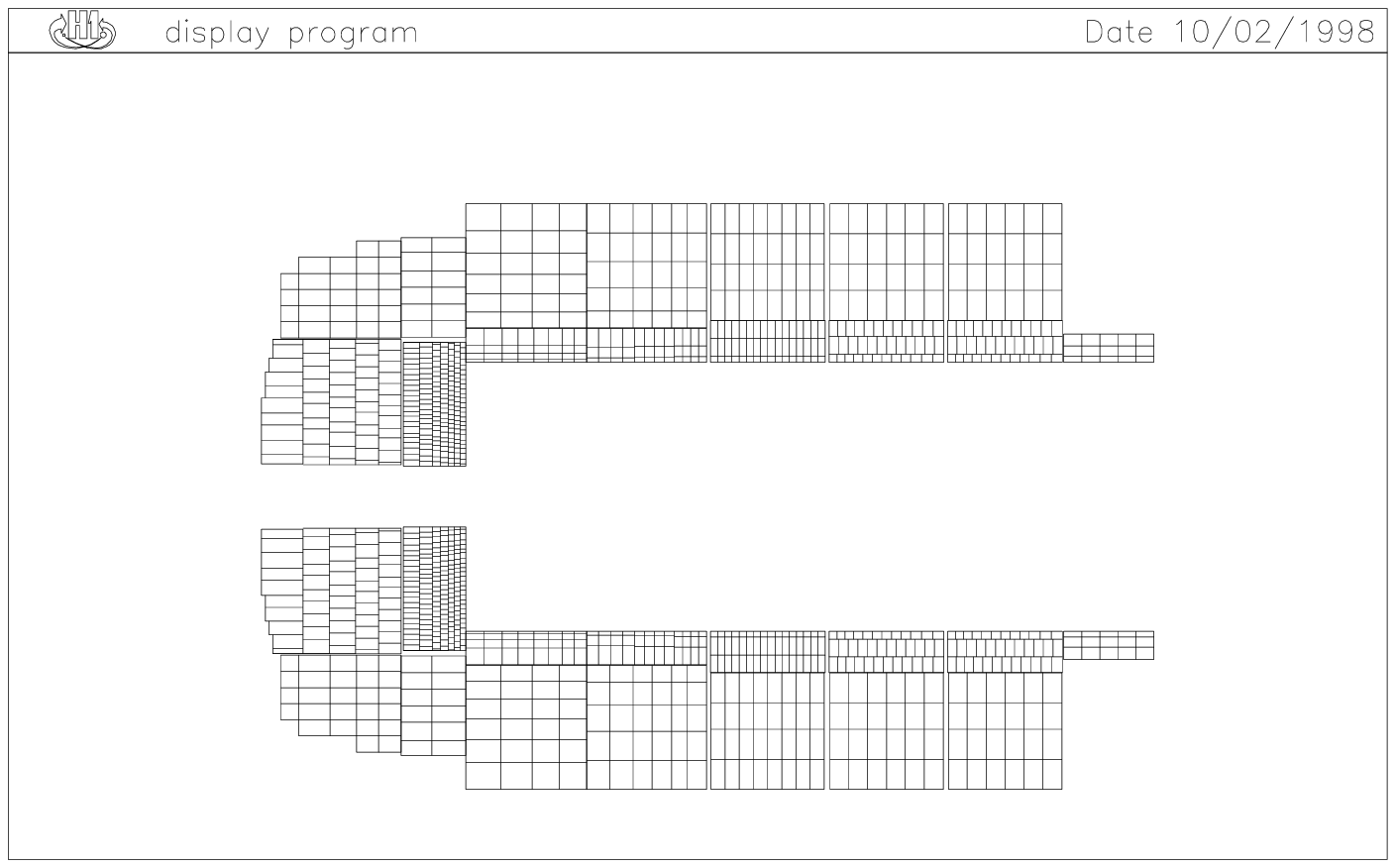}
    \caption[The orientation of the absorber plates and the cell
    structure of the LAr in $(r,z)$-view.]  {The orientation of the
      absorber plates (top) and the cell structure of the LAr in
      $(r,z)$-view (bottom).}
    \label{fig:LAr}
  \end{minipage}
\end{figure}

Concluding, the CTD, LAr and BEMC are the most important subdetectors
for the observables considered and will therefore be described in more
detail~\cite{H1:NIMTCM}:

\begin{itemize}
\item The central tracking device CTD:
  
  The CTD is subdivided into two major units, the {\bf C}entral {\bf
    J}et {\bf C}hambers CJC1 and CJC2 as shown in fig.~\ref{fig:CTD}
  in a view perpendicular to the beam. The track reconstruction in the
  central region, $25\grad < \theta < 155\grad$, depends primarily on
  these concentric drift chambers with wires strung parallel to the
  beam pipe. They provide a good space point resolution in the
  $(r,\phi)$-plane of $170\,{\rm \mu m}$ and, by comparing the signals
  at both ends, furthermore supply information about the
  $z$-coordinate of the hits. To avoid completely insensible paths for
  almost straight tracks, the sequence of the sense wires is inclined
  with respect to the radial direction.
  
  The CJC1 is sandwiched between the {\bf C}entral {\bf I}nner and
  {\bf O}uter {\bf Z}-chambers CIZ and COZ with wires oriented in the
  $(r,\phi)$-plane.  They complement the measurement in the CJC's with
  data on the $z$-coordinate of track elements with a precision of
  $\approx 300\,{\rm \mu m}$.
  
  Mainly for the purpose of fast timing ($< 96\,{\rm ns}$) and
  triggering, multiwire proportional chambers are added for polar
  angles between $5\grad$ and $175\grad$. Two of them, the {\bf
    C}entral {\bf I}nner and {\bf O}uter {\bf P}roportional chambers
  CIP and COP, are indicated in fig.~\ref{fig:CTD}. For electrons in
  the backward direction, only short track segments can be seen in the
  CTD\@. The {\bf B}ackward {\bf P}roportional {\bf C}hamber BPC,
  consisting of four sensible layers of wires oriented along azimuthal
  angles of $0\grad$, $45\grad$, $90\grad$ and $135\grad$,
  substantially improves the tracking for polar angles of $155.5\grad
  < \theta \leq 174.5\grad$. Installed directly in front of the BEMC,
  fig.~\ref{fig:BEMC}, it also helps to distinguish between photons
  and electrons in this region.
  
  In general, the tracking system was designed to determine the
  momentum and angles of charged particles to a precision of
  $\sigma_p/p^2 \approx 0.003\gev^{-1}$ and $\sigma_\theta \approx
  1\,{\rm mrad}$.
  
\item The liquid argon calorimeter LAr:
  
  As mentioned, the LAr, covering polar angles of $4\grad < \theta <
  154\grad$, is a sampling calorimeter where layers of absorber
  material and sensitive gaps, filled with liquid argon, alternate.
  Since only a sample of all energy deposits can be measured that way,
  the energy resolution of this technique is worse than that of
  calorimeters built of only one medium. To achieve nevertheless a
  maximal precision, care has been taken that the orientation of the
  absorber plates, shown in the upper part of fig.~\ref{fig:LAr},
  ensures angles near the normal direction of the stacks for particles
  originating from the vertex.  Another drawback is the difference in
  response on electromagnetic and hadronic showers produced by impacts
  of particles of the same original energy.  The LAr is {\em
    non-compensating}.
  
  This is counterbalanced by stable calibration and homogeneous
  response properties as well as the possibility of a compact but
  still finely segmented construction. The structure of the read-out
  cells, again in an $(r,z)$-view, is presented in the lower half of
  fig.~\ref{fig:LAr}.
  
  Apart from the general subdivision in an {\bf E}lectromagnetic and a
  {\bf H}adronic section, the LAr is composed of eight wheels each
  further partitioned into eight octants in~$\phi$.  Due to the use of
  weak lead for the ECAL, a pointing geometry, blind for hits straight
  on the border between two octants, could not be avoided in contrast
  to the CTD\@.  Particles reaching these $\phi$-cracks are detected
  only in the non-pointing HCAL\@. All wheels are, with respect to the
  nominal interaction point, combined to four aggregates: a {\bf
    B}ackward, {\bf C}entral and {\bf F}orward {\bf B}arrel and the
  {\bf F}orward end cap with an {\bf I}nner and {\bf O}uter ring.  The
  abbreviations in fig.~\ref{fig:LAr} translate e.g.\ as \qi outer
  hadronic ring of forward end cap wheel 1\qo\ for OF1H.
  
  Test beam results yield energy resolutions of $\approx
  11\%/\sqrt{E}\,\oplus\,1\%$ for electromagnetic and $\approx
  50\%/\sqrt{E}\,\oplus\,2\%$ for hadronic showers.\footnote{The \qi
    $\oplus$\qo\ indicates quadratic addition.}  The absolute energy
  scales are known to $1$--$3\%$~\cite{H1:ICHEP98hq} and $4\%$ for
  electrons and hadrons respectively.
  
\item The backward electromagnetic calorimeter BEMC:
  
  A transverse view of the warm lead-scintillator calorimeter and its
  segmentation into $88$ stacks aligned parallel to the beam pipe can
  be seen in fig.~\ref{fig:BEMC}. The stacks are multi-layer sandwich
  structures with active sampling units made of plastic scintillators.
  The front face directly behind the BPC is located at $z = -144\cm$.
  The angular region covered extends from polar angles of $151\grad$
  up to $176\grad$.
  
  The electromagnetic energy resolution was derived from test beam
  measurements to be $\approx 10\%/\sqrt{E}\,\oplus\,1.7\%$, the
  absolute energy scale is known to a precision of $1\%$.
\end{itemize}

\chapter{Deep-Inelastic Scattering}
\label{chap:DIS}
\section{The Kinematics}
\label{sec:kine}

\begin{figure} 
  \centering \psfrag{s}{$\hat{\sigma}$} \psfrag{S}{$\hat{\rm s}$}
  \psfrag{|}{$\left . \mbox{\rule{0.pt}{48.pt}} \right \} $}
  \includegraphics{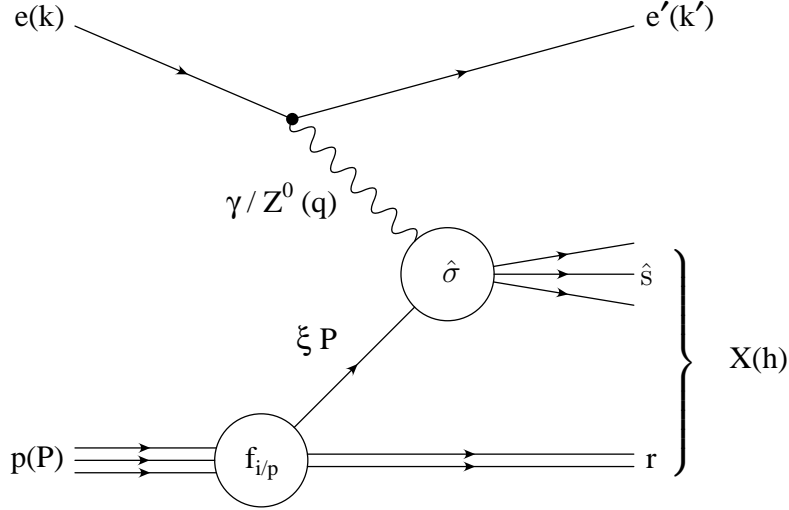}
  \caption[Diagram of the basic $ep$ DIS process via neutral currents.]
  {Diagram of the basic $ep$ DIS process via neutral currents.  The letters in
    parentheses label the corresponding four-momenta.}
  \label{fig:epDIS}
\end{figure}

Due to the high beam energies of $E_e = 27.5\gev$ and $E_p = 820\gev$
available at HERA, the masses of the electron and proton can safely be
neglected for most purposes. This will be done throughout this study.  Their
four-momenta,\footnote{To disentangle four-momenta like $p$ from
  three-momenta, the latter are set in Roman font with arrows on top:
  $\rmvec{p}$.}  indicated in fig.~\ref{fig:epDIS}, can therefore be written
as $k = (E_e, 0, 0, p_{z_e}) = (27.5, 0, 0, -27.5)\gev$ and $P = (E_p, 0, 0,
p_{z_p}) = (820, 0, 0, 820)\gev$ with $k^2 = P^2 = 0$.  Thus, the center of
mass energy $\sqrt{s}$ follows from
\begin{equation}
  s := (k+P)^2 = 2k\cdot P = 4E_eE_p = 90200\gevq
\end{equation}
to be $300.33\gev$.

The {\bf N}eutral {\bf C}urrent (NC) interaction displayed in
fig.~\ref{fig:epDIS} is mediated via the exchange of a virtual $\gamma$ or
$Z^0$ boson with four-momentum $q = k - k'$ and mass $q^2 < 0$ where $k'$, the
four-momentum of the outgoing electron, may be chosen to be $k' = (E_{e'},
E_{e'}\sin \theta_{e'}, 0, E_{e'} \cos \theta_{e'})$, i.e.~$\phi_{e'}:=0$.
{\bf C}harged {\bf C}urrent (CC) reactions involving $W^\pm$ bosons would
yield an outgoing neutrino $\nu_e$ and will not be considered here.  In the
case of elastic scattering $ep \rightarrow ep$, the Lorentz invariant momentum
transfer squared
\begin{equation}
  Q^2 := -q^2 = -(k-k')^2 = 2k\cdot k' = 2E_eE_{e'} (1+\cos{\theta_{e'}}) 
  \label{eqn:defq2}
\end{equation}
would suffice to characterize the process. Protons, however, are not
point-like particles, but have a complex internal structure revealing itself
at distances of ${\scriptstyle\lesssim}\,1\,{\rm fm}$, equivalent to
$Q^2\,{\scriptstyle\gtrsim}\,0.04\gevq$.  As consequences the proton,
represented by the lower blob in fig.~\ref{fig:epDIS}, on the one hand has to
be described in terms of {\em structure functions}, and on the other hand it
usually breaks up since elastic reactions are strongly suppressed compared to
inelastic ones with increasing $Q^2$~\cite{Berger}.  This inevitably leads to
the need of another quantity to define the global outcome of an event, i.e.\ 
without differentiating the hadronic final state.  Looking at the
photon-proton ($\gamma^*p$) interaction by itself, we have only one additional
invariant at our disposal:
\begin{equation}
  W^2 := (q+P)^2 = 2P\cdot q - Q^2,
\end{equation}
which is limited by $0 \leq W^2 \leq s$.

Other popular choices are the two dimensionless variables\footnote{The {\em
    scaling}\/ variable $x$ is also called Bj{\o}rken $x$ and $y$ may be
  denoted as {\em inelasticity}.}
\begin{eqnarray}
\label{eqn:defx}
  x & := & \frac{Q^2}{2P\cdot q}\,,\\
  y & := & \frac{P\cdot q}{P\cdot k}
\label{eqn:defy}
\end{eqnarray}
with the neat property that $0 \leq (x,y) \leq 1$.  In the context of the {\em
  infinite momentum frame}, where the proton is conceived of as a collinear
stream of fast moving {\em partons}\/ and masses are negligible, $x$ can be
interpreted as the fraction of the total momentum carried by the struck
constituent as seen from the virtual boson. This reference frame has
implicitly been adopted in fig.~\ref{fig:epDIS} as the hard scattering with a
parton, labelled $\hat{\sigma}$ with mass $\sqrt{\hat{s}}$, is assumed to be
incoherent and well separated from ensuing soft processes. In the simplest
situation of a boson-parton collision where
\begin{equation}
  \hat{s} := ( q + \xi P ) ^2 = 2 \xi q \cdot P - Q^2 =
  \left ( \frac{\xi}{x} -1 \right ) Q^2
\end{equation}
vanishes, $x$ is identical to $\xi$ of fig.~\ref{fig:epDIS} according to
\begin{equation}
\xi = \left ( 1 + \frac{\hat{s}}{Q^2} \right ) x\,.
\end{equation}
Under more complex circumstances with $\hat{s} > 0$, it can only be concluded
that $x \leq \xi \leq 1$.

In fixed-target experiments, $y$ is easily interpreted as relative energy loss
of the scattered lepton because in the rest frame of the target denoted by
$^\bullet$'s:
\begin{equation}
  y = \frac{m_p (E^\bullet_e - E^\bullet_{e'})}{m_p E^\bullet_e} =
  \frac{E^\bullet_e - E^\bullet_{e'}}{E^\bullet_e}\,.
\end{equation}

Of course, only two of the four introduced kinematic quantities are
independent of each other. The conversion formulae for the pairs $(Q^2,W^2)
\leftrightarrow (x,y)$ are:
\begin{eqnarray}
  \nonumber x = \frac{Q^2}{Q^2+W^2}\,, &\hspace{2cm}&
  \quad y = \frac{Q^2+W^2}{s}\,,\\
  \label{eqn:conversion}&&\\
  \nonumber Q^2 = sxy\,,\qquad\;\: &\hspace{2cm}& W^2 = s\,(1-x)\,y\,.
\end{eqnarray}

\section{Reconstruction of the Kinematical Quantities}
\label{sec:kinereco}

When confronted with real data, one has to reconstruct the kinematical
quantities of an event from the measured energy depositions.  Basically, we
have four measurements at our disposal for the determination of two unknowns:
the energies and polar angles of the scattered electron and the {\em current
  jet}, i.e.\ a collimated shower of hadrons produced by the struck parton:
$E_{e'}$, $\theta_{e'}$, $E_j$, $\theta_j$.  Depending on the choice of input
variables, several reconstruction methods exist, each with specific advantages
and drawbacks.

\begin{figure} 
  \centering \includegraphics[height=10.cm]{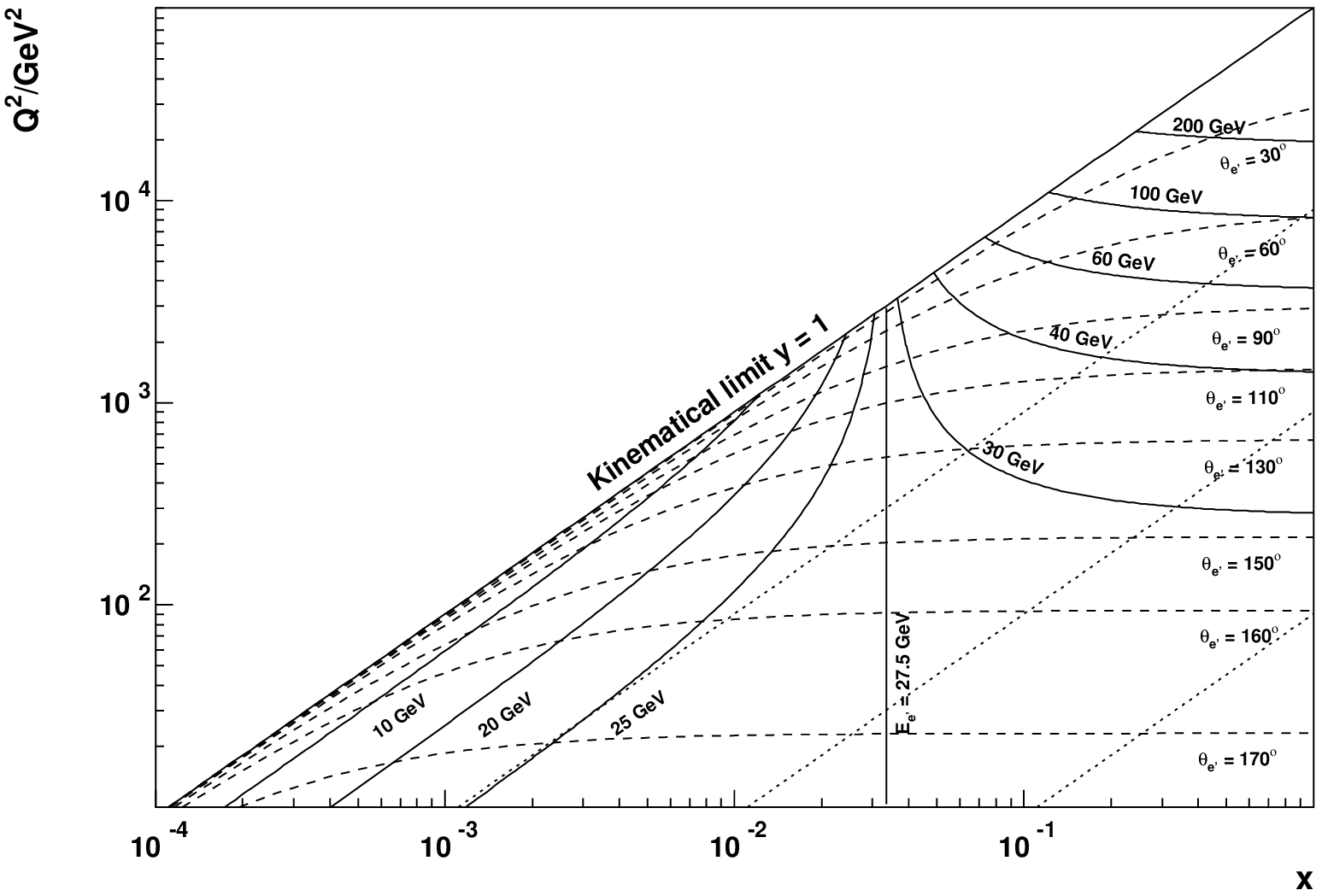}
  \includegraphics[height=10.cm]{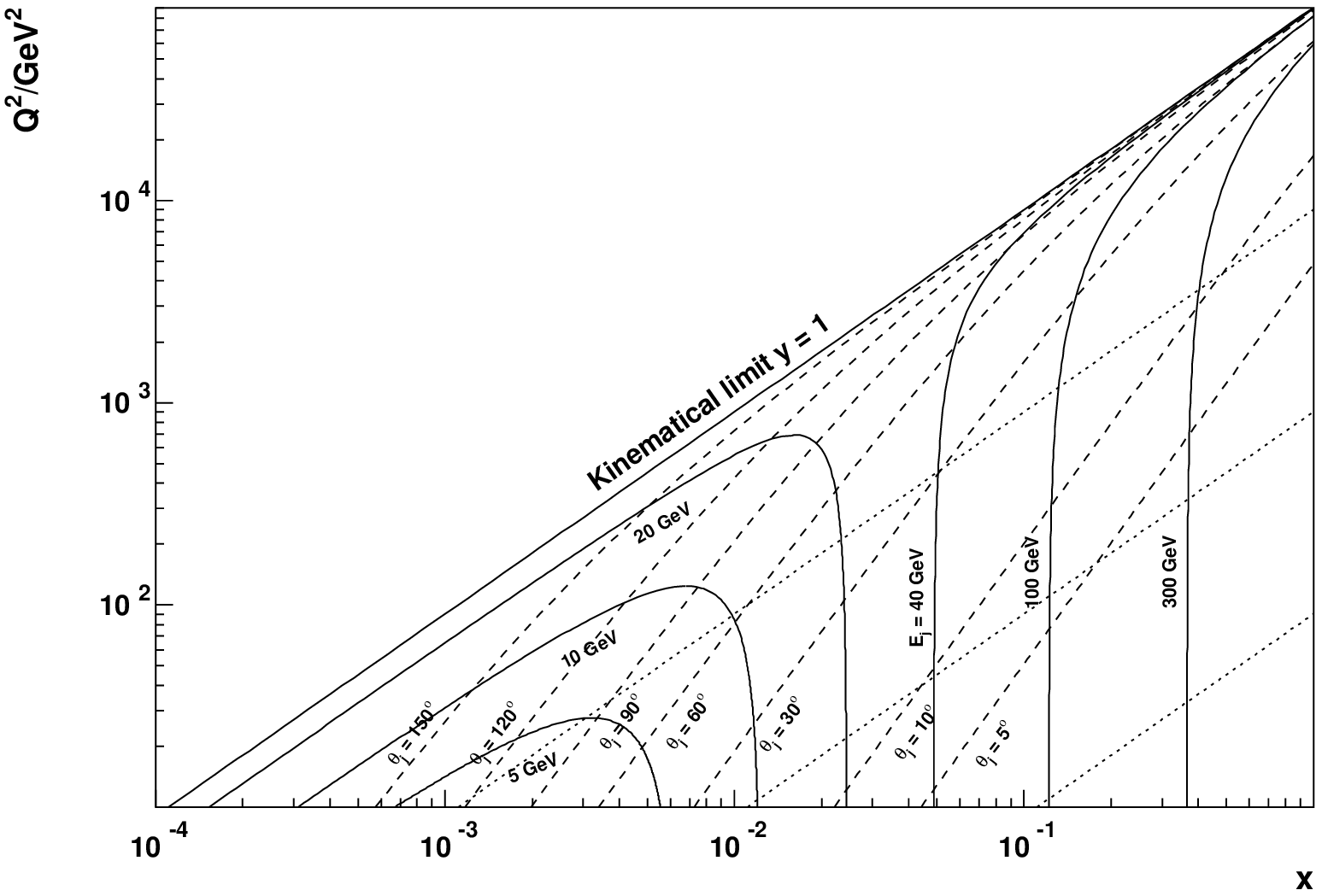}
  \caption[Lines of constant energy and polar angle
  for the scattered electron and the current jet in the $(x,Q^2)$-plane.]
  {The two plots show the lines of constant energy (full lines) and polar
    angle (dashed) for the scattered electron (top) and the current jet
    (bottom) in an $(x,Q^2)$-plane.  Lines of constant $y$ (dotted except for
    $y=1$) are drawn for $y = 1, 10^{-1}, 10^{-2}$ and $10^{-3}$.}
  \label{fig:isolines}
\end{figure}

Fig.~\ref{fig:isolines} presents the lines of constant energy and polar angle
for the scattered electron and the current jet in the $(x,Q^2)$-plane. Lines
of constant $y = Q^2/(sx)$ are displayed for $y=1, 10^{-1}, 10^{-2}$ and
$10^{-3}$.  For reconstruction purposes it is best if the isolines of a
quantity are closely staggered and intersect the isolines of the corresponding
second variable predominantly under large angles.  Henceforth, the electron
data alone are sufficient for a good determination of $Q^2$ in the complete
range shown. The large gaps, however, between energies $E_{e'} = 25\gev$ and
$E_{e'} = 40\gev$ and angles $\theta_{e'}$ larger than $90\grad$ demonstrate
that the electron method is not very reliable for large $x$ or low $y$
respectively.  Despite the fact that hadronic energies and angles have much
larger uncertainties than electromagnetic ones, the current jet data can
provide better estimates for $x$ or $y$ in that region.

Yet, because we are interested in the shape of the hadronic final state, we
have to restrict ourselves mainly to the electron quantities in order not to
bias our results. The application of a {\em jet algorithm}\/ like the ones
described in section~\ref{sec:shapesjets} to obtain $E_j$ and $\theta_j$ is
not advisable.  Fortunately, there is also an inclusive method to derive the
kinematics from hadronic measurements first proposed by Jacquet and
Blondel~\cite{DIS:jbkine}.  Alternatively, one could rely on the angular
measurements of the electron and the hadronic system alone~\cite{DIS:dakine}.
These three methods will be presented in the next sections.  For a comparison
ref.~\cite{DIS:sgkine} may be consulted.

\subsection{The Electron Method}
\label{sec:ekine}

The electron method is the simplest possibility to extract $y$ and $Q^2$ and
employs the scattered electron only.  $x$ and $W^2$ can then be calculated
according to eqs.~(\ref{eqn:conversion}). Using the
definitions~(\ref{eqn:defq2}) and~(\ref{eqn:defy}), it follows directly:
\begin{equation}
  \label{eqn:defye}
  y_e = 1 - \frac{P\cdot k'}{P\cdot k} =
  1 - \frac{E_{e'}}{E_e} \frac{1- \cos \theta_{e'}}{2} =
  1 - \frac{E_{e'} - p_{z_{e'}}}{2E_e}\,,
\end{equation}
\begin{equation}
  \label{eqn:defq2e}
  Q^2_e = 2k\cdot k' = 2E_eE_{e'} (1+\cos{\theta_{e'}}) =
  2E_e (E_{e'} + p_{z_{e'}}) = \frac{p_{t_{e'}}^2}{1-y_e}\,.
\end{equation}

\subsection{The Jacquet-Blondel Method}
\label{sec:jbkine}

With $h$ representing the four-momentum of the complete hadronic final state,
it is clear from fig.~\ref{fig:epDIS} that $q = k-k' = h-P$.  Furthermore,
$p_t$-balance enforces $p_{t_h}=p_{t_e}$ so that
\begin{equation}
  \label{eqn:defyh}
  y_h = \frac{P\cdot (h-P)}{P\cdot k} =
  \frac{\sum\limits_i \left ( E_i - p_{z_i} \right )}{2E_e}\,,
\end{equation}
\begin{equation}
  Q^2_h = \frac{p^2_{t_h}}{1-y_h} =
  \frac{\left ( \sum\limits_i p_{x_i} \right ) ^2 +
    \left ( \sum\limits_i p_{y_i} \right ) ^2}{1-y_h}
\end{equation}
where $i$ loops over all hadronic objects.

\subsection{The Double Angle Method}
\label{sec:dakine}

Requiring $y_h = y_e$ and $p_{t_h}=p_{t_{e'}}$, the dependence on $E_{e'}$ and
$\sum_i E_i$ for all hadrons $i$ can be eliminated such that:
\begin{equation}
  y_{da} = \frac{\sin\theta_{e'} (1-\cos\theta_h)}
  {\sin\theta_{e'} (1-\cos\theta_h) + 
    \sin\theta_h (1-\cos\theta_{e'})}\,,
\end{equation}
\begin{equation}
  Q^2_{da} = 4 E_e^2 \frac{\sin\theta_h (1+\cos\theta_{e'})}
  {\sin\theta_{e'} (1-\cos\theta_h) + 
    \sin\theta_h (1-\cos\theta_{e'})}\,.
\end{equation}

\section{The Born Cross Section}
\label{sec:Borncross}

To derive a cross section formula for deep-inelastic $ep \rightarrow eX$
scattering, it will be required that the total process can be separated into a
two step procedure. In the first part, involving very small space-time scales
of $\order(1/Q)$ where the strong interaction is weak, the basic kinematic
outcome of a reaction is fixed by the incoherent elastic scattering of the
boson probe and a proton constituent.  The ensuing soft hadronization of the
struck parton and the proton remnant takes place at a time scale typically of
the order of inverse hadron masses $1/M$ with $M \approx 200\,{\rm MeV} \ll Q$
instead and merely affects the detailed structure of the hadronic final state.

\begin{figure} 
  \centering
  \includegraphics{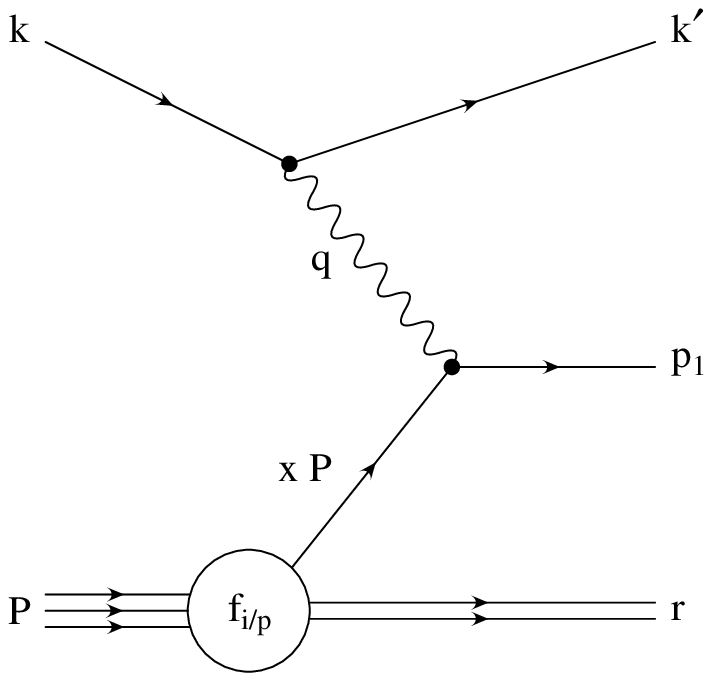}\hfill%
  \includegraphics{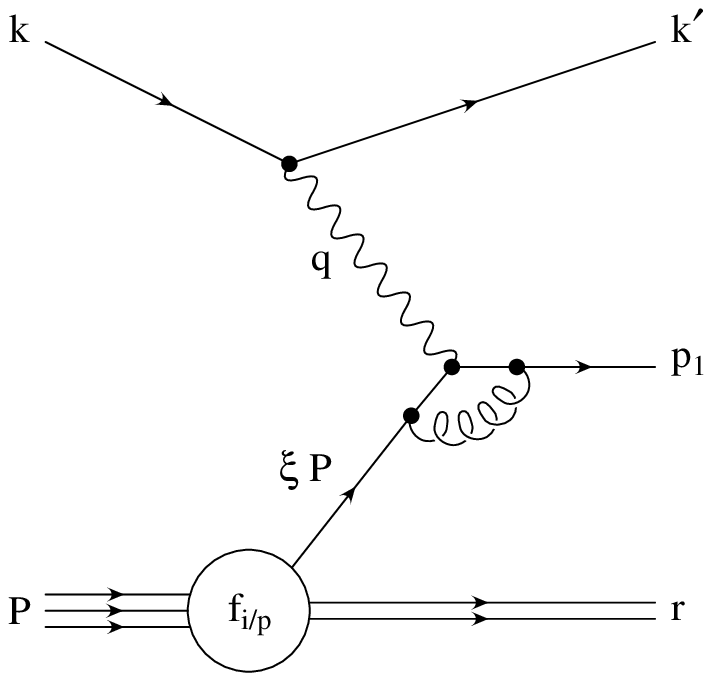}
  \caption[Quark-Parton-Model Feynman graph and a virtual correction to it.]
  {Quark-Parton-Model Feynman graph (left) and an example of a virtual
    correction to it (right).}
  \label{fig:QPM}
\end{figure}

Without resolving such details, it is therefore possible to calculate a cross
section by applying perturbation theory in lowest or {\bf L}eading {\bf O}rder
(LO, here: $\order(\alpha^2\as^0)$) to the hard subprocess $\hat{\sigma}$ of
fig.~\ref{fig:epDIS}. The blob is then replaced by a single boson-parton
vertex as shown in the {\bf Q}uark-{\bf P}arton-{\bf M}odel (QPM) Feynman
diagram of fig.~\ref{fig:QPM}.  If the proton itself is left out, essentially
an elastic two-body scattering reaction remains for which the cross sections
have been calculated, s.\ e.g.~\cite{Berger}.  One possibility to derive a
general cross section formula is to assume partons with spin $1/2$ and
$0$~\cite{Berger}.  Depending on the spin, one takes over the result from
$e\mu$ respectively $e\pi$ scattering:
\begin{eqnarray}
  \frac{d\sigma^{e\mu}}{dQ^2} &=& \frac{2\pi \alpha^2}{Q^4} (1+(1-y)^2)\,,\\
  \frac{d\sigma^{e\pi}}{dQ^2} &=& \frac{4\pi \alpha^2}{Q^4} (1-y)\,.
\end{eqnarray}
Here, $\alpha$ denotes the electromagnetic coupling strength and NC
contributions from $Z^0$ exchange, which are suppressed at least $\propto
Q^2/(Q^2+M_Z^2)$, are neglected.

At this stage, we must again consider the internal arrangement of partons in
the proton that can be parameterized in the form of {\bf p}arton {\bf d}ensity
{\bf f}unctions (pdfs) $f_{i/p}(x)$ which represent the probability to find a
constituent $i$ with a momentum fraction in the interval $[x,x+dx]$. Denoting
the spin-$1/2$ pdfs with $f_{i/p}(x)$ and the spin-$0$ ones with $f_{j/p}(x)$,
the differential cross section can be written as
\begin{equation}
  \frac{d\sigma^{ep}}{dQ^2} = \frac{4\pi\alpha^2}{Q^4} \left (
    \frac{1+(1-y)^2}{2} \sum\limits_i q_i^2 f_{i/p}(x) dx +
    (1-y) \sum\limits_j q_j^2 f_{j/p}(x) dx \right ) \,,
  \label{eqn:Born0}
\end{equation}
where $q_i$ and $q_j$ are the corresponding electromagnetic charges of the
constituent flavours $i$ and $j$ in units of the positron charge.
Historically, the derivation of the cross section implied the evaluation of a
leptonic tensor $L^{\mu\nu}$ prescribed by QED and its contraction with the
most general form of a hadronic tensor $W_{\mu\nu}$ describing the hadronic
vertex.  This approach, employed e.g.\ in~\cite{HM,Quigg}, led to the
definition of structure functions $F_1(x)$ and $F_2(x)$ which here translate
to
\begin{eqnarray}
  F_1(x) &=& \sum\limits_i \frac{1}{2} q_i^2 f_{i/p}(x)\,,\\
  F_2(x) &=& \sum\limits_i q_i^2 x f_{i/p}(x) +
  \sum\limits_j q_j^2 x f_{j/p}(x)\,.
\end{eqnarray}
The double differential cross section~(\ref{eqn:Born0}) now reads
\begin{equation}
  \frac{d^2\sigma^{ep}}{dQ^2dx} = \frac{4\pi\alpha^2}{xQ^4}
  \left ( (1-y) F_2(x) + xy^2 F_1(x) \right ) \,.
  \label{eqn:Born}
\end{equation}

So far the partons served as a term for point-like constituents within the
proton that lead to the experimentally observed $1/Q^4$-dependence of the
inelastic cross section in contrast to the dramatic $1/Q^{12}$-decrease for
elastic reactions.  Combining this with the quark model by identifying partons
and spin-$1/2$ quarks, one immediately concludes that $f_{j/p}(x) \equiv 0 \,
\forall j$ and henceforth
\begin{equation}
  F_2(x) = 2xF_1(x)\,.
  \label{eqn:Callan}
\end{equation}
This is the experimentally well established Callan-Gross relation which
demonstrates the fermionic nature of the charged proton constituents.  In
addition, one can conclude from eq.~(\ref{eqn:Born}) that the cross section
normalized to the corresponding one for point-like particles depends on the
scaling variable $x$ only.  The observation of deviations can be attributed to
the neglect of masses, intrinsic transverse momenta and ---~most
importantly~--- the strong interaction.

\section{The NLO Cross Section}
\label{sec:NLOcross}

Up to now, the proton was treated like a stream of collinear non-interacting
quarks. However, in {\bf N}ext-to-{\bf L}eading {\bf O}rder (NLO), i.e.\ 
$\order(\alpha^2\as^1)$, quarks can emit and absorb gluons which again may
split up into two gluons or $q\bar{q}$ pairs and so forth.  The rather simple
picture involving structure functions $F_1(x)$, $F_2(x)$ to describe the
composition of the proton in terms of quark densities alone has to be modified
accordingly to include a gluon density.  Two of the most important
consequences are {\em scaling violations}, i.e.\ a~dependence of $F_1(x),
F_2(x)$ on $Q^2$, and non-vanishing contributions to the longitudinal
structure function
\begin{equation}
  F_L := F_2 - 2x F_1\,.
\end{equation}
For low $x$, i.e.\ $x\,\lesssim\,10^{-3}$, approximations reveal the effects
to be proportional to $\as(Q^2)$ times the gluon density $g(x,Q^2)$:
\begin{equation}
  \frac{dF(x,Q^2)}{d\log Q^2} \propto \as(Q^2)\cdot g(x,Q^2)\,,
\end{equation}
\begin{equation}
  F_L(x,Q^2) \propto \as(Q^2)\cdot g(x,Q^2)\,.
\end{equation}
Both are intensively studied in DIS experiments and can be exploited in
inclusive measurements to gain information on the strong coupling constant as
well as the gluon density.  For H1 publications on this topic consult
refs.~\cite{H1:F2,H1:FL}.

In this analysis all hadronic final states are considered but will be
differentiated with respect to suitably chosen characteristic variables (s.\ 
section~\ref{sec:shapesdef}), i.e.\ the measurement is semi-inclusive.

\begin{figure} 
  \centering
  \includegraphics{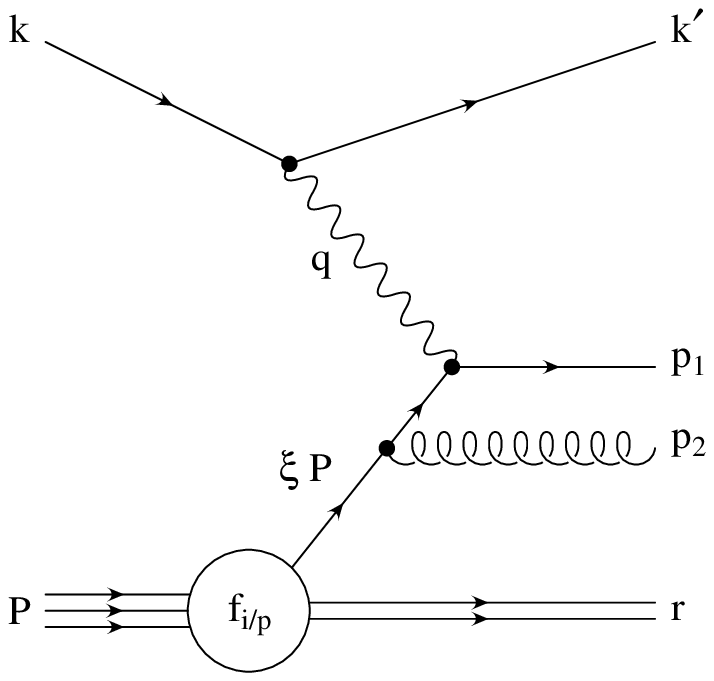}\hfill%
  \includegraphics{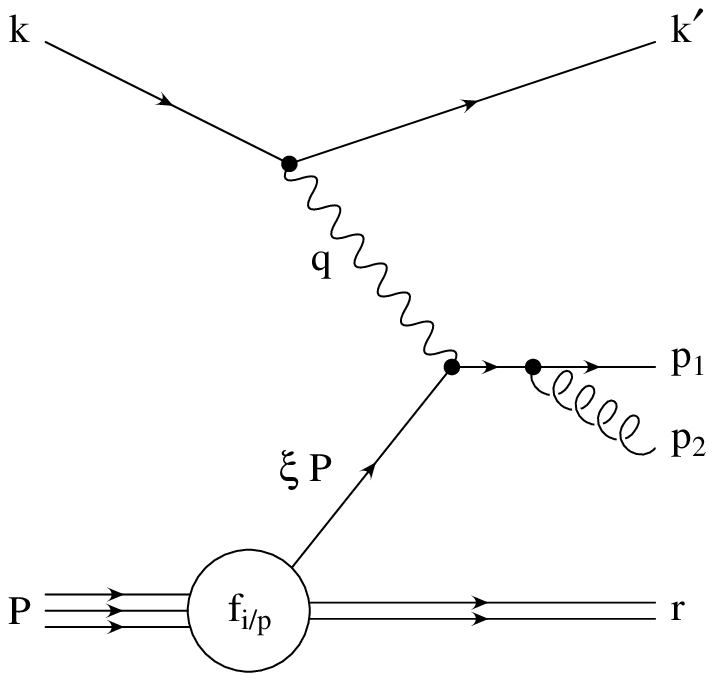}
  \caption{QCD-Compton Feynman graphs.}
  \label{fig:QCDC}
\end{figure}

\begin{figure} 
  \centering
  \includegraphics{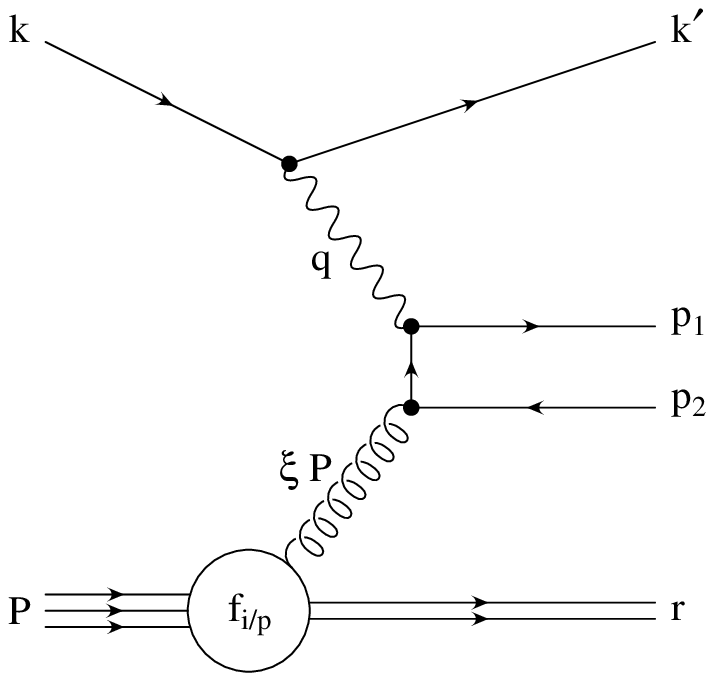}\hfill%
  \includegraphics{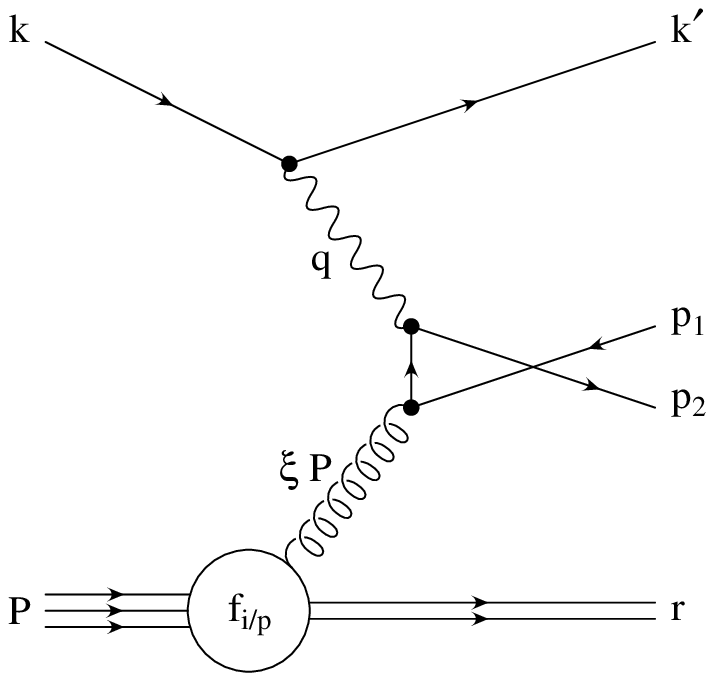}
  \caption{Boson-Gluon-Fusion Feynman graphs.}
  \label{fig:BGF}
\end{figure}

To $\order(\as^1)$ there are two kinds of processes resulting in two instead
of one final state parton, where \qi parton\qo\ from now on is used
synonymously for quarks, anti-quarks and gluons.  Since they may lead to \qi
real\qo\ effects like an additional {\em jet}\/ they are called {\em real
  corrections}.
In the first case of the {\bf QCD}-{\bf C}ompton graphs (QCDC), an additional
gluon is emitted by the struck quark. The two possible Feynman diagrams are
presented in fig.~\ref{fig:QCDC}.
The second kind with diagrams shown in fig.~\ref{fig:BGF} involves the
production of a $q\bar{q}$ pair from a reaction between the boson and a gluon
emitted from the proton. They are expressively labelled {\bf B}oson-{\bf
  G}luon-{\bf F}usion graphs (BGF).

To complete the set of contributions to $ep$ DIS to $\order(\as^1)$, one also
has to consider loop diagrams.  One example is given on the right-hand side of
fig.~\ref{fig:QPM}. The diagram itself results in an $\order(\as^2)$ add-on to
the matrix element squared.  However, since the final states of both graphs in
fig.~\ref{fig:QPM} are indistinguishable, an interference term of
$\order(\as^1)$ has to be taken into account.  For the purpose of
differentiating the hadronic final state, this {\em virtual correction}\/ is,
of course, irrelevant and merely changes the total cross section.

Focusing on the two-parton QCDC and BGF processes, three new degrees of
freedom arise in the matrix elements corresponding to energy, azimuthal and
polar angle of one parton. The other parton is then determined by
energy-momentum conservation. Usually, the now five-fold differential cross
section is written employing the variables
\begin{eqnarray}
  x_p & := & \frac{Q^2}{2\,(\xi P) \cdot q} = \frac{x}{\xi}\,,\\
  \nonumber&&\\
  z_p & := & \frac{P \cdot p_1}{P \cdot q} = 1-\frac{P \cdot p_2}{P \cdot q}\,,
\end{eqnarray}
and $\phi$, where $\phi$ denotes the angle between the planes fixed by the two
outgoing partons on the one hand, and the scattered electron on the other hand
in a suitable reference frame. This may be e.g.\ the {\em Breit frame}\/
defined in the next chapter or any other reference system connected to it by a
Lorentz boost along its $z$-direction.  Concerning the characterization of the
hadronic final state, no reference will be made to the electron.  Therefore,
the integration over $\phi$ can be performed beforehand such that for our
purposes the NLO cross section may be labelled as
\begin{equation}
  \frac{d^4\sigma^{ep}}{dQ^2dxdx_pdz_p}\,.
  \label{eqn:Born1}
\end{equation}
The allowed ranges for $x_p$ and $z_p$ are $x \leq x_p \leq 1$ and $0 \leq z_p
\leq 1$.

Yet, it does contain singularities!  For the QCDC and BGF processes
they can be extracted from ref.~\cite{calc:NLOcrosssection}, which presents
the complete $\order(\as^1)$ corrections to electroweak NC and CC $ep$ DIS
cross sections, to be:
\pagebreak
\begin{eqnarray}
  \label{eqn:singQCDC}
  d\sigma^{\rm QCDC} &\propto& \frac{1+x_p^2z_p^2}{(1-x_p)(1-z_p)}\,,\\
  \nonumber &&\\
  d\sigma^{\rm BGF}  &\propto& \frac{[x_p^2+(1-x_p)^2][z_p^2+(1-z_p)^2]}
  {z_p(1-z_p)}\,.
  \label{eqn:singBGF}
\end{eqnarray}
Partially, they cancel against corresponding divergent terms of the virtual
corrections. The remaining initial state mass singularities can be absorbed in
a proper redefinition of the parton densities.  Nevertheless, one has to be
careful about the precise definition of quantities that shall be calculated by
pQCD\@. In order to allow the divergences to compensate each other, the
investigated variables must agree in the limits where the real and virtual
corrections go to infinity. The {\em number of final state particles}\/ for
example would always be two in the first case and one in the latter.

To be more definite, we make use of eqs.~(\ref{eqn:z0})--(\ref{eqn:barz3}) and
can refer the limits $x_p \rightarrow 1$ and $z_p \rightarrow 0,1$ to
configurations where either $p_1$ or $p_2$ are soft ({\em infrared
  divergence}\/) or any pair of $\{p_1,p_2,r\}$ is collinear ({\em collinear
  divergence}\/).  Hence, any quantity $F_n(\rmvec{p}_1,\ldots,\rmvec{p}_n)$
to be meaningful in pQCD has to fulfil the conditions
\begin{eqnarray}
  \nonumber F_{n}(\rmvec{p}_1,\ldots,\lambda\rmvec{p}_i,\ldots,
  \rmvec{p}_j=(1-\lambda)\rmvec{p}_i,\ldots,\rmvec{p}_n) &=&
  F_{n-1}(\rmvec{p}_1,\ldots,\rmvec{p}_i,\ldots,\rmvec{p}_{j-1},
  \rmvec{p}_{j+1},\ldots,\rmvec{p}_n)\\
  \label{eqn:safe} &{\rm and}&\\
  \nonumber \lim_{\lambda\rightarrow 0}F_{n}(\rmvec{p}_1,\ldots,
  \lambda\rmvec{p}_i,\ldots,\rmvec{p}_n) &=&
  F_{n-1}(\rmvec{p}_1,\ldots,\rmvec{p}_{i-1},\rmvec{p}_{i+1},
  \ldots,\rmvec{p}_n)
\end{eqnarray}
for $0<\lambda<1$.  That is, collinear splittings and soft particles do not
affect $F$; it is {\em collinear and infrared safe}.  One example for such a
variable is {\em thrust}, or rather $1-$thrust, defined by
eq.~(\ref{eqn:tau}). It is invariant with respect to collinear splittings and
varies smoothly for one momentum approaching zero.


\chapter{Event Shapes}
\label{chap:shapes}
\section{The Breit Frame}
\label{sec:Breit}
\subsection{Introduction}
\label{sec:Breitintro}

\begin{figure}[tb]
  \centering \psfrag{+zb}{$+z\Bf$} \psfrag{-zb}{$-z\Bf$}
  \psfrag{thetaq}{$\theta_q$} \psfrag{q2=3934}{$Q^2 = 3934\gevq$}
  \psfrag{y=0.318}{$y = 0.318$} \psfrag{thetaq=37.3}{$\theta_q = 37.3\grad$}
  \psfrag{Ee=54.5}{$E_{e'} = 54.5\gev$} \psfrag{phie=291.3}{$\phi_{e'} =
    291.3\grad$} \psfrag{thetae=71.9}{$\theta_{e'} = 71.9\grad$}
  \includegraphics[width=\textwidth]{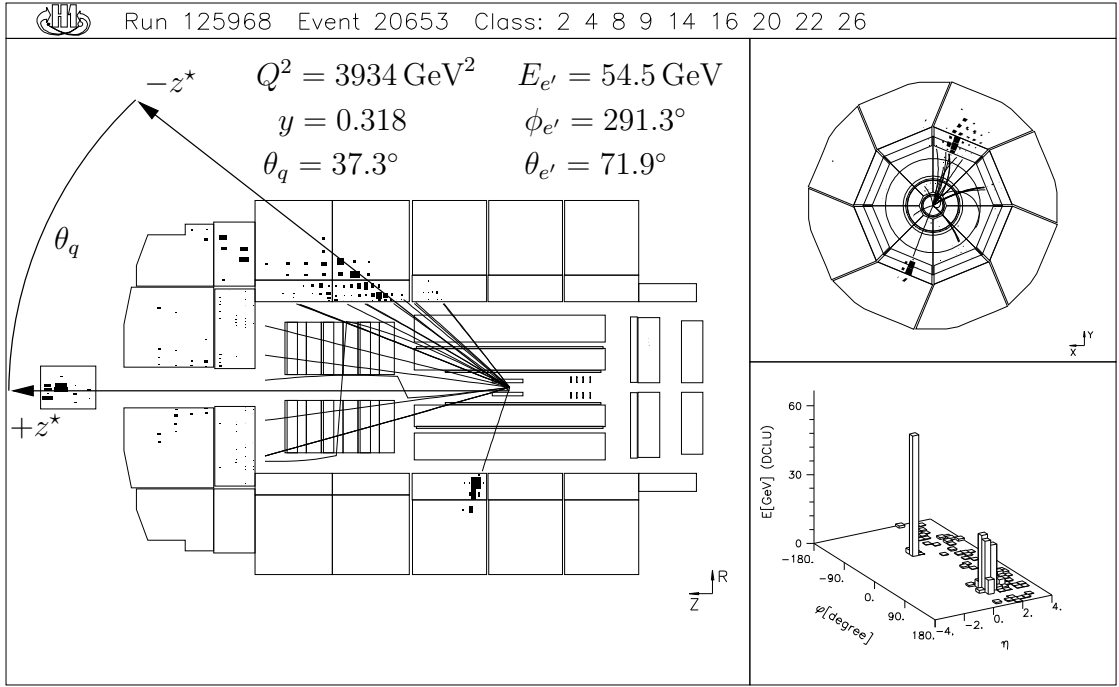}
  \caption[Example of an $ep$ collision measured with the H1 detector.]
  {Example of an $ep$ collision measured with the H1 detector which resembles
    a QPM-like configuration. The isolated electromagnetic cluster in the LAr
    with one track linked to it represents the scattered electron, whereas the
    broader energy deposits in the upper half of the LAr balance the electron
    $p_t$ and correspond to a jet of hadrons produced by the struck quark.
    Some energy measured in the very forward direction indicates the proton
    remnant.  The basic kinematic quantities deduced from the electron are
    given.}
  \label{fig:NC1}
\end{figure}

Initially, all energy deposits measured with the H1 detector are given in the
laboratory system. Since we are, however, merely interested in specific
properties of the hadronic final state caused by the hard interaction, two
problems arise:
\begin{enumerate}
\item The transverse momentum of the scattered electron is balanced by the
  hadronic system. Hence, from longitudinal and transverse momenta of the
  hadrons one can reconstruct the global event kinematics, but they are not
  characteristic of the underlying hadronic process.
  
\item Somehow we have to differentiate between the products of the hard
  reaction and the proton remnant. A maximal separation, which is not given in
  the laboratory system, is desirable.
\end{enumerate}
For illustration fig.~\ref{fig:NC1} shows a comparatively simple NC event with
a clearly identifiable electron and a lot of hadronic activity on the opposite
side with respect to $\phi$. In addition, the proton rest manifests itself in
the form of some clusters in the forward direction.

\begin{figure} 
  \centering \includegraphics{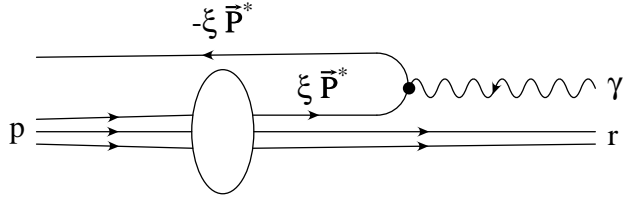}
  \caption{Diagram of a QPM-type $ep$ collision as seen in the Breit frame.}
  \label{fig:QPMBreit}
\end{figure}

Here already, it is not too obvious how to define characteristic properties of
the hadronic energy deposits. In the case of much more complicated average
events, it becomes forbidding. The solution to the two problems is to apply a
Lorentz transformation into the Breit frame of reference~\cite{prop:Breit}.
For an event of QPM type as shown in fig.~\ref{fig:QPMBreit}, it is defined as
the reference system where the incoming parton with momentum $xP\Bf$ in
$+z\Bf$-direction\footnote{Note that this is in contrast to
  e.g.~\cite{prop:Breit} and~\cite{pc:WDDIS} where the $+z\Bf$-axis has been
  chosen for $q\Bf$.}  is back-scattered by a purely space-like boson of
momentum $q\Bf$.\footnote{To distinguish non-invariant quantities in the Breit
  system from those in the laboratory, they will be marked by a $\Bf$.}

\pagebreak\noindent From $q\BF^2 = q^2 = -Q^2$ and $xP\Bf + q\Bf = -xP\Bf$, it
follows that
\begin{equation}
q\Bf = (0, 0, 0, -Q)
\label{eqn:qstern}
\end{equation}
and
\begin{equation}
  P\Bf = \left (\frac{Q}{2x}, 0, 0, \frac{Q}{2x}\right )
  \label{eqn:Pstern}
\end{equation}
where $Q$ is defined to be
\begin{equation}
  \label{eqn:Q}
  Q:=\sqrt{-q^2}=\sqrt{Q^2}\,.
\end{equation}

Thereby, the event and also the \qi soft\qo\ and \qi hard\qo\ physics is
separated by the $(x\Bf,y\Bf)$-plane into a $+z\Bf$ {\bf R}emnant {\bf
  H}emisphere (RH) and a $-z\Bf$ {\bf C}urrent {\bf H}emisphere (CH).
Simultaneously, the transverse momentum of the scattered quark has been
eliminated such that the CH with its available energy of $Q/2$ is very similar
to one half of an $e^+e^- \rightarrow q\bar{q}$ event with purely \qi
time-like\qo\ energy $\sqrt{s}/2$, relating the \qi relevant energies\qo\ $Q$
and $\sqrt{s}$.

Technically, the necessary Lorentz transformation is decomposed into a pure
boost demanding that
\begin{equation}
  2xP\Bf + q\Bf = \left (
    \begin{array}{c}
      Q\\
      \rmvec{0}
    \end{array}
  \right )
\end{equation}
and a rotation afterwards to readjust $P\Bf$ into the $+z\Bf$-direction.  In
addition, the incoming and outgoing electron is usually required to lie in the
$(x\Bf,z\Bf)$-plane.  Their boosted and rotated four-momenta read
\begin{equation}
  k\Bf = \left ( \frac{Q}{2y}(2-y), \frac{Q}{y}\sqrt{1-y},
    0, -\frac{Q}{2} \right )
\end{equation}
and
\begin{equation}
  k\BF' = \left ( \frac{Q}{2y}(2-y), \frac{Q}{y}\sqrt{1-y},
    0, \frac{Q}{2} \right )\,.
\end{equation}
Thus, another feature of the Breit frame is that the energy loss of the
lepton, $E_e\Bf-E_e\BF'$, vanishes.

\subsection{Properties}
\label{sec:Breitprop}

For a better clarification of what happens to the clusters in
fig.~\ref{fig:NC1}, we introduce another variable $\theta_q$ which corresponds
to the polar angle of the scattered quark as seen by the electron:
\begin{equation}
  \label{eqn:thetaq}
  \cos \theta_q = \frac{Q_e^2\,(1-y_e) - 4y_e^2E_e^2}
  {Q_e^2\,(1-y_e) + 4y_e^2E_e^2}\,.
\end{equation}
A comparison of $\theta_q$, drawn also in fig.~\ref{fig:NC1}, with the polar
angle of the most energetic hadronic energy deposition in the LAr obviously
demonstrates that they are approximately equal.  When looked at it from the
Breit frame, the $+z$-axis remains at its position, but the new
$-z\Bf$-direction is given by $\theta_q$!  As a result, the angular region of
$0\grad \leq \theta \leq \theta_q$ is stretched until $0\grad \leq \theta \Bf
\leq 180\grad$ and, together with the remaining $180\grad - \theta_q$, the
complete polar angular range for opposite $\phi$, including the electron and
most of the proton remnant, is squeezed. Intermediate $\phi$ angles other from
$\phi_q$ and $\phi_e$ with $\phi_q - \phi_e = 180\grad$ constitute the
transitional domain between the two cases.

\begin{figure} 
  \centering
  \includegraphics{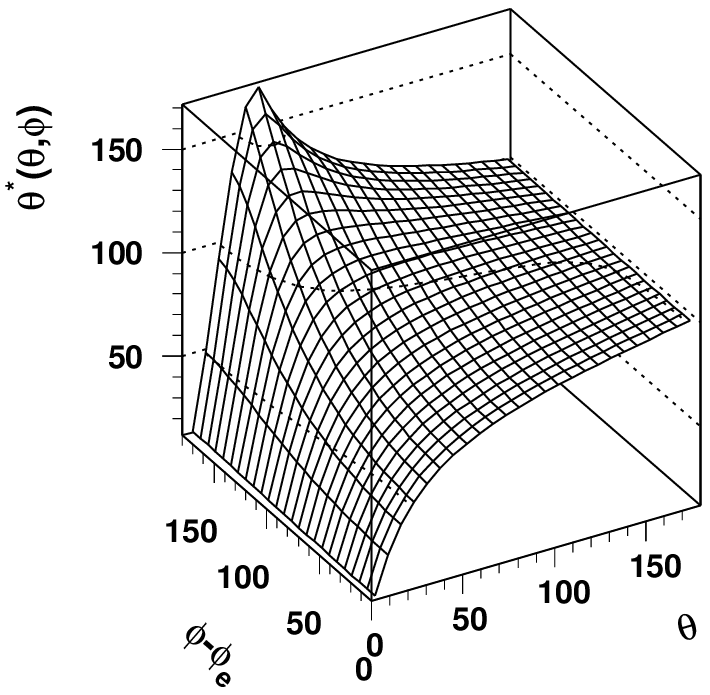}%
  \includegraphics{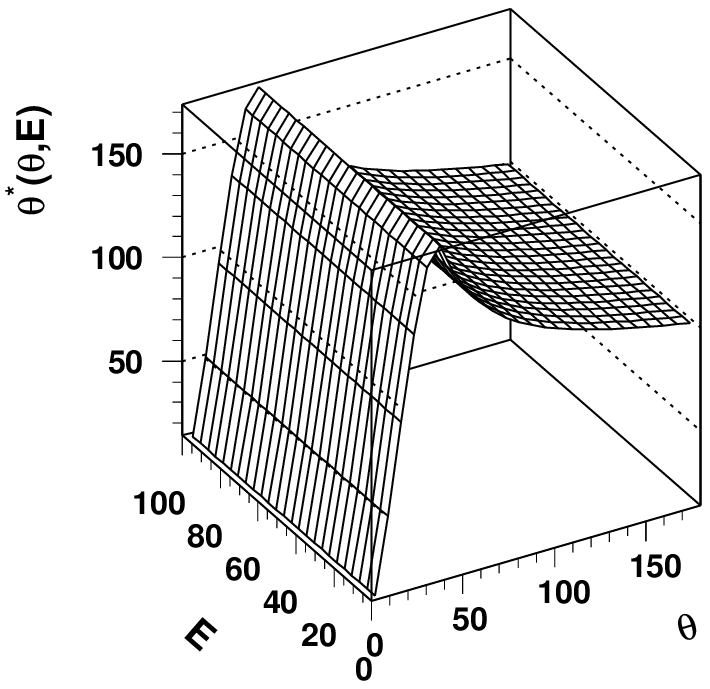}
  \includegraphics{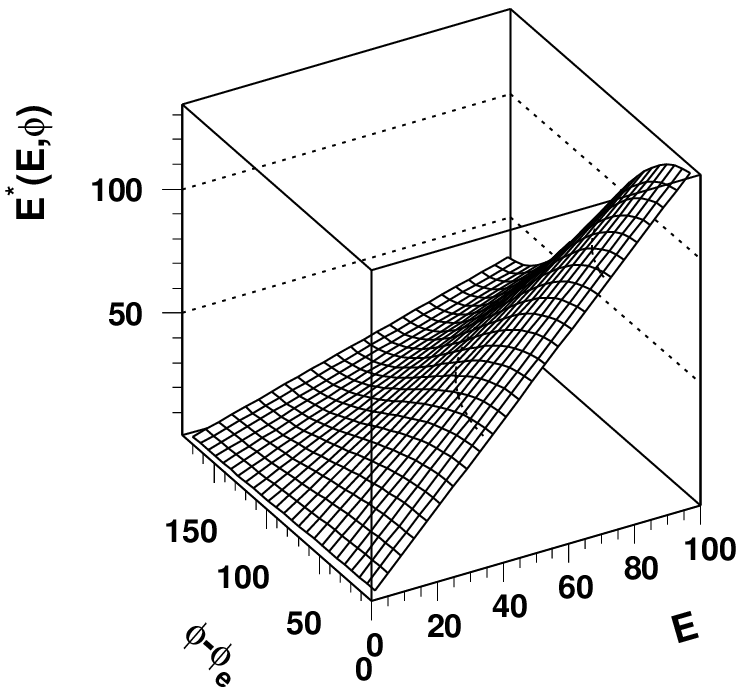}%
  \includegraphics{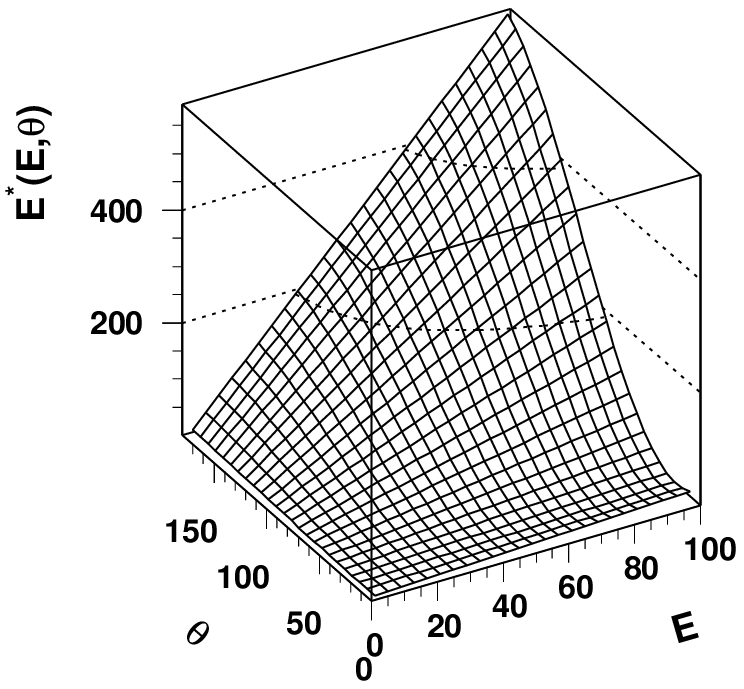}
  \caption[Polar angles and energies in the Breit frame in dependence
  of energy and angles of a four-vector in the laboratory system.]  {The
    $\theta\Bf (\theta ,\phi)$- (top left), $\theta\Bf (\theta ,E)$- (top
    right), $E\Bf (E,\phi)$- (bottom left) and $E\Bf (E,\theta)$-functions
    (bottom right) for the sample boost of fig.~\ref{fig:NC1}. The third input
    variable is set to $E=85.5\gev$, $\phi = \phi_q = 111.3\grad$, $\theta =
    \theta_q = 37.3\grad$ and again $\phi = \phi_q$ respectively.}
  \label{fig:trafo}
\end{figure}

For all variables defined in the next section, the energies and polar angles
of the transformed four-vectors are the important input quantities. Azimuthal
angles with respect to that of the scattered electron, i.e. $\phi\Bf_e =
0\grad$, are rather insignificant. In fig.~\ref{fig:trafo} we therefore
restrict ourselves to four plots showing the $\theta\Bf (\theta ,\phi)$-,
$\theta\Bf (\theta ,E)$-, $E\Bf (E,\phi)$- and $E\Bf (E,\theta)$-functions for
the sample boost of fig.~\ref{fig:NC1}. For simplicity, an exact balance in
$p_t$ of the electron and the hadronic final state is assumed so that $E =
85.5\gev$ is taken as default.

\noindent From these plots it can be concluded that:
\begin{enumerate}
\item $\theta\Bf (\theta ,\phi)$ peaks sharply at $(\theta_q,\phi_q)$.
\item Only clusters opposite of the electron hemisphere in the laboratory
  system appear in the CH, i.e. $\theta\Bf > 90\grad$.
\item $\theta\Bf$ is independent of the energy of the boosted vector.
\item $E\Bf \propto E$ with the slope strongest at $\phi_e$ and $\theta =
  180\grad$.
\end{enumerate}
The first property may lead to a serious deterioration of the resolution in
polar angle in the Breit frame and is the motivation for the cut-off
no.~\ref{cut:thetaq} introduced in section~\ref{sec:fincut}.

\section{Definition of the Event Shapes}
\label{sec:shapesdef}

According to fig.~\ref{fig:QPMBreit}, the hadronic final state in the simplest
case consists of merely one parton with longitudinal momentum only and no
mass. The hard interaction is of a purely electromagnetic nature. QCD induces
deviations from this constellation.  By investigating variables that are
sensitive to these deviations, it is possible to learn more about perturbative
and non-perturbative aspects of QCD.

All quantities, introduced in the next sections and generically labelled as
$F$, are {\em event shapes}\/ and have the following properties in common:

\begin{enumerate}
\item They are dimensionless.
\item For $ep$ DIS they are defined in the Breit frame of reference.
\item In the limit of a QPM-like event $F=0$, otherwise $F\geq 0$.
\item They are infrared and collinear safe, which is most important for a
  valid comparison with pQCD.
\item Soft fragmentation and hadronization processes cause discrepancies
  between fixed-order pQCD calculations and measured data that generally can
  be parameterized to be $\propto 1/Q^p$ with $Q$ being the relevant energy
  scale and $p=1$ or $p=2$ in our circumstances.
\end{enumerate}

The event shapes discussed in the next two subsections assume that the hard
subprocess takes place in the CH alone. However, for $\xi > 2x$ it is possible
for the CH to be completely empty, although experimentally this is improbable
due to hadronization, backscattering, noise, etc. In order to be insensitive
to such effects and to remain infrared safe~\cite{prc:GrauSey1}, one has to
specify what is meant by an \qi empty\qo\ or, vice versa, a \qi full\qo\ CH\@.
We adopt the prescription that the total energy available in the CH $E\Bf :=
\sum_\ich E_i^{\star}$ has to exceed $20\%$ of the value it should have
according to QPM:
\begin{equation}
E\Bf > 0.2 \cdot Q/2 = Q/10\,.
\end{equation}
Otherwise the event is ignored. The precise value of the cut-off is motivated
by a study of the measured energy flow in both hemispheres performed for
ref.~\cite{H1:Shapes}.

To keep event shapes dimensionless, it was originally suggested in
\cite{pc:WDIS95} to normalize energies, momenta and masses to $Q/2$.  \qi
Empty\qo\ events would then lead to $F=0$.  Following the proposal
in~\cite{KR:DIS97} to use for experimental reasons the actually present total
energy $E\Bf$ or momentum $P\Bf := \sum_\ich |\rmvec{p}_i\BF|$ instead, one
would get ill-defined expressions.  Therefore, $F$ is set to zero
in~\cite{pc:WDDIS} analogously for this normalization in contrast to our
definition, thereby affecting the total cross section $\sigma$ and the
left-most bin of the differential cross section $d\sigma /dF$.  Except for the
difference in $\sigma$, mean values $\fmean$, however, are not altered.  In
this work, we will mainly restrict ourselves to the study of the event shapes
normalized to $E\Bf$ or $P\Bf$.

Note that in contrast to the first published experimental results on event
shapes in $ep$ DIS~\cite{KR:DIS97,H1:Shapes} we adopt the modified naming
scheme from ref.~\cite{pc:WDDIS}. Except for $\tau_C$ defined below, the
subscript indicates the quantity ($E$, $P$ or $Q$) used for the normalization.
The event shapes $1-T_C$, $1-T_Z$, $B_C$ and $\rho_C$ investigated in the
above-mentioned publications will be labelled $\tau_C$, $\tau_P$, $B_P$ and
$\rho_Q$.

\subsection[Event Shapes employing $z\Bf$ as Event Axis]{Event Shapes
  employing \boldmath$z\Bf$\unboldmath\ as Event Axis}
\label{sec:shapesaxis}

Choosing the direction of the exchanged boson $\rmvec{q}_{~}\BF = (0,0,-Q)$ as
event axis $\rmvec{n}$, one can, with fig.~\ref{fig:QPMBreit} in mind, easily
deduce quantities $F$ being zero for QPM-like configurations and $F>0$
otherwise.  The simplest event shapes thrust (or rather $1-$thrust $\tau :=
1-T$) and jet broadening $B$ can be written as~\cite{def:B,pc:WDIS95,KR:DIS97}
\begin{eqnarray}
  \label{eqn:tau}
  \tau_P := 1 - \frac{\sum\limits_\ich \left | \rmvec{p}_i\BF \cdot
      \rmvec{n} \right |}{\sum\limits_\ich \left | \rmvec{p}_i\BF \right |} =
  1 - \frac{\sum\limits_\ich \left | p_{li}\Bf \right |}{P\Bf}\,,
  &\qquad&
  \tau_Q := 1-\frac{\sum\limits_\ich \left| p_{li}\Bf \right |}{Q/2}\,,\\
  \nonumber &&\\
  B_P := \frac{\sum\limits_\ich \left | \rmvec{p}_i\BF \times
      \rmvec{n} \right |}{2\sum\limits_\ich \left | \rmvec{p}_i\BF \right |} =
  \frac{\sum\limits_\ich \left | p_{ti}\Bf \right |}{2P\Bf}\,,
  \qquad\quad
  &\qquad&
  B_Q : = \frac{\sum\limits_\ich \left | p_{ti}\Bf \right |}{Q}
\end{eqnarray}
where $p_l$ and $p_t$ denote the longitudinal respectively the transversal
momentum components of $\rmvec{p}$. The factor of $1/2$ for $B$ is
conventional.

\subsection[Event Shapes without Reference to $z\Bf$ as Event Axis]{Event
  Shapes without Reference to \boldmath$z\Bf$\unboldmath\ as Event Axis}
\label{sec:shapesnoaxis}

A possibility for differentiation without reference to $z\Bf$ as event axis is
given by the jet mass $\rho$~\cite{def:rho}:
\begin{eqnarray}
  \rho_E := \frac{\left ( \sum\limits_\ich p_i\Bf \right ) ^2}
  {4 \left ( \sum\limits_\ich E_i\Bf \right ) ^2} =
  \frac{M^2}{4E\BF^2}\,,
  \qquad\qquad\,
  &\qquad& \rho_Q := \frac{M^2}{Q^2}
\end{eqnarray}
with $M$ being the total mass in the CH\@.

Another way is the evaluation with respect to a new axis to be defined.
Originally, event shapes were introduced in the context of \ee~annihilations,
where, in distinction to $ep$ DIS, a preferred direction like $z\Bf$ is not
given beforehand. The first definition deals with the {\em thrust axis}\/
$\rmvec{n}_T$ which is characterized as the normalized vector that maximizes
the sum of all projections of momenta (absolute values) onto it. With regard
to that, $\tau = 1-$thrust can now be described as
\begin{equation}
  \tau_C := 1 - \max_{\rmvec{n}, \rmvec{n}^2=1}
  \frac{\sum\limits_\ich \left | \rmvec{p}_i\BF \cdot \rmvec{n} \right |}
  {\sum\limits_\ich \left | \rmvec{p}_i\BF \right |} =
  1 - \frac{\sum\limits_\ich \left | \rmvec{p}_i\BF \cdot \rmvec{n}_T \right |}
  {P\Bf}\,.
\end{equation}
In contrast to \ee physics, the maximizing procedure is applied in the CH
alone. For merely one momentum vector, $\tau_C$ always equals zero.

A second kind of event axis brings the momentum tensor
\begin{equation}
  \Theta_{jk}\Bf := \frac{\sum\limits_\ich \frac{p_{j_i}\Bf p_{k_i}\Bf}
    {\left | \rmvec{p}_i\Bf \right |}}{\sum\limits_\ich
    \left | \rmvec{p}_i\Bf \right |}
\end{equation}
into play. This real symmetric matrix is positive semi-definite with trace
$Tr(\Theta) = \lambda_1 + \lambda_2 + \lambda_3 = 1$.  For $0 < \lambda_3 \leq
\lambda_2 \leq \lambda_1 < 1$ it describes an ellipso\"{\i}d with pairwise
orthogonal axes named minor, semi-major and major with increasing eigenvalue.
The major axis is similar but not identical to $\rmvec{n}_T$.  If $\lambda_3 =
0$, then the ellipso\"{\i}d degenerates into an elliptical cylinder with all
momentum vectors in one plane. Is $\lambda_2 = 0$ as well, then all momenta
are collinear and the corresponding normal area consists of one or two
parallel planes.  Utilizing the eigenvalues, we can define the $C$
parameter~\cite{def:C}:
\begin{eqnarray}
  C_P := 3( \lambda_1 \lambda_2 + \lambda_2 \lambda_3 + \lambda_3 \lambda_1)\,,
  &\qquad&
  C_Q := \frac{4P\BF^2}{Q^2} C_P\,.
\end{eqnarray}
The conventional factor of three ensures a maximal value of one for $C_P$.

\subsection{Event Shapes employing Jet Algorithms for the Separation
  of the Remnant}
\label{sec:shapesjets}

Another approach of characterizing an event with regard to deviations from the
QPM type does not make use of the CH\@. As depicted in section~\ref{sec:kine},
the elementary reaction yields one hard parton moving along the
$-z\Bf$-direction. The ensuing production of soft partons during fragmentation
and the final hadronization distort this picture to a limited extent only
since no large transverse momenta with respect to the original one are
involved.  Basically, the one parton gets transformed into a tight stream of
hadrons, flying along the original direction, which one refers to as a {\em
  jet}.

The inclusion of more complex processes into the pQCD calculation facilitates
the production of two or more hard partons (s.\ section~\ref{sec:NLOcross}),
and henceforth the events can acquire more than one jet in addition to the
remnant jet. QPM-like events may also be called to be of a $(1+1)$-jet type in
contrast to $(n+1)$-jet events with $n \geq 2$. In order to decide to which
category a given constellation belongs, precise instructions on how to combine
jets from an assortment of four-momenta are needed. The most basic of these
{\em jet algorithms}~\cite{def:jet} makes use of angular {\em cones}\/ around
seeds given by the input four-vectors.  The two schemes we employ are of
another type called {\em cluster}\/ algorithms. Both, the Durham- or
$k_t$-~\cite{def:kt,def:epkt} and the JADE-algorithm~\cite{def:JADE,def:fJADE}
are applied in a modified form adapted to $ep$ DIS in the Breit frame.

The central procedure is almost the same for both.  Two distance measures are
defined, one for distances between two four-vectors, $y_{ij}$, and another one
for the separation of each from the remnant, $y_{ir}$. When all combinations
are evaluated, the minimal value determines which are the closest two in $y$.
If an $y_{ij}$ was smallest, then these two are recombined to one new
four-vector. In case of $y_{ir}$ to be minimal, $i$ is ascribed to the
remnant. The whole routine is repeated until either all $y$'s are larger than
a lower bound $y_{\rm cut}$, or until a certain number of jets is reached. The
first prescription is used to divide a sample of reactions into sets of
$(1+1)$-, $(2+1)$- and so forth events. The second approach is taken to employ
$y$ as an event shape variable.  Here, $y_{k_t}$ and $y_{fJ}$ always denote
the $y$-value where the transition $(2+1) \rightarrow (1+1)$ occurs.  The
respective distance measures are
\begin{eqnarray}
  \label{eqn:fjdefij}
  y_{ij} & := & \frac{2 E_i\Bf E_j\Bf (1-\cos \theta_{ij}\Bf)}{Q^2}\,, \\
  y_{ir} & := & \frac{2 E_i\Bf x E_p\Bf (1-\cos \theta_i\Bf)}{Q^2}
  \label{eqn:fjdefir}
\end{eqnarray}
for the factorizable JADE-algorithm and
\begin{eqnarray}
  \label{eqn:ktdefij}
  y_{ij} & := &
  \frac{2 \min(E_i\BF^2,E_j\BF^2) (1-\cos \theta_{ij}\Bf)}{Q^2}\,, \\
  y_{ir} & := &
  \frac{2 E_i\BF^2 (1-\cos \theta_i\Bf)}{Q^2}
  \label{eqn:ktdefir}
\end{eqnarray}
for the Durham-algorithm.

\section[Event Shapes to $\order(\as)$]{Event Shapes
  to \boldmath$\order(\as)$\unboldmath}
\label{sec:shapesas1}

As explained above, the event shapes are designed to distinguish between a
QPM-like topology of the hadronic final state with the relevant cross section
given by eq.~(\ref{eqn:Born}) and deviations from it that are described in
lowest order pQCD by eq.~(\ref{eqn:Born1}).  In the first case, $F$ always
equals zero, whereas in the latter, $F$ depends on the two additional degrees
of freedom $x_p$ and $z_p$.

To derive explicit formulae, we first determine $p\Bf_1$ and $p\Bf_2$ from
\begin{equation}
  p\Bf_1 + p\Bf_2 = \xi P\Bf + q\BF = \left (
    \begin{array}{c}
      \frac{Q}{2x_p}\\ 0 \\ 0 \\ \frac{Q}{2x_p} - Q
    \end {array}
  \right )
\end{equation}
according to eqs.~(\ref{eqn:qstern}) and~(\ref{eqn:Pstern}).
Remembering that azimuthal angles are irrelevant, we define
the $y$-components to be zero such that
\begin{eqnarray}
  p\Bf_1 := \frac{Q}{2} \left (
    \begin{array}{c}
      z_0 \\ z_t \\ 0 \\ z_3
    \end{array}
  \right )\,, & \quad\quad\quad &
  p\Bf_2 := \frac{Q}{2} \left (
    \begin{array}{c}
      \bar{z}_0 \\ - z_t \\ 0 \\ \bar{z}_3
    \end{array}
  \right )\,.
\end{eqnarray}
Using $p^{\star 2}_1 = p^{\star 2}_2 = 0$ and
\begin{equation}
  z_p = \frac{p\Bf_{1_0}-p\Bf_{1_3}}{Q} = \frac{z_0-z_3}{2} =
  1 - \frac{\bar{z}_0-\bar{z}_3}{2}\,,
\end{equation}
one can calculate $z_0$, $\bar{z}_0$, $z_t$, $z_3$ and $\bar{z}_3$ 
to be
\begin{eqnarray}
  \label{eqn:z0}
  z_0       & = & \frac{2x_pz_p - x_p - z_p +1}{x_p}\,,\\
  \bar{z}_0 & = & \frac{-2x_pz_p + x_p + z_p}{x_p}\,,\\
  z_t       & = & 2 \sqrt{\frac{z_p}{x_p}(1-x_p)(1-z_p)}\,,\\
  z_3       & = & \frac{1 - x_p - z_p}{x_p}\,,\\
  \bar{z}_3 & = & \frac{z_p-x_p}{x_p}\,.
  \label{eqn:barz3}
\end{eqnarray}
In the limit of $x_p \rightarrow 1$ and $z_p \rightarrow 1$
they evaluate to $z_0 = 1$, $z_3 = -1$ and $\bar{z}_0 = \bar{z}_3 = z_t = 0$
corresponding to a QPM-like event with only one final parton.
Taking into account the mismatch in the sign of the $z\Bf$-direction
and noting that $x_p$ is named $\xi$ in~\cite{pc:WDDIS},
the formulae coincide with those of~\cite{prop:Breit} and~\cite{pc:WDDIS}.

\begin{figure} 
  \centering
  \hftwo\includegraphics{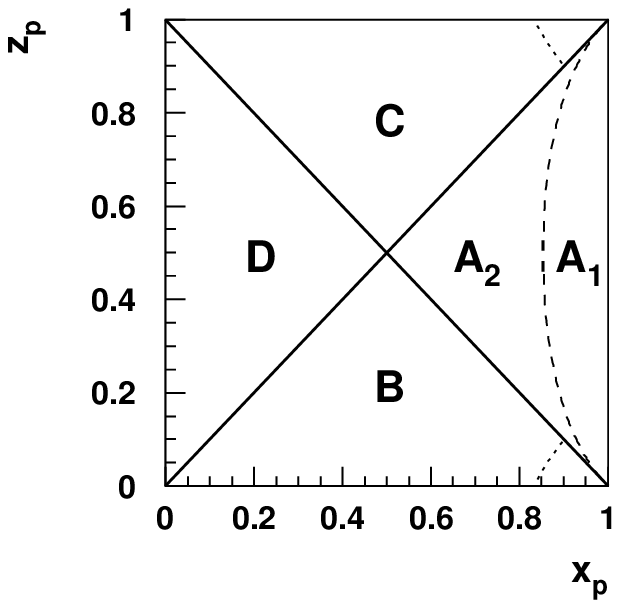}\hftwo%
  \includegraphics{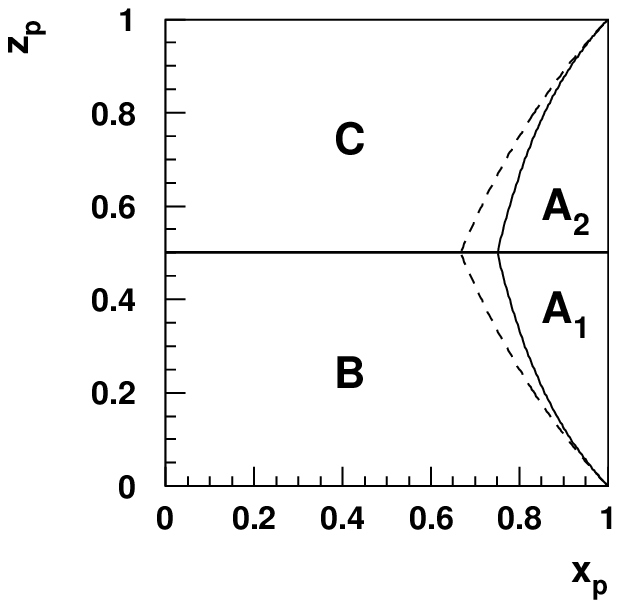}\hftwo
  \caption[$(x_p,z_p)$ phase space region for $\order(\as)$ corrections
  to DIS\@.]{$(x_p,z_p)$ phase space region for $\order(\as)$ corrections to
    DIS\@. See text for an explanation of the labels.}
  \label{fig:xpzp}
\end{figure}

Finally, to achieve results for the event shapes, the available
phase space in $(x_p,z_p)$ has to be subdivided. Fig.~\ref{fig:xpzp}
shows on the left-hand side the appropriate subregions for the
definitions of sections~\ref{sec:shapesaxis} and~\ref{sec:shapesnoaxis}
involving the CH only. The four triangles A to D correspond to:
\begin{description}
\item[A:] Both partons are in the CH: $z_3,\bar{z}_3 \leq 0$.
\item[B:] Only parton $p\Bf_1$ is in the CH: $z_3 \leq 0, \bar{z}_3 > 0$.
\item[C:] Only parton $p\Bf_2$ is in the CH: $z_3 > 0, \bar{z}_3 \leq 0$.
\item[D:] The CH is empty: $z_3,\bar{z}_3 > 0$.
\end{description}

Except for subregion D which is excluded from our event shape
definitions by the explicit requirement of a minimal energy of
$Q/10$ in the CH, all results are listed in table~\ref{tab:alphasshapes}. 
The dotted lines in fig.~\ref{fig:xpzp} point out two parts of
regions B and C rejected in addition to D. Note that
$\rho_E$, $\rho_Q$, $C_P$ and $C_Q$ vanish throughout B and C.

The additional separation into ${\rm A}_1$ and ${\rm A}_2$ indicated
by the dashed line is necessary for $\tau_C$. Depending on the
angle enclosed by $\rmvec{p}_1\BF$ and $\rmvec{p}_2\BF$, the thrust axis
$\rmvec{n}_T$ is represented by
\begin{equation}
  \rmvec{n}_T = \left\{
      \begin{array}{ll}
        \frac{\rmvec{p}_1\BF + \rmvec{p}_2\BF}
        {\left| \rmvec{p}_1\BF + \rmvec{p}_2\BF \right|}\,, &
        {\rm for}\ \angle_{12} \leq 90\grad\, ({\rm A}_1)\, {\rm and}\\
        &\\
        \frac{\rmvec{p}_1\BF - \rmvec{p}_2\BF}
        {\left| \rmvec{p}_1\BF - \rmvec{p}_2\BF \right|}\,, &
        {\rm for}\ \angle_{12} > 90\grad\, ({\rm A}_2)\,.
      \end{array}
    \right .
\end{equation}
In the first case, $\tau_C$ is equal to $\tau_P$, in the latter,
$\tau_C$ also measures momentum components perpendicular to the boson
axis $\rmvec{n}$ as can be seen from the appearance of $z_t$ in the
formula for ${\rm A}_2$.

For the event shapes employing jet algorithms as explained in
section~\ref{sec:shapesjets}, the complete phase space is accessed.
As displayed on the right-hand side of fig.~\ref{fig:xpzp},
it is split up into three main regions:
\begin{description}
\item[A:] The two partons are merged together.
\item[B:] Parton $p\Bf_1$ is clustered to the remnant.
\item[C:] Parton $p\Bf_2$ is clustered to the remnant.
\end{description}
The full line designates the border between A, B and C for $y_{k_t}$, whereas
the dashed line is valid for $y_{fJ}$. Here, the subdivision into ${\rm A}_1$
and ${\rm A}_2$ reflects the $\min(E_i\BF^2,E_j\BF^2)$ condition
of eq.~(\ref{eqn:ktdefij}).

\begin{table}[p]
  \centering
  \begin{tabular}{|c||cc|c|c|}
    \hline
    \rbthr$F$ & ${\rm A}_1$ & ${\rm A}_2$ & ${\rm B}$ & ${\rm C}$\\\hline\hline
    $\tau_P$ & \multicolumn{2}{|c|}{$2(1-x_p)$} &
    $1+z_3/z_0$ & $1+\bar{z}_3/\bar{z}_0$ \rbtrm\\\hline
    $\tau_Q$ & \multicolumn{2}{|c|}{$(1-x_p)/x_p$} &
    $1+z_3$ & $1+\bar{z}_3$ \rbtrm\\\hline
    $B_P$ & \multicolumn{2}{|c|}{$x_pz_t$} &
    $z_t/2/z_0$ & $z_t/2/\bar{z}_0$ \rbtrm\\\hline
    $B_Q$ & \multicolumn{2}{|c|}{$z_t$} &
    $z_t/2$ & $z_t/2$ \rbtrm\\\hline
    $\rho_E$ & \multicolumn{2}{|c|}{$x_p(1-x_p)$} &
    $0$ & $0$ \rbtrm\\\hline
    $\rho_Q$ & \multicolumn{2}{|c|}{$(1-x_p)/x_p$} &
    $0$ & $0$ \rbtrm\\\hline
    $\tau_c$ & $2(1-x_p)$ & $1-\sqrt{4x_p^2z_t^2 + (1-2z_p)^2}$ &
    $0$ & $0$ \rbtrm\\\hline
    $C_P$ & \multicolumn{2}{|c|}{$3(2x_p-1)^2z_t^2/(z_0\bar{z}_0)$} &
    $0$ & $0$ \rbtrm\\\hline
    $C_Q$ & \multicolumn{2}{|c|}{$3(2x_p-1)^2z_t^2/(x_p^2z_0\bar{z}_0)$} &
    $0$ & $0$ \rbtrm\\\hline
    $y_{fJ}$ & \multicolumn{2}{|c|}{$(1-x_p)/x_p$} &
    $z_p$ & $(1-z_p)$ \rbtrm\\\hline
    $y_{k_t}$ & $z_0/\bar{z}_0 \cdot (1-x_p)/x_p$ & 
    $\bar{z}_0/z_0 \cdot (1-x_p)/x_p$ &
    $z_pz_0$ & $(1-z_p)\bar{z}_0$ \rbtrm\\\hline
  \end{tabular}
  \caption[Formulae for the event shapes in $\order(\as)$.]
  {Formulae for the event shapes $F$ in $\order(\as)$.
    For the definition of the phase space subdivisions ${\rm A}_1$,
    ${\rm A}_2$, B and C see fig.~\ref{fig:xpzp} and the explanations
    in the text.}
  \label{tab:alphasshapes}
\end{table}

It is remarkable that to $\order(\as)$ some of the definitions
from section~\ref{sec:shapesdef} lead to the same formulae, e.g.\
for $\tau_Q$, $\rho_Q$ and $y_{fJ}$. However, considering the complete
phase space in $(x_p,z_p)$, discrepancies appear. When higher
orders are included, all event shapes will differ from each other.
Table~\ref{tab:shaperange} gives an overview of the allowed ranges for the
defined event shapes.

\begin{table}[p]
  \centering
  \begin{tabular}{|c||c|c||c||c|c|}
    \hline & \multicolumn{2}{|c||}{Upper bounds} & &
    \multicolumn{2}{|c|}{Upper bounds} \\
    \raisebox{2.0ex}{$F$} &
    \multicolumn{1}{|c}{to $\order(\as)$} &
    \multicolumn{1}{c||}{absolute} &
    \raisebox{2.0ex}{$F$} &
    \multicolumn{1}{|c}{to $\order(\as)$} &
    \multicolumn{1}{c|}{absolute}\\\hline\hline
    $\tau_P$ & $1$ & $1$ &
    $\tau_Q$ & $1$ & $1$ \rbtrm\\\hline
    $B_P$ & $1/2$ & $1/2$ &
    $B_Q$ & $1$ & $1^*$ \rbtrm\\\hline
    $\rho_E$ & $1/4$ & $1/4$ &
    $\rho_Q$ & $1$ & $1^*$ \rbtrm\\\hline
    $\tau_C$ & $1-\sqrt{2}/2$ & $1/2$ & & & \rbtrm\\\hline
    $C_P$ & $3/4$ & $1$ &
    $C_Q$ & $3(\sqrt{5}-1)^3(3-\sqrt{5})/4$ & $2.89^*$ \rbtrm\\\hline
    $y_{fJ}$ & $1/2$ & $0.84^*$ &
    $y_{k_t}$ & $\leq1/(4x)$ & $276^*$\rbtrm\\\hline
  \end{tabular}
  \caption[Upper bounds for the defined event shapes.]
  {Upper bounds for the defined event shapes.
    The starred numbers are not absolute values, but
    the largest ones encountered.}
  \label{tab:shaperange}
\end{table}


\chapter{Data Selection}
\label{chap:datsel}

The very basis of every collider experiment is formed by the equation
\begin{equation}
  \frac{dN}{dt} = {\cal L} \sigma\,.
\end{equation}
It relates the observed event rate $dN/dt$ with the corresponding cross
section $\sigma$. The machine dependent proportional factor ${\cal L}$ is
called {\em luminosity}\/ and has to be measured e.g.\ via comparison to a
theoretically well-known reaction like the Bethe-Heitler process $ep \,
\rightarrow \, ep\gamma$ employed in H1 (s.~section~\ref{sec:H1overview}).

\begin{figure} 
  \centering \includegraphics{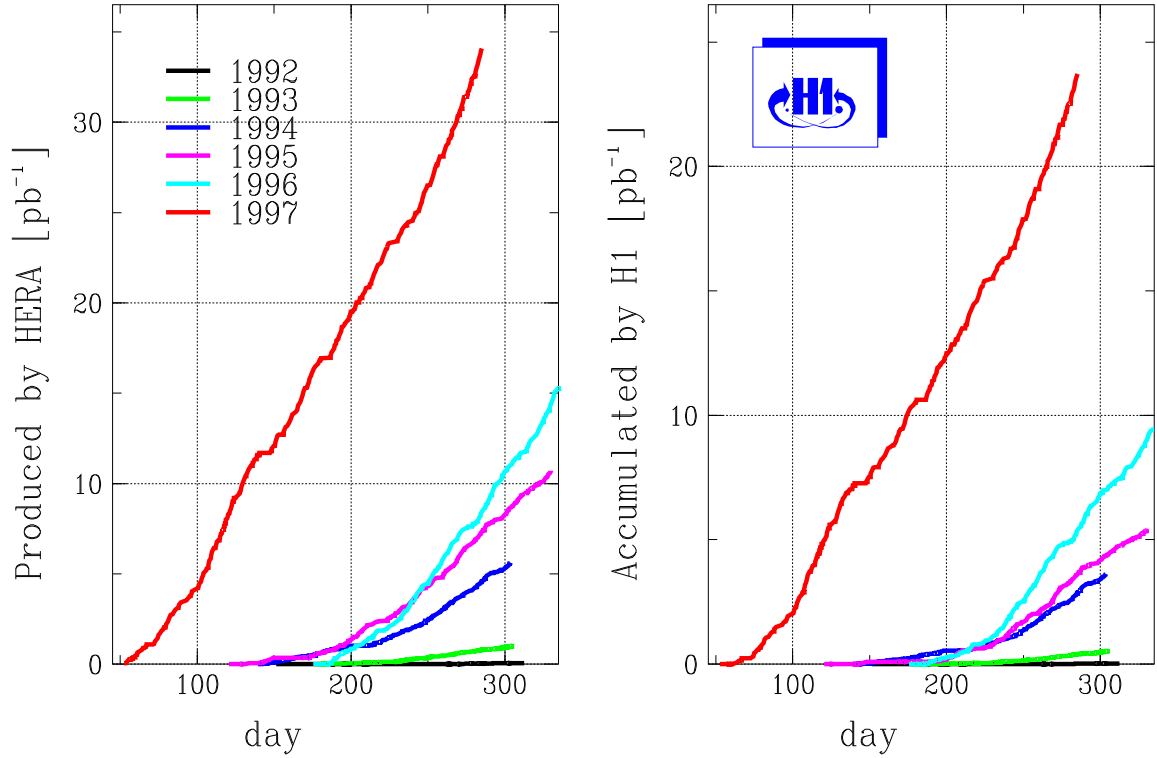}
  \caption{Integrated luminosity produced and accumulated during the
    running periods from $1992$ up to $1997$.}
  \label{fig:INTLUMI}
\end{figure}

In order to gather as many events of a certain kind as possible, one would
like to have a large luminosity. For Gaussian beam profiles with horizontal
and vertical widths $\sigma_{x_1}^*$, $\sigma_{x_2}^*$, $\sigma_{y_1}^*$ and
$\sigma_{y_2}^*$, it is given by
\begin{equation}
  {\cal L} = \frac{1}{e^2f_b}
  \frac{I_1 I_2}{2\pi\left(\sigma_{x_1}^*\sigma_{y_1}^* +
      \sigma_{x_2}^*\sigma_{y_2}^*\right)}
\end{equation}
where $I_1$, $I_2$ are the beam currents and $f_b$ is the bunch frequency.  By
increasing the currents, the HERA crew was able to improve the performance
considerably over the years.  Fig.~\ref{fig:INTLUMI} gives an overview of the
integrated luminosity ${\cal L}_{\rm int}$ that was produced and accumulated
during the running periods from $1992$ up to $1997$.\footnote{Cross sections
  $\sigma$ are usually measured in barns b, where $1\,{\rm b} =
  10^{-24}\cm^{-2}$.  The luminosity integrated over the time ${\cal L}_{\rm
    int}$ may therefore be given in e.g.  ${\rm pb}^{-1}$.}

\section{Background Sources}
\label{sec:back}

Nevertheless, it should be kept in mind that background processes are enhanced
right along. In fact, interactions with atoms of the rest gas in the beam pipe
are even dominating! Table~\ref{tab:rates} gives an impression of the rates to
expect. Therefore, a careful consideration of possible background sources is
necessary. Basically, they can be subdivided into \qi true\qo\ background not
related to $ep$ collisions and misinterpreted competing $ep$ reactions.  The
first are common to all analyses, whereas the latter depend on the topic under
study.

\begin{table}
  \centering
  \begin{tabular}{|l|r|r|}
    \hline
    \multicolumn{3}{|l|}{\rbthr Cross sections and rates
      (at design luminosity)}
    \\\hline\hline
    Beam gas interactions & & $50000\,{\rm Hz}$ \rbtrr\\\hline
    Cosmic myons          & & $  700\,{\rm Hz}$ \rbtrr\\\hline
    Photoproduction       & $1600\,{\rm nb}$ & $25\,{\rm Hz}$ \rbtrr\\\hline
    NC DIS, $Q^2<100\gevq$& $150\,{\rm nb}$  & $2.2\,{\rm Hz}$ \rbtrr\\\hline
    NC DIS, $Q^2>100\gevq$& $1.5\,{\rm nb}$  & $0.022\,{\rm Hz}$ \rbtrr\\\hline
    CC DIS, $P_t>25\gev$  & $0.05\,{\rm nb}$ & $0.001\,{\rm Hz}$ \rbtrr\\\hline
  \end{tabular}
  \caption[Cross sections and rates at design luminosity.]
  {Cross sections and rates at a design luminosity of
    $1.5\times 10^{31} \cm^{-2}{\rm s}^{-1}$~\cite{H1:NIMH1}.}
  \label{tab:rates}
\end{table}

\subsection[Non-$ep$ Background]{Non-\boldmath$ep$\unboldmath\ Background}
\label{sec:backnep}
\subsubsection{Beam Gas Interactions}

Instead of colliding with particles from the other beam, it is also possible
(and even probable) to hit residual gas atoms. Especially the high energetic
protons are able to produce large numbers of particles that may be scattered
into the detector. Most of these events can be rejected because there is no
vertex or the tracks point to vertices outside the interaction region.

\subsubsection{Cosmic Myons}

The surface of the earth and henceforth the detector are constantly hit by
myons of cosmic origin. Most of them cross the experiment out of time with
respect to $ep$ collisions and out of place, i.e.\ nowhere near the
interaction region. Owing to their high rate, however, sometimes they pass in
time right through the CTD and fake DIS events, despite the fact that they
usually deposit little energy in the LAr. Still, they can be identified
topologically by looking for back to back tracks in the IRON and CTD\@. In
addition, the produced clusters are low energetic and imbalanced in transverse
momentum.

\subsubsection{Halo Myons}

A second source of myons are stray protons interacting with material around
the beam line. They always surround the proton beam (halo) and occasionally
give rise to electromagnetic showers in the calorimeters, but since they are
moving along the proton direction, they usually do not cause high $p_t$ tracks
in the CJC.

\subsection[$ep$ Background]{\boldmath$ep$\unboldmath\ Background}
\label{sec:backep}

Concerning our aim to investigate NC DIS, the following processes have to be
considered as background.

\subsubsection{Photoproduction}

The term \qi photoproduction\qo\ derives from the picture of a quasi-real
photon interacting with the proton, i.e. $Q^2 \approx 0$.  Hence, the electron
is only slightly deflected from its original direction and escapes the central
detectors. Instead, it can be found with a certain efficiency in the electron
tagger of the luminosity system.  Nevertheless, the $\gamma p$ reaction may
result in a hadronic final state involving large transverse momenta.  The
misidentification of a particle as the scattered electron is then, due to the
very high rate (s.~table~\ref{tab:rates}), a main source of background.

\subsubsection{Diffractive Events}

Another name for these reactions is \qi{\bf L}arge {\bf R}apidity {\bf G}ap
events\qo\ (LRG's), where the rapidity is defined as
\begin{equation}
  \label{eqn:rapidity}
  y_R := \frac{1}{2} \ln\frac{E+p_z}{E-p_z}\,.
\end{equation}
In the case of massless four-momenta, it is identical to the pseudo-rapidity
\begin{equation}
  \label{eqn:pseudorapidity}
  \eta := - \ln \left( \tan \frac{\theta}{2} \right)\,. 
\end{equation}
In the region between the proton remnant and the current jet, which are
normally connected via colour forces, an unusual gap without hadronic activity
is exhibited.  These events constitute around $5$--$10\%$ of the NC DIS
sample, yet they are not described by the pQCD calculations invoked and have
to be excluded.

\subsubsection{CC Reactions}

At last, it may happen that clusters of the hadronic final state in one of the
seldom CC events with a neutrino as scattered lepton are misidentified as
electron. Since the neutrino, however, escapes detection completely, the total
transverse momentum is imbalanced.

\section{Trigger Scheme}
\label{sec:trigger}

To keep notwithstanding the high rates the dead time of the experiment low and
specifically select true $ep$ collisions, a trigger system has been set up in
four stages~\cite{H1:NIMH1}.  The total read-out time of the detector of
$\approx 1\,{\rm ms}$ is four orders of magnitude larger than that between two
bunch-crossings of $\approx 0.1 \mu{\rm s}$. Therefore, only a small part of
the measured data, especially those from detector parts with short response
times, is available for a fast decision.

The first level trigger L1 collects information from nine trigger systems
attached to one subdetector each. These {\em trigger elements}\/ are combined
to form various subtriggers which provide a {\em KEEP}\/ or {\em REJECT}\/
signal within $2.5\,\mu{\rm s}$.  A pipelining system stores the full data at
the front end during the delay caused by L1 and ensures a dead time free
running at this stage.

For future requirements intermediate trigger levels L2\footnote{L2 was
  commissioned in 1996.} and L3 are foreseen to operate during primary dead
time of the read-out and are based on the same input as L1.  However, they are
able to evaluate a larger number of signals and their complex correlations.
The decision times of these systems are designed to be around $20\,\mu{\rm s}$
and $800\,\mu{\rm s}$ respectively.

The last stage consists of the L4 software trigger, which has the raw data of
the full event at its disposal. Depending on the time consumption, either fast
algorithms specifically adapted to the requirements of L4 or parts of the
standard offline reconstruction program H1REC~\cite{H1REC} are applied to
reach a quick decision. All events accepted by L4 are finally stored on tape.
In addition, a small fraction of {\bf L4} {\bf R}e{\bf J}ec{\bf T}ed events
(L4RJT's) of $\approx 1\%$ are kept for monitoring purposes.

\begin{figure} 
  \centering \includegraphics[width=0.80\textwidth]{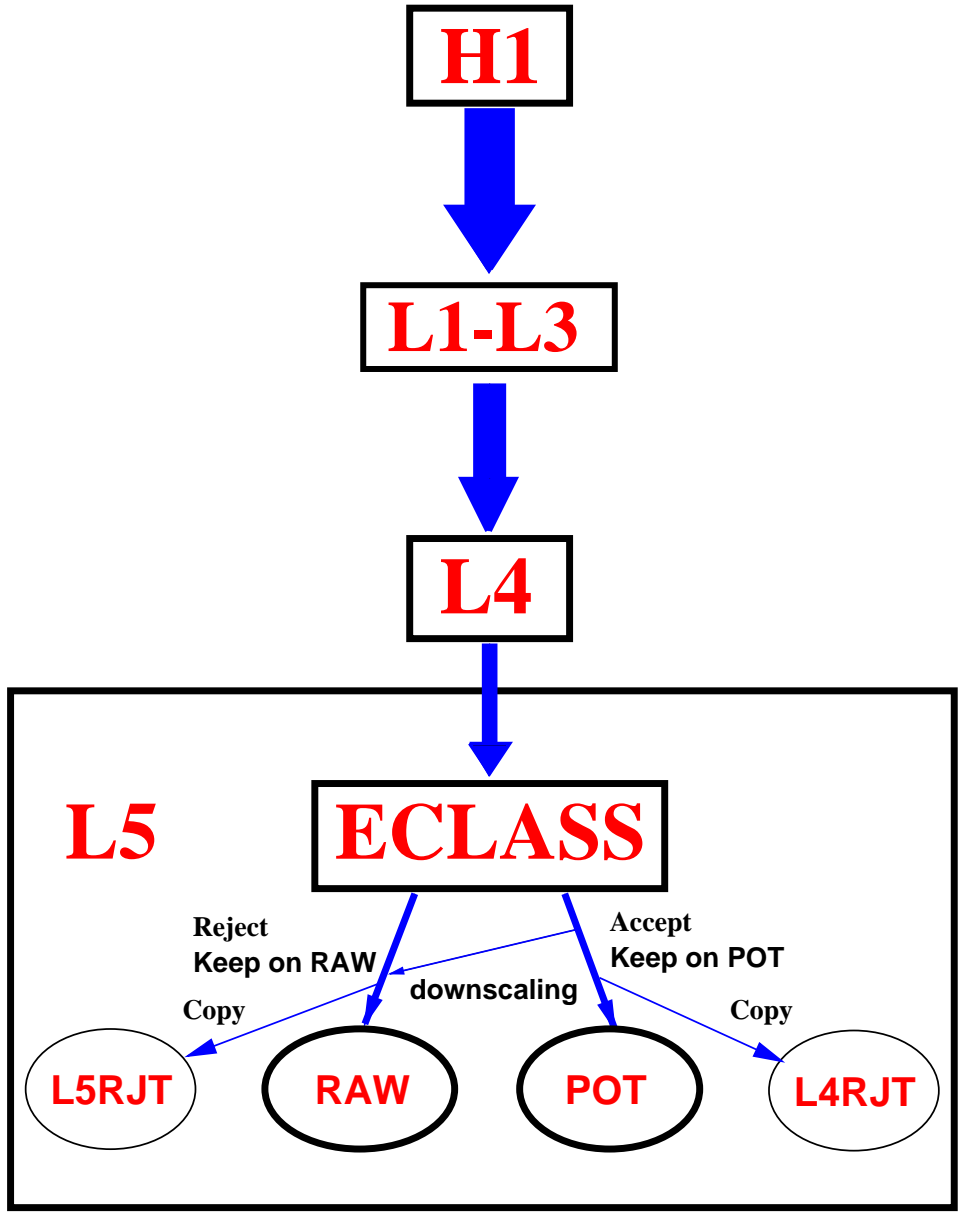}
  \caption[A principal diagram of the data stream in H1.]
  {A principal diagram of the data stream in H1. Three levels of triggers
    operating with fast signals from nine detector branches and one software
    trigger evaluating the complete raw data are responsible for the rejection
    of background. A fifth level L5 works on fully reconstructed events and
    categorizes them into physics classes that are stored as POT's.  To avoid
    overflowing them, the classes may be downscaled.  Events accepted by L4
    but falling in no physics class at all are kept as raw data only.  Small
    samples of L4 and L5 rejected data are retained for monitoring purposes.}
  \label{fig:datastream}
\end{figure}

The complete scheme including a fifth level L5 discussed in the next section
is illustrated in fig.~\ref{fig:datastream}.

\section{Event Classification}
\label{sec:eclass}

Even after all four trigger stages, the reconstructed events kept for physics
analysis contain to a large part unwanted or background reactions for most
investigations. To save tape, network and computing resources, the program
FPACK~\cite{FPACK} used for platform independent data access has the
possibility to skip events according to a classification word stored directly
at the beginning of an event.  Thereby, only the properly marked data tracks
are completely read.  Depending on the aim of a study, different classes, each
matching to a set bit in the classification word, can be selected. Events
falling in no physics class at all are {\bf L5} {\bf R}e{\bf J}ec{\bf T}ed
(L5RJT).  They are kept as raw data, but do not appear fully reconstructed on
{\bf P}roduction {\bf O}utput {\bf T}apes (POT's), except for a small fraction
for monitoring purposes as in the case of L4RJT's.

The chief requirement distinguishing NC DIS from background processes is the
presence of a candidate for the scattered electron in the BEMC or LAr
calorimeter. Therefore, loose criteria such as the existence of at least one
compact mainly electromagnetic cluster with a minimal energy of several $\gev$
are applied to define two DIS classes, a \lowq\ class (no.~$11$) in case of
the BEMC and a \highq\ one (no.~$9$) for the LAr.

Using additionally data from the tracking and myon systems, beam-induced and
cosmic background is rejected more effectively.

\section{Preselection}
\label{sec:presel}

Due to the soft cuts applied in the event classification, the resulting
classes are usually still too contaminated with background.  In addition, some
analyses might not need all parts of the {\bf intended} content. Yet, physics
studies are refined and iterated several times before being finished so it is
recommendable to reduce the data sample further. This is done by a
preselection which copies the selected events to a local disk for fast access.

Since the operation of such a complicated machinery like HERA and H1 is no
easy task, running conditions are subject to variations.  Any time a
significant change in status like the (temporary) failure of an important
detector component occurs, a new {\em run}\/ is initiated.  Only runs
qualified as \qi medium\qo\ or \qi good\qo\ are allowed by the preselection.
Additionally, the most important detector components for this analysis, i.e.\ 
the CJC1/2 and LAr for class $9$ and also the BEMC and BPC for class $11$, are
required to be operational.

Relying on the fact that at least the scattered electron should cause a well
measured track, the existence of a reconstructed vertex is mandatory.

As mentioned above, it is evident for NC DIS that finally a clear candidate
for the scattered electron should be found. In distinction to the event
classification, stricter cuts have to be fulfilled now.  In {\bf H1} a program
package for {\bf PH}ysics {\bf AN}alysis (H1PHAN~\cite{H1PHAN}) is available
containing several algorithms for that purpose.  Two of them, QFSELH/M and
QESCAT, are applied. Both exploit the properties that the scattered electron
should be isolated from the jet of the struck parton due to $p_t$-balance and
that it deposits its energy in form of a compact electromagnetic shower mainly
in the ECAL part of the LAr or in the BEMC\@.

In the case of QFSELH/M, some additional requirements are imposed on the found
candidate.  A cone of radius $0.5$ in azimuthal angle $\phi$ and
pseudo-rapidity $\eta$ is drawn around the candidate. At most $5\%$ of its
energy is allowed to be measured within this cone.  Additionally, a matching
track has to be found within $5\grad$ for \highq.  For electrons in the BEMC,
a hit in the BPC with a maximal distance of $5\cm$ from the cluster center
projected onto the BPC $(x,y)$-plane has to be measured and the cluster radius
may not exceed $5\cm$.

For the purpose of the preselection, events are accepted if a candidate is
determined either way.  In the final analysis QFSELH/M is utilized for the
\lowq\ sample and QESCAT for the \highq\ events. A comparison revealed
marginal differences between the two electron finders.

At last, a minimal energy of the candidate of $9\gev$ is called for, and it is
ensured that safe regions of the calorimeters are hit by cuts on the polar
angle of $10\grad < \theta < 150\grad$ for the LAr and $157\grad < \theta <
173\grad$ for the BEMC respectively.  Because the minimal $Q^2$ demanded later
is $49\gevq$, a slightly smaller value of $45\gevq$, calculated according to
eq.~(\ref{eqn:defq2e}), has to be surpassed by the \lowq\ events.

\section{Final Cut Scenario}
\label{sec:fincut}

Before the last selection is applied, an additional routine rejects residual
events due to cosmic myons. Furthermore, there may be no photons detected in
the PD\@.

In order to achieve a clean DIS sample, the final cuts have to take into
account detector acceptances, efficiencies and resolutions as well as the
diverse background sources described in section~\ref{sec:back}. The latter
have already been suppressed to some extent in the selection procedure above.
Additionally, possible constraints of the theoretical model to compare with
have to be considered.\vspace{0.5cm}

\noindent The final cuts are subdivided into two basic categories:

\begin{itemize}
\item
  {\bf Phase space cuts:}\\
  Due to unavoidable limitations either of the experimental apparatus or the
  theoretical model, it is not possible to choose at will a phase space to
  investigate. Hence, common kinematic requirements have to be imposed on the
  data as well as on the theory.
\item
  {\bf Data quality cuts:}\\
  These are necessary to ensure that background is suppressed and the
  measurement is of good quality. Then, the data can be corrected for detector
  effects and may be compared to theoretical predictions.
\end{itemize}

\noindent First, the phase space cuts will be defined. For a quick overview
see table~\ref{tab:pscuts}.\vspace{0.5cm}

\noindent{\bf Phase space cuts:}

\begin{figure}[t] 
  \centering \includegraphics{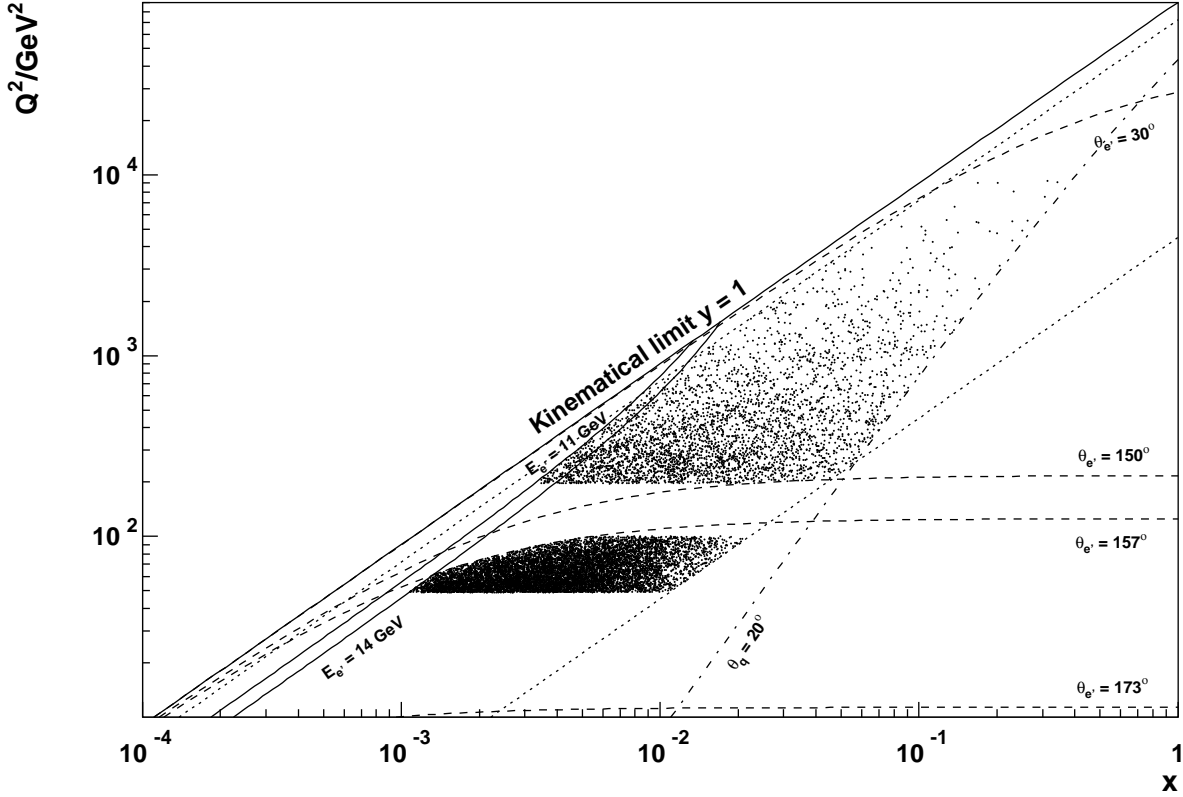}
  \caption[Distribution of finally selected events in the $(x,Q^2)$-plane.]
  {Distribution of finally selected events in the $(x,Q^2)$-plane.  For
    clarity, $1994$ data only are shown. The curves mark out the phase space
    cuts nos.~\ref{cut:y}--\ref{cut:thetaq} in $y$, $E_{e'}$, $\theta_{e'}$
    and $\theta_q$ as indicated; the cuts in $y$ correspond to the
    (unlabelled) dotted lines.}
  \label{fig:kinescat}
\end{figure}

\begin{itemize}
\item \cut\label{cut:Q2} \boldmath {\bf \lowq:} $49 < Q^2/\gevq < 100$,\quad
  {\bf \highq:} $196 < Q^2/\gevq < 10000$:\unboldmath\\
  To examine the $Q$-dependence of the event shapes, all events are grouped
  into eight bins in $Q$: $7$--$8\gev$, $8$--$10\gev$, $14$--$16\gev$,
  $16$--$20\gev$, $20$--$30\gev$, $30$--$50\gev$, $50$--$70\gev$ and
  $70$--$100\gev$. Note that there is a gap from $10\gev$ to $14\gev$ which
  corresponds to the excluded transition region between the BEMC and LAr
  calorimeters. The lower bound of $7\gev$ is motivated by the fact that an
  energy of $Q/2 = 3.5\gev$ should be available in the CH of the Breit frame.
  For $Q > 100\gev$ there is not enough statistics at hand.\\
  In case of data, $Q^2$ is identified with $Q_e^2$ according to
  eq.~(\ref{eqn:defq2e}).
  
\item \cut\label{cut:y}
  \boldmath $0.05 < y < 0.8$:\unboldmath\\
  As explained in section~\ref{sec:kinereco}, the electron method is not well
  suited for the reconstruction of $y$-values as low as $0.05$. Hence, these
  events are rejected. At high~$y$, radiative corrections due to an additional
  real photon emission of the scattered electron
  (s.~section~\ref{sec:simpar}) are enormous and have to be avoided.\\
  For data, $y$ is identified with $y_e$ according to eq.~(\ref{eqn:defye})
  down to $0.15$. Below that value, $y_h$ from eq.~(\ref{eqn:defyh}) is taken
  instead.\footnote{Concerning the boost into the Breit frame, it follows that
    $x=Q_e^2/(sy_h)$ for $0.05<y_e<0.15$!}  Note that nevertheless $y_e>0.05$
  is asked for.
  
\item \cut\label{cut:Ee} \boldmath {\bf \lowq:} $E_{e'} > 14\gev$,\quad
  {\bf \highq:} $E_{e'} > 11\gev$:\unboldmath\\
  Lower limits on the electron energy are imposed for several reasons: First,
  the trigger efficiency is above $99\%$~\cite{H1:NIMBEMC,H1:ICHEP98hq}.
  Moreover, there are numerous clusters of hadronic origin, e.g.\ pions, which
  are low energetic and may fake electrons, especially when the real one did
  not even hit the main detectors e.g.\ in photoproduction.  At last, true
  leptons which lost a large part of their original energy due to radiative
  effects are excluded. Thereby, a good measurement of $E_{e'}$ should be
  reached, which is most important for the boost into the Breit frame.
    
\item \cut\label{cut:thetae} \boldmath {\bf \lowq:} $157\grad < \theta_{e'} <
  173\grad$,\quad
  {\bf \highq:} $30\grad < \theta_{e'} < 150\grad$:\unboldmath\\
  These cuts reflect the coverage in polar angle of the BEMC and LAr
  calorimeters, although the upper limit of $173\grad$ for the \lowq\ sample
  is redundant owing to \mbox{$Q^2>49\gevq$}.  For the \highq\ events, the
  forward region $\theta_{e'} \leq 30\grad$ with its high hadronic activity is
  left out to avoid misidentifications of the scattered electron.
  
\item \cut\label{cut:thetaq}
  \boldmath $20\grad < \theta_q$:\unboldmath\\
  $\theta_q$, eq.~(\ref{eqn:thetaq}), indicates the direction of the
  $-z\Bf$-axis in the laboratory system. The requirement $\theta_q > 20\grad$
  ensures a sufficient detector resolution in polar angle (s.\ the discussion
  in section~\ref{sec:Breitprop}). In the \lowq\ regime, this cut-off is
  automatically fulfilled owing to the previous selection in $Q^2$ and $y$; a
  small number of \highq\ events, however, is discarded, cf.\ 
  fig.~\ref{fig:kinescat}.
  
\item \cut\label{cut:QBreit} \boldmath {\bf $\tau_P$, $B_P$, $\rho_E$,
    $\tau_C$ and $C_P$ only:}
  $E\Bf > Q/10$:\unboldmath\\
  As explained in section~\ref{sec:shapesdef}, this cut is essential to keep
  the event shapes $\tau_P$, $B_P$, $\rho_E$, $\tau_C$ and $C_P$ infrared
  safe. In this sense, it is part of their {\bf definition}.  Experimentally,
  it ensures a minimum of hadronic activity in the CH and suppresses events
  substantially influenced by noise and leakage out of the RH.
\end{itemize}

\begin{table}
  \centering
  \begin{tabular}{|l||c|c|}
    \hline\multicolumn{3}{|c|}{\rbthr\bf\large Phase space cuts}\\\hline
    & \multicolumn{1}{|c|}{\rbthr\bf\boldmath \lowq\ sample (BEMC)} &
    {\bf\boldmath \highq\ sample (LAr)}\\\hline\hline
    {\bf Cut~\ref{cut:Q2}}& $49\gevq < Q^2 < 100\gevq$ &
    $196\gevq < Q^2 < 10000\gevq$ \rbtrm\\\hline
    {\bf Cut~\ref{cut:y}}& \multicolumn{2}{|c|}{$0.05 < y < 0.8$}\rbtrm\\\hline
    {\bf Cut~\ref{cut:Ee}}& $E_{e'} > 14\gev$ & $E_{e'} > 11\gev$\rbtrm\\\hline
    {\bf Cut~\ref{cut:thetae}} & $157\grad < \theta_{e'} < 173\grad$ &
    $30\grad < \theta_{e'} < 150\grad$\rbtrm\\\hline
    {\bf Cut~\ref{cut:thetaq}}& \multicolumn{2}{|c|}
    {$20\grad < \theta_q$}\rbtrm\\\hline
    {\bf Cut~\ref{cut:QBreit}$^*$}& \multicolumn{2}{|c|}
    {$E\Bf > Q/10$}\rbtrm\\\hline
  \end{tabular}
  \caption[Final phase space cuts.]
  {Final phase space cuts. Note that the starred cut no.~\ref{cut:QBreit}
    is applied for $\tau_P$, $B_P$, $\rho_E$, $\tau_C$ and $C_P$ only.}
  \label{tab:pscuts}
\end{table}

\noindent For an overview of the data quality cuts
see table~\ref{tab:dqcuts}.\vspace{0.5cm}

\noindent{\bf Data quality cuts:}

\begin{itemize}
\item \cut\label{cut:NCH} \boldmath {\bf $\tau_P$, $B_P$, $\rho_E$, $\tau_C$
    and $C_P$ only:} $N_{\rm CH}\geq 2\,;\;
  (F-F_{\rm min})\,,(F_{\rm max}-F)>5\cdot10^{-5}$:\unboldmath\\
  Asking for at least two objects in the CH, \qi unnatural\qo\ peaks at zero
  for $\rho_E$, $\tau_C$ and $C_P$ in the \lowq\ region are removed.  \qi
  Unnatural\qo\ means that they are caused either by leakage out of the
  remnant fragmentation region or by cutting off \qi regular\qo\ jets just at
  the border between the two hemispheres.  In fact, $B_P$ does exhibit a
  pronounced peak at $0.5$ for the rejected events showing the
  hadrons/clusters to have polar angles marginally larger than the required
  $90\grad$ in the Breit frame.  For the same reason, the extreme values
  $F_{\rm min}$ and $F_{\rm max}$ are excluded by very small cut-offs.
  
\item \cut\label{cut:Eforw}
  \boldmath $E_{\rm forw} > 0.5\gev$:\unboldmath\\
  Requiring a minimal energy deposition in the forward region defined by
  \mbox{$4\grad < \theta < 15\grad$}, diffractive events not described by
  usual pQCD calculations are discarded.
  
\item \cut\label{cut:acc}
  \boldmath $5.7\grad \leq \theta_{\rm cl} \leq 170\grad$:\unboldmath\\
  This cut reflects the angular acceptance for clusters that are completely
  contained in the LAr or BEMC calorimeters.
  
\item \cut\label{cut:EPz}
  \boldmath $30\gev < (E-P_z) < 65\gev$:\unboldmath\\
  If all emerging particles of an $ep$ reaction could be measured perfectly,
  then
  \begin{equation}
    \label{eqn:EPz}
    (E-P_z) := \sum_i (E_i - p_{z_i})
  \end{equation}
  would be $E_p - p_{z_p} + E_e - p_{z_e} = 2\cdot E_e = 55\gev$.  Losses due
  to limited acceptance, e.g.\ around the beam pipe, transform this peak into
  a broad distribution with tails down to very low values. Restricting the
  range of $(E-P_z)$ effectively reduces the photoproduction background and the
  size of radiative corrections.
  
\item \cut\label{cut:Pt} \boldmath {\bf \lowq:} $P_t < 7.5\gev$,\quad
  {\bf \highq:} $P_t < 15\gev$:\unboldmath\\
  The total transverse momentum
  \begin{equation}
    \label{eqn:Pt}
    P_t := \sqrt{\left (\sum_i p_{x_i}\right )^2 +
      \left (\sum_i p_{y_i}\right )^2}
  \end{equation}
  of a NC event should normally be zero. To suppress remaining background from
  CC reactions or badly measured events, maximal $P_t$'s of $7.5\gev$ and
  $15\gev$ are allowed in the \lowq\ and \highq\ samples respectively.
  
\item \cut\label{cut:Eda}
  \boldmath $\left|(E_{e'}-E_{da})/E_{da}\right| < 0.25$:\unboldmath\\
  According to
  \begin{equation}
    E_{da} := \frac{Q_{da}^2}{4E_e}+E_e(1-y_{da})\,,
  \end{equation}
  the energy of the scattered electron can be derived from angular information
  only. This fact is exploited e.g.\ for the energy calibration of the
  calorimeter. Asking for both values to be compatible within $\approx
  3\sigma$ suppresses events strongly affected by QED radiation.
  
\item \cut\label{cut:zv}
  \boldmath $-35\cm < z_{\rm v} - \mean{z_{\rm v}} < 35\cm$:\unboldmath\\
  Here, it is enforced that an interaction vertex could be determined which
  lies within $\pm 35\cm \approx 3\sigma$ around the average $z$-position for
  the corresponding run period. In $1994$, $\mean{z_{\rm v}} \approx 3\cm$,
  and in $1995$--$1997$, $\mean{z_{\rm v}} \approx -1\cm$.
  
\item \cut\label{cut:Ebackw}
  \boldmath $E_{\rm backw} < 10\gev$:\unboldmath\\
  Due to the very limited capability of the BEMC calorimeter to measure
  hadronic energies, events with significant activity there are discarded. In
  addition, unidentified scattered electrons in the BEMC may lead to a
  rejection.
  
\item \cut\label{cut:crack} \boldmath {\bf LAr only:}
  $\left|\phi_{\rm imp}-\phi_{\rm cr}\right|\geq 2\grad$,\\
  $\neg\,(-65\cm<z_{\rm imp}<-55\cm)\wedge\neg\,(15\cm<z_{\rm imp}<25\cm)$:
  \unboldmath\\
  In order to ensure a reliable measurement of the scattered electron,
  partially inefficient regions such as cracks between calorimeter modules
  ($\phi$-cracks) or wheels ($z$-cracks) have to be avoided.  $\phi_{\rm imp}$
  and $z_{\rm imp}$ denote the impact coordinates of the electron.
\end{itemize}

\begin{table}
  \centering
  \begin{tabular}{|l||c|c|}
    \hline\multicolumn{3}{|c|}{\rbthr\bf\large Data quality cuts}\\\hline
    & \multicolumn{1}{|c|}{\rbthr\bf\boldmath \lowq\ sample (BEMC)} &
    {\bf\boldmath \highq\ sample (LAr)}\\\hline\hline
    {\bf Cut~\ref{cut:NCH}$^*$}&
    \multicolumn{2}{|c|}{$N_{\rm CH}\geq 2\,,\quad 
      F-F_{\rm min}> 5\cdot10^{-5}\,,\quad
      F_{\rm max}-F > 5\cdot10^{-5}$}\rbtrm\\\hline
    {\bf Cut~\ref{cut:Eforw}}&
    \multicolumn{2}{|c|}{$E_{\rm forw}>0.5\gev$}\rbtrm\\\hline
    {\bf Cut~\ref{cut:acc}}&
    \multicolumn{2}{|c|}{$5.7\grad \leq \theta_{\rm cl} \leq 170\grad$}
    \rbtrm\\\hline
    {\bf Cut~\ref{cut:EPz}}&\multicolumn{2}{|c|}{$30\gev <(E-P_z)< 65\gev$}
    \rbtrm\\\hline
    {\bf Cut~\ref{cut:Pt}}& $P_t < 7.5\gev$ & $P_t < 15\gev$\rbtrm\\\hline
    {\bf Cut~\ref{cut:Eda}}& \multicolumn{2}{|c|}
    {$\left|(E_{e'}-E_{da})/E_{da}\right| < 0.25$}\rbtrm\\\hline
    {\bf Cut~\ref{cut:zv}}& \multicolumn{2}{|c|}
    {$-35\cm < z_{\rm v} - \mean{z_{\rm v}} < 35\cm$}\rbtrm\\\hline
    {\bf Cut~\ref{cut:Ebackw}}&
    \multicolumn{2}{|c|}{$E_{\rm backw} < 10\gev$}\rbtrm\\\hline
    {\bf Cut~\ref{cut:crack}}&&
    $\left|\phi_{\rm imp}-\phi_{\rm cr}\right|\geq 2\grad$\rbtrm\\
    && $\left|z_{\rm imp}-z_{\rm cr}\right|\geq 5\cm$\rbtrm\\\hline
  \end{tabular}
  \caption[Final data quality cuts.]
  {Final data quality cuts. Note that the starred cut
    no.~\ref{cut:NCH} is applied for $\tau_P$, $B_P$, $\rho_E$, $\tau_C$
    and $C_P$ only. Depending on the
    theory to compare with, the cuts nos.~\ref{cut:NCH} and~\ref{cut:Eforw}
    may have to be considered as phase space cuts, cf.\ the discussion
    in section~\ref{sec:corrrad}.}
  \label{tab:dqcuts}
\end{table}

The only considerable background remaining hereafter stems from
photoproduction. In the \lowq\ sample it is estimated to be less than $3\%$,
for the \highq\ region it is negligible~\cite{H1:Shapes}.  Residual radiative
effects are accounted for by the correction procedure described in
ch.~\ref{chap:corrproc}.

\begin{figure}[b]
  \centering
  \psfrag{L int * pb}{\boldmath ${\cal L}_{\rm int}/{\rm pb}^%
    {\scriptscriptstyle -1}$\unboldmath} \psfrag{Nr. of
    events}{\sffamily\bfseries No.\ of events} \psfrag{1994 lq}
  {\scriptsize{\sffamily\bfseries 1994 low }\boldmath$Q^2$\unboldmath}
  \hftwo\includegraphics{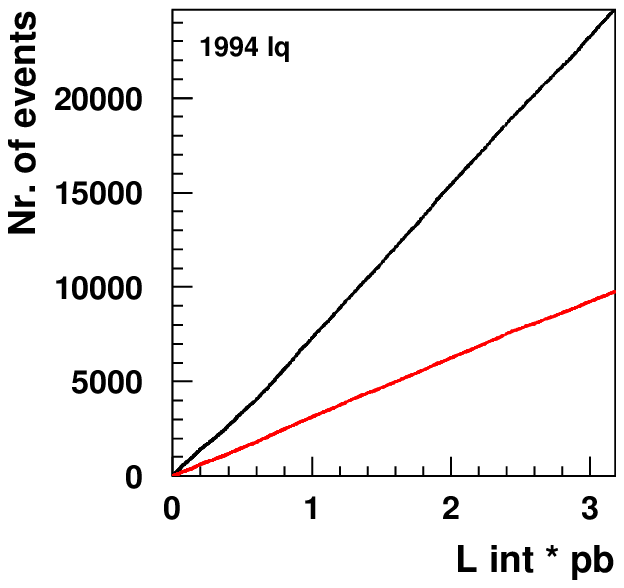}\hftwo%
  \psfrag{L int * pb}{\boldmath ${\cal L}_{\rm int}/{\rm pb}^%
    {\scriptscriptstyle -1}$\unboldmath}%
  \psfrag{Nr. of events}{\sffamily\bfseries No.\ of events}%
  \psfrag{1994 hq} {\scriptsize{\sffamily\bfseries 1994 high
      }\boldmath$Q^2$\unboldmath}
  \includegraphics{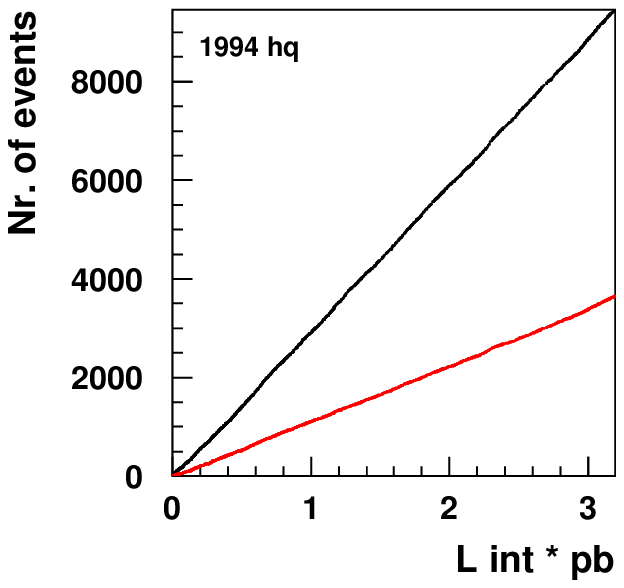}\hftwo\vspace{0.5cm}\linebreak \psfrag{L int
    * pb} {\boldmath ${\cal L}_{\rm int}/{\rm pb}^ {\scriptscriptstyle
      -1}$\unboldmath} \psfrag{Number of events/Lint/pb} {{\sffamily\bfseries
      No.\ of events}\boldmath$/{\cal L}_{\rm int}\, /{\rm pb}$\unboldmath}
  \includegraphics{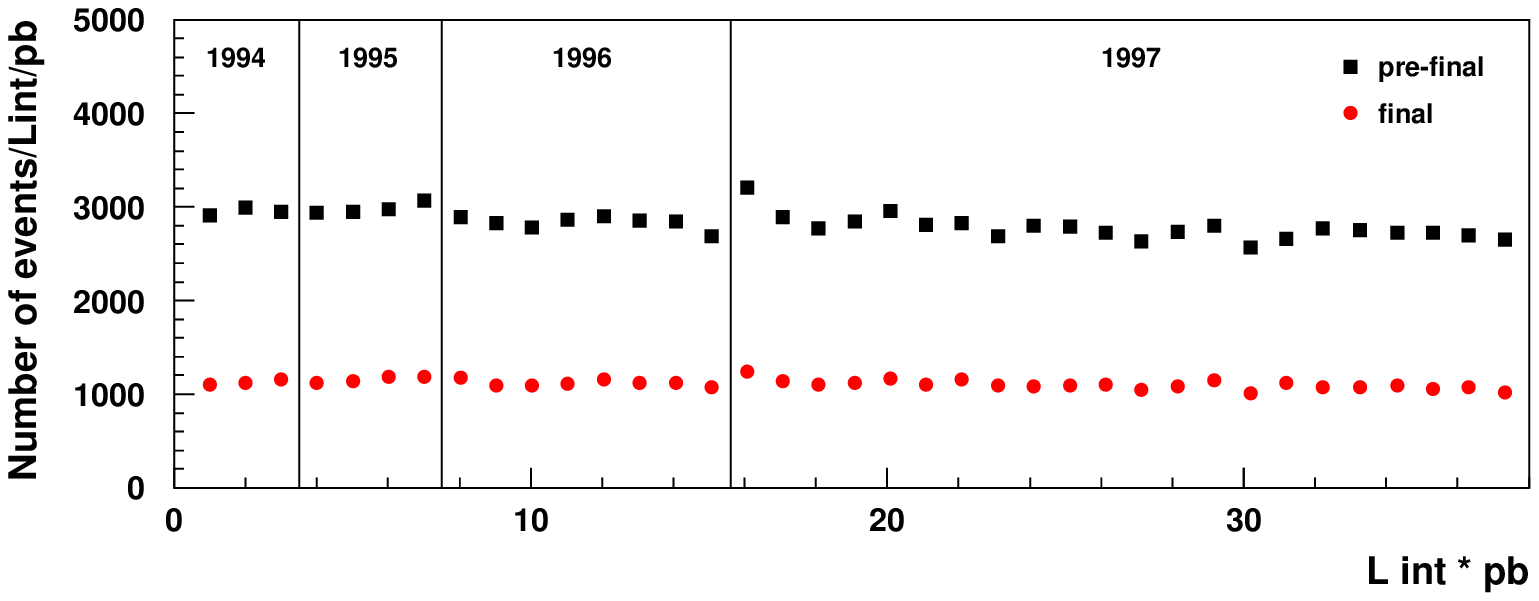}
  \caption[Number of selected events in 1994 and the rate of selected
  events from 1994--1997 versus the integrated luminosity.]  {In the two plots
    on top, the accumulated events (upper curve: pre-final, lower curve:
    final) are drawn against ${\cal L}_{int}$ for the low and \highq\ samples
    of 1994. A constant rate of incoming events is demonstrated.  The lower
    plot presents the constancy of the number of events gathered per ${\cal
      L}_{int}$ of $\approx 1\,{\rm pb}^{-1}$ for all four years contributing
    to the \highq\ sample.  The statistical uncertainties are smaller than the
    marker sizes.}
  \label{fig:lumiplot}
\end{figure}

\pagebreak
To illustrate the stability of the described selection procedure,
the number of events accumulated in 1994 for the \lowq\ as well as the \highq\ 
sample is plotted versus the integrated luminosity ${\cal L}_{int}$ in
fig.~\ref{fig:lumiplot}.  Also demonstrated is the constancy of the number of
events gathered per ${\cal L}_{int}$ of $\approx 1\,{\rm pb}^{-1}$ for all
four years contributing to the \highq\ sample.  The term \qi pre-final\qo\ 
refers to the production of data n-tuples where additionally to the
preselection the cosmic filter, the anti-photon tag and the cuts
nos.~\ref{cut:Eforw},~\ref{cut:zv} and~\ref{cut:Ebackw} are in effect already.
QFSELM/QESCAT is employed for the determination of the scattered electron.
Statistics on the selection procedure can be looked up in
table~\ref{tab:data}.

\begin{table}[t]
  \centering
  \begin{tabular}{|l||r|r|r||r|}
    \hline
    \multicolumn{1}{|c||}{\rbthr year} &
    \multicolumn{1}{c|}{${\cal L}_{\rm int}/{\rm pb}^{-1}$} &
    \multicolumn{1}{c|}{$\#$ preselected} &
    \multicolumn{1}{c||}{$\#$ pre-final} &
    \multicolumn{1}{c|}{$\#$ final}\\\hline\hline
    1994 \highq &  $3.20$ &  $12177$ &   $9467$ &  $3646$ \rbtrr\\\hline
    1995 \highq &  $4.29$ &  $16199$ &  $12817$ &  $5006$ \rbtrr\\\hline
    1996 \highq &  $8.49$ &  $30319$ &  $23924$ &  $9478$ \rbtrr\\\hline
    1997 \highq & $22.26$ & $100466$ &  $61711$ & $24477$ \rbtrr\\\hline\hline
    total \highq& $38.24$ & $159161$ & $107919$ & $42607$ \rbtrr\\\hline\hline
    1994 \lowq  &  $3.18$ &  $35051$ &  $24690$ &  $9761$ \rbtrr\\\hline
  \end{tabular}
  \caption{Integrated luminosities and events gathered in the
    selection steps.}
  \label{tab:data}
\end{table}

\chapter{Event Simulation}
\label{chap:evsim}

Measurements with the H1 detector essentially comprise clusters, i.e.\ energy
depositions in the calorimeters, and tracks in the tracking devices that are
caused by long-lived\footnote{Here, long-lived means lifetimes $\tau >
  10^{-8}\,{\rm s}$.} particles, mainly hadrons.  Yet, the objects dealt with
in theoretical considerations of pQCD are partons or, equivalently, quarks,
anti-quarks and gluons. Due to the complexity of the measuring apparatus and
the underlying physical processes, a direct link from clusters and tracks
backwards to hadrons or even partons can not be established.  Nevertheless,
information can be drawn from the selected data by comparing with model
assumptions on a statistical basis.  The first task to be carried out in this
analysis chain consists in simulating the detector response to a given physics
model.  This simulation procedure is the subject of the next two sections.

\section{Simulation}
\label{sec:sim}
\subsection{Parton Level}
\label{sec:simpar}

Starting with the calculation of the matrix element of the hard scattering to
$\order(\alpha^2\as)$, complicated integrals arise. For the purpose of
generating \qi events,\qo\ they are solved by employing a {\em Monte Carlo
  integration}\/ technique, s.~e.g.~\cite{VEGAS}.  Subsequently, the result
has to be folded with pdfs describing the proton structure. For the purpose of
easy access, the available sets of pdfs are compiled in the PDFLIB program
library~\cite{PDFLIB}.  According to the probability distribution derived from
the matrix element and the available phase space, a limited number of final
state partons is \qi produced\qo\ within the framework of dedicated computer
programs called {\bf M}onte {\bf C}arlo (MC) {\em generators}.  For $ep$ DIS
two such programs, LEPTO~\cite{MC:LEPTO} and HERWIG~\cite{MC:HERWIG}, are
available.

To account for higher orders in a {\bf L}eading {\bf L}ogarithmic {\bf
  A}pproximation (LLA), both offer an implementation of {\bf P}arton {\bf
  S}howers (PS) including coherence effects.  These branching algorithms may
be attached to incoming ({\bf I}nitial {\bf S}tate PS, ISPS) and outgoing
partons ({\bf F}inal {\bf S}tate PS, FSPS) as long as their four-momentum
squared $t$ ({\em virtuality}\/) is above some adjustable threshold, typically
around $t_0 \approx 1$--$4\gevq$.  Otherwise, the PS is terminated.

Alternatively, a {\bf C}olour {\bf D}ipole {\bf M}odel (CDM) can be invoked to
describe gluon radiation including the first emission in the QCDC reaction.
ARIADNE~\cite{MC:ARIADNE} supplies an implementation of the CDM, but is not
intended to be a stand-alone program.  Instead, it provides an interface to
LEPTO so that it may be used within its framework.

\begin{figure}
  \centering
  \includegraphics{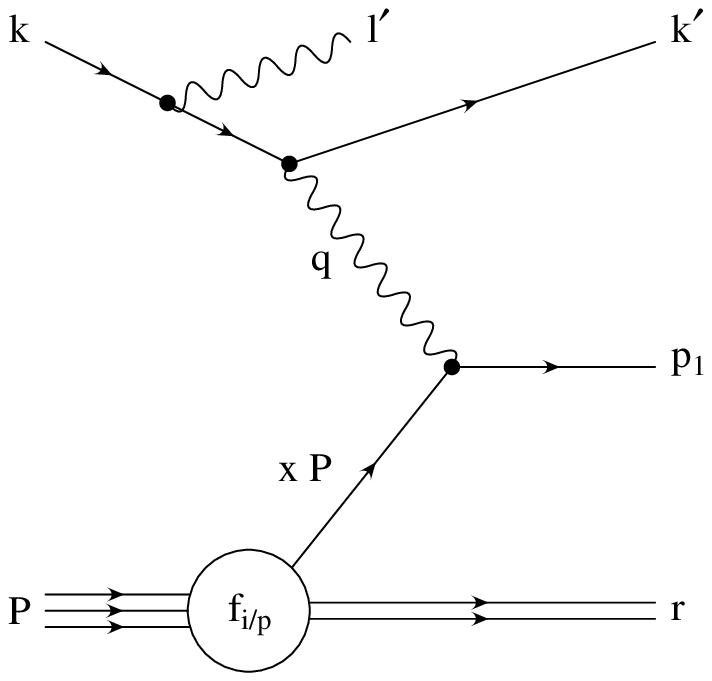}\hfill%
  \includegraphics{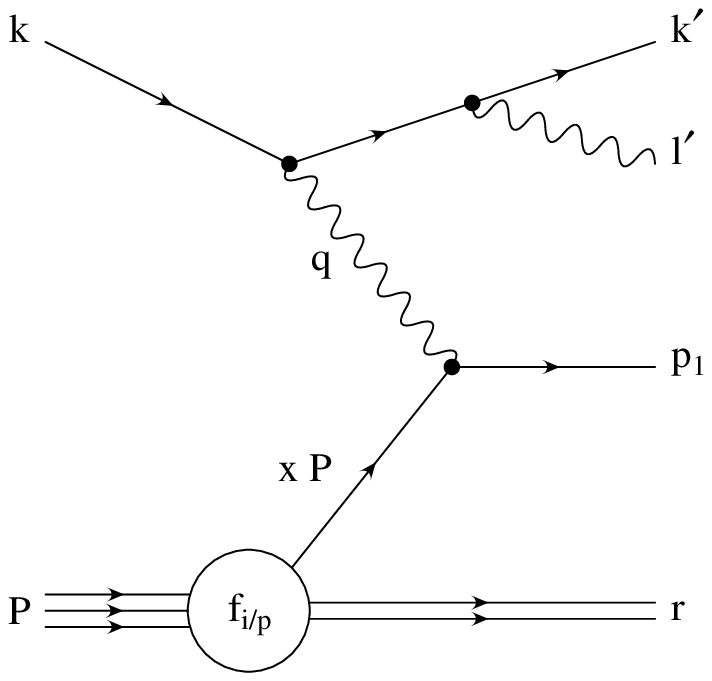}
  \caption[$\order(\alpha)$ real corrections to the QPM Feynman graph.]
  {$\order(\alpha)$ real corrections to the QPM Feynman graph
    fig.~\ref{fig:QPM}: ISR (left) and FSR (right) on the lepton side.}
  \label{fig:QPMrad}
\end{figure}

Apart from $\order(\as)$ QCD corrections to the Born cross
section~(\ref{eqn:Born}), also QED corrections of $\order(\alpha)$ may be
sizable depending on the phase space. This is especially true for high $y$ and
low $x$. Following the singularities proportional to $1/(k\cdot l')$,
$1/(k'\cdot l')$ and $1/(k-k'-l')^2$ that appear in the calculation of the
real diagrams, they can instructively be labelled {\bf I}nitial {\bf S}tate
{\bf R}adiation (ISR), {\bf F}inal {\bf S}tate {\bf R}adiation (FSR) and
Compton contribution, where the latter plays only a minor role.
Fig.~\ref{fig:QPMrad} depicts the first two of them where the photons are
radiated predominantly collinear to the incoming respectively outgoing lepton.
Nevertheless, all three parts are defined in the whole phase space such that
e.g.\ \qi ISR\qo\ photons may also be directed along the scattered lepton!  In
addition, this simple picture is only valid in LLA; beyond, the separation is
not unique.

The event generator DJANGO~\cite{MC:DJANGO} combines the abilities of LEPTO
and HERACLES~\cite{MC:HERACLES}, which provides the QED radiative effects
including virtual corrections due to 1-loop diagrams, into one software
package offering the most complete description of $ep$ DIS events available.

\subsection{Hadron Level}
\label{sec:simhad}

Neglecting leptons and photons at this stage, we are left with partons of low
virtualities \mbox{$t \approx t_0$}, where perturbation theory ceases to be
applicable.  Lacking better theoretical means, phenomenological models have to
be invoked to perform the necessary fragmentation of partons into hadrons.
Since the PS cut-off $t_0$ is an arbitrary parameter, the hadronization model
should employ a similar scale from which to start in such a way that the
dependence on $t_0$ largely cancels between the PS and the fragmentation
process.

The two most popular models existing for this purpose are the {\em string}\/
and the {\em cluster}\/ fragmentation, both of which also consider colour
coherence effects.  The first one is implemented in JETSET~\cite{MC:JETSET},
which is applied for this task by the MC generators of the last section except
for HERWIG, which makes use of the second approach.  Taking into account
decays of unstable particles as well, the final outcome of this step consists
of all particles traversing through a real detector.

\subsection{Detector Simulation}
\label{sec:simdet}

So far the involved processes were of a basically theoretical nature and could
be described by general purpose MC programs.  Since now the detector response
to the passing particles has to be reproduced, software specifically adapted
to the measuring device is required. The software package employed by H1 is
called H1SIM~\cite{H1SIM}.

It is responsible for tracking the MC particles through a virtual H1 detector
and simulating the response signals in detail. The \qi events\qo\ produced
that way look like real data.  For comparison purposes it is desirable to have
as many MC events as possible, at least about the same amount as data are
available.  However, the simulation step is very time consuming and takes
several seconds of computing time per event on the computer systems at
disposal.  Therefore, the MC models to use have to be chosen carefully.

\section{Reconstruction}
\label{sec:reco}
\subsection{Cluster Level}
\label{sec:reccls}

Because the \qi raw\qo\ information (real and simulated) in the form of wire
hits, cell voltages, etc.\ is not very intuitive with respect to the physics
of $ep$ scattering, the data have to be refined.  This {\em reconstruction}\/
process is the task of another H1 software program, H1REC~\cite{H1REC}, which
has to be identical for real and simulated events.  As a result, it provides
i.a.\ particle tracks and the calorimeter clusters extensively used in this
study. For that reason, this stage is called \qi cluster level.\qo

\section{Comparison to Data}
\label{sec:compdat}

Once the simulation and reconstruction have been completed, the MC models must
be confronted with data.  Ideally, they should give a good account of both,
\qi standard\qo\ distributions where selection cuts are applied (the energy
spectrum of the scattered electron, $dn/dE_{e'}$, for example), as well as
event shape distributions $dn/dF$ that are of special interest here.

In the previous publication~\cite{H1:Shapes}, LEPTO~6.1 was shown to give the
best description of the data. This is still true. However, it does not contain
radiative corrections. Therefore, DJANGO~6.2 together with ARIADNE has been
chosen as a replacement. In the following, the term \qi DJANGO~6.2\qo\ {\bf
  always} refers to this combination!

For testing purposes a newer LEPTO version, LEPTO~6.5, and HERWIG~5.8 have
been used. The basic differences to the old LEPTO version are changed default
settings, a~new cut-off scheme for the divergences of the matrix elements, an
improved target remnant treatment and the introduction of soft colour
interactions to facilitate the production of diffractive events within DIS
samples. As will be seen, this has a considerable impact on the event shape
distributions.  Statistics of the employed MC files are given in
table~\ref{tab:MCstat}.  In case of DJANGO~6.2 \highq, it should be noted that
an extra very \highq\ MC file was produced to increase the statistics in the
$Q$-bins seven and eight.  When combined with the other \highq\ data sets, it
would lead to unnatural steps in some of the distributions presented in the
next section.  Therefore, it was excluded there.

\begin{table}
  \centering
  \begin{tabular}{|c||r|r|r|r|}
    \hline
    \multicolumn{1}{|c||}{\rbthr MC model} &
    \multicolumn{1}{c|}{DJANGO~6.2} &
    \multicolumn{1}{c|}{LEPTO~6.1} &
    \multicolumn{1}{c|}{LEPTO~6.5} &
    \multicolumn{1}{c|}{HERWIG~5.8}\\\hline\hline
    \lowq\ statistics &   $4446$ & $4253$ &
    \multicolumn{1}{c|}{$-$} & \multicolumn{1}{c|}{$-$} \rbtrr\\\hline
    \highq\ statistics & $121483$ & $8906$ & $84561$ & $12617$ \rbtrr\\\hline
  \end{tabular}
  \caption{Cluster level MC statistics.}
  \label{tab:MCstat}
\end{table}

\subsection{Standard Distributions}
\label{sec:compstand}

\begin{figure} 
  \centering
  \includegraphics{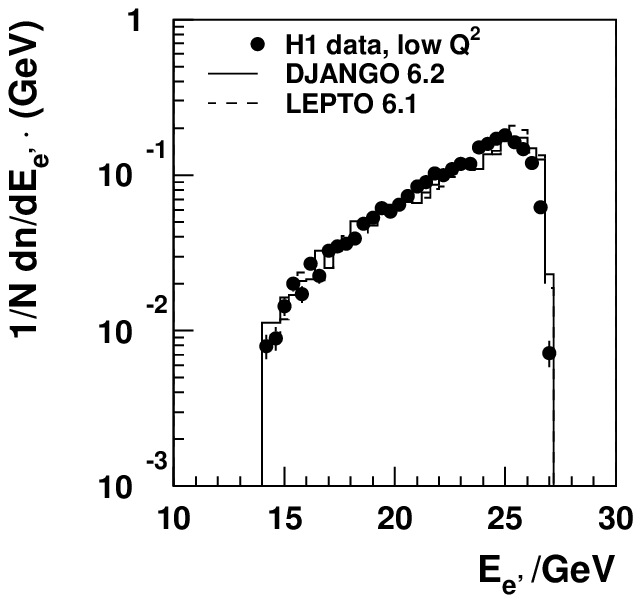}\hftwo%
  \includegraphics{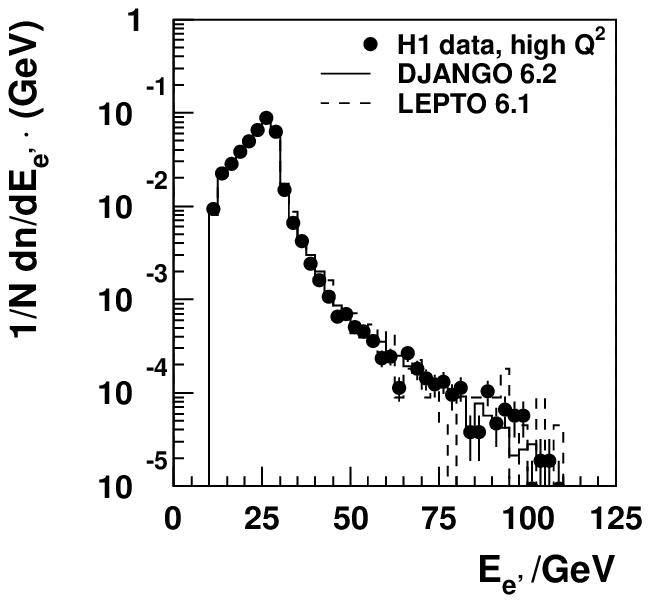}
  \includegraphics{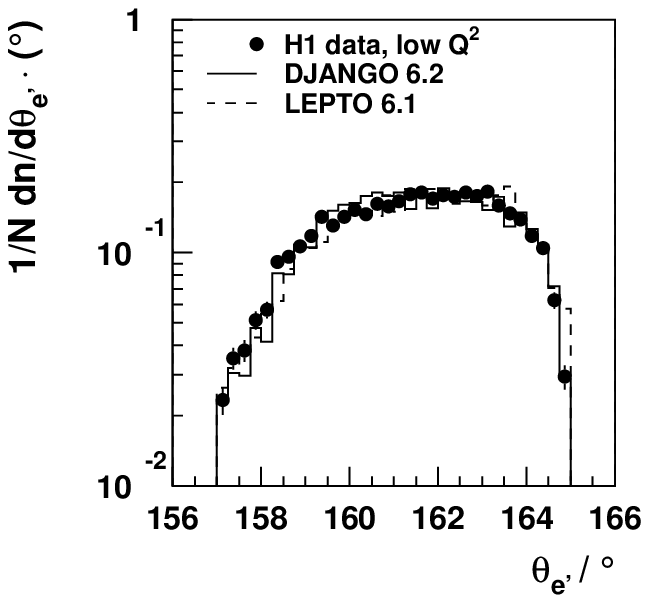}\hftwo%
  \includegraphics{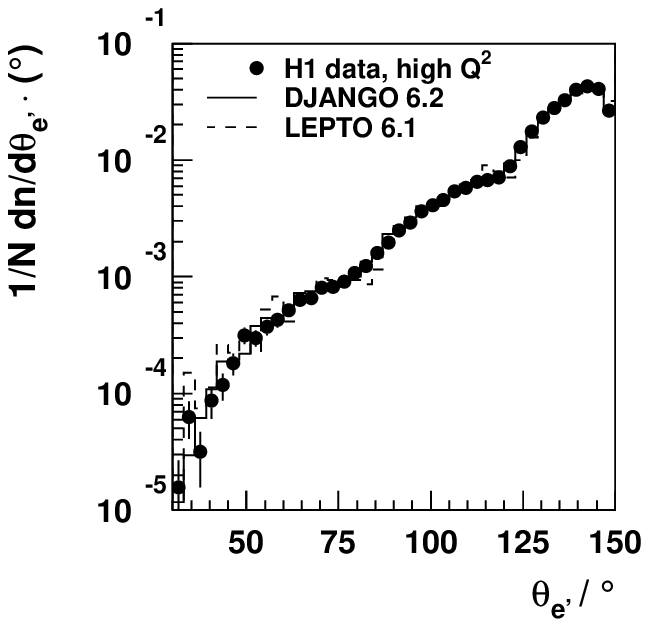}
  \includegraphics{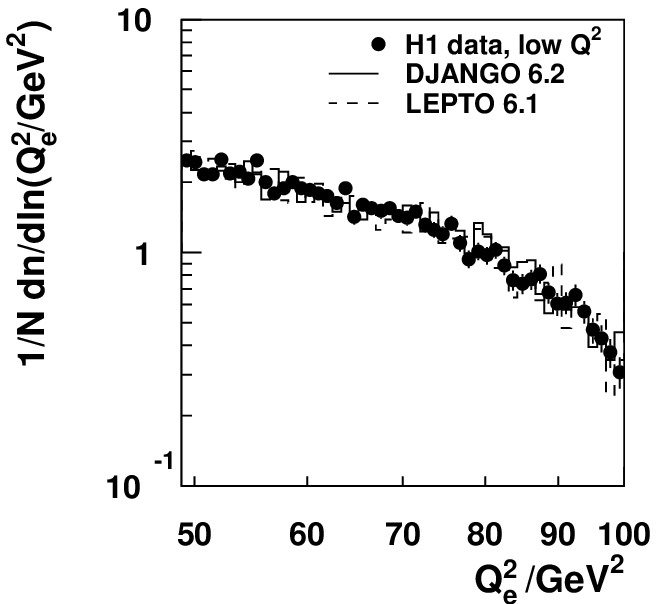}\hftwo%
  \includegraphics{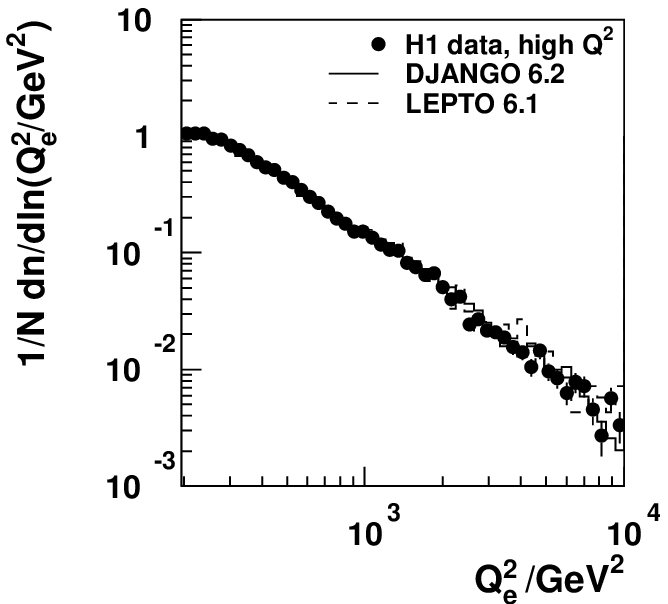}
  \caption[Normalized differential distributions of $E_{e'}$, $\theta_{e'}$
  and $\ln (Q_e^2/\gevq)$ on cluster level.]  {Normalized differential
    distributions of $E_{e'}$, $\theta_{e'}$ and $\ln (Q_e^2/\gevq)$ on
    cluster level for the \lowq\ (left) and \highq\ (right) data sample (full
    symbols) in comparison with DJANGO~6.2 (full line) and LEPTO~6.1 (dashed
    line). The error bars represent statistical uncertainties only.}
  \label{fig:dndstdcl1}
\end{figure}

\begin{figure} 
  \centering
  \includegraphics{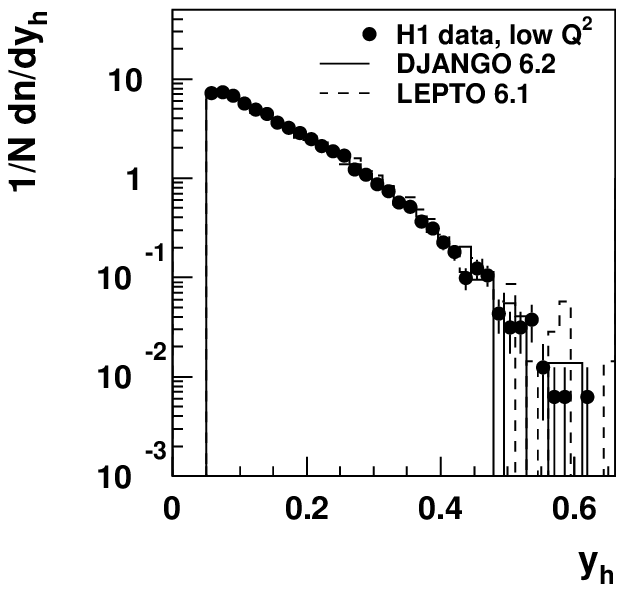}\hftwo%
  \includegraphics{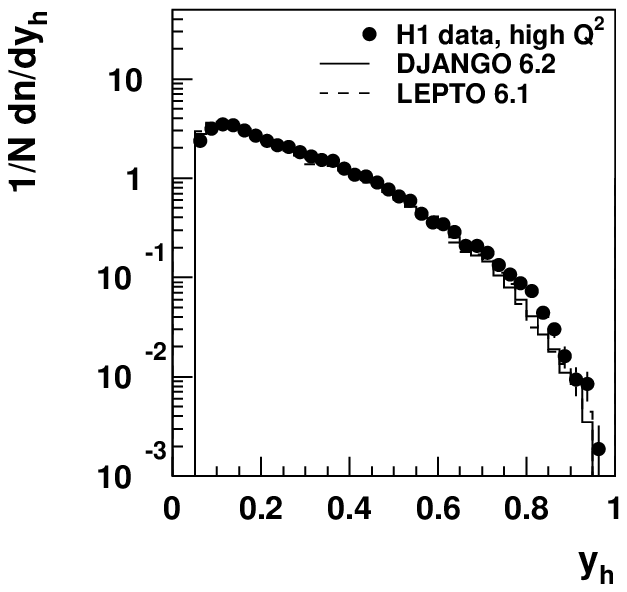}
  \includegraphics{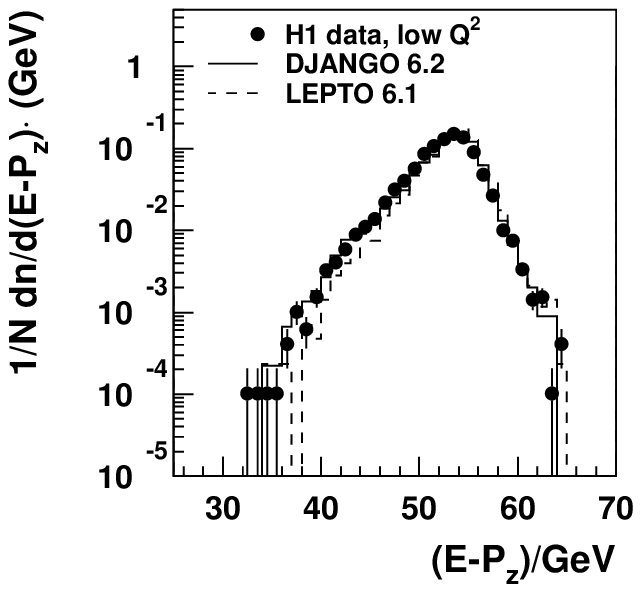}\hftwo%
  \includegraphics{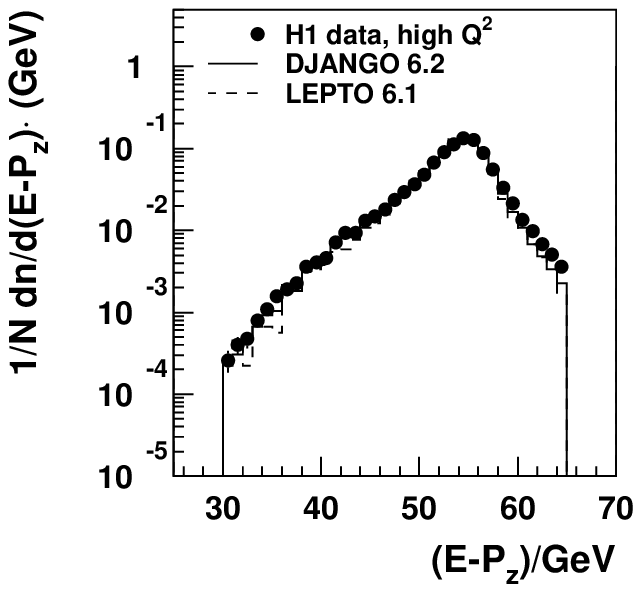}
  \includegraphics{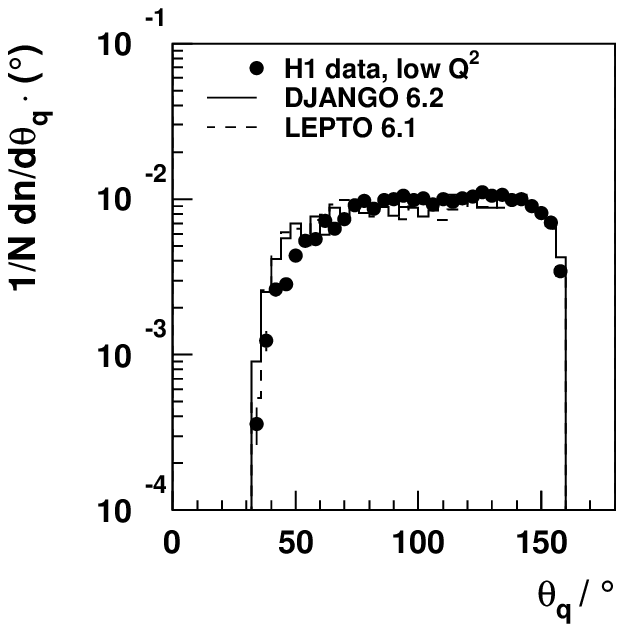}\hftwo%
  \includegraphics{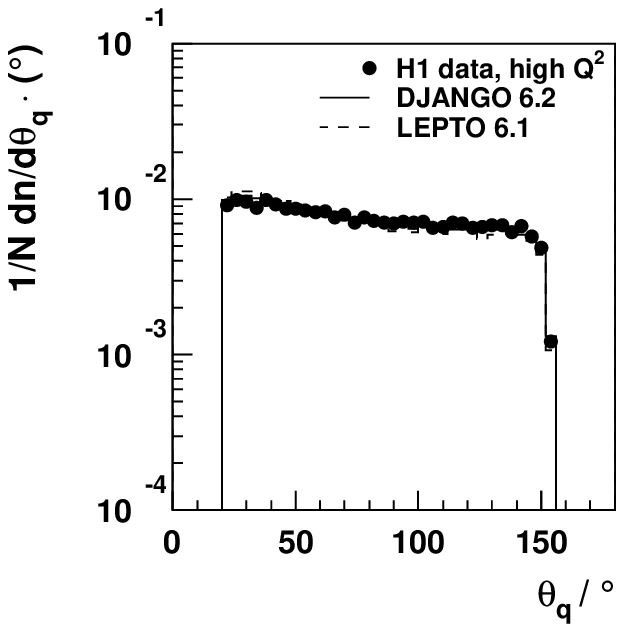}
  \caption[Normalized differential distributions of $y_h$, $(E-P_z)$
  and $\theta_q$ on cluster level.]  {Normalized differential distributions of
    $y_h$, $(E-P_z)$ and $\theta_q$ on cluster level for the \lowq\ (left) and
    \highq\ (right) data sample (full symbols) in comparison with DJANGO~6.2
    (full line) and LEPTO~6.1 (dashed line). The error bars represent
    statistical uncertainties only.}
  \label{fig:dndstdcl2}
\end{figure}

In figs.~\ref{fig:dndstdcl1} and~\ref{fig:dndstdcl2} the normalized
differential distributions of $E_{e'}$, $\theta_{e'}$, $\ln
(Q_e^2/\gevq)$, $y_h$, $(E-P_z)$ and $\theta_q$ are presented separately
for the \lowq\ and \highq\ sample. The comparison with DJANGO~6.2 as
well as LEPTO~6.1 reveals a good overall agreement with the exception
of low $\theta_q$-values in case of \lowq\ events.  Yet, the imposed
cut-off no.~\ref{cut:thetaq} of $20\grad < \theta_q$ is well below the
smallest occurring angle of $\theta_{q,{\rm min}}\approx 30\grad$,
rendering the deviation harmless.

The dips in the \highq\ $\theta_{e'}$-distribution stem from the
rejection of events with the scattered lepton found in areas of the
LAr where the energy measurement deteriorates (cut
no.~\ref{cut:crack}).

LEPTO~6.5 and HERWIG~5.8 are not shown here. Their agreement with data
is similar.

\subsection{Event Shape Distributions}
\label{sec:compshapes}

\begin{figure} 
  \centering
  \includegraphics{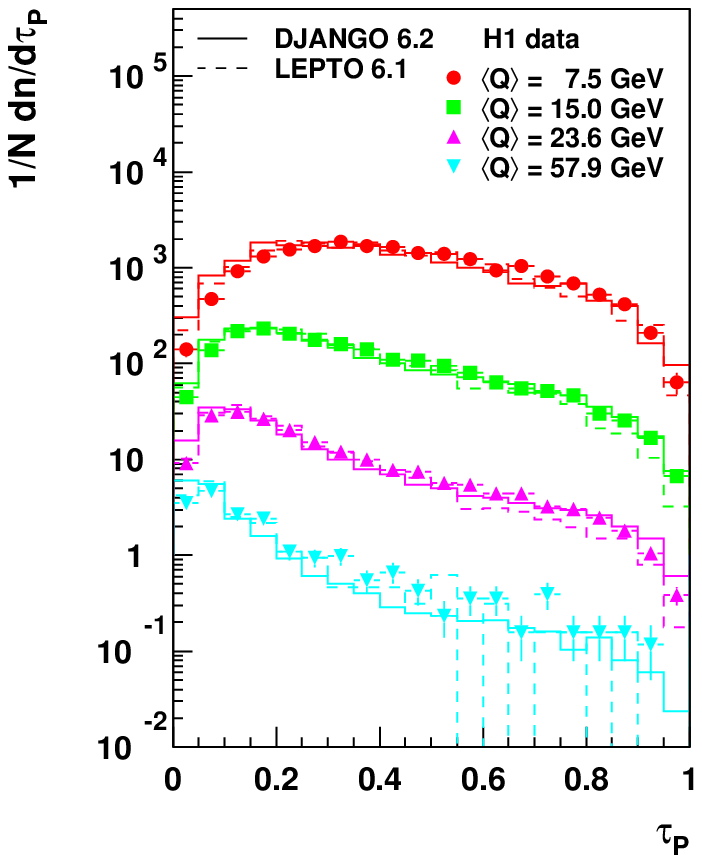}\hftwo%
  \includegraphics{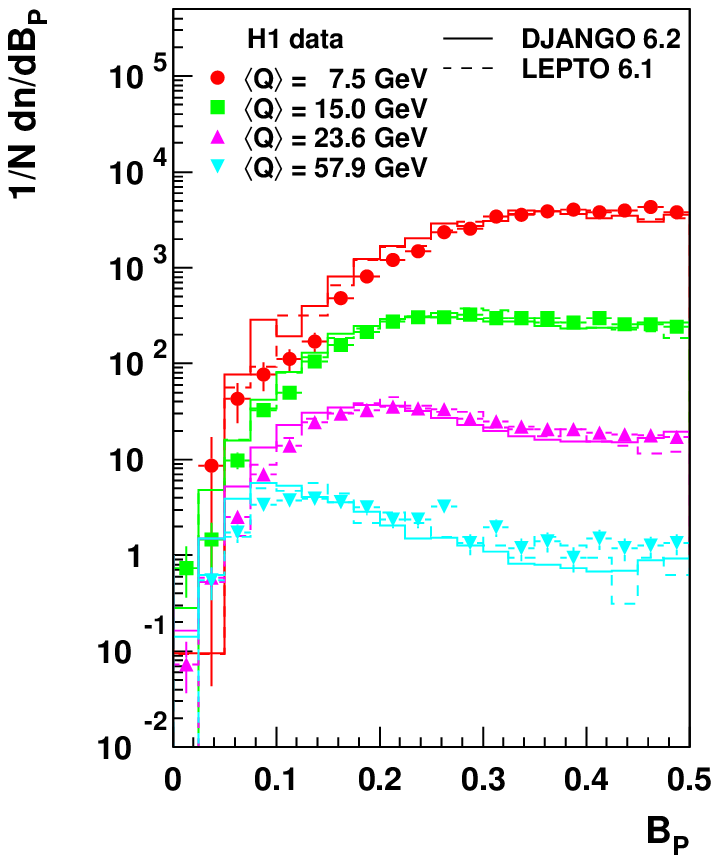}
  \includegraphics{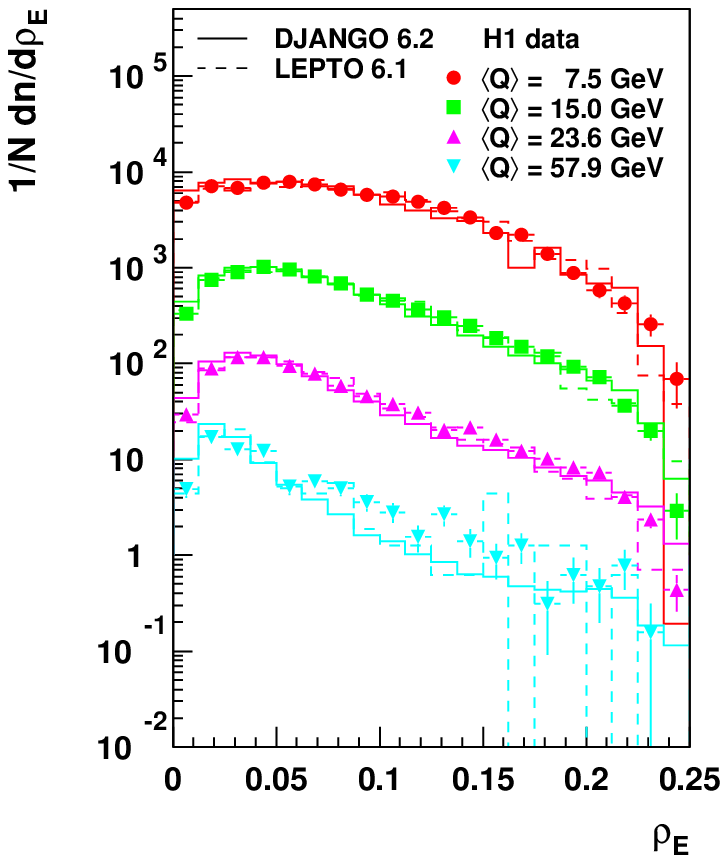}\hftwo%
  \includegraphics{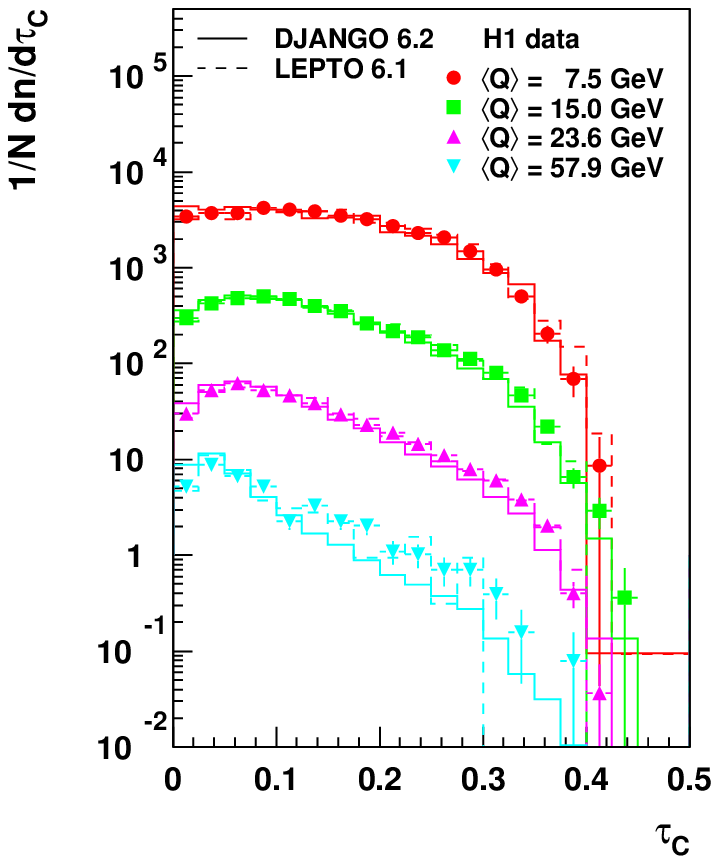}
  \caption[Normalized differential distributions of
  $\tau_P$, $B_P$, $\rho_E$ and $\tau_C$ on cluster level.]  {Normalized
    differential distributions of the event shapes $\tau_P$, $B_P$, $\rho_E$
    and $\tau_C$ on cluster level.  H1 data (full symbols) are compared with
    DJANGO~6.2 (full lines) and LEPTO~6.1 (dashed lines) for four out of eight
    investigated bins in $Q$. The spectra for $\mean{Q} = 7.5$--$57.9\gev$ are
    multiplied by factors of $10^n$, $n=0,1,2,3$.  The error bars represent
    statistical uncertainties only.}
  \label{fig:dndFcl1}
\end{figure}

\begin{figure} 
  \centering \includegraphics{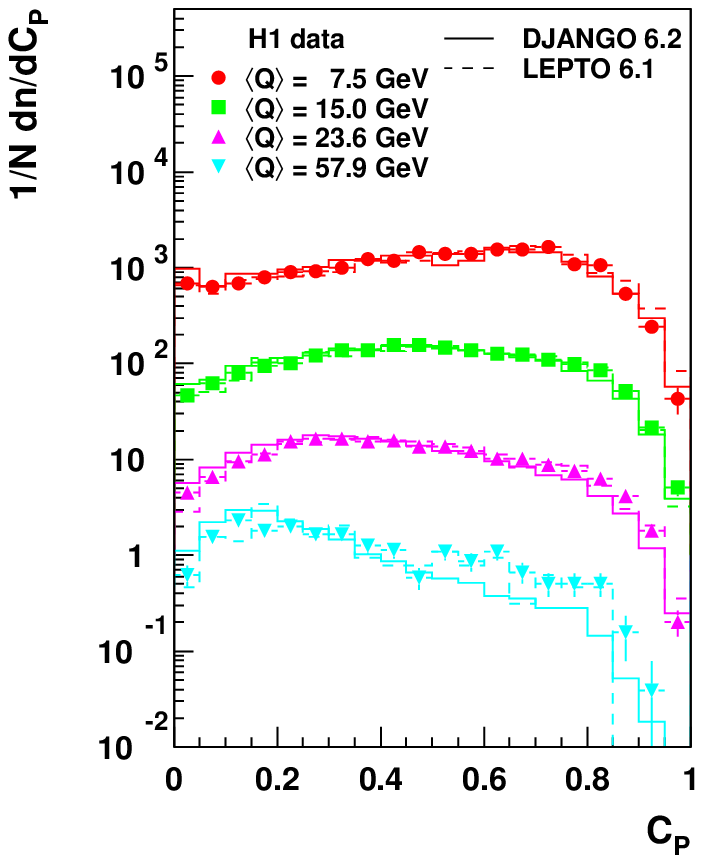}
  \includegraphics{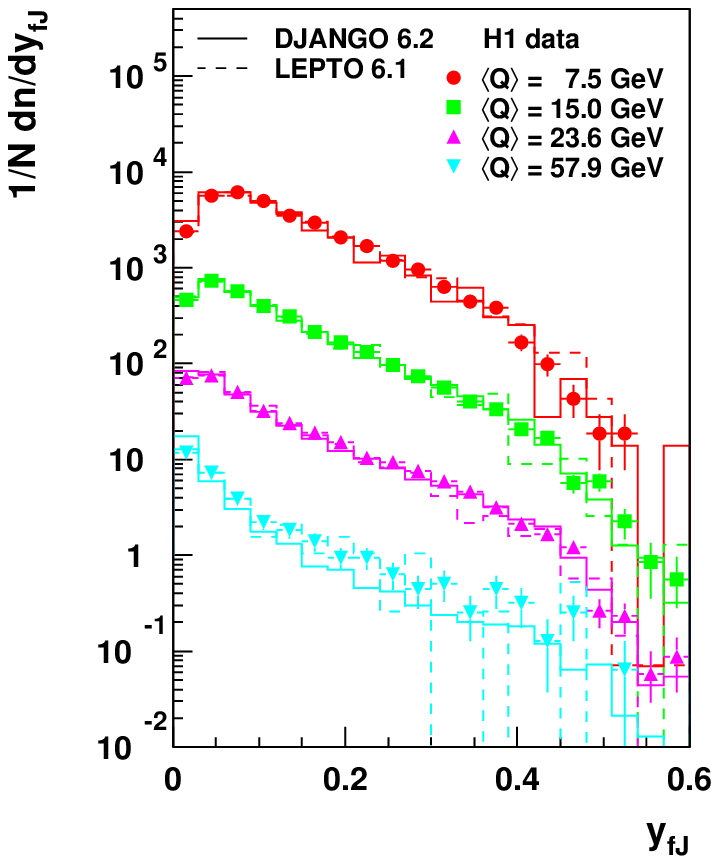}\hftwo%
  \includegraphics{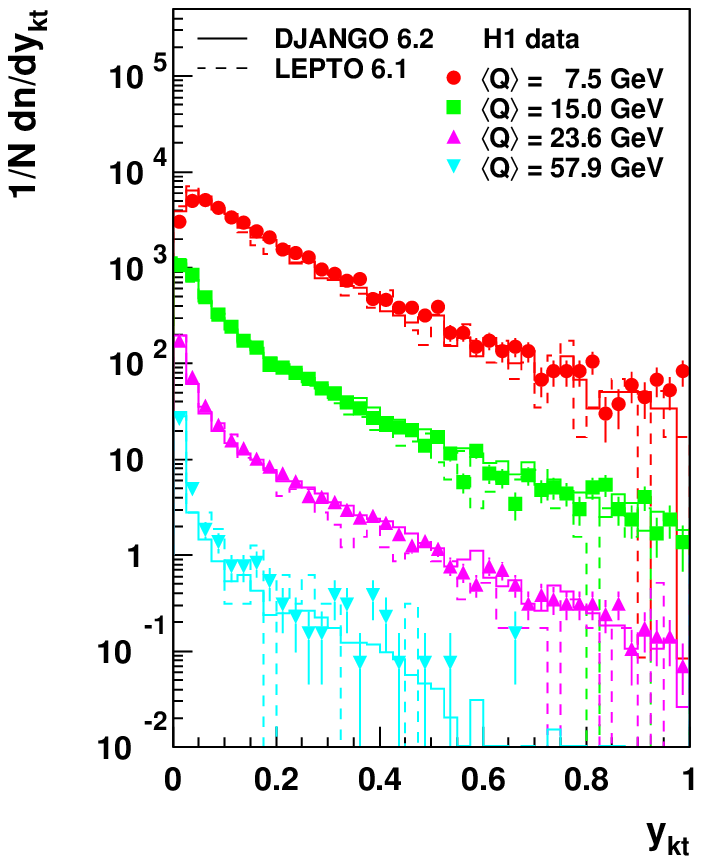}
  \caption[Normalized differential distributions of
  $C_P$, $y_{fJ}$ and $y_{k_t}$ on cluster level.]  {Normalized differential
    distributions of the event shapes $C_P$, $y_{fJ}$ and $y_{k_t}$ on cluster
    level.  H1 data (full symbols) are compared with DJANGO~6.2 (full lines)
    and LEPTO~6.1 (dashed lines) for four out of eight investigated bins in
    $Q$. The spectra for $\mean{Q} = 7.5$--$57.9\gev$ are multiplied by
    factors of $10^n$, $n=0,1,2,3$.  The error bars represent statistical
    uncertainties only.}
  \label{fig:dndFcl2}
\end{figure}

The next step is to check the description of the event shape distributions.
They are shown for four out of eight investigated bins in $Q$ in
figs.~\ref{fig:dndFcl1} and~\ref{fig:dndFcl2}.  Here, the similarity is not as
satisfactory as before.  DJANGO~6.2 systematically tends to overshoot the data
in the low $F$-region which is compensated for by underestimating them for
high values in $F$.  This is worst for $\tau_P$ and especially $B_P$, whose
mean values come out too low in the MC\@.  The uncorrected data means $\fmean$
are given in table~\ref{tab:rawmeans}.

Within the much lower statistics available for LEPTO~6.1, it seems to do a
better job in reproducing the data distributions.  Nevertheless, DJANGO~6.2 in
combination with ARIADNE provides an acceptable account of the available data
and, since it includes radiative corrections, is employed for the primary
correction procedure discussed in the next chapter.

\begin{figure}[t]
  \centering
  \includegraphics{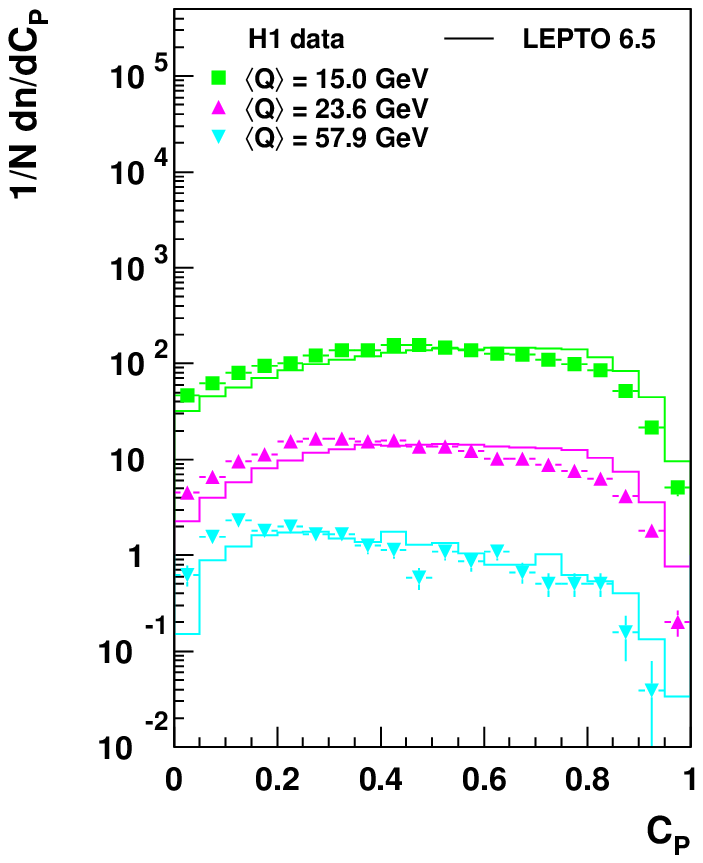}\hftwo%
  \includegraphics{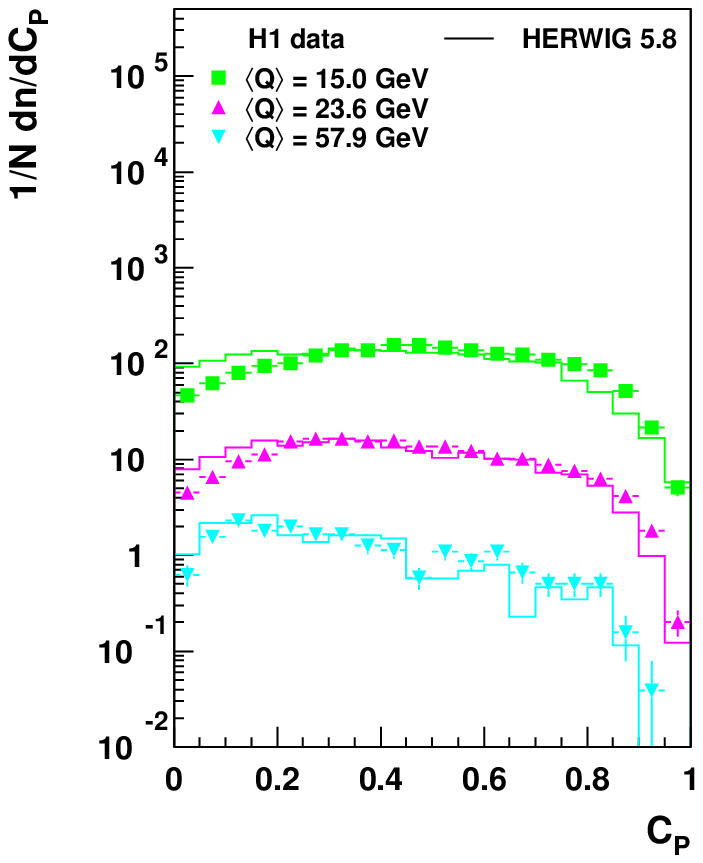}
  \caption[Normalized differential distributions of $C_P$
  on cluster level compared with LEPTO~6.5 and HERWIG~5.8.]  {Normalized
    differential distributions of $C_P$ on cluster level. H1 data (full
    symbols) are compared with LEPTO~6.5 (full lines) on the left-hand side
    and HERWIG~5.8 (full lines) on the right-hand side for four out of eight
    investigated bins in $Q$. The spectra for $\mean{Q} = 7.5$--$57.9\gev$ are
    multiplied by factors of $10^n$, $n=0,1,2,3$.  The error bars represent
    statistical uncertainties only.}
  \label{fig:dndCPcl}
\end{figure}

As an example of a rather bad performance, fig.~\ref{fig:dndCPcl} presents the
normalized differential distributions of $C_P$ for three bins of the \highq\ 
sample in comparison with LEPTO~6.5 on the left-hand side and HERWIG~5.8 on
the right-hand side.  Severe deviations contrary to each other are observed.

\begin{table}
  \centering
  \hftwo\begin{tabular}{|c||l|l|}
    \hline
    \multicolumn{1}{|c||}{\rbthm$\mean{Q}/\gev$}&
    \multicolumn{1}{|c|}{$\mean{\tau_P}$}&
    \multicolumn{1}{|c|}{$\mean{B_P}$}
    \\\hline\hline
    $7.46$ &
    $0.4351\pm 0.0031$&
    $0.3651\pm 0.0013$
    \rbtrr\\\hline 
    $8.74$ &
    $0.4110\pm 0.0034$&
    $0.3543\pm 0.0014$
    \rbtrr\\\hline
    $14.97$&
    $0.3456\pm 0.0021$&
    $0.3186\pm 0.0010$
    \rbtrr\\\hline
    $17.75$&
    $0.3203\pm 0.0018$&
    $0.3039\pm 0.0009$
    \rbtrr\\\hline
    $23.62$&
    $0.2782\pm 0.0020$&
    $0.2790\pm 0.0010$
    \rbtrr\\\hline
    $36.72$&
    $0.2367\pm 0.0034$&
    $0.2502\pm 0.0019$
    \rbtrr\\\hline
    $57.93$&
    $0.2079\pm 0.0090$&
    $0.2271\pm 0.0053$
    \rbtrr\\\hline
    $81.32$&
    $0.1598\pm 0.0134$&
    $0.1891\pm 0.0082$\rbtrr\\\hline
  \end{tabular}\hftwo\vspace{0.5cm}
  \hftwo\begin{tabular}{|c||l|l|l|}
    \hline
    \multicolumn{1}{|c||}{\rbthm$\mean{Q}/\gev$}&
    \multicolumn{1}{|c|}{$\mean{\rho_E}$}&
    \multicolumn{1}{|c|}{$\mean{\tau_C}$}&
    \multicolumn{1}{|c|}{$\mean{C_P}$}
    \\\hline\hline
    $7.46$ &
    $0.0797\pm 0.0007$&
    $0.1401\pm 0.0013$&
    $0.4960\pm 0.0034$
    \rbtrr\\\hline 
    $8.74$ &
    $0.0805\pm 0.0008$&
    $0.1429\pm 0.0014$&
    $0.5048\pm 0.0038$
    \rbtrr\\\hline
    $14.97$&
    $0.0737\pm 0.0005$&
    $0.1297\pm 0.0008$&
    $0.4701\pm 0.0022$
    \rbtrr\\\hline
    $17.75$&
    $0.0725\pm 0.0004$&
    $0.1267\pm 0.0007$&
    $0.4603\pm 0.0019$
    \rbtrr\\\hline
    $23.62$&
    $0.0685\pm 0.0004$&
    $0.1179\pm 0.0008$&
    $0.4324\pm 0.0021$
    \rbtrr\\\hline
    $36.72$&
    $0.0626\pm 0.0008$&
    $0.1041\pm 0.0013$&
    $0.3847\pm 0.0036$
    \rbtrr\\\hline
    $57.93$&
    $0.0578\pm 0.0021$&
    $0.0950\pm 0.0033$&
    $0.3495\pm 0.0098$
    \rbtrr\\\hline
    $81.32$&
    $0.0482\pm 0.0030$&
    $0.0804\pm 0.0050$&
    $0.3041\pm 0.0147$\rbtrr\\\hline
  \end{tabular}\hftwo\vspace{0.5cm}
  \hftwo\begin{tabular}{|c||l|l|}
    \hline
    \multicolumn{1}{|c||}{\rbthm$\mean{Q}/\gev$}&
    \multicolumn{1}{|c|}{$\mean{y_{fJ}}$}&
    \multicolumn{1}{|c|}{$\mean{y_{k_t}}$}
    \\\hline\hline
    $7.46$ &
    $0.1273\pm 0.0012$&
    $0.1746\pm 0.0023$
    \rbtrr\\\hline 
    $8.74$ &
    $0.1274\pm 0.0013$&
    $0.1471\pm 0.0023$
    \rbtrr\\\hline
    $14.97$&
    $0.1131\pm 0.0009$&
    $0.1077\pm 0.0013$
    \rbtrr\\\hline
    $17.75$&
    $0.1115\pm 0.0008$&
    $0.0975\pm 0.0011$
    \rbtrr\\\hline
    $23.62$&
    $0.1047\pm 0.0009$&
    $0.0826\pm 0.0012$
    \rbtrr\\\hline
    $36.72$&
    $0.0911\pm 0.0015$&
    $0.0603\pm 0.0018$
    \rbtrr\\\hline
    $57.93$&
    $0.0874\pm 0.0044$&
    $0.0483\pm 0.0040$
    \rbtrr\\\hline
    $81.32$&
    $0.0681\pm 0.0060$&
    $0.0341\pm 0.0050$\rbtrr\\\hline 
  \end{tabular}\hftwo
  \caption[Uncorrected mean values of the event shapes as a function of $Q$.]
  {Uncorrected mean values of the event shapes and their
    statistical uncertainty as a function of $Q$.}
  \label{tab:rawmeans}
\end{table}


\chapter{Correction Procedure}
\label{chap:corrproc}

After having established a sufficient description of data by MC models, the
most important task remaining is disentangling the underlying physics from
mere detector effects due to limited efficiencies and resolutions.
Additionally, radiative corrections are taken into account.

In principle, it is feasible to do this in a one-step procedure.  Compensating
effects, however, may mislead to the conclusion that the corrections are
small, even if indeed they are not. Moreover, one would like to identify the
dominant influence. Another disadvantage is the complete neglect of migrations
enforced by the lack of correlations.

For those reasons, the procedure applied here tries to differentiate between
these contributions, and the unfolding of the data is correspondingly
performed in two stages.  The different correction methods available are
explained in the next section.

\section{Unfolding Methods}
\label{sec:unf}
\subsection{Factor Method}
\label{sec:unffactor}

When only mean values do matter, the easiest thing to think of is to invoke
one correction factor for each mean:
\begin{equation}
  \label{eqn:factor}
  \fmean_{\rm corr} = c_{\rm had/sim}\cdot\fmean_{\rm dat}\,,\qquad\qquad
  c_{\rm had/sim} :=
  \frac{\fmean_{\rm had}^{\rm MC}}{\fmean_{\rm sim}^{\rm MC}}\,.
\end{equation}

\subsection{Bin-to-bin Correction}
\label{sec:unfbtb}

By applying eq.~(\ref{eqn:factor}) not to the mean values alone, but to each
bin $i$ of a differential distribution $F_i$,
\begin{equation}
  \label{eqn:bintobin}
  F_{i,{\rm corr}} = c_{i,{\rm had/sim}}\cdot F_{i,{\rm dat}}\,,\qquad\qquad
  c_{i,{\rm had/sim}}
  := \frac{\frac{1}{N_{\rm had}^{\rm MC}}F_{i,{\rm had}}^{\rm MC}}
  {\frac{1}{N_{\rm sim}^{\rm MC}}F_{i,{\rm sim}}^{\rm MC}}\,,
\end{equation}
the factor method can easily be extended to unfold complete distributions
which again can be reevaluated to give a corrected mean $\fmean_{\rm corr}$.

As can be seen from eq.~(\ref{eqn:bintobin}), however, this simple approach
completely ignores the possibility of migrations from one bin $i$ on hadron
level to another bin $j$ on cluster level. Depending on the quantity to deal
with, this constitutes a severe disadvantage. A measure to improve the
situation is to choose bin sizes in such a way that migrations are minimized.

\subsection{Matrix Method}
\label{sec:unfmatrix}

It is still better to take these migrations explicitly into consideration by
employing a correction matrix $C_{ij}$:
\begin{equation}
  F_{i,{\rm corr}} = \sum\limits_j C_{ij}F_{j,{\rm dat}}\,.
\end{equation}
The matrix $C_{ij}$ can in principle be obtained from MC by inverting the
transfer matrix $T_{ij}$ which transforms the \qi true\qo\ values $F_{i,{\rm
    had}}^{\rm MC}$ into the observed ones
\begin{equation}
  F_{i,{\rm sim}}^{\rm MC} = \sum\limits_j T_{ij}F_{j,{\rm had}}^{\rm MC}\,.
\end{equation}
In practice, however, the inversion leads to instabilities and oscillations
unless extremely large MC statistics is available.  Instead, we follow the
strategy employed in~\cite{H1:Shapes} to define the correction matrix to be
\begin{equation}
  C_{ij} := \frac{\rho_{ij}}{\sum\limits_k \rho_{kj}}
\end{equation}
where $\rho_{ij}$ represents a probability density derived from MC
correlations as
\begin{equation}
  F_{i,{\rm had}}^{\rm MC} = \sum\limits_j \rho_{ij}F_{j,{\rm sim}}^{\rm MC}\,.
\end{equation}

As a drawback of this scheme, there may remain some model dependence which can
be overcome by an iterative procedure.

\subsection{Bayesian Unfolding}
\label{sec:unfBayes}

The last method employs a program developed in~\cite{Bayes}.  It exploits
Bayes' theorem on conditional probabilities to extract information from the
observed distributions.  To state it in a more general form, we label the fact
of an event to have a \qi true\qo\ value $F_{\rm corr}$ allocated to bin $i$
as \qi cause\qo\ $C_i$ and the observed value $F_{\rm dat}$ in bin $j$ as \qi
effect\qo\ $E_j$.  Since in general the binning and also the {\bf domain of
  definition} may be different, we have a number of $N_C$ causes and $N_E$
effects.  Denoting further the initial probability for cause $C_i$ to happen
with $P_0(C_i)$ and the conditional probability for the effect $E_j$ to occur
if $C_i$ has already taken place with $P(E_j|C_i)$, Bayes' theorem can be
written as
\begin{equation}
  \label{eqn:Bayes}
  P(C_i|E_j) = \frac{P(E_j|C_i)\cdot P_0(C_i)}
  {\sum\limits_{k=1}^{N_C} P(E_j|C_k)\cdot P_0(C_k)}\,.
\end{equation}
That is, given an observation of $E_j$, we get a probability that it was
produced by cause~$C_i$.

The result, however, depends on two ingredients: the conditional probabilities
$P(E_j|C_i)$, which can be estimated from MC, and the initial probabilities
$P_0(C_i)$, that are {\em a priori}\/ unknown.  At first, the latter have to
be assumed by the observer according to his prejudices which may even result
in a uniform distribution in the case of complete ignorance.  But, given a
number of experimental observations with frequencies $N(E) :=
\{N(E_1),\ldots,N(E_{N_E})\}$, Bayes' theorem can then be applied to gain
information on $P_0(C_i)$.  Based on these observed event numbers for the
effects $E_j$, the expected number of events to be assigned to the different
causes $C_i$ can be inferred to be
\begin{equation}
  \hat{N}(C_i) = \sum\limits_{j=1}^{N_E} P(C_i|E_j)\cdot N(E_j)\,.
\end{equation}
Ignoring the possibility of inefficiencies, i.e.\ the case when a cause does
not produce one of the effects under consideration, the initial probability is
estimated by
\begin{equation}
  \hat{P}(C_i) := P(C_i|N(E)) = \frac{\hat{N}(C_i)}
  {\sum\limits_{i=1}^{N_C}\hat{N}(C_i)}\,.
\end{equation}
If the initial distribution $P_0(C_i)$ is not consistent with the data, it
will in general differ from the final one $\hat{P}(C_i)$.  In an iterative
procedure where the assumption on $P_0(C_i)$ is updated in each step, the
observed events $N(E)$ can now be exploited to unfold the \qi true\qo\ 
distribution $\hat{N}(C)$.

\section{Correction for Detector Effects}
\label{sec:corrdet}

The first step consists in deriving a correction to hadron level from the
simulated DJANGO~6.2 MC files. Here, special care has to be taken when
specifying what is meant by \qi hadron level\qo\ since there may be radiative
photons. But we are interested in detector effects only.  Thus, the main
influence of these photons has to be accounted for in a later step.  This is
accomplished by first enforcing a limitation in acceptance according to cut
no.~\ref{cut:acc}, thereby throwing away radiation collinear to the incoming
lepton. Furthermore, with regard to the finite angular resolution of the LAr
calorimeter, the scattered electron is redefined to be merged with radiative
photons if they are closer in angle than $5\grad$. This value is motivated by
a study in~\cite{Wobisch}.  As a consequence, the cuts
nos.~\ref{cut:EPz}--\ref{cut:Eda} may reject events with respect to this \qi
radiative\qo\ hadron level.

Now one is free to draw a connection between the cluster and hadron level of
the simulation and apply this to the data according to the four unfolding
schemes described.  Note that the selection criteria
nos.~\ref{cut:zv}--\ref{cut:crack} are imposed on the cluster level only such
that they are accounted for in this first step.

\begin{figure} 
  \centering
  \includegraphics{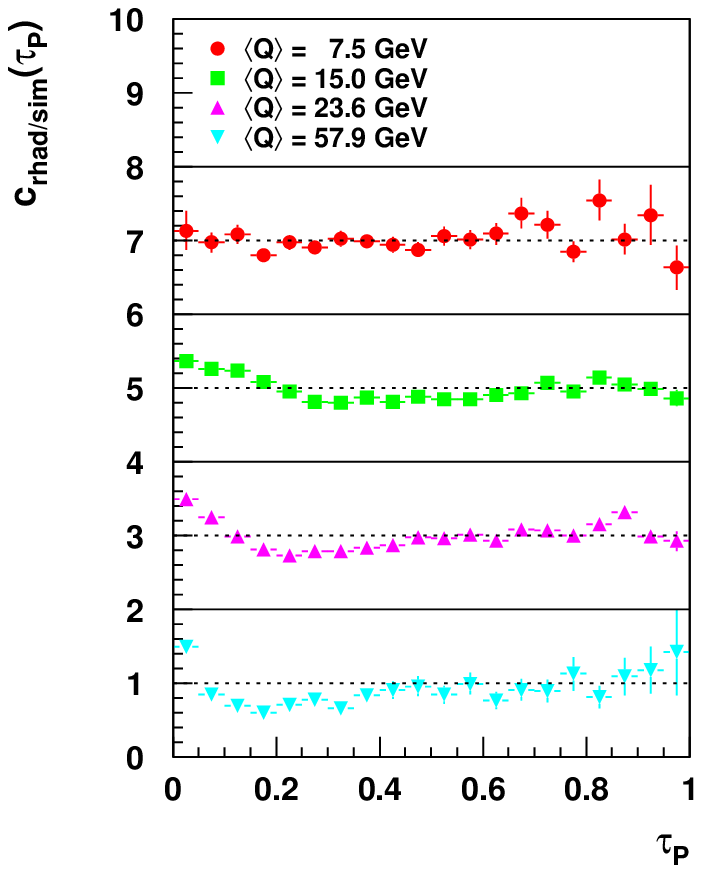}\hftwo%
  \includegraphics{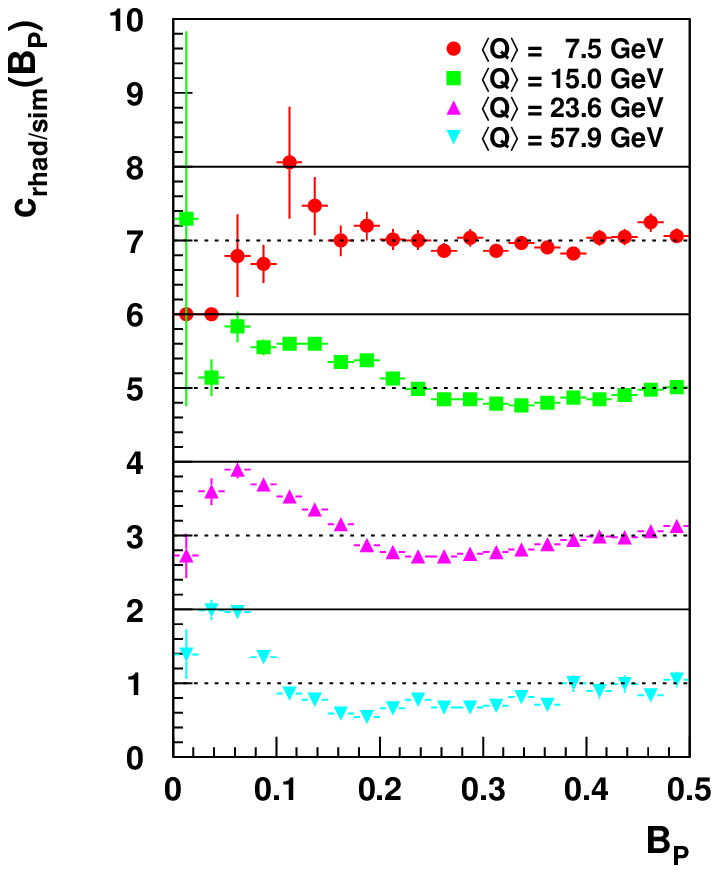}
  \includegraphics{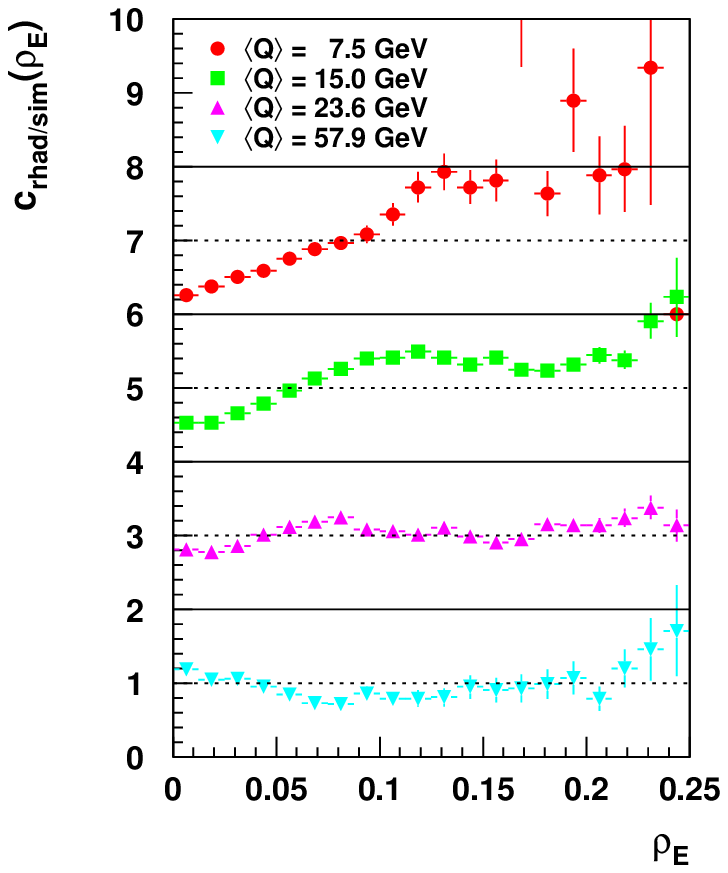}\hftwo%
  \includegraphics{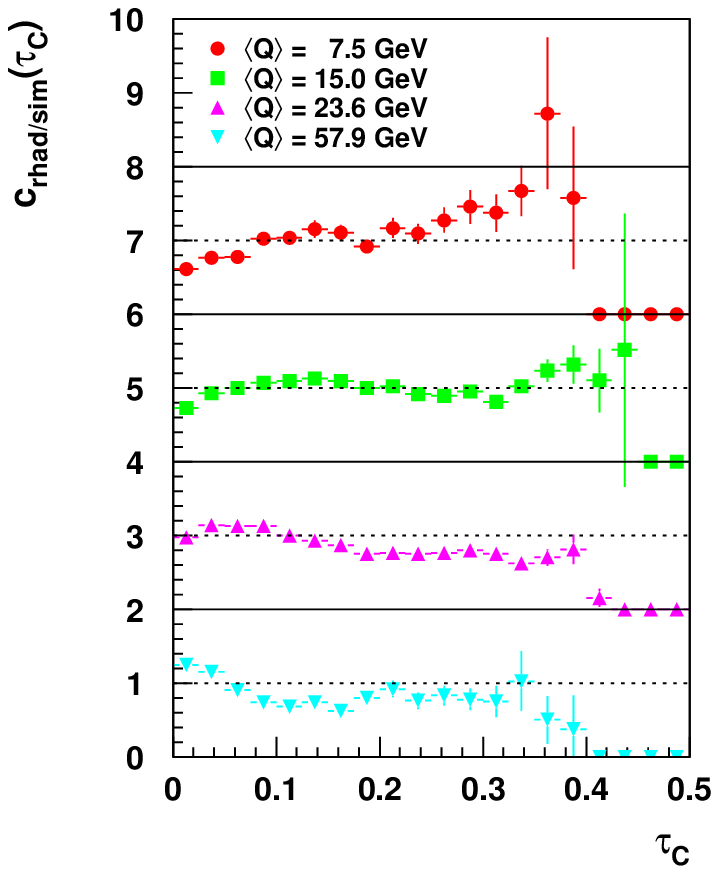}
  \caption[Bin-wise detector corrections for 
  $\tau_P$, $B_P$, $\rho_E$ and $\tau_C$.]  {Bin-wise detector corrections for
    the event shapes $\tau_P$, $B_P$, $\rho_E$ and $\tau_C$ as derived from
    DJANGO~6.2 for four out of eight investigated bins in $Q$.  The factors
    for $\mean{Q}=7.5$--$57.9\gev$ are shifted by offsets of $2\cdot n$, $n =
    0,1,2,3$. The dashed lines delineate the corresponding position of unity.
    Entries exactly at zero (or 2, 4, 6) mean that no events were found on
    either level.  The error bars represent the statistical uncertainty.}
  \label{fig:cdet1}
\end{figure}

\begin{figure} 
  \centering \includegraphics{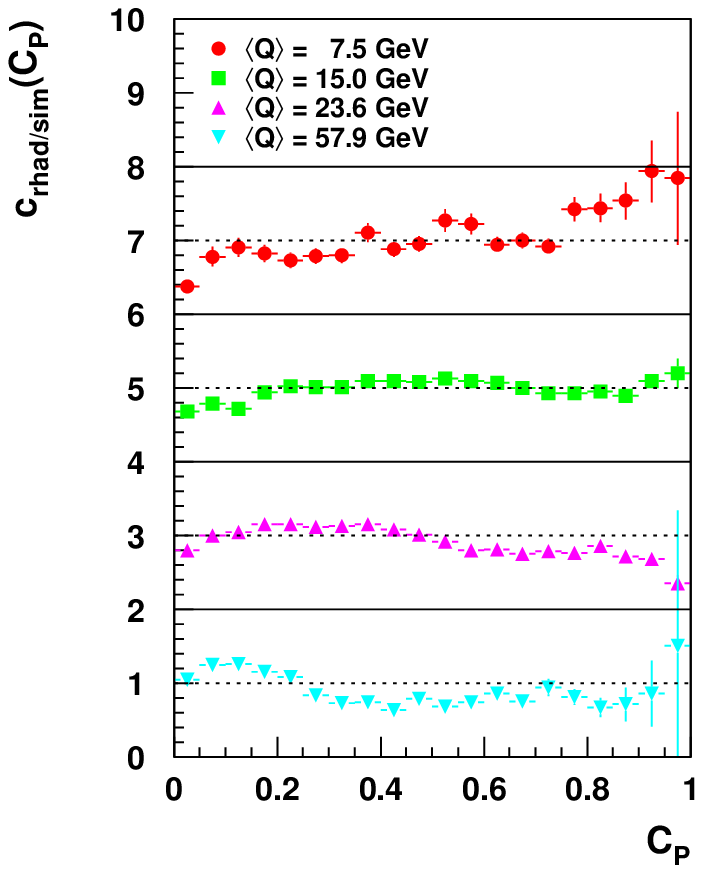}
  \includegraphics{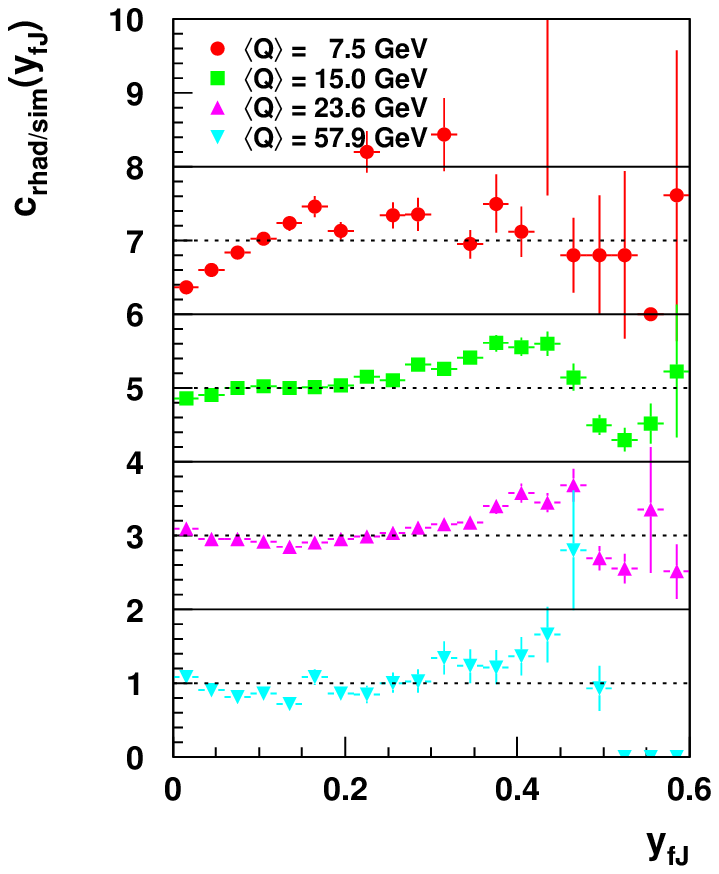}\hftwo%
  \includegraphics{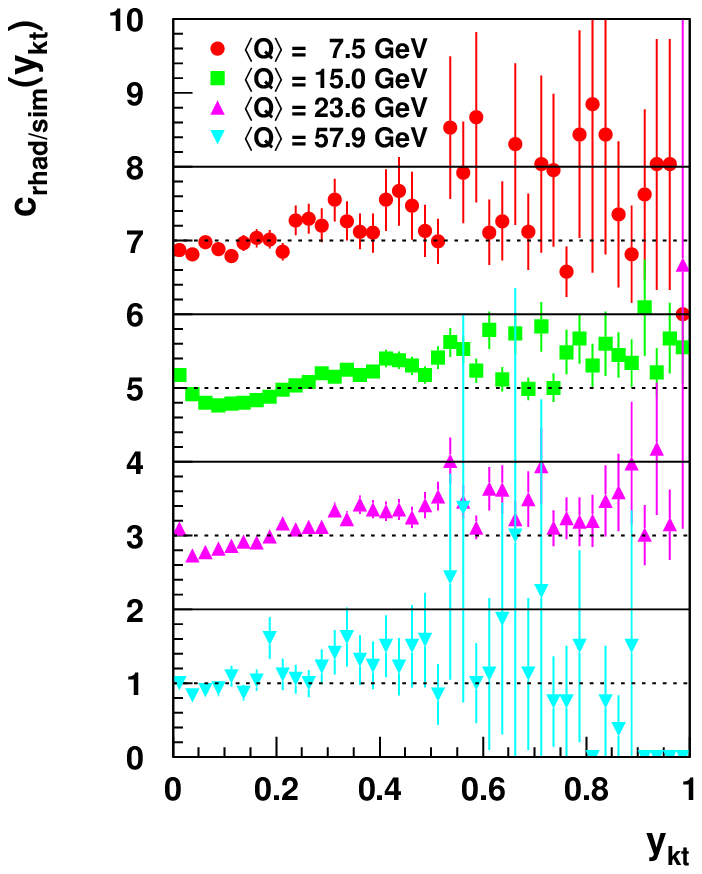}
  \caption[Bin-wise detector corrections for
  $C_P$, $y_{fJ}$ and $y_{k_t}$.]  {Bin-wise detector corrections for the
    event shapes $C_P$, $y_{fJ}$ and $y_{k_t}$ as derived from DJANGO~6.2 for
    four out of eight investigated bins in $Q$.  The factors for
    $\mean{Q}=7.5$--$57.9\gev$ are shifted by offsets of $2\cdot n$, $n =
    0,1,2,3$. The dashed lines delineate the corresponding position of unity.
    Entries exactly at zero (or 2, 4, 6) mean that no events were found on
    either level.  The error bars represent the statistical uncertainty.}
  \label{fig:cdet2}
\end{figure}

Figs.~\ref{fig:cdet1} and~\ref{fig:cdet2} show the bin-wise detector
corrections as derived from DJANGO~6.2 for four out of eight investigated bins
in~$Q$. Entries of zero (or correspondingly two, four or six) occurring at the
edges mean that no events were found at either level. Concentrating on the
statistically relevant points, it must be concluded that the corrections
sometimes are larger than agreeable. Most often they are below $30\%$--$50\%$,
but may rise higher in some bins, especially for $B_P$ and $\rho_E$.

\begin{figure} 
  \centering
  \includegraphics{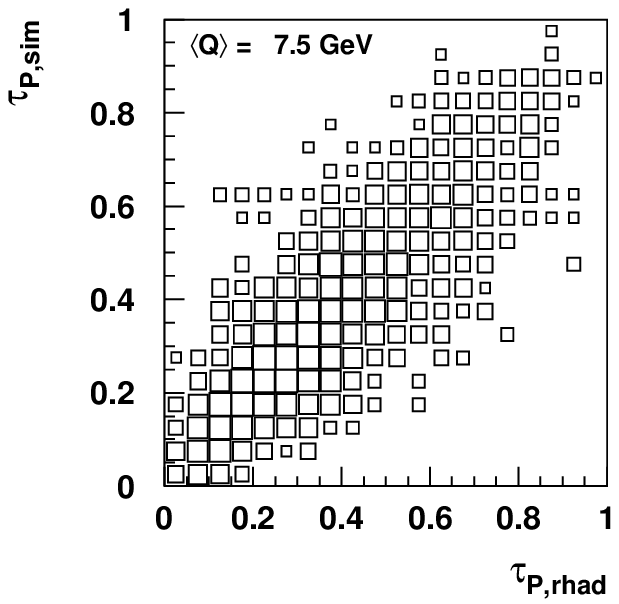}\hftwo%
  \includegraphics{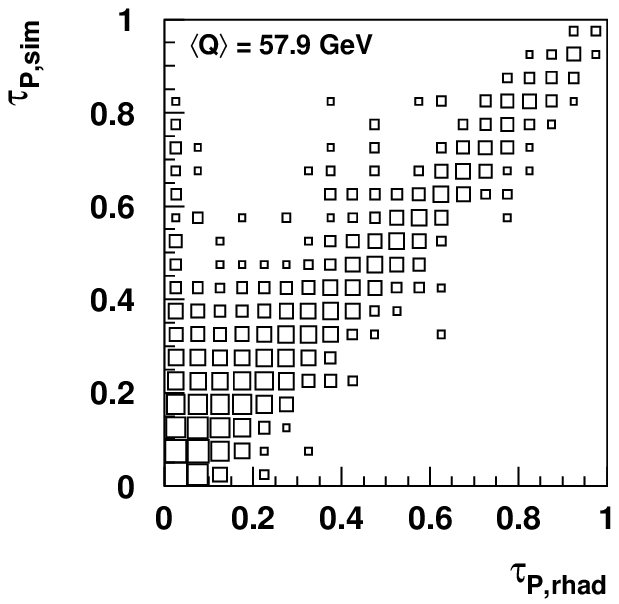}
  \includegraphics{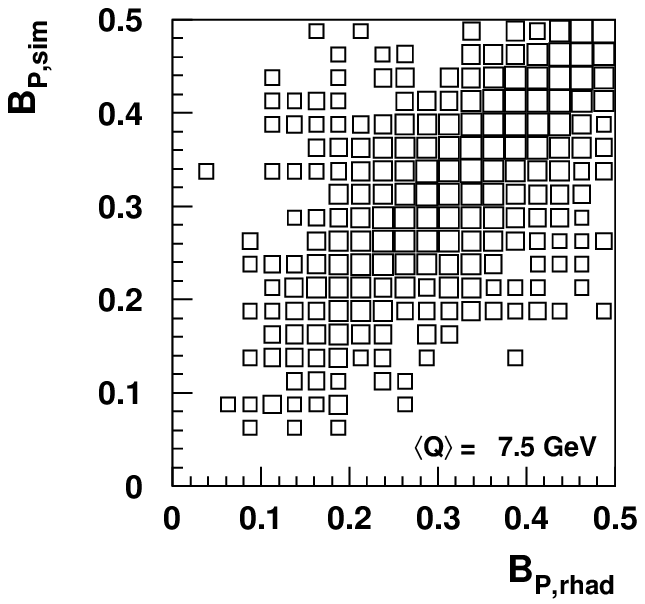}\hftwo%
  \includegraphics{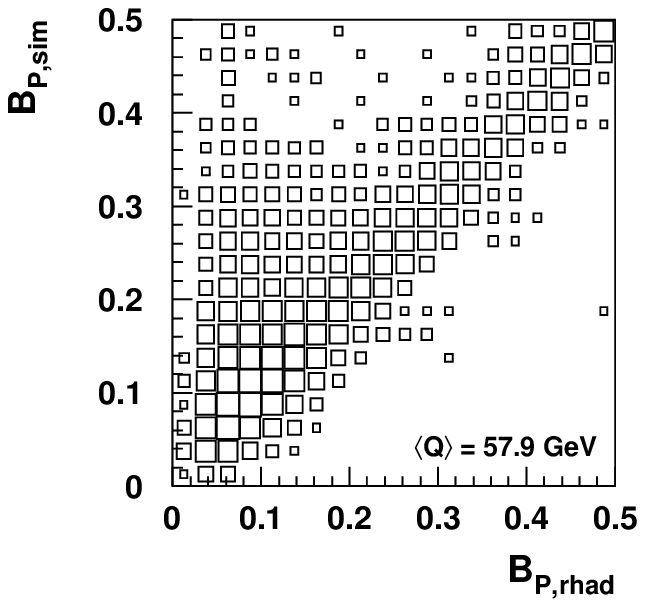}
  \includegraphics{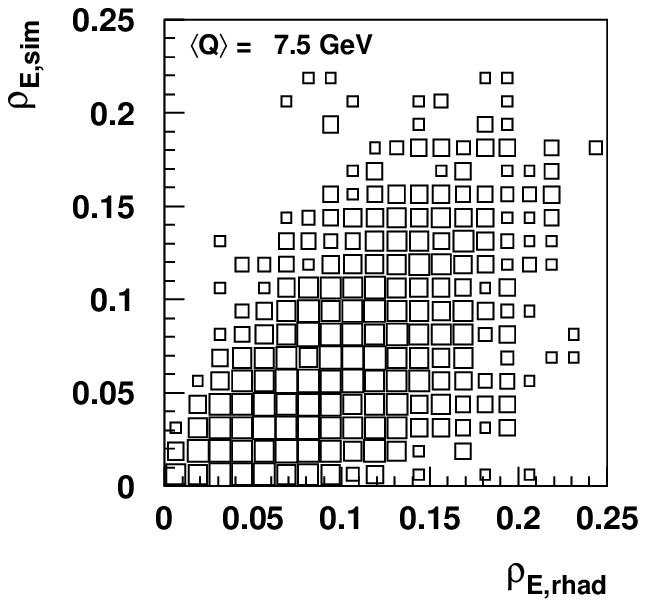}\hftwo%
  \includegraphics{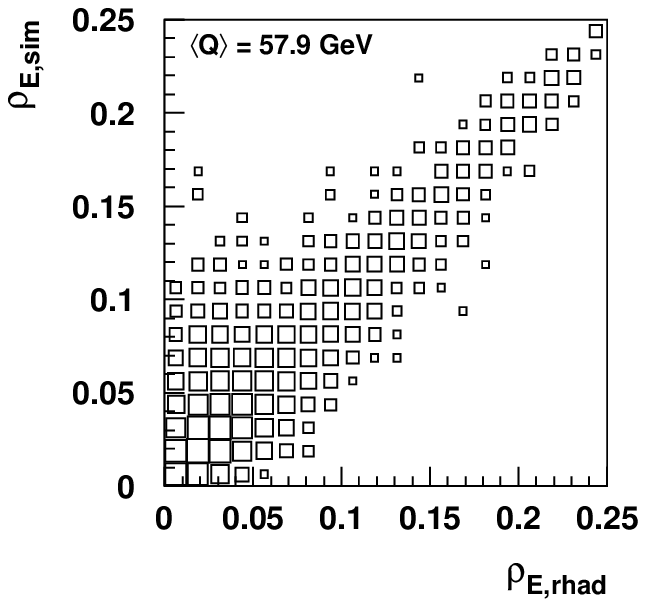}
  \caption[Correlation matrices for $\tau_P$, $B_P$ and $\rho_E$.]
  {Correlation matrices for the event shapes $\tau_P$, $B_P$ and $\rho_E$ as
    derived from DJANGO~6.2 for two out of eight investigated bins in $Q$.
    Note that the box sizes are scaled {\bf logarithmically} with the number
    of entries.}
  \label{fig:corrm1}
\end{figure}

\begin{figure} 
  \centering
  \includegraphics{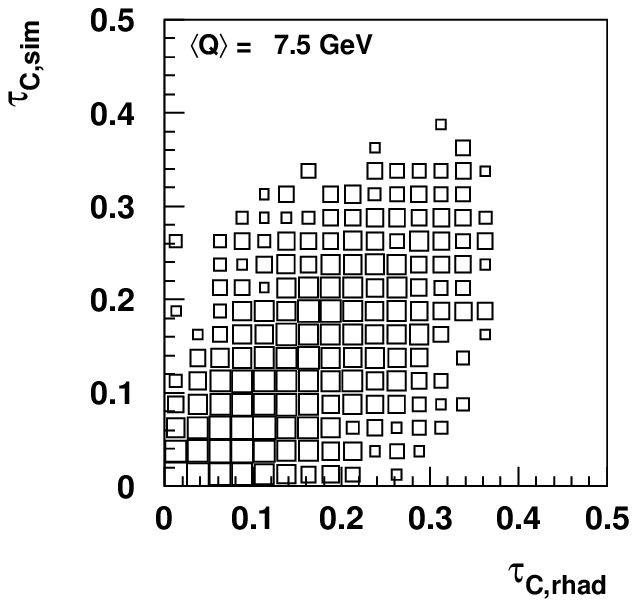}\hftwo%
  \includegraphics{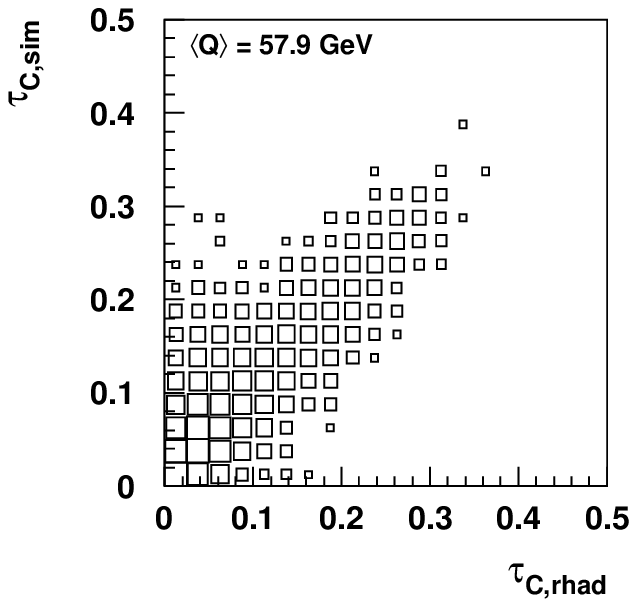}
  \includegraphics{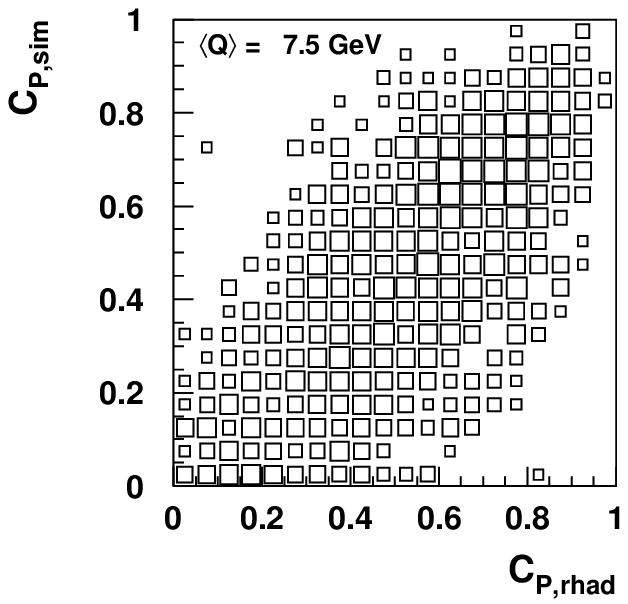}\hftwo%
  \includegraphics{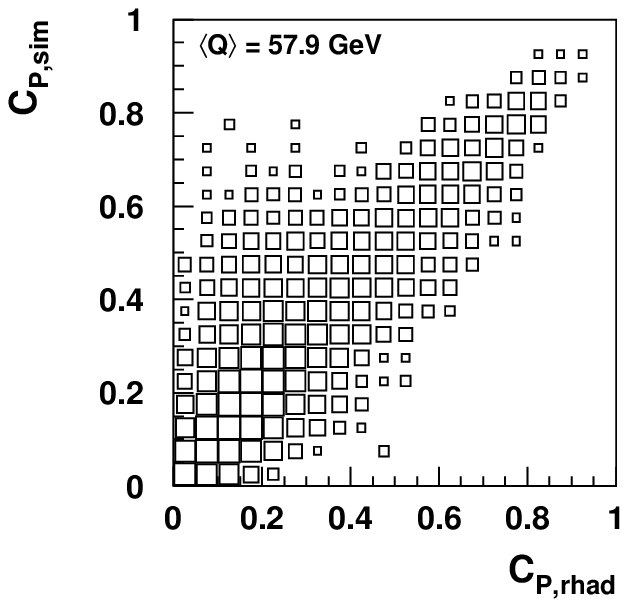}
  \includegraphics{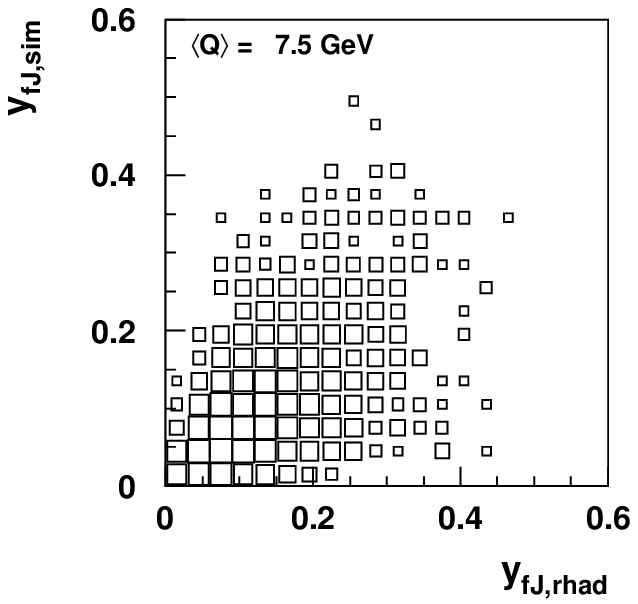}\hftwo%
  \includegraphics{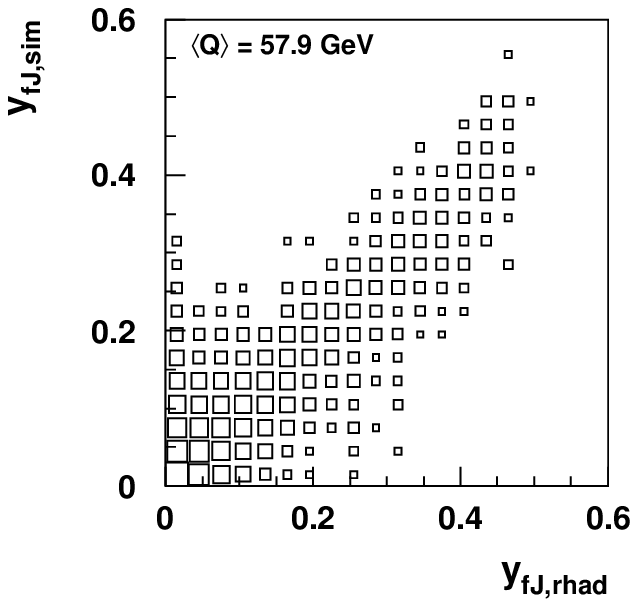}
  \caption[Correlation matrices for $\tau_C$, $C_P$ and $y_{fJ}$.] 
  {Correlation matrices for the event shapes $\tau_C$, $C_P$ and $y_{fJ}$ as
    derived from DJANGO~6.2 for two out of eight investigated bins in $Q$.
    Note that the box sizes are scaled {\bf logarithmically} with the number
    of entries.}
  \label{fig:corrm2}
\end{figure}

\begin{figure} 
  \centering
  \includegraphics{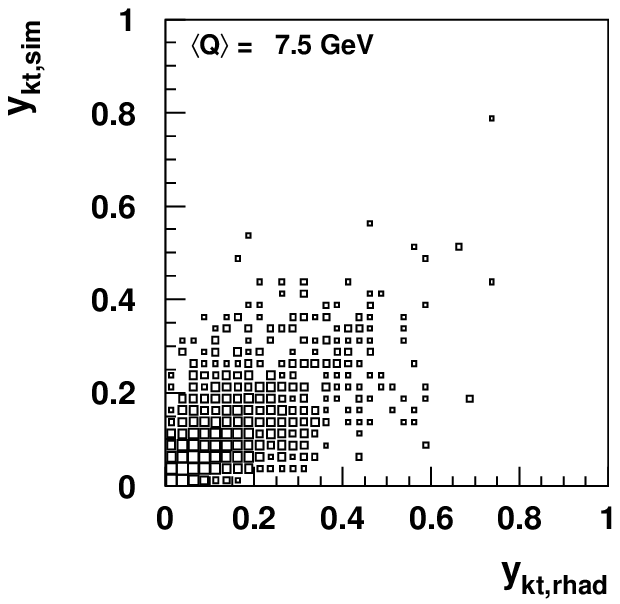}\hftwo%
  \includegraphics{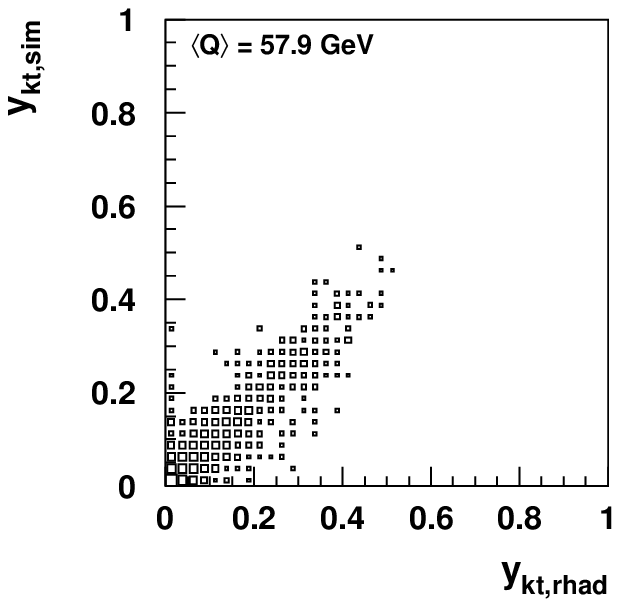}
  \caption[Correlation matrices for $y_{k_t}$.]
  {Correlation matrices for the event shape $y_{k_t}$ as derived from
    DJANGO~6.2 for two out of eight investigated bins in $Q$.  Note that the
    box sizes are scaled {\bf logarithmically} with the number of entries.}
  \label{fig:corrm3}
\end{figure}

Both, the matrix method as well as the Bayesian unfolding need the correlation
matrices between hadron and cluster level as input. They are presented for
each event shape in figs.~\ref{fig:corrm1}--\ref{fig:corrm3} for two bins, one
at low and one at \highq.  Note that the box sizes are scaled {\bf
  logarithmically} with the number of entries.  From these we can learn mainly
two things: First, correlations, which improve with rising $Q$, are clearly
demonstrated, and second, with respect to the resolution, the bin sizes are
too small for the bin-to-bin correction.  Nevertheless, the binning is kept as
a compromise, because the matrix and the Bayes unfolding account for
migrations as well as resolution effects and the corrected means should not be
affected by too coarse bin sizes.

With these ingredients given, two supplementary remarks are due.  Concerning
the matrix approach, the effect of iterations was revealed to be small, except
for the fact that oscillations are gathered up.  Hence, no iteration is
performed here.

With respect to the Bayesian unfolding, it is recommended in~\cite{Bayes} to
avoid building up large fluctuations with an increasing number of iterations
by smoothing a new \qi initial\qo\ distribution $\hat{P}(C_i)$ in the
intermediate steps. This proved to be difficult for the event shapes and
finally was not done.  To start from uniformly distributed initial
probabilities did work, but required some iterations before converging, and
consequently fluctuations showed up. It could be circumvented by choosing the
MC distribution for $P_0(C_i)$. Then, three iterations are fully sufficient.

\section{Radiative Corrections}
\label{sec:corrrad}

At this stage we have to perform the remaining correction from the previously
defined radiative to a non-radiative hadron level.  For that purpose a second
kind of DJANGO~6.2 MC files is necessary.  They have to be generated with
exactly the same settings as the ones before, except for the radiative effects
which have to be switched off. A simulation is not required.  Because there
are no more ISR or FSR photons, the acceptance cut no.~\ref{cut:acc} may be
dropped leaving the cuts nos.~\ref{cut:EPz}--\ref{cut:Eda} ineffective.  In
that way, the acceptance holes of the detector near the beam pipe are
additionally compensated for.  But due to our limited knowledge of that
region, the extrapolation from MC possibly is not very precise.  Notice that
the cuts nos.~\ref{cut:NCH} and~\ref{cut:Eforw} have been applied to {\bf all}
levels up to now and consequently they have not been accounted for by the
unfolding procedure.  Depending on the theory to compare with, e.g.\ the
parton level of some MC, they may have to be considered as {\bf phase space
  cuts}.  Yet, they are not suitable in the framework of a pQCD calculation to
NLO and hence are regarded as data quality cuts in this study.

For the factor and bin-to-bin method, the combined correction steps are fully
equivalent to a one-step procedure.  Yet, as a matter of fact, we now have to
deal with two different MC files.  That means we do not have any correlations
at our disposal such that the unfolding schemes number three and four are not
applicable.  The bin-to-bin correction has to serve as the only available
replacement.

\begin{figure} 
  \centering
  \includegraphics{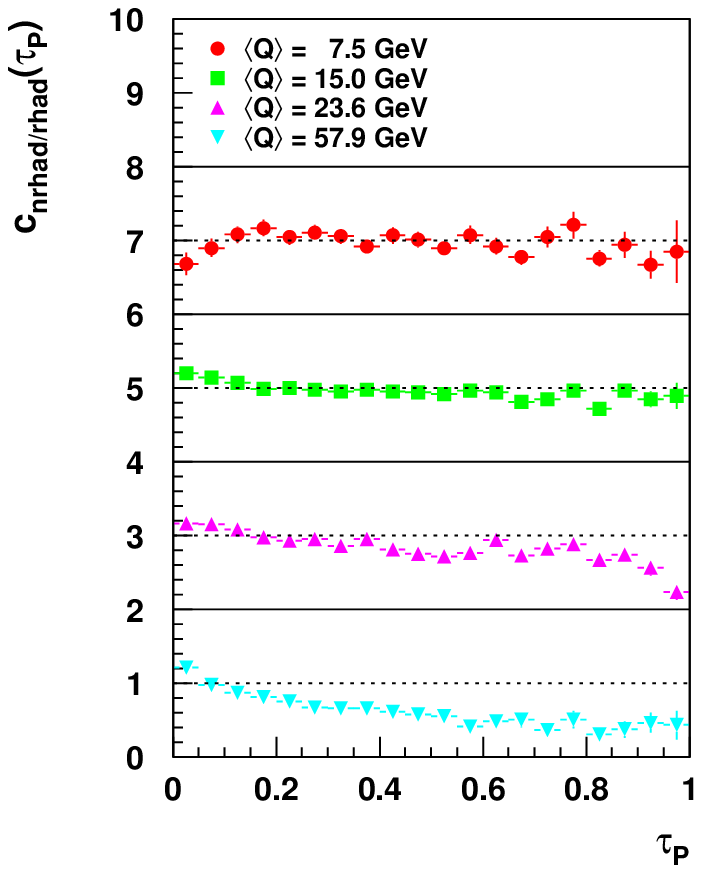}\hftwo%
  \includegraphics{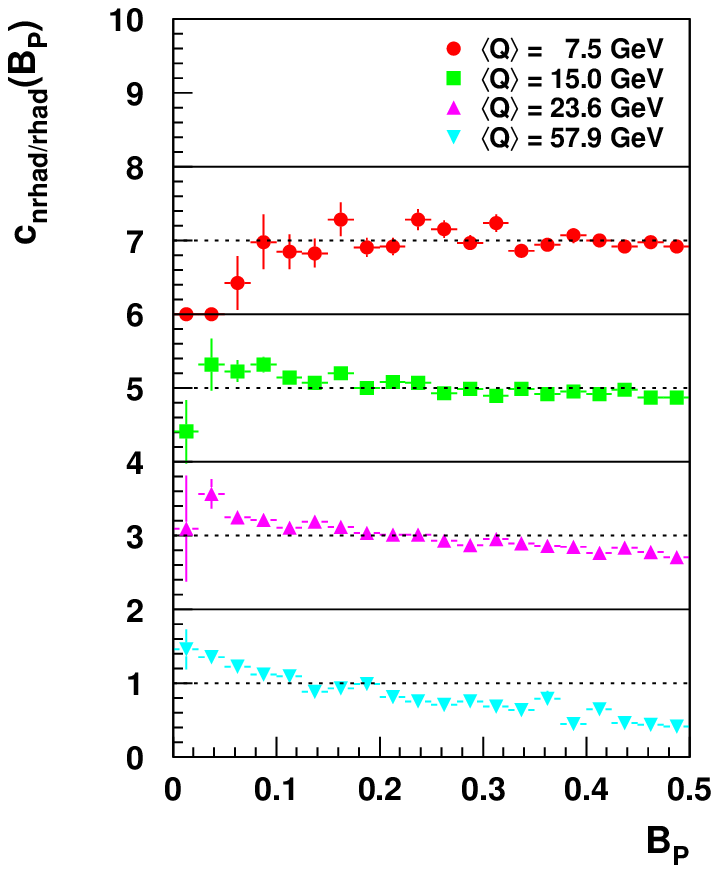}
  \includegraphics{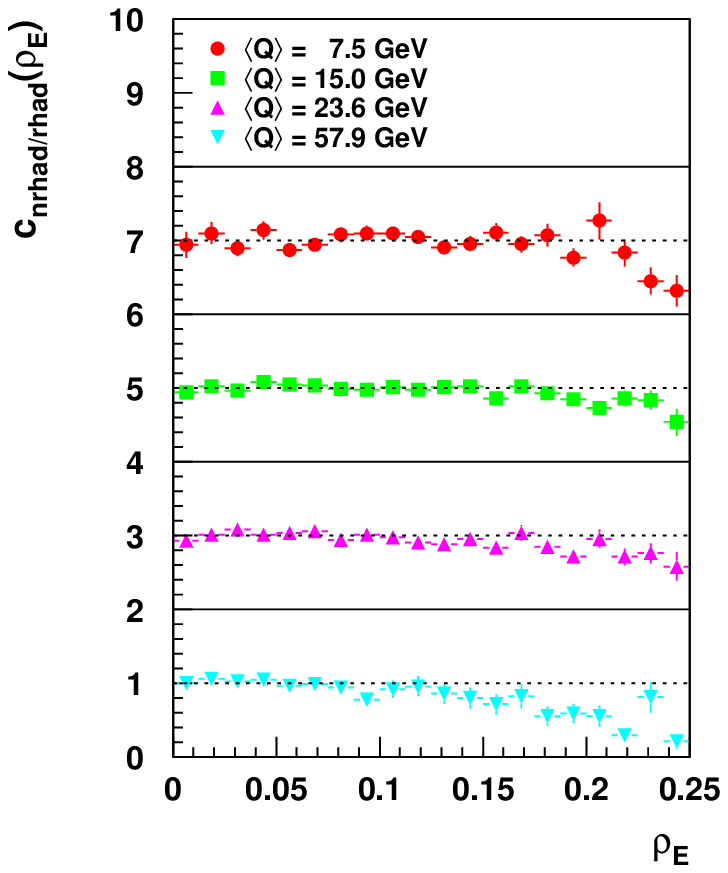}\hftwo%
  \includegraphics{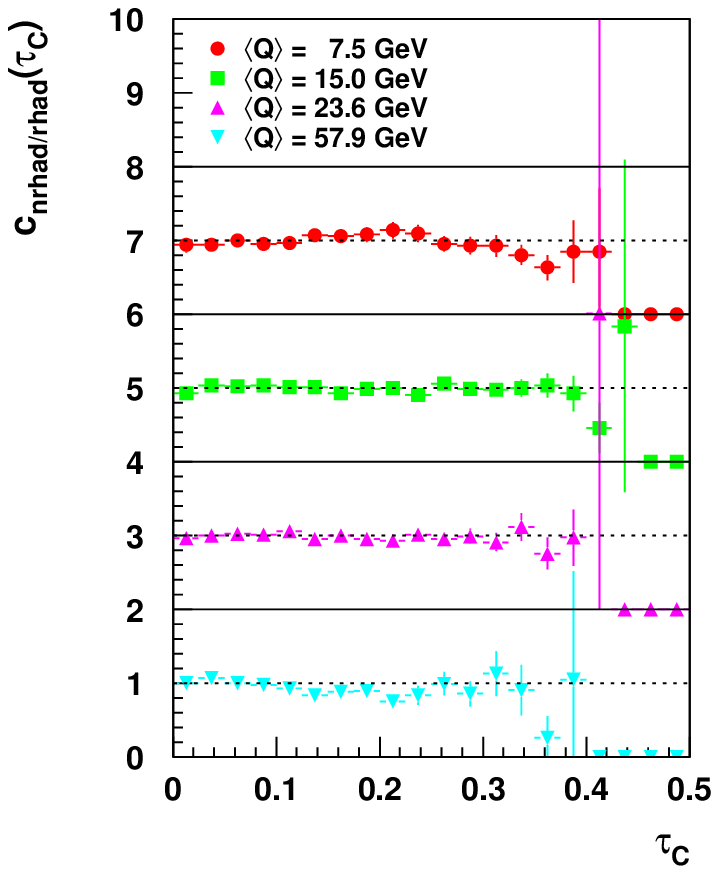}
  \caption[Bin-wise radiative corrections for
  $\tau_P$, $B_P$, $\rho_E$ and $\tau_C$.]  {Bin-wise radiative corrections
    for the event shapes $\tau_P$, $B_P$, $\rho_E$ and $\tau_C$ as derived
    from DJANGO~6.2 for four out of eight investigated bins in $Q$.  The
    factors for $\mean{Q}=7.5$--$57.9\gev$ are shifted by offsets of $2\cdot
    n$, $n = 0,1,2,3$. The dashed lines delineate the corresponding position
    of unity.  Entries exactly at zero (or 2, 4, 6) mean that no events were
    found on either level.  The error bars represent the statistical
    uncertainty.}
  \label{fig:crad1}
\end{figure}

\begin{figure} 
  \centering \includegraphics{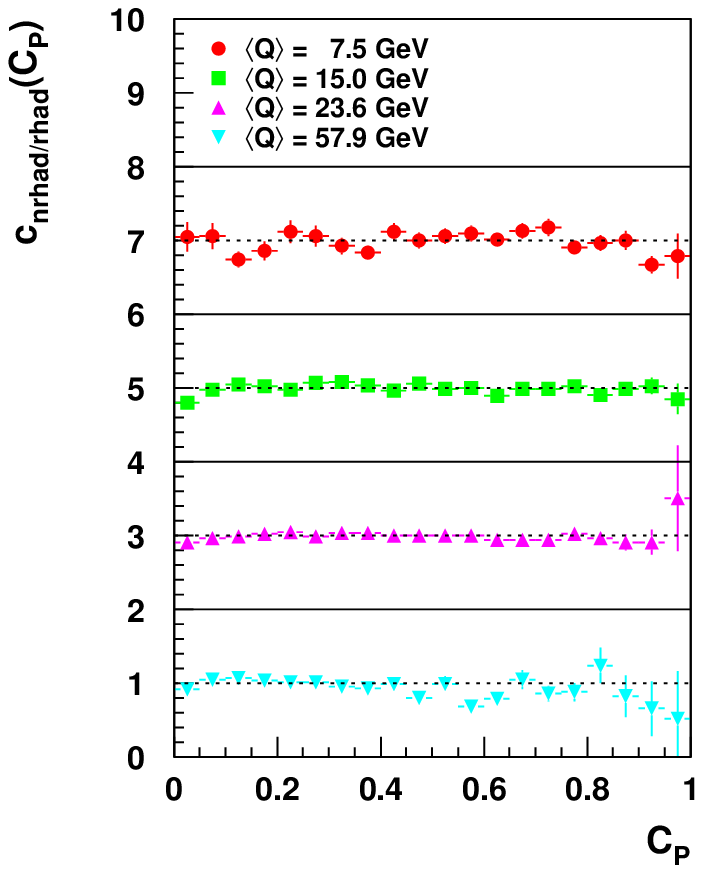}
  \includegraphics{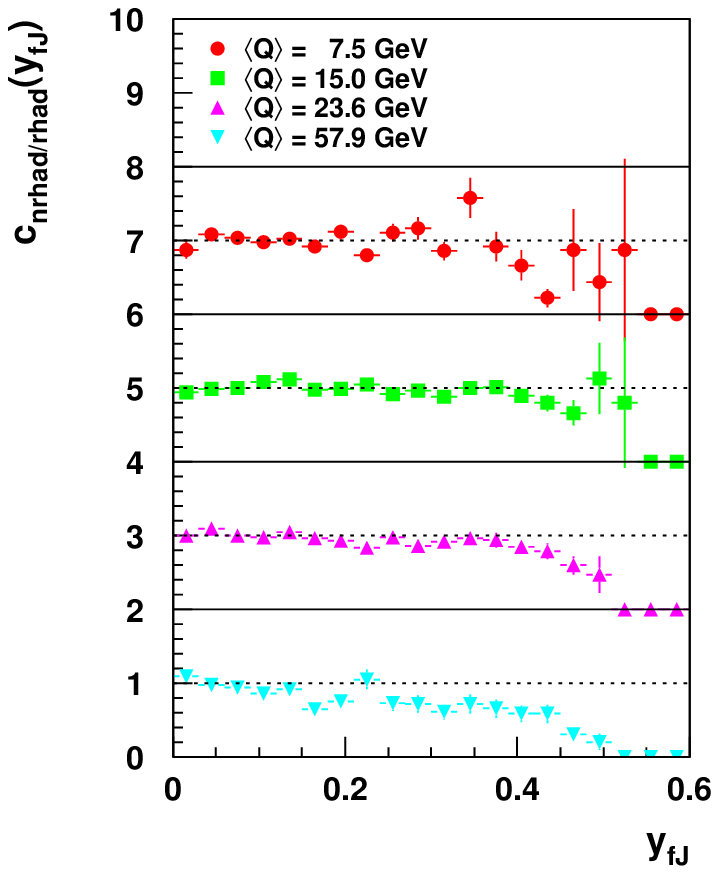}\hftwo%
  \includegraphics{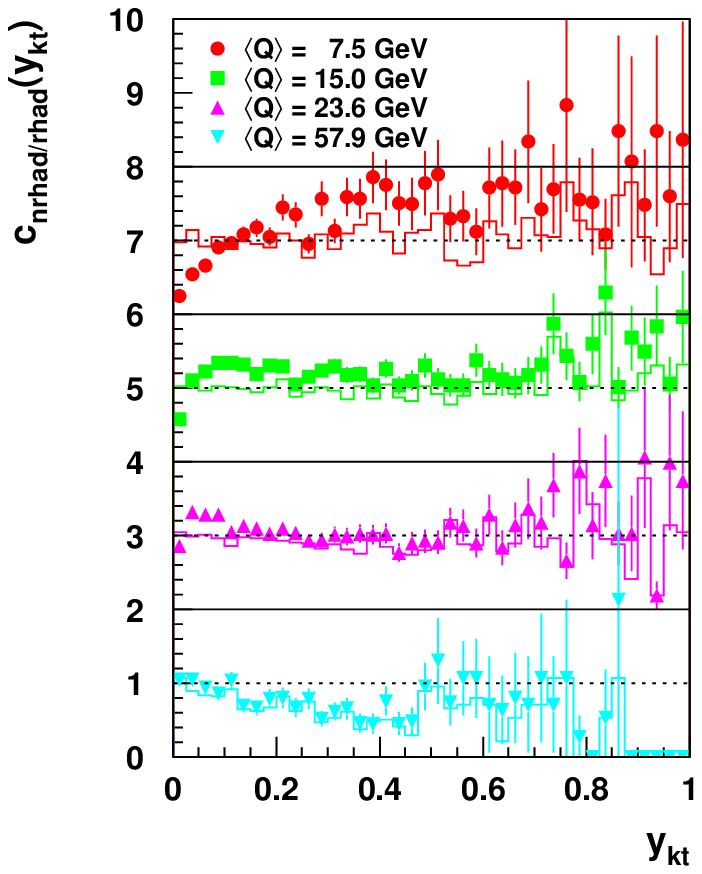}
  \caption[Bin-wise radiative corrections for
  $C_P$, $y_{fJ}$ and $y_{k_t}$.]  {Bin-wise radiative corrections for the
    event shapes $C_P$, $y_{fJ}$ and $y_{k_t}$ as derived from DJANGO~6.2 for
    four out of eight investigated bins in $Q$.  The factors for
    $\mean{Q}=7.5$--$57.9\gev$ are shifted by offsets of $2\cdot n$, $n =
    0,1,2,3$. The dashed lines delineate the corresponding position of unity.
    Entries exactly at zero (or 2, 4, 6) mean that no events were found on
    either level.  The error bars represent the statistical uncertainty.  For
    $y_{k_t}$ the additional histogram (full line) shows the correction factor
    without acceptance extrapolation.}
  \label{fig:crad2}
\end{figure}

Figs.~\ref{fig:crad1} and~\ref{fig:crad2} display the radiative correction
factors for four out of eight investigated bins in $Q$.  Except for $\tau_P$
and $B_P$ where radiative effects are not negligible, they are very close to
one.  $y_{k_t}$ is special. The additionally shown histogram with the full
line usually near \qi one\qo\ corresponds to the correction factors where the
acceptance cut no.~\ref{cut:acc} was kept!  This demonstrates that it is not a
radiative effect.  $y_{k_t}$ as applied here seems to be affected by the MC
extrapolation, which is unfortunate and not understood at the moment. For all
other event shapes including $y_{fJ}$, almost nothing changes whether cut
no.~\ref{cut:acc} is active or not. In the case of $\tau_P$ up to $C_P$ this
could be expected since they only deal with the CH opposite to the remnant in
the Breit frame.

\section{Performance Check}
\label{sec:unfcheck}

\begin{figure}
  \centering
  \includegraphics{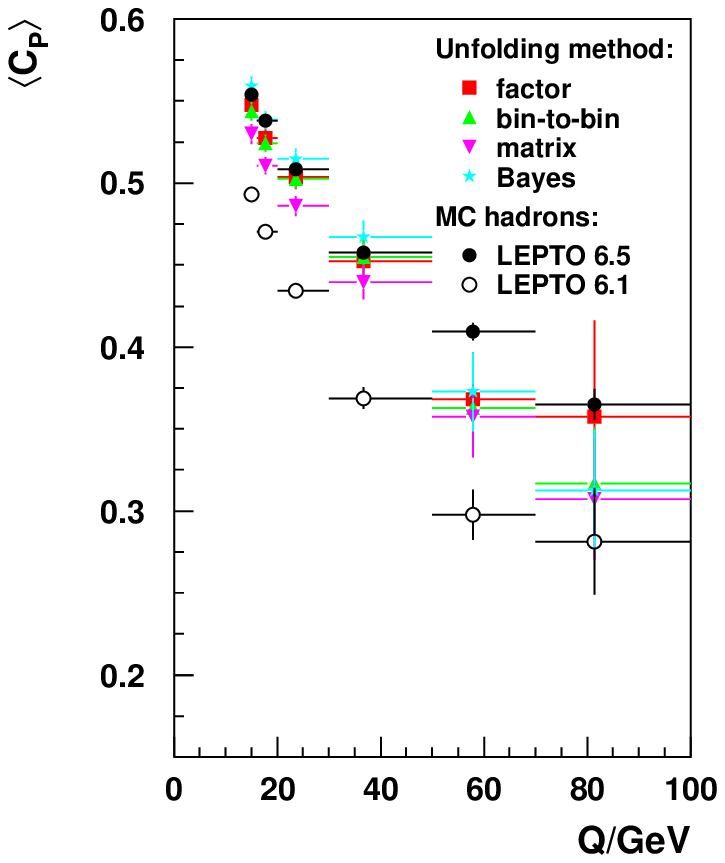}\hftwo%
  \includegraphics{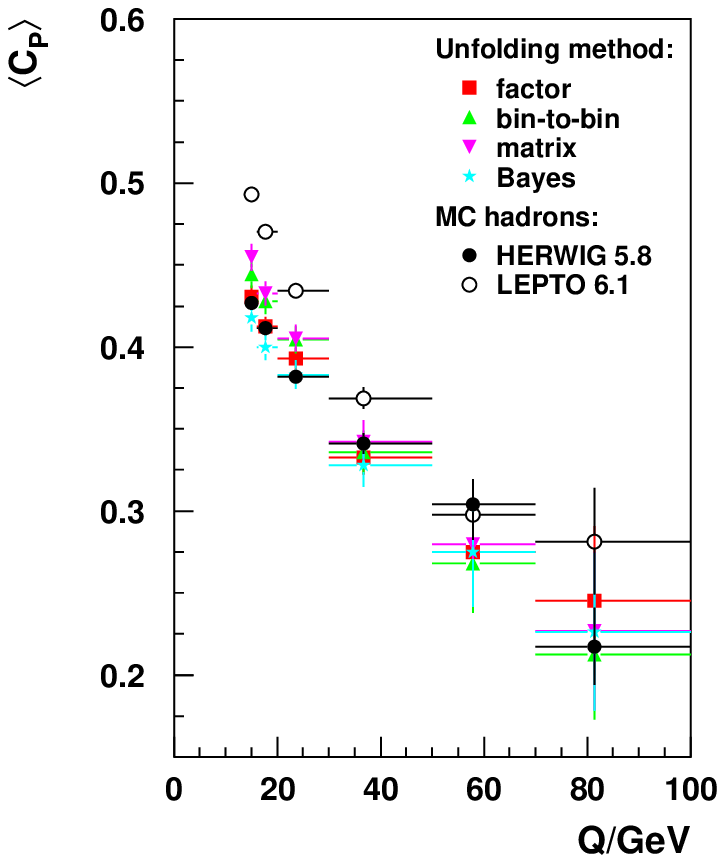}
  \includegraphics{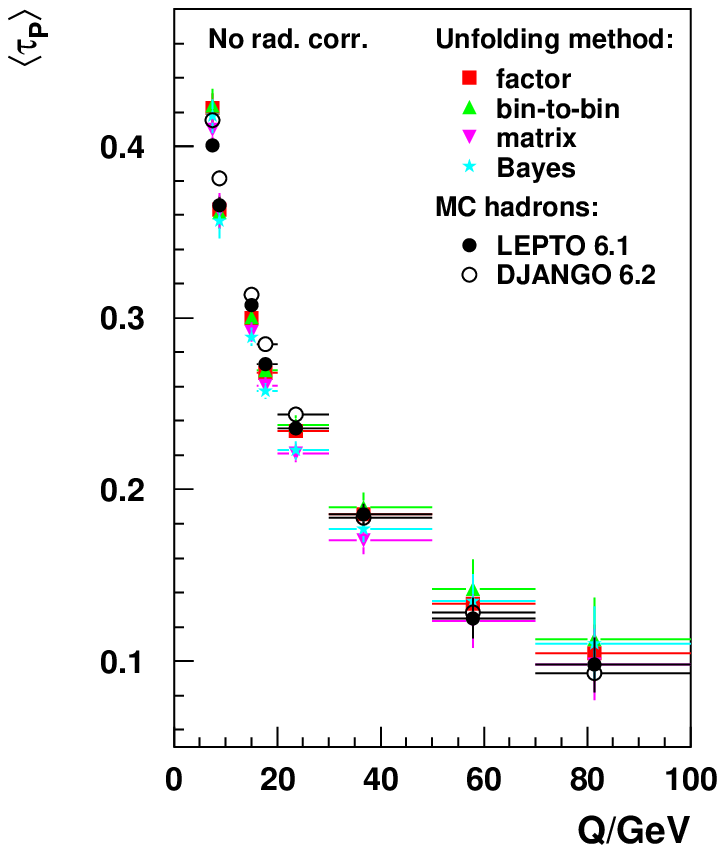}\hftwo%
  \includegraphics{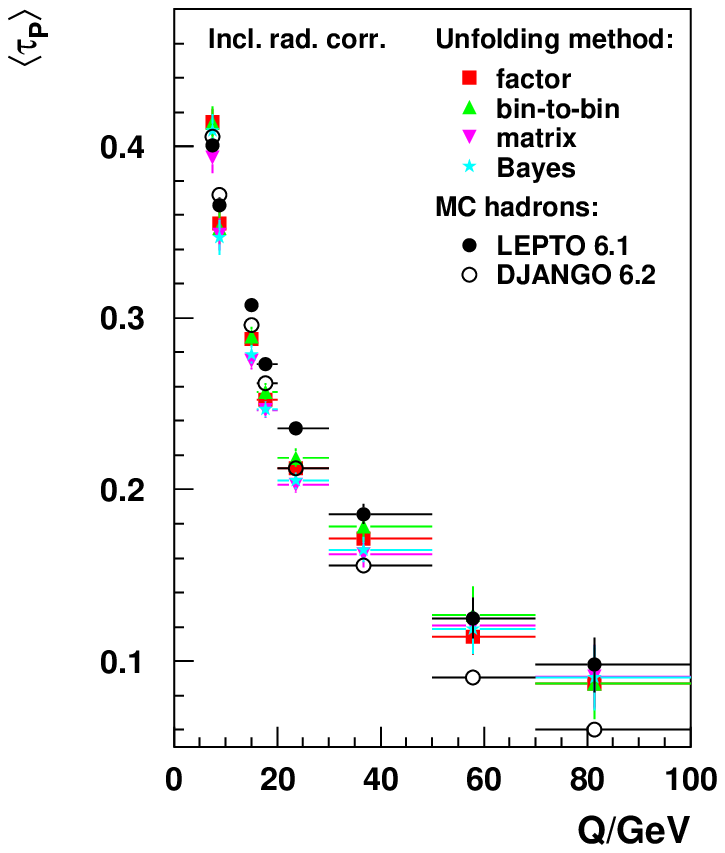}
  \caption[Means of $C_P$ and $\tau_P$ as a function of $Q$
  for simulated event samples corrected with a second MC to check the
  unfolding procedure.]  {Top: Unfolded means of $C_P$ as a function of $Q$
    for simulated \highq\ LEPTO~6.5 (left) and HERWIG~5.8 (right) event
    samples. LEPTO~6.1 was employed as unfolding MC\@.  Bottom: Unfolded means
    of $\tau_P$ as a function of $Q$ for simulated LEPTO~6.1 event samples
    with (right) and without (left) radiative corrections derived from
    DJANGO~6.2.  In addition, the corresponding hadron levels of the MC's are
    shown. The error bars represent statistical uncertainties only.}
  \label{fig:l65h58l61means}
\end{figure}

A mandatory check of correction methods is to employ one MC to unfold another.
As a first test, this was done for models without QED radiation.  LEPTO~6.1
served as the MC, whereas LEPTO~6.5 and HERWIG~5.8 were treated like real
data.  Consequently, the unfolded means should predominantly lie close to the
hadron level means of the corresponding MC (almost) regardless of what
LEPTO~6.1 would prefer.  To be able to discriminate between them, the models
have to make sufficiently different predictions for the quantity under study.
From fig.~\ref{fig:dndCPcl} it is clear that for $C_P$ this is certainly true.
The outcome is presented in the upper plots of fig.~\ref{fig:l65h58l61means}
proving all four correction schemes to work properly.  Nevertheless, all other
event shapes were tested, too, yielding similar results, although not as
evident.

In a second test, LEPTO~6.1 events were smuggled in as data for DJANGO~6.2.
Here, the additional complication arises that the latter does include
radiative effects in contrast to the \qi data.\qo\, As expected, the first
correction step alone suffices since the radiative hadron level is designed to
separate detector and radiative effects as much as possible.  If the radiative
factors are applied, too, then the sensible variables, i.e.\ $\tau_P$ and
$B_P$, are even overdone with.  This is demonstrated in the lower plots of
fig.~\ref{fig:l65h58l61means}, where the \qi fully\qo\ corrected means of the
right plot deviate considerably from the hadron level means of LEPTO~6.1 as
opposed to the plot on the left-hand side.  Using only DJANGO~6.2 events
without radiative photons, the discrepancy disappeared.

Concluding, it can be stated that the unfolding works satisfactorily.

\section{Final Means}
\label{sec:finmeans}

\begin{figure} 
  \centering
  \includegraphics{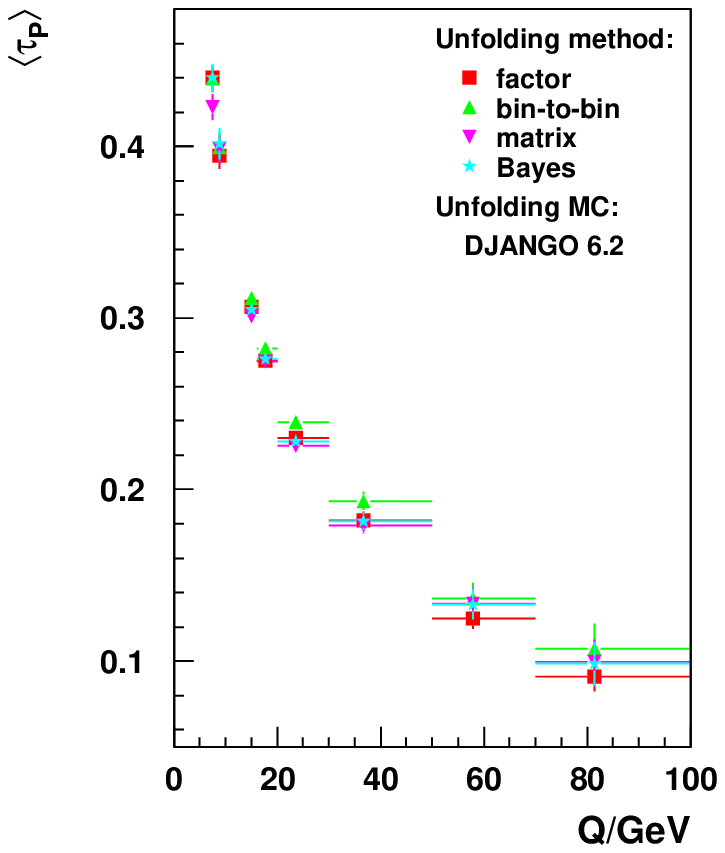}\hftwo%
  \includegraphics{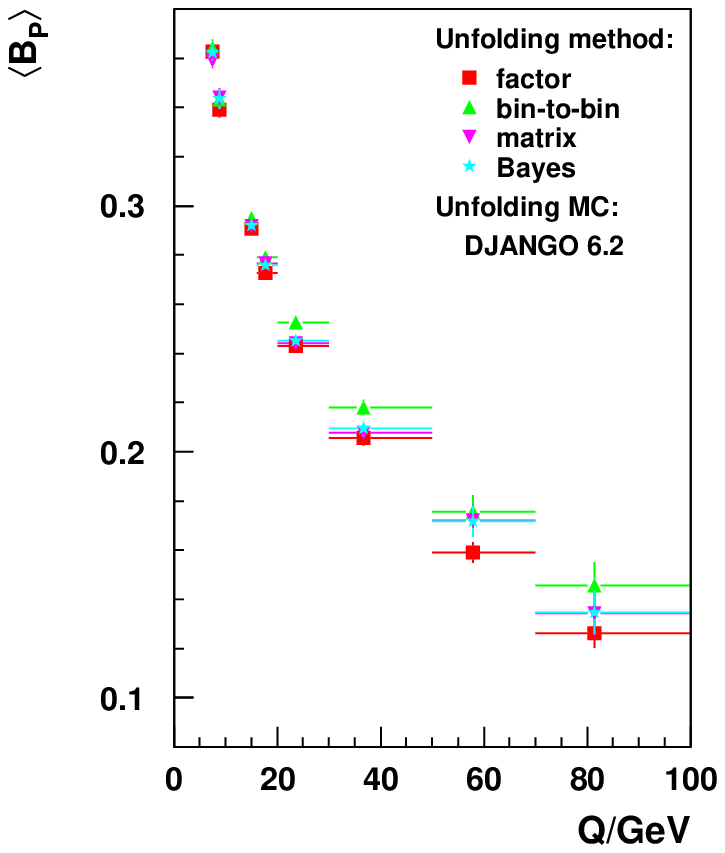}
  \includegraphics{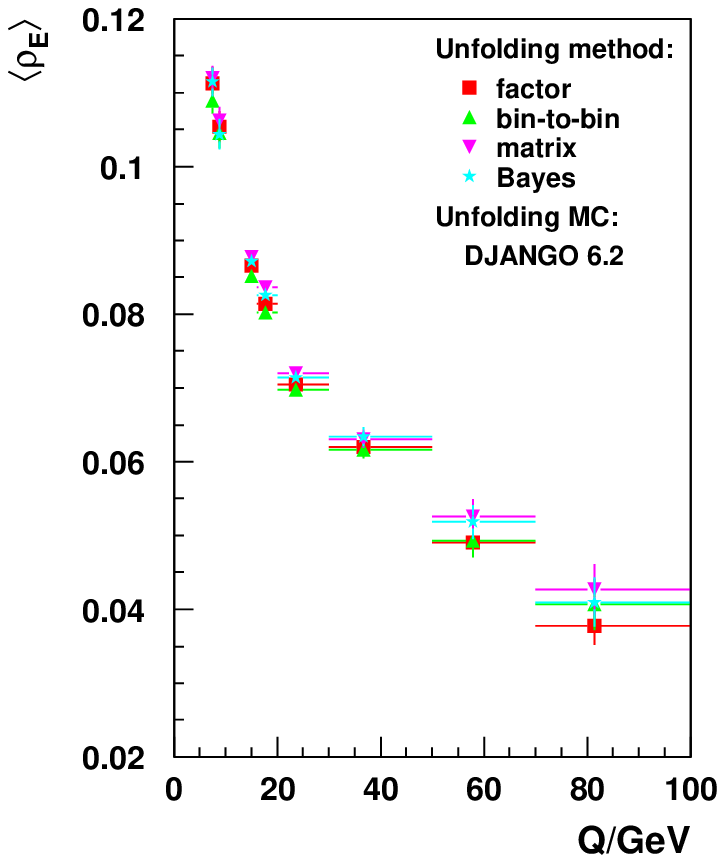}\hftwo%
  \includegraphics{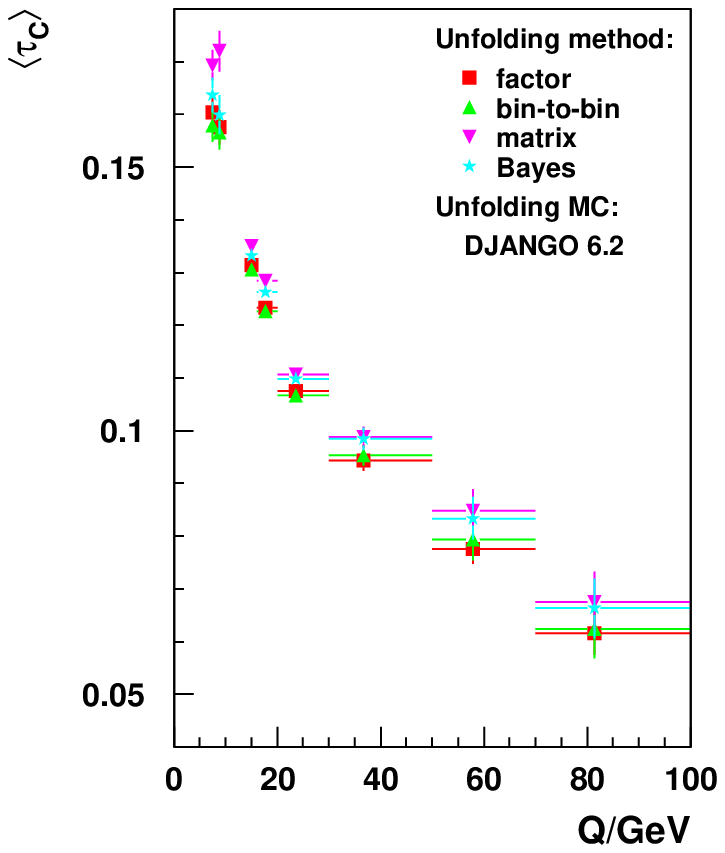}
  \caption[Unfolded means of $\tau_P$, $B_P$, $\rho_E$ and $\tau_C$
  as a function of $Q$ for four correction schemes.]  {Unfolded means of
    $\tau_P$, $B_P$, $\rho_E$ and $\tau_C$ as a function of $Q$ for the four
    described correction schemes.  DJANGO~6.2 served as unfolding MC\@.  Note
    that the step from radiative to non-radiative hadron level can be
    accomplished by the factor or bin-to-bin methods only.  The latter was
    employed for this purpose in case of the matrix and Bayes procedures.  The
    error bars represent statistical uncertainties only.}
  \label{fig:d62unfmeans1}
\end{figure}

\begin{figure} 
  \centering \includegraphics{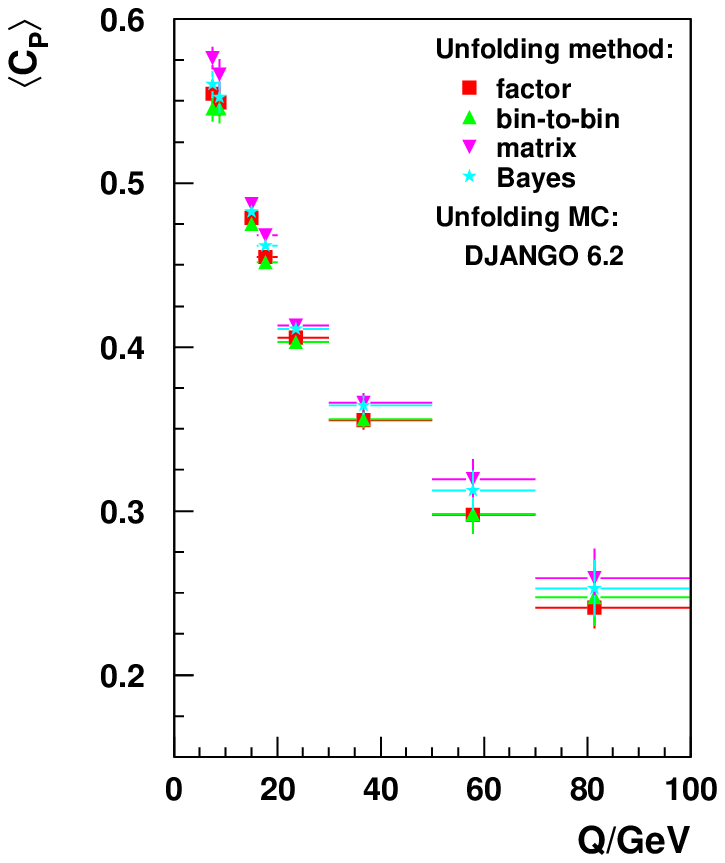}
  \includegraphics{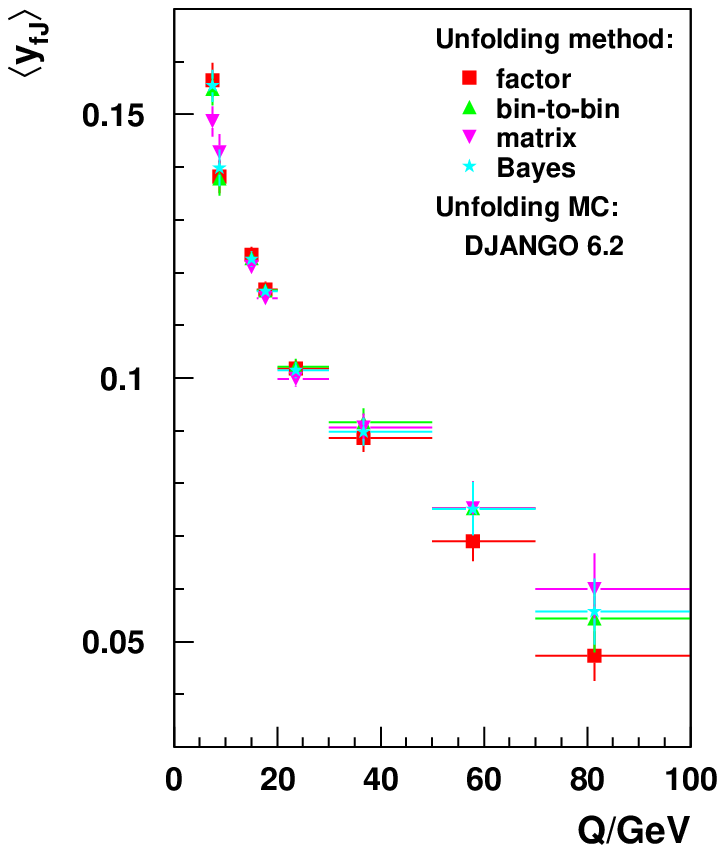}\hftwo%
  \includegraphics{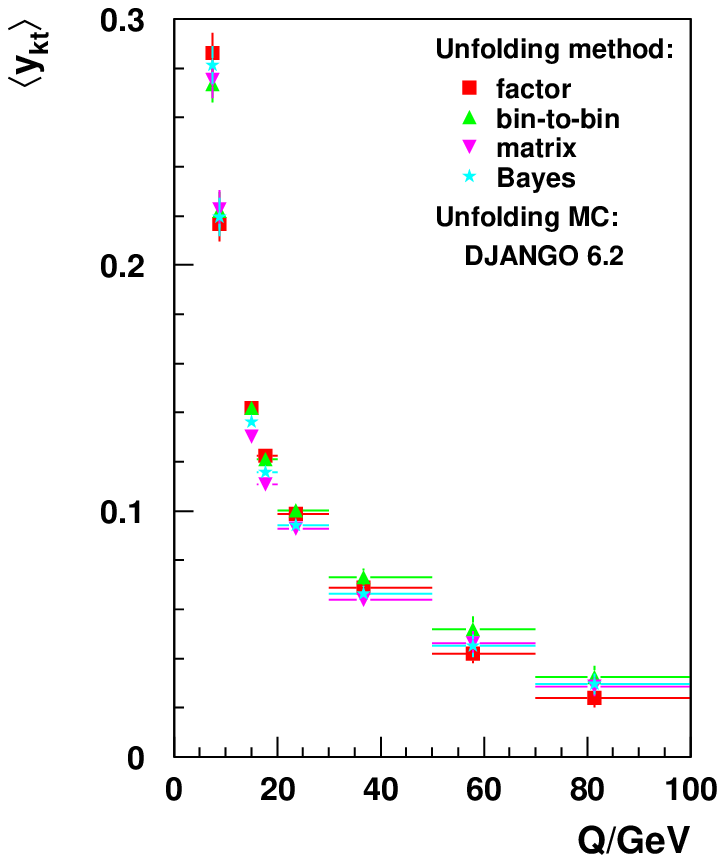}
  \caption[Unfolded means of $C_P$, $y_{fJ}$ and $y_{k_t}$
  as a function of $Q$ for four correction schemes.]  {Unfolded means of
    $C_P$, $y_{fJ}$ and $y_{k_t}$ as a function of $Q$ for the four described
    correction schemes.  DJANGO~6.2 served as unfolding MC\@.  Note that the
    step from radiative to non-radiative hadron level can be accomplished by
    the factor or bin-to-bin methods only.  The latter was employed for this
    purpose in case of the matrix and Bayes procedures.  The error bars
    represent statistical uncertainties only.}
  \label{fig:d62unfmeans2}
\end{figure}

To illustrate possible systematic effects due to the different unfolding
schemes, all resulting corrected means are shown in
figs.~\ref{fig:d62unfmeans1} and~\ref{fig:d62unfmeans2}.  The statistical
uncertainties encompass data as well as MC statistics. For the factor method,
they are obtained properly by error propagation. In case of an evaluation from
unfolded distributions, they are estimated by
\begin{equation}
  \Delta\fmean = \sigma(\fmean)\cdot\left(\frac{1}{\sqrt{N_{\rm dat}}}
    + \frac{1}{\sqrt{N_{\rm MC}}}\right)\,,
\end{equation}
where $N_{\rm dat}$, $N_{\rm MC}$ are the appropriate event numbers and
$\sigma$ denotes the standard deviation. Note that $1/\sqrt{N_{\rm dat}}$ and
$1/\sqrt{N_{\rm MC}}$ are not added conventionally in quadrature. This is
motivated by the fact that the uncertainties derived this way are closer in
value to those calculated for the factor method.

For the final results the differential distributions as well as the central
values are taken from the most sophisticated approach, i.e.\ the Bayesian
unfolding with subsequent bin-to-bin radiative correction. In case of the
event shape spectra, no further error evaluation beyond statistical ones was
attempted. For a comparison with pQCD calculations, their presentation is
postponed to ch.~\ref{chap:powcorr}, figs.~\ref{fig:dndFhl1}
and~\ref{fig:dndFhl2}.

Concerning the mean values, the other methods serve to judge systematic
effects.  The due asymmetric uncertainty is estimated to be half of the spread
caused by the maximal deviation to larger respectively smaller values.  The
sum of both then corresponds to half the total spread which is somewhat
smaller than $2/\sqrt{12} = 1/\sqrt{3}$ as appropriate for a uniform
distribution.

A second source of systematic uncertainties are the electromagnetic and
hadronic energy scales of the calorimeters.  Due to recent
improvements~\cite{H1:ICHEP98hq}, the electron energies now can be measured to
a precision of $1\%$, $2\%$ and $3\%$ up to $z_{\rm imp}$-coordinates of
$24\cm$, $110\cm$ and the forward end of the LAr calorimeter respectively.
Hadronic energies are known only to about $4\%$.  In order to gain information
on the influence of this uncertainty, the whole analysis is repeated scaling
up and down separately for both scales.  The discrepancies with respect to the
normal central values are then attributed to two further asymmetric systematic
uncertainties.  All three, i.e.\ including the unfolding, added in quadrature,
yield the total systematics given together with the final results in
table~\ref{tab:finalmeans}.

Concerning individual contributions, the impact of the electron energy, which
directly affects the boost into the Breit frame, is almost always largest for
$\tau_P$ up to $C_P$, followed by unfolding effects. Realizing that hadronic
systematics cancel between numerator and denominator for these event shapes,
this is understandable. The situation is reversed for $y_{fJ}$ and $y_{k_t}$.
Here, no cancellation occurs and the systematic uncertainty due to hadronic
energies dominates.

\begin{figure} 
  \centering
  \includegraphics{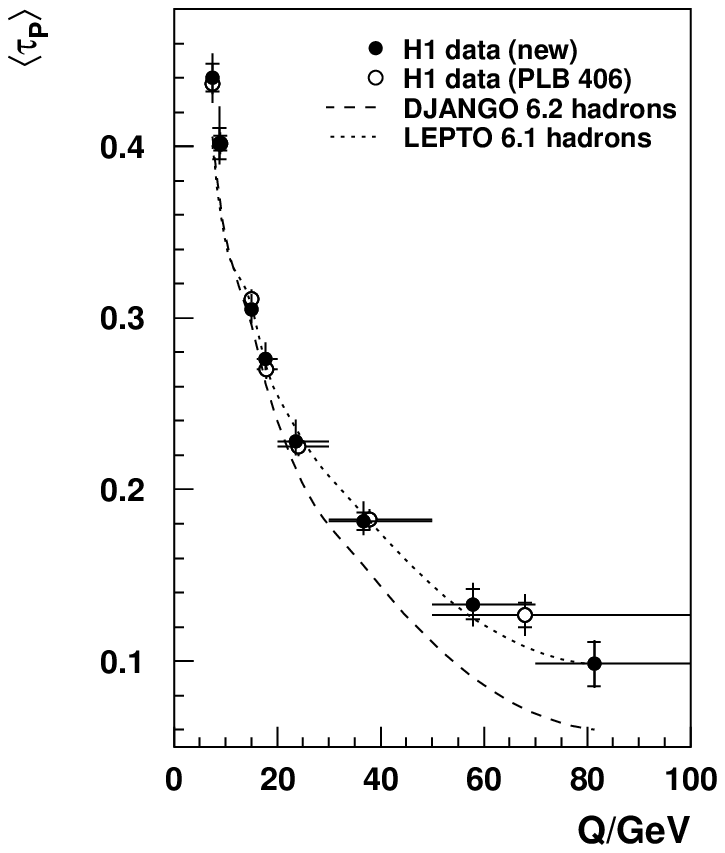}\hftwo%
  \includegraphics{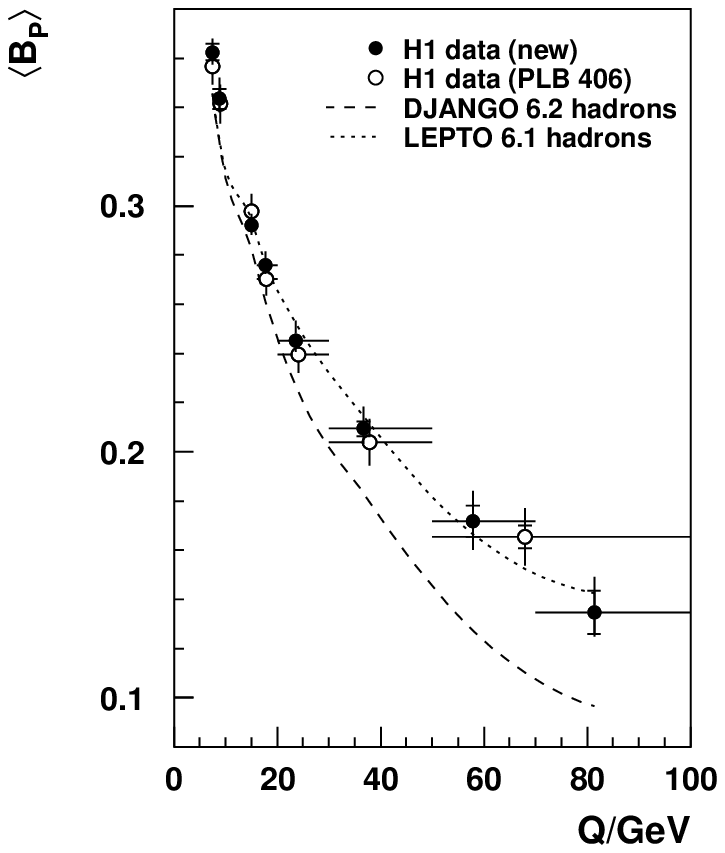}
  \includegraphics{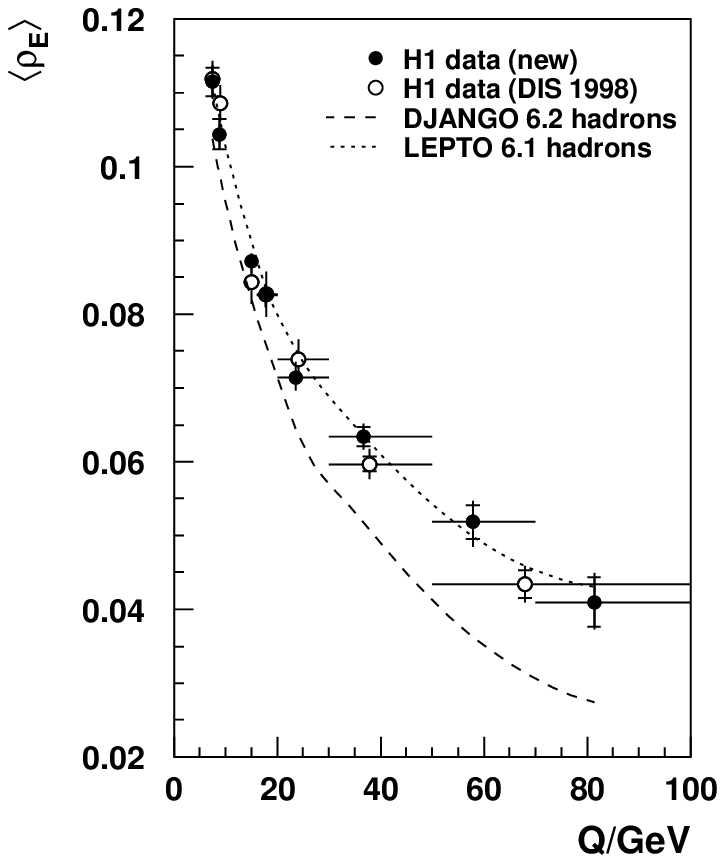}\hftwo%
  \includegraphics{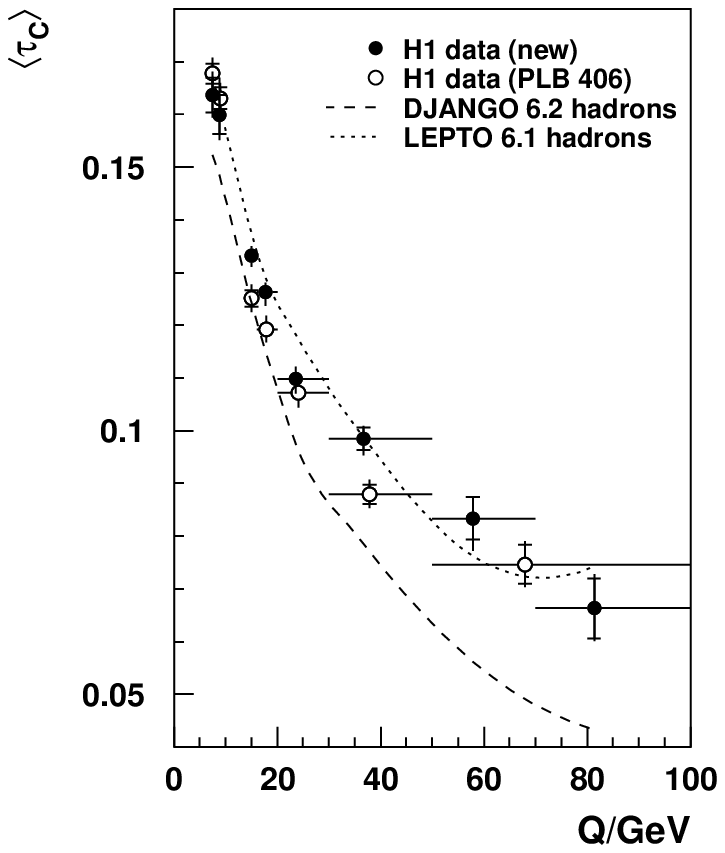}
  \caption[Corrected means of $\tau_P$, $B_P$, $\rho_E$ and $\tau_C$
  as a function of $Q$.]  {Corrected means (full symbols) of $\tau_P$, $B_P$,
    $\rho_E$ and $\tau_C$ as a function of $Q$. The inner error bars represent
    the statistical uncertainty, the outer ones the total statistic and
    systematic uncertainty. They are compared with the hadron level mean
    values of the DJANGO~6.2 (dashed) and LEPTO~6.1 (dotted) MC and with
    already published results by H1 (hollow symbols).}
  \label{fig:finalmeans1}
\end{figure}

\begin{figure} 
  \centering \includegraphics{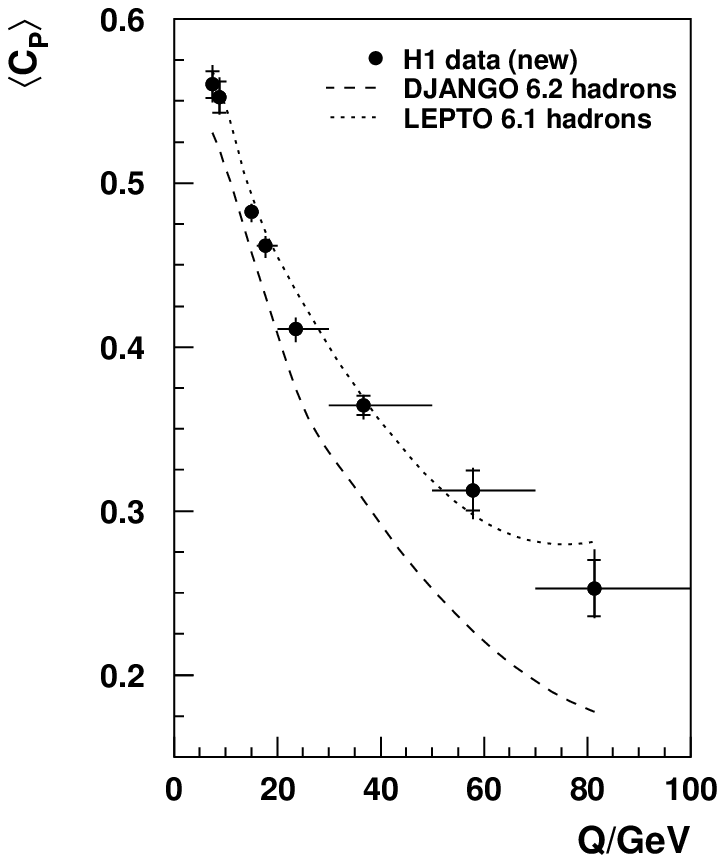}
  \includegraphics{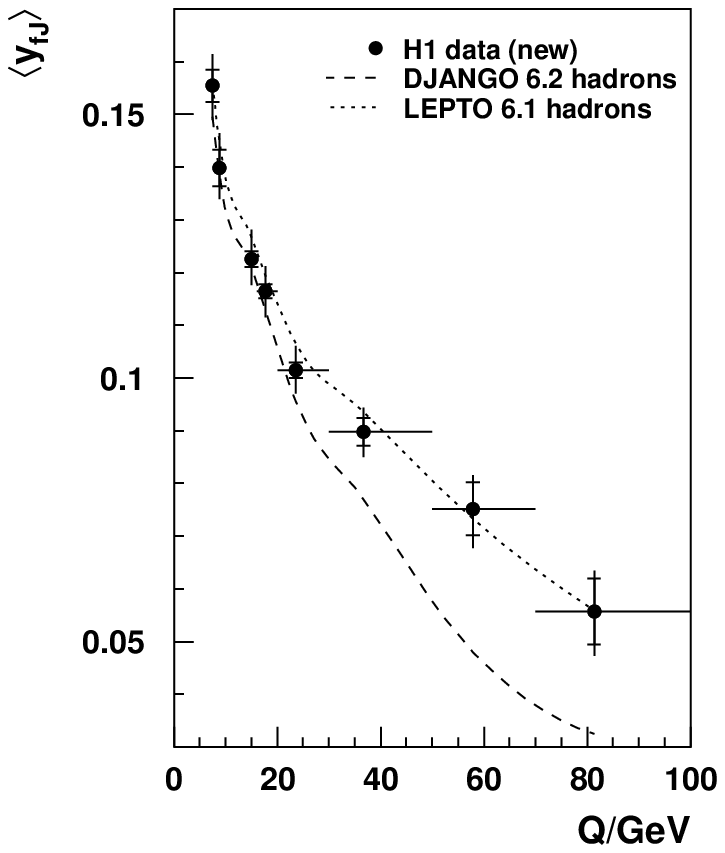}\hftwo%
  \includegraphics{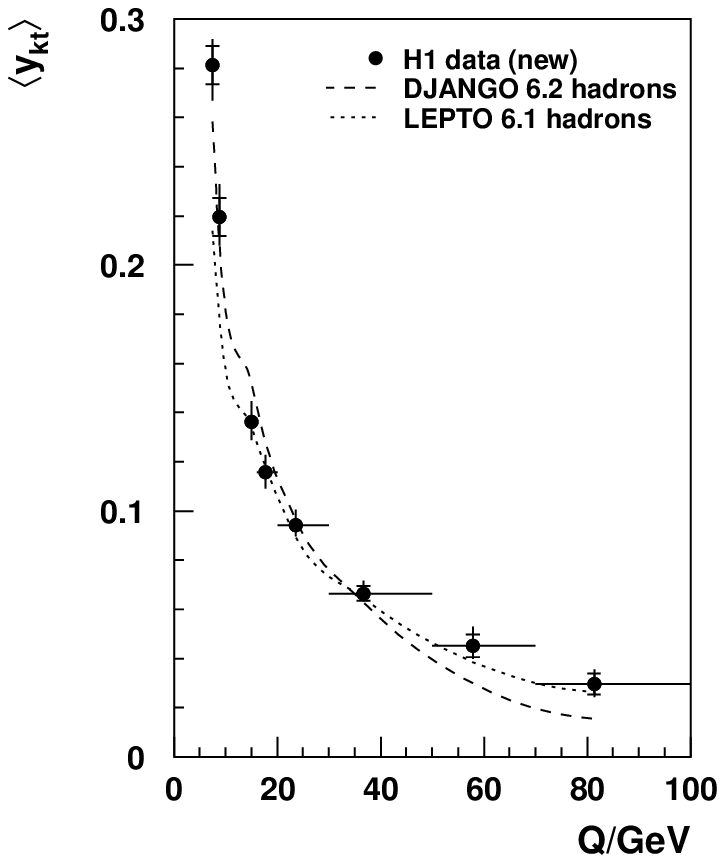}
  \caption[Corrected means of $C_P$, $y_{fJ}$ and $y_{k_t}$
  as a function of $Q$.]  {Corrected means (full symbols) of $C_P$, $y_{fJ}$
    and $y_{k_t}$ as a function of $Q$. The inner error bars represent the
    statistical uncertainty, the outer ones the total statistic and systematic
    uncertainty. They are compared with the hadron level mean values of the
    DJANGO~6.2 (dashed) and LEPTO~6.1 (dotted) MC\@.}
  \label{fig:finalmeans2}
\end{figure}

At last, figs.~\ref{fig:finalmeans1} and~\ref{fig:finalmeans2} present the
final means including all uncertainties as well as the hadron level means of
DJANGO~6.2 and LEPTO~6.1.  Where available, already published results
from~\cite{H1:Shapes} and~\cite{HUM:DIS98} are shown for
comparison.\footnote{$C_P$ is also given in~\cite{HUM:DIS98}, but the values
  are believed to be erroneous and therefore are not reproduced here.}
$\tau_P$ and $B_P$ are in good agreement, but they should not because
radiative corrections leading to smaller values were previously neglected.
Seemingly, this was compensated by an overestimation of the electron energy
that happened before the new calibration was introduced into the analysis.
Since $\rho_E$ and $\tau_C$ are much less affected by QED radiation, the new
data are shifted somewhat towards higher means.

\begin{table}
  \centering
  \hftwo\begin{tabular}{|c||l|l|}
    \hline
    \multicolumn{1}{|c||}{$\rbthm\mean{Q}/\gev$}&
    \multicolumn{1}{|c|}{$\mean{\tau_P}$}&
    \multicolumn{1}{|c|}{$\mean{B_P}$}
    \\\hline\hline
    $7.46$ &
    $0.4402\pm 0.0082~^{+0.0111}_{-0.0122}$&
    $0.3624\pm 0.0034~^{+0.0046}_{-0.0038}$
    \rbtrr\\\hline 
    $8.74$ &
    $0.4017\pm 0.0090~^{+0.0196}_{-0.0080}$&
    $0.3435\pm 0.0040~^{+0.0076}_{-0.0037}$
    \rbtrr\\\hline
    $14.97$&
    $0.3052\pm 0.0034~^{+0.0081}_{-0.0075}$&
    $0.2921\pm 0.0017~^{+0.0044}_{-0.0033}$
    \rbtrr\\\hline
    $17.75$&
    $0.2762\pm 0.0029~^{+0.0091}_{-0.0059}$&
    $0.2760\pm 0.0016~^{+0.0054}_{-0.0034}$
    \rbtrr\\\hline
    $23.62$&
    $0.2279\pm 0.0031~^{+0.0125}_{-0.0071}$&
    $0.2452\pm 0.0018~^{+0.0078}_{-0.0042}$
    \rbtrr\\\hline
    $36.72$&
    $0.1814\pm 0.0049~^{+0.0107}_{-0.0065}$&
    $0.2094\pm 0.0031~^{+0.0083}_{-0.0053}$
    \rbtrr\\\hline
    $57.93$&
    $0.1330\pm 0.0089~^{+0.0092}_{-0.0092}$&
    $0.1717\pm 0.0062~^{+0.0109}_{-0.0096}$
    \rbtrr\\\hline
    $81.32$
    &$0.0984\pm 0.0130~^{+0.0045}_{-0.0051}$&
    $0.1346\pm 0.0088~^{+0.0117}_{-0.0041}$\rbtrr\\\hline
  \end{tabular}\hftwo\vspace{0.5cm}
  \hftwo\begin{tabular}{|c||l|l|l|}
    \hline
    \multicolumn{1}{|c||}{$\rbthm\mean{Q}/\gev$}&
    \multicolumn{1}{|c|}{$\mean{\rho_E}$}&
    \multicolumn{1}{|c|}{$\mean{\tau_C}$}&
    \multicolumn{1}{|c|}{$\mean{C_P}$}
    \\\hline\hline
    $7.46$ &
    $0.1115\pm 0.0019~^{+0.0011}_{-0.0013}$&
    $0.1637\pm 0.0032~^{+0.0030}_{-0.0029}$&
    $0.5601\pm 0.0082~^{+0.0083}_{-0.0073}$
    \rbtrr\\\hline 
    $8.74$ &
    $0.1044\pm 0.0021~^{+0.0017}_{-0.0000}$&
    $0.1600\pm 0.0037~^{+0.0060}_{-0.0031}$&
    $0.5524\pm 0.0094~^{+0.0074}_{-0.0054}$
    \rbtrr\\\hline
    $14.97$&
    $0.0872\pm 0.0007~^{+0.0007}_{-0.0014}$&
    $0.1333\pm 0.0013~^{+0.0013}_{-0.0019}$&
    $0.4824\pm 0.0034~^{+0.0036}_{-0.0051}$
    \rbtrr\\\hline
    $17.75$&
    $0.0826\pm 0.0007~^{+0.0013}_{-0.0015}$&
    $0.1263\pm 0.0011~^{+0.0014}_{-0.0024}$&
    $0.4621\pm 0.0030~^{+0.0045}_{-0.0069}$
    \rbtrr\\\hline
    $23.62$&
    $0.0714\pm 0.0007~^{+0.0019}_{-0.0016}$&
    $0.1098\pm 0.0012~^{+0.0020}_{-0.0026}$&
    $0.4112\pm 0.0033~^{+0.0056}_{-0.0076}$
    \rbtrr\\\hline
    $36.72$&
    $0.0634\pm 0.0012~^{+0.0013}_{-0.0013}$&
    $0.0985\pm 0.0021~^{+0.0012}_{-0.0023}$&
    $0.3644\pm 0.0058~^{+0.0029}_{-0.0058}$
    \rbtrr\\\hline
    $57.93$&
    $0.0518\pm 0.0023~^{+0.0016}_{-0.0025}$&
    $0.0834\pm 0.0040~^{+0.0015}_{-0.0046}$&
    $0.3127\pm 0.0122~^{+0.0065}_{-0.0131}$
    \rbtrr\\\hline
    $81.32$&
    $0.0410\pm 0.0034~^{+0.0022}_{-0.0016}$&
    $0.0663\pm 0.0057~^{+0.0031}_{-0.0025}$&
    $0.2529\pm 0.0173~^{+0.0160}_{-0.0065}$\rbtrr\\\hline 
  \end{tabular}\hftwo\vspace{0.5cm}
  \hftwo\begin{tabular}{|c||l|l|}
    \hline
    \multicolumn{1}{|c||}{$\rbthm\mean{Q}/\gev$}&
    \multicolumn{1}{|c|}{$\mean{y_{fJ}}$}&
    \multicolumn{1}{|c|}{$\mean{y_{k_t}}$}
    \\\hline\hline
    $7.46$ &
    $0.1554\pm 0.0031~^{+0.0052}_{-0.0057}$&
    $0.2814\pm 0.0077~^{+0.0072}_{-0.0123}$
    \rbtrr\\\hline 
    $8.74$ &
    $0.1399\pm 0.0034~^{+0.0056}_{-0.0048}$&
    $0.2196\pm 0.0078~^{+0.0109}_{-0.0088}$
    \rbtrr\\\hline
    $14.97$&
    $0.1226\pm 0.0014~^{+0.0053}_{-0.0047}$&
    $0.1364\pm 0.0026~^{+0.0080}_{-0.0071}$
    \rbtrr\\\hline
    $17.75$&
    $0.1164\pm 0.0014~^{+0.0047}_{-0.0047}$&
    $0.1156\pm 0.0021~^{+0.0068}_{-0.0061}$
    \rbtrr\\\hline
    $23.62$&
    $0.1015\pm 0.0015~^{+0.0044}_{-0.0042}$&
    $0.0943\pm 0.0022~^{+0.0059}_{-0.0041}$
    \rbtrr\\\hline
    $36.72$&
    $0.0898\pm 0.0026~^{+0.0037}_{-0.0041}$&
    $0.0665\pm 0.0030~^{+0.0041}_{-0.0015}$
    \rbtrr\\\hline
    $57.93$&
    $0.0752\pm 0.0050~^{+0.0039}_{-0.0056}$&
    $0.0451\pm 0.0046~^{+0.0064}_{-0.0024}$
    \rbtrr\\\hline
    $81.32$&
    $0.0557\pm 0.0063~^{+0.0045}_{-0.0056}$&
    $0.0295\pm 0.0042~^{+0.0046}_{-0.0028}$\rbtrr\\\hline 
  \end{tabular}\hftwo
  \caption[Corrected mean values of the event shapes as a function of $Q$.]
  {Corrected mean values of the event shapes as a function of $Q$.
    The first error is statistical, the second systematic and comprises
    unfolding as well as energy scale uncertainties added quadratically.}
  \label{tab:finalmeans}
\end{table}

\chapter{NLO Integration Programs}
\label{chap:NLOint}

According to the definition of the event shapes $F$ in
section~\ref{sec:shapesdef}, they will take on non-trivial values, i.e.\ $F
\neq 0$, when derived to $\order(\as)$ at least.  In terms of cross sections,
the distributions $dn/dF$ may be written to LO accuracy as
\begin{equation}
  \frac{1}{\sigma_{\rm tot}} \frac{d\sigma}{dF} = C_1(F)\as\,.
\end{equation}
The first moment or mean value, characterizing the distribution in $F$ of an
event sample in a simple way, now reads
\begin{equation}
  \fmean := \frac{\int\limits_0^{F_{\rm max}} F\frac{d\sigma}{dF}dF}
  {\int\limits_0^{F_{\rm max}} \frac{d\sigma}{dF}dF} =
  \frac{1}{\sigma_{\rm tot}} \int\limits_0^{F_{\rm max}} F\frac{d\sigma}{dF}dF
  \,,
  \label{pertmean}
\end{equation}
which can be expanded to LO in $\as$ as well:
\begin{equation}
  \fmean = c_{1,F}\as\,.
\end{equation}

However, theoretical arguments show that the achieved precision of LO
calculations is not sufficient to extract much information on pQCD from
experimental data.  Especially the concept of a {\em running coupling
  constant}\/ $\as(\mr)$, explained in the next section, requires the
distributions and mean values to be known at least to NLO,
i.e.~$\order(\as^2)$.

\section{The Running Coupling Constant}
\label{sec:runcoupl}

Following an approach from~\cite{ESW}, one may assume that a dimensionless
quantity $\fmean$ depends on a single energy scale $Q$ which is much larger
than any other parameter measured in units of energy such that quark masses
may be set to zero.\footnote{Note that this step requires a more careful
  consideration, cf.\ the ref.\ given above.}  Then, the conclusion is that
$\fmean$ has a constant value independent of $Q$; there is no other
possibility to remove the dimension of $Q$. However, this is not true in a
renormalizable quantum field theory such as QCD.

Perturbative calculations operating with \qi naked\qo\ charges $e_0$ (or
masses $m_0$) of point particles have to deal with ultraviolet divergences
occurring in the integrations of loop momenta $d^4k/(2\pi)^4$. The procedure
allowing us to absorb these divergences in physical charges $e$ is called {\em
  renormalization}.  Results concerning the (measured) physical charges are
then all finite.

There exist several recipes how this can be done~\cite{Collins}. One of the
most popular, also adopted in this analysis, is the {\em modified minimal
  subtraction}\/ scheme $\overline{\rm MS}$, where the singularities are
regulated by a reduction of the space-time dimensions to $n<4$:
\begin{equation}
  \frac{d^4k}{(2\pi)^4} \rightarrow
  \mu^{2\epsilon}\frac{d^{4-2\epsilon}k}{(2\pi)^{4-2\epsilon}}\,.
\end{equation}
At an arbitrary scale called {\em renormalization scale}\/ $\mr$, the
integrals are subdivided into finite and divergent parts for $\epsilon
\rightarrow 0$. The singular terms (and some finite ones depending on the
renormalization scheme) are then attributed to a renormalized charge $e$.
However, the remaining finite terms also modify the charge $e$ or equivalently
the coupling constant $\alpha := e^2/4\pi$, which now explicitly varies with
$\mr$!

Relating this to QCD, we have a strong coupling constant $\as := g^2/4\pi$
containing $\mr$ ---~a second mass scale that enters the game.  $\fmean$ may
therefore depend on the ratio $Q^2/\mr^2$. But since $\mr$ is an arbitrary
parameter of the renormalization procedure, it should not make an appearance
in physical observables like $\fmean$.  This can be expressed mathematically
by
\begin{equation}
  \mr^2 \frac{d}{d\mr^2} \fmean \left( Q^2/\mr^2, \as\right) =
  \left[ \mr^2 \frac{\partial}{\partial \mr^2} + \mr^2
    \frac{\partial \as}{\partial \mr^2}\frac{\partial}{\partial\as}\right]
  \fmean \left( Q^2/\mr^2, \as\right) = 0
\end{equation}
or, using the notation
\begin{eqnarray}
  t &=& \ln\left( \frac{Q^2}{\mr^2}\right)\,,\\
  \nonumber&&\\
  \beta(\as) &=& \mr^2\frac{\partial \as}{\partial\mr^2}\,,
  \label{eqn:RGE}
\end{eqnarray}
as
\begin{equation}
  \left[ -\frac{\partial}{\partial t} + \beta(\as) \frac{\partial}{\partial\as}
    \right] \fmean (e^t,\as) = 0\,.
    \label{eqn:mrind2}
\end{equation}
A solution to this partial differential equation can be given implicitly by
defining the running coupling $\as(Q^2)$ as
\begin{equation}
  t = \int\limits_{\as}^{\as(Q^2)} \frac{dx}{\beta(x)}\,,\quad
  \as := \as(\mr^2)
  \label{eqn:asq2}
\end{equation}
such that $\fmean(1,\as(Q^2))$ fulfils eq.~(\ref{eqn:mrind2}).  The scale
dependence of any $\fmean$ calculated perturbatively now follows from
$\as(Q^2)$ provided eq.~(\ref{eqn:asq2}) could be solved.  Since QCD is an
asymptotically free theory, i.e.\ $\lim\limits_{Q\rightarrow\infty}\as(Q^2) =
0$, this can also be achieved by expanding the $\beta$-function of the {\bf
  R}enormalization {\bf G}roup {\bf E}quation~(\ref{eqn:RGE}) (RGE) in powers
of $\as$:
\begin{equation}
  \frac{\beta(\as)}{4\pi} = - \sum\limits_{n=0}^{\infty} \beta_n
  \left( \frac{\as}{4\pi}\right)^{n+2}\,.
\end{equation}
The first two coefficients defined that way are
\begin{eqnarray}
  \label{eqn:beta0}
  \beta_0 &=& \frac{33-2N_f}{3}\,,\\
  \nonumber&&\\
  \beta_1 &=& \frac{306-38N_f}{3}\,,
  \label{eqn:beta1}
\end{eqnarray}
where $N_f$ is the number of active flavours. $\beta_2$ and further
coefficients are scheme dependent.

Retaining only the first term, the 1-loop solution for $\as(Q)$ can be written
as
\begin{equation}
  \as(Q) = \frac{\as(\mr)}{1+\frac{\beta_0}{2\pi}\as(\mr)
    \ln\left( \frac{Q}{\mr} \right)}\,.
\end{equation}
Using this equation, $\as$ can be evaluated at any (sufficiently high) scale
provided it is known at one point, say $M_Z$. Historically, a dimensional
parameter $\Lambda$ derived from
\begin{equation}
  \ln\frac{Q^2}{\Lambda^2} =
  -\int\limits_{\as(Q^2)}^{\infty}\frac{dx}{\beta(x)}
\end{equation}
was determined instead of $\asmz$.  The 1-loop formula then reads
\begin{equation}
  \as(Q) = \frac{2\pi}{\beta_0\ln\left(\frac{Q}{\Lambda}\right)}
\end{equation}
and demonstrates that $\Lambda$ indicates the scale at which the coupling
would diverge if extrapolated outside the perturbative regime.

In principle, both approaches are equivalent and any measurement of $\asmz$
can be converted into a corresponding value for $\Lambda$ and vice versa.  The
latter, however, has some disadvantages: It is not dimensionless, it depends
on $N_f$ and on the renormalization scheme\footnote{That is, it has to be
  labelled e.g.\ $\Lambda_{N_f,\overline{\rm MS}}$.}  and ---~most
important~--- there exist two slightly different definitions for $\Lambda$ in
the literature~\cite{PDG,ESW}.

\pagebreak\noindent
Except for the consistency check in section~\ref{sec:xdep}
employing $\lmsb$, we therefore follow ref.~\cite{DIS:as2loop} and apply the
2-loop equation according to
\begin{equation}
  \as(Q) = \frac{\asmz}{1+\asmz\cdot L^{(n)}\left(\frac{Q}{M_Z}\right)}
\end{equation}
where
\begin{equation}
  L^{(1)}\left(\frac{Q}{M_Z}\right) =
  \frac{\beta_0}{2\pi}\ln\frac{Q}{M_Z}
\end{equation}
reproduces the 1-loop formula above and
\begin{equation}
  L^{(2)}\left(\frac{Q}{M_Z}\right) =
  \left(\frac{\beta_0}{2\pi} + \frac{\beta_1}{8\pi^2}\asmz\right)
  \ln\frac{Q}{M_Z}
\end{equation}
gives the 2-loop result.

Returning to $\fmean$, the perturbative expression to all orders
\begin{equation}
  \fmean = \sum\limits_{n=1}^{\infty}c_n(\mr)\as^n(\mr)
  \label{eqn:expansion}
\end{equation}
does not depend on $\mr$. But we do not know all coefficients, and therefore
any approximation by a truncated series does.  To be more precise,
\begin{equation}
  \mr^2\frac{d}{d\mr^2} \sum\limits_{n=1}^N c_n(\mr)\as^n(\mr)
  \propto \order(\as^{N+1}(\mr))\,.
\end{equation}
According to current knowledge, the dependence on $\mr$ is much reduced if
more terms are inserted.  Hence, the theoretical uncertainty usually
associated with the choice of scale, whose variation reflects the unknown
higher orders, is diminished. Up to now, we merely considered $c_1$, which
itself does not contain $\mr$, and the complete $\mr$-dependence stems from
$\as(\mr)$ and is rather large. It is reduced with the inclusion of
$c_2(\mr)\as^2(\mr)$ with
\begin{equation}
  \label{eqn:c2mrdep}
  c_2(Q) = c_2(1) + \frac{\beta_0}{2\pi}\ln\frac{\mr}{Q}c_1\,,
\end{equation}
which can be calculated by the three NLO integration programs presented in the
next sections.

\section{NLO Integration Techniques}
\label{sec:NLOtech}

\begin{figure} 
  \centering
  \includegraphics{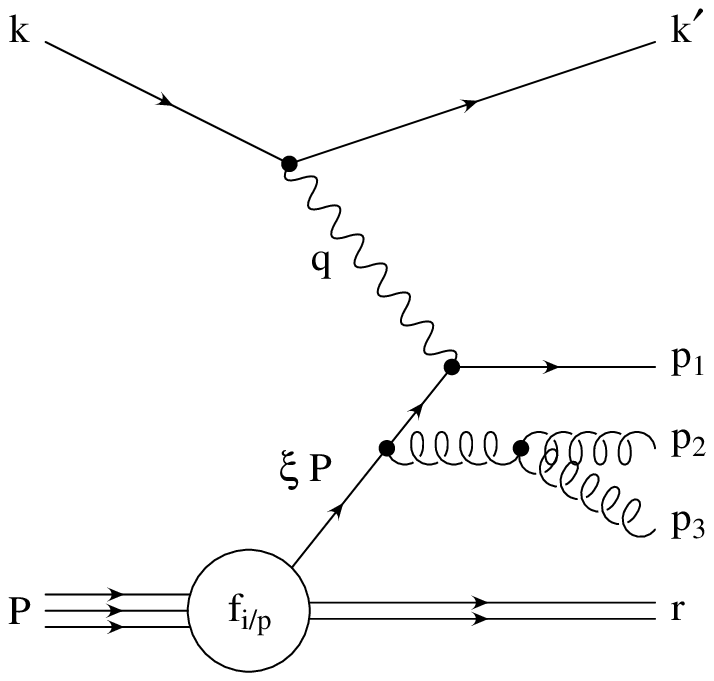}\hftwo%
  \includegraphics{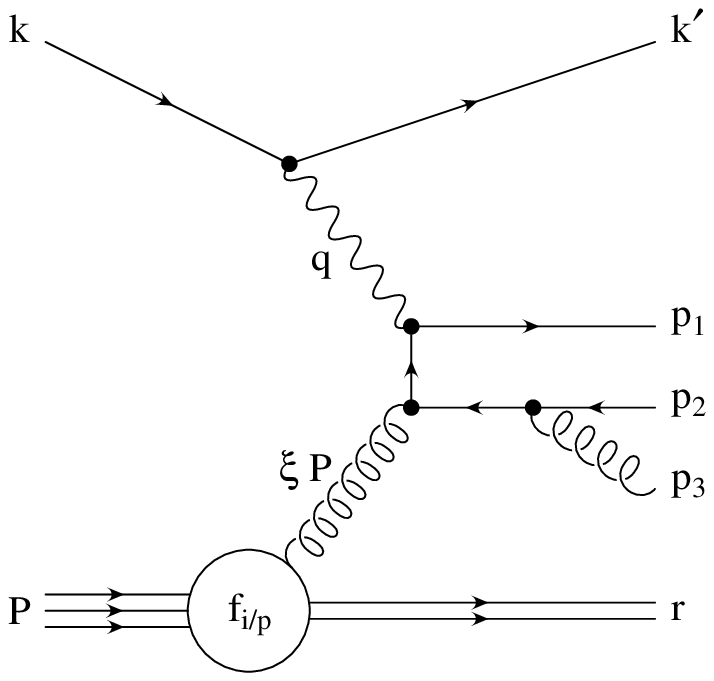}
  \caption[Examples of real corrections to QCDC and
  BGF Feynman graphs.]  {Examples of real corrections to QCDC (left) and BGF
    (right) Feynman graphs.}
  \label{fig:realcorr}
\end{figure}

As mentioned in section~\ref{sec:NLOcross} on the cross section to
$\order(\as)$, again infrared and collinear divergences are present in
calculations to $\order(\as^2)$ which cancel between real and virtual
diagrams. Examples of real corrections to the QCDC and BGF processes with now
three final state partons are given in fig.~\ref{fig:realcorr}.

In order to eliminate the occurring singularities and obtain numerically
stable results, the necessary cancellations were performed analytically for a
certain jet algorithm in the programs DISJET~\cite{NLO:DISJET} and
PROJET~\cite{NLO:PROJET}.  However, these programs lack the flexibility to
evaluate NLO corrections for arbitrary infrared-safe observables and they
contain approximations that are inappropriate for certain phase space regions.
Therefore, more elaborate methods have been developed that ensure the
cancellations and evaluate numerical results for arbitrary observables.
Basically, two such schemes are known: the {\em phase space slicing}\/
method~\cite{NLO:slicing} and the {\em subtraction}\/ method~\cite{def:C}.

In the phase space slicing method, the separation of singular from
non-singular phase space regions is done by surrounding partons with a small
cone. Real emissions of partons outside the cone defined by $s_{ij} = (p_i +
p_j)^2 > s_{\rm min}$, where $s_{\rm min} \approx 0.1 \gev^2$, are computed
exactly. Inside the cone, soft and collinear approximations are used. A
drawback of this procedure is the possibility of a residual dependence on the
technical parameter $s_{\rm min}$.  For any calculation it has to be
established that a plateau has been reached for small enough cut-offs.  In
addition, this scheme prevents the integration of the event shape
distributions $Fd\sigma/dF$ to be carried out over the whole phase space from
zero to $F_{\rm max}$.  Instead, a lower bound $F_{\rm cut} > 0$ has to be
imposed and $s_{\rm min}$ has to be adjusted accordingly.  Moreover, the
introduction of {\em crossing functions}~\cite{NLO:crossfunc}, that have to be
evaluated for every parton density parameterization, is required.

In case of the subtraction method, the singularities are cancelled
point-by-point in phase space. A cut-off like $s_{\rm min}$ is not formally
required by this scheme, although a tiny value of $s_{\rm min} \approx
10^{-8}$--$10^{-10}\gev^2$ is applied to set the number of significant digits.
Hence, the event shape distributions $Fd\sigma/dF$ can be integrated down to
zero~\cite{prc:Sey1}.

\begin{figure} 
  \centering
  \includegraphics{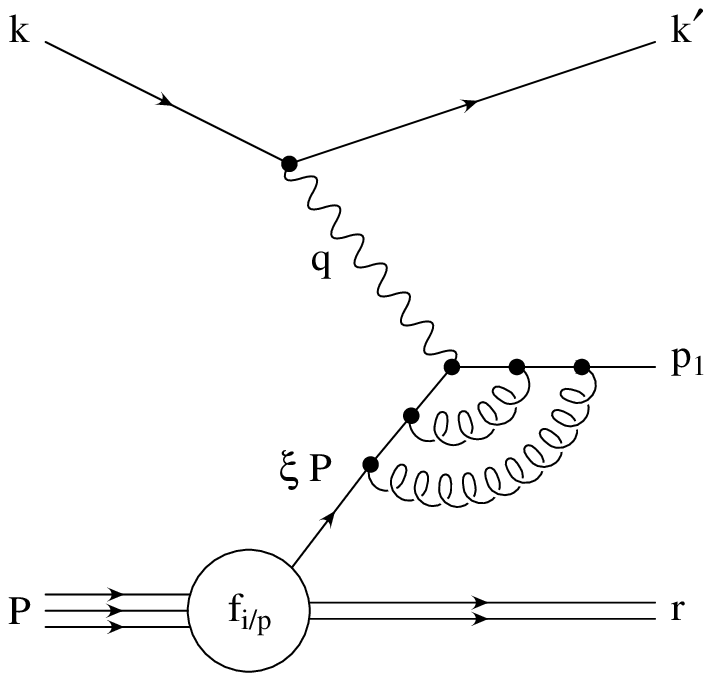}\hftwo%
  \includegraphics{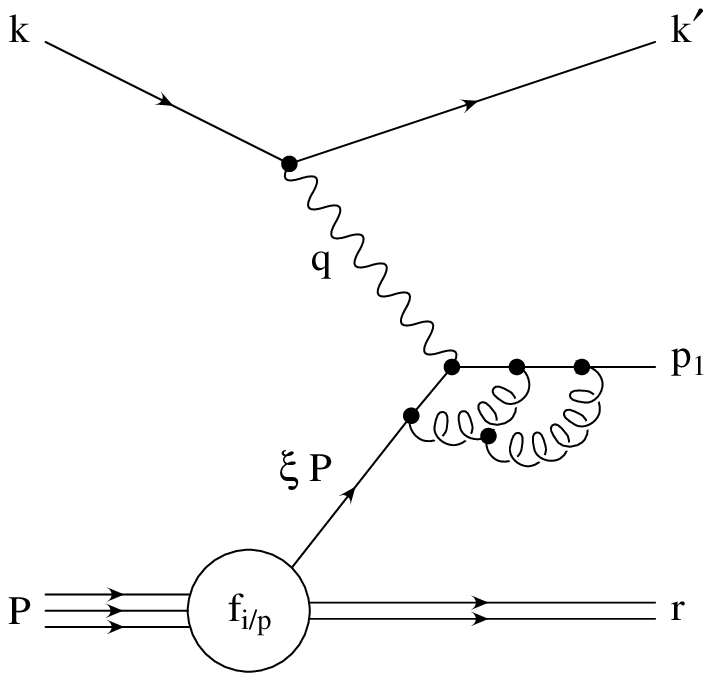}
  \caption{Examples of 2-loop virtual correction Feynman graphs.}
  \label{fig:2loop}
\end{figure}

The first method is employed in the NLO integration program
MEPJET~\cite{NLO:MEPJET}, whereas the subtraction procedure is used in
DISENT~\cite{NLO:DISENT} and DISASTER++~\cite{NLO:DISASTER}.  All three
programs, however, do not contain the full set of diagrams to $\order(\as^2)$.
Additional 2-loop virtual corrections from graphs like the ones in
fig.~\ref{fig:2loop} appear and are difficult to handle.  Yet, they contribute
as interference term to the Born process only and therefore do not affect the
event shape distributions except for the bin at zero. The integral
$\int_0^{F_{\rm max}} F\frac{d\sigma}{dF} dF$ still contains integrable
divergences, but they are not harmful for DISENT or
DISASTER++~\cite{prc:GrauSey2}. A calculation of the total cross section
$\int_0^{F_{\rm max}}\frac{d\sigma}{dF} dF$, needed for the normalization, to
$\order(\as^2)$ (i.e.\ NNLO), however, is not possible.  A Taylor expansion of
eq.~(\ref{pertmean})
\begin{eqnarray}
  \fmean &=& \frac{\int\limits_0^{F_{\rm max}} F\frac{d\sigma}{dF}dF}
  {\int\limits_0^{F_{\rm max}} \frac{d\sigma}{dF}dF}\\
  &\approx& \frac{a_1\as^1 + a_2\as^2}{b_0\as^0 + b_1\as^1 + b_2\as^2}\\
  &=& \frac{a_1}{b_0}\as^1 + \frac{a_2-\frac{a_1b_1}{b_0}}{b_0} \as^2
    + \order(\as^3)
\end{eqnarray}
shows that, owing to the property of $F$ to be zero in the QPM limit and hence
$a_0=0$, $\sigma_{\rm tot}$ needs only be known to $\order(\as)$ for $\fmean$
to be still correct to $\order(\as^2)$.

\section{First Comparisons of NLO Programs}
\label{sec:NLOcomp}

\begin{figure} 
  \centering
  \includegraphics{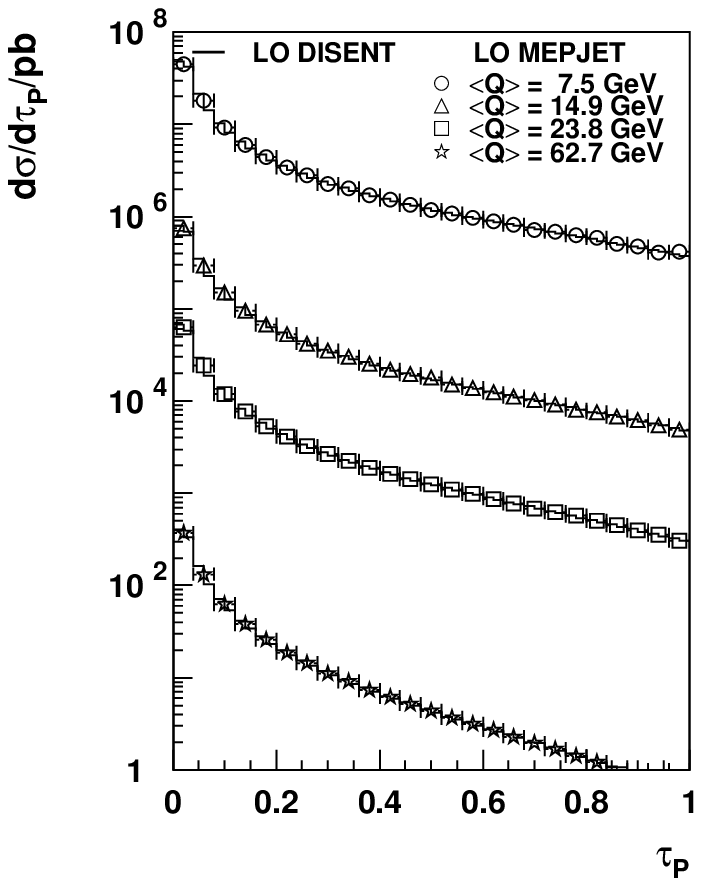}\hftwo%
  \includegraphics{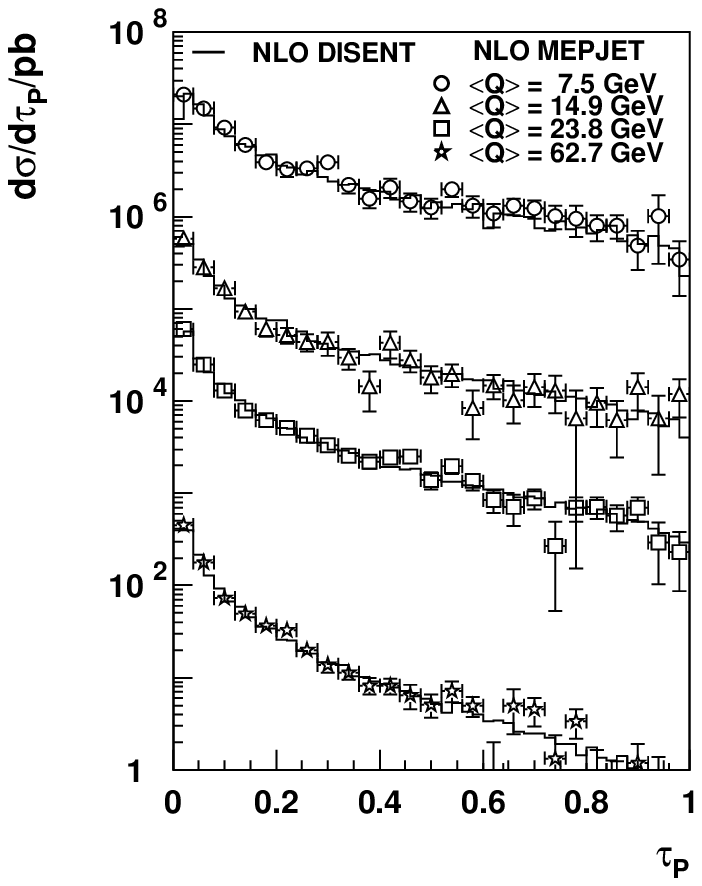}
  \caption[Differential distributions of $\tau_P$ at LO and NLO
  for MEPJET~1.4 and DISENT~0.0.]  {Differential distributions of $\tau_P$ for
    four out of seven investigated bins in $Q$. LO (left) and NLO (right) pQCD
    calculations of MEPJET~1.4 (hollow symbols) and DISENT~0.0 (full lines)
    are compared.  The spectra for $\mean{Q} = 7.5$--$62.7\gev$ are multiplied
    by factors of $10^n$, $n=0,1,2,3$.}
  \label{fig:NLOcomparison}
\end{figure}

In view of the complexity of the calculations and their implementations into
program code, it is mandatory to compare the results where possible. A first
comparison of MEPJET~1.4 and DISENT~0.0 was performed in~\cite{KR:DIS97}. Due
to the phase space slicing method applied in MEPJET, the differential
distributions for $F > F_{\rm cut} > 0$ only could be considered.  In order to
keep $F_{\rm cut}$ small, a cut-off of $s_{\rm min} = 0.01\, \gev^2$ instead
of $0.1\,\gev^2$ was used. Nevertheless, the left-most bins extending down to
zero should not be taken seriously, even if they accidentally do agree.
Fig.~\ref{fig:NLOcomparison} shows a perfect agreement of the LO spectra and,
within the calculational precision of $3$--$5\%$, compatible results for the
NLO distributions.

With the advent of DISASTER++, a check of all three programs against each
other was carried out and some discrepancies have been
revealed~\cite{NLO:DISASTER}.  Partially, these are understood and meanwhile
an improved version DISENT~0.1 has been released.  In event shape spectra like
fig.~\ref{fig:NLOcomparison}, effects are hardly visible, but the mean values
have increased considerably in the \lowq\ region compared to the calculations
that entered in~\cite{H1:Shapes}.  The worst case was the jet broadening
$B_P$, whose mean value at $Q\approx 7.5\gev$ rose by about $14\%$.

The most recent program versions are MEPJET~2.2, DISENT~0.1 and
DISASTER++~1.0.1. In this analysis only DISENT will be used, but more
elaborate comparisons are under way in the DESY workshop on {\em Monte Carlo
  Generators for HERA Physics}.


\chapter{DISENT Results}
\label{chap:DISENT}

DISENT, the NLO integration program mainly employed here, is structured
similarly to a MC event generator. A main routine generates configurations of
parton momenta in the allowed phase space region ---~for our purposes defined
by the cuts nos.~\ref{cut:Q2}--\ref{cut:QBreit} of
section~\ref{sec:fincut}~---, stores them in an event record and calculates
the weight according to the matrix elements. Consequently, a user routine may
analyze these events with respect to arbitrary infrared and collinear safe
observables.  For physical cross sections, the weights have to be multiplied
by the strong coupling constant raised to the proper power and by weights
extracted from the pdfs of the proton.  These are not inherent to DISENT but
must be gathered from somewhere else, e.g.\ the PDFLIB~\cite{PDFLIB}. As
standard options we use the set MRSA'-115~\cite{QCD:MRSAp} and $N_f=5$ as
number of active flavours.  Note that DISENT currently neglects quark masses
as well as electroweak corrections proportional to $Q^2/(Q^2+M_Z^2)$ and
$(Q^2/(Q^2+M_Z^2))^2$.  Fortunately, only the highest $Q$-bin investigated
with $\mean{Q}=81.32\gev$ may be affected noticeably~\cite{H1:ICHEP98hq}.  In
the near future, DISENT may be extended to include the electroweak
terms~\cite{prc:Sey2}.

In distinction to MC generators, however, several different but correlated
event records, in total $14$, are produced for {\bf one} DISENT \qi
event.\qo\, They correspond to contributions of real and virtual diagrams plus
counter terms that ensure the necessary cancellation of divergences. These
contributions have coupled weights that are positive as well as negative. For
the error evaluation, it is mandatory that these weights are allowed to
counterbalance each other before finally e.g.\ being histogrammed.  This may
be done by creating an intermediate array or histogram, where only the
contributions of one event are entered. All bins not empty at the end are then
transferred to the final array with their respective weights. A more detailed
program description can be found in~\cite{NLO:DISENTmanual}.

\section{Evaluation of the Statistical Uncertainty}
\label{sec:staterr}

Given the framework of the DISENT program, the calculation of differential
distributions $d\sigma/dF$ and mean values $\fmean$ is straightforward.
Positive and negative weights of the different contributions are summed up.
Yet, the evaluation of the statistical uncertainties, where the technique of
intermediate histograms is presupposed, is more involved.

First, we want to consider cross sections that can be written as a sum of
weights $w_i$, given by DISENT, which are already normalized to the number of
events $N$:
\begin{equation}
  \sigma = \sum\limits_{i=1}^N w_i\,.
\end{equation}
By reintroducing unnormalized weights $v_i$, one obtains
\begin{equation}
  \sigma = \frac{1}{N}\sum\limits_{i=1}^N v_i = \mean{v}\,,
\end{equation}
demonstrating that $\sigma$ is computed analogously to a mean value.  Its
statistical uncertainty can be estimated by
\begin{eqnarray}
  \nonumber\Delta\sigma &=& \sqrt{\frac{1}{N-1} \left( \mean{v^2}
      - \mean{v}^2 \right)}\\
  \nonumber&&\\
  &=& \sqrt{\frac{1}{N-1} \left( \frac{1}{N} \sum\limits_{i=1}^N v_i^2 -
      \frac{1}{N^2} \left( \sum\limits_{i=1}^N v_i \right)^2 \,\right)}\\
  \nonumber&&\\
  \nonumber&=& \sqrt{\frac{1}{N-1} \left( N\sum\limits_{i=1}^N w_i^2 -
      \left( \sum\limits_{i=1}^N w_i \right)^2 \,\right)}\,.
  \label{eqn:error}
\end{eqnarray}
With $N\approx N-1$, this finally gives
\begin{equation}
  \Delta\sigma = \sqrt{\sum\limits_{i=1}^N w_i^2 - 
    \frac{\left( \sum\limits_{i=1}^N w_i \right)^2}{N}}\,,
  \label{eqn:sigmaerror}
\end{equation}
which in the limit of very large $N$ becomes
\begin{equation}
  \Delta\sigma = \sqrt{\sum\limits_{i=1}^N w_i^2}\,.
  \label{eqn:sigmaerror2}
\end{equation}
For our purposes the second term is not negligible and
formula~(\ref{eqn:sigmaerror}) has to be applied.  Turning to the event shape
means
\begin{equation}
  \fmean = \frac{\sum\limits_{i=1}^N F_iw_i}{\sum\limits_{i=1}^Nw_i^{'}}
  \label{eqn:mean}
\end{equation}
one could be misled to the assumption that a similar uncertainty as in
eq.~(\ref{eqn:error}) could be derived here.  Yet, formula~(\ref{eqn:mean})
emphasizes that the weights $w_i$ and $w_i^{'}$ are not identical because the
numerator is evaluated to $\order(\as^2)$ and the cross section in the
denominator to $\order(\as)$ only. Instead, the uncertainties have to be
estimated separately, and afterwards the error of the quotient can be obtained
by the usual formulae. The final result reads:
\begin{equation}
  \Delta\fmean = \fmean\sqrt{\frac{\sum\limits_{i=1}^N\left(F_iw_i\right)^2}
    {\left(\sum\limits_{i=1}^NF_iw_i\right)^2} +
    \frac{\sum\limits_{i=1}^Nw_i^{'2}}{\left(\sum\limits_{i=1}^N
        w_i^{'}\right)^2}-\frac{2}{N}}\,.
\end{equation}

\begin{figure}
  \centering
  \begin{minipage}{0.47\textwidth}
    \includegraphics{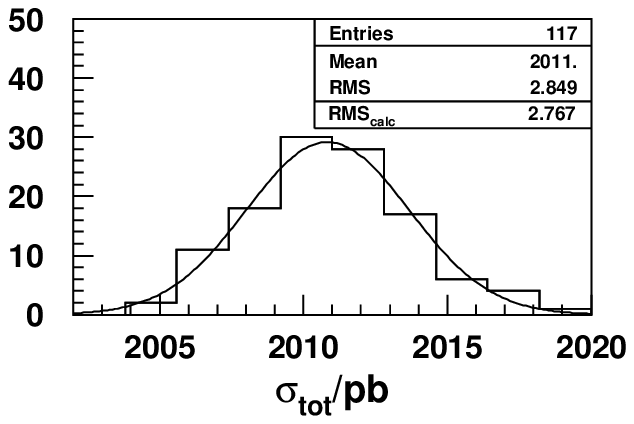} \includegraphics{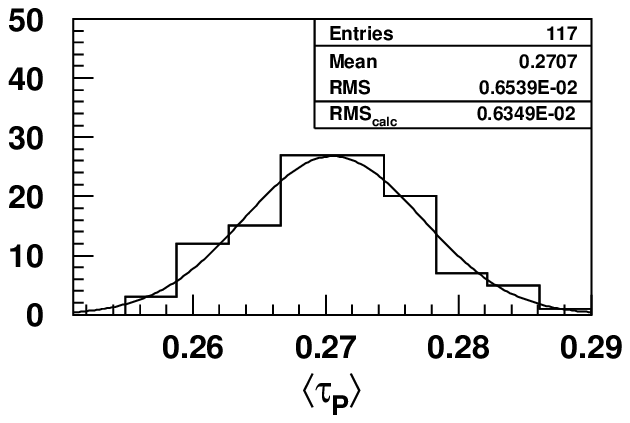}
  \end{minipage}\hftwo%
  \begin{minipage}{0.47\textwidth}%
    \includegraphics{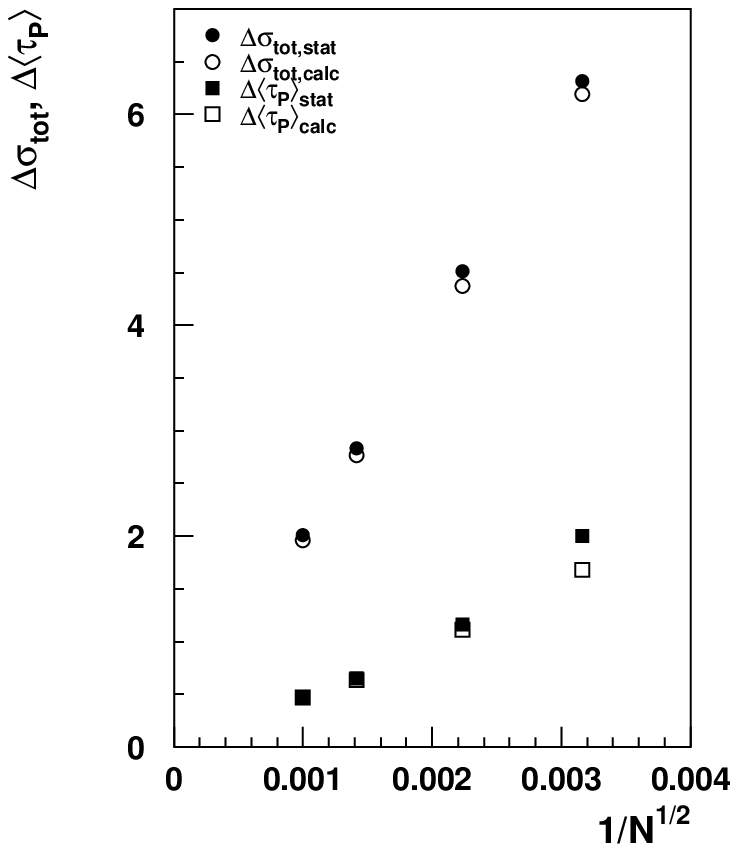}
  \end{minipage}
  \caption[Histograms of $\sigma_{\rm tot}$ and $\mean{\tau_P}$
  resulting from repeated DISENT~0.0 calculations and the dependence of their
  statistical uncertainties on $1/\sqrt{N}$.]  {Left: Histograms of
    $\sigma_{\rm tot}$ (top) and $\mean{\tau_P}$ (bottom) plus Gaussian fit
    resulting from $117$ DISENT~0.0 calculations with half a million events
    each.  ${\rm RMS}_{\rm calc}$ denotes the on average estimated standard
    deviation.  Right: Dependence on $1/\sqrt{N}$ of the statistical (full
    symbols) and computational (hollow symbols) uncertainty estimates for
    $\sigma_{\rm tot}$ and $\mean{\tau_P}$.}
  \label{fig:statuncertainty}
\end{figure}

As a cross-check, the same DISENT~0.0 calculation has been repeated about a
hundred times with different seeds of the random number generator for $0.1$,
$0.2$, $0.5$ and $1.0$ million generated parton events each.
Fig.~\ref{fig:statuncertainty} shows on the left-hand side histograms of the
obtained cross sections $\sigma_{\rm tot}$ and mean values $\mean{\tau_P}$
together with the fit of a Gaussian distribution for half a million events.
It describes the distribution of the results well and its standard deviation
is compatible with the deduced uncertainty ${\rm RMS}_{\rm calc}$. The
right-hand side presents the dependence of the uncertainties for both, the
statistical and the computational derivation, on the number of events
produced. It nicely exhibits errors shrinking proportional to $1/\sqrt{N}$.
The computation underestimates only slightly the statistically achieved
variance.

The error bars in differential distributions are evaluated according to
eq.~(\ref{eqn:sigmaerror2}).

\section[$\symbol{120}$-Dependence]
{\boldmath$\symbol{120}$\unboldmath\ -Dependence}
\label{sec:xdep}

$\asmz$ is the fundamental parameter of QCD that has to be determined by
experiment. An assumption on it, however, is already contained in our cross
section and mean value calculations. Therefore, they cannot directly be used
to extract information from our data.  In~\cite{KR:DIS97} and~\cite{H1:Shapes}
this problem was avoided by fitting eq.~(\ref{eqn:expansion}) up to
$\order(\as^2)$ to the obtained mean values of each event shape. The resulting
coefficients $c_{1,F}$ and $c_{2,F}$ are subsequently employed for an
$\asmz$-dependent parameterization of the event shape means.

\begin{figure} 
  \centering
  \includegraphics{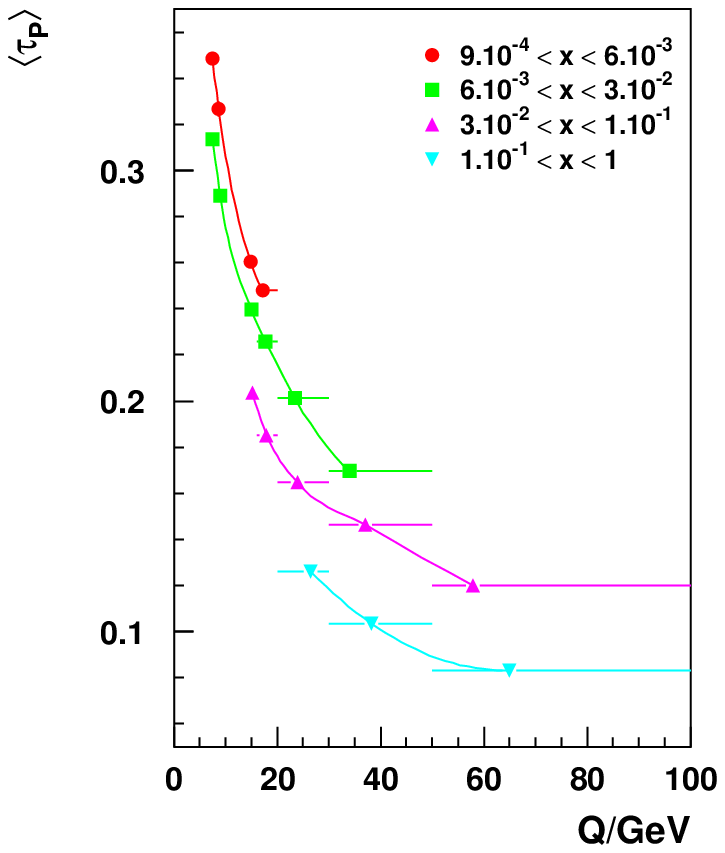}\hftwo%
  \includegraphics{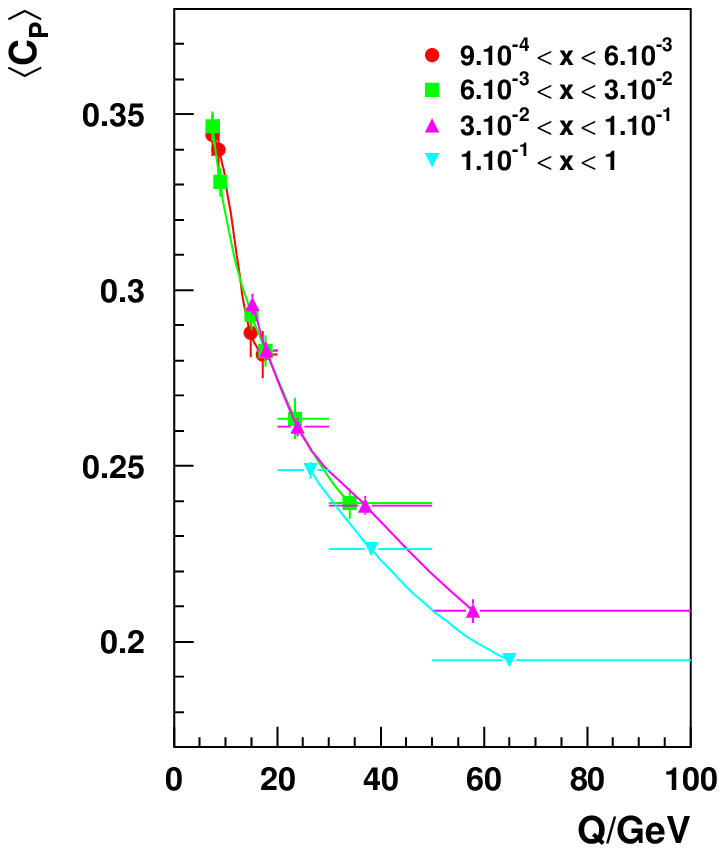}
  \caption[Mean values of $\tau_P$ and $C_P$ versus $Q$
  in four different bins of $x$.]  {Mean values of $\tau_P$ (left) and $C_P$
    (right) versus $Q$ in four different bins of $x$.}
  \label{fig:xdep}
\end{figure}

In contrast to $e^+e^-$ physics, where $c_{1,F}$ and $c_{2,F}$ truly are pure
numbers (s.~e.g.~\cite{LEP:YRLEP1}), this is not the case in $ep$ DIS because
in addition to the matrix elements $x$-dependent pdfs enter their evaluation.
In general, the coefficients will therefore be \boldmath{\bf functions of
  $x$}\unboldmath\ which was taken into account in an approximate way only by
the above ansatz.  Fig.~\ref{fig:xdep} presents as examples for a strong
---~the \qi worst\qo\ case being $y_{k_t}$~--- and a weak variation of the
mean values with $x$ curves of $\mean{\tau_P}$ and $\mean{C_P}$ versus $Q$ for
four different bins in $x$.  Note that in this investigation the two highest
$Q$-bins were merged.

A solution to this problem is the direct calculation of $c_{1,F}$ and
$c_{2,F}$ for every $Q$-bin with its specific $x$-range separately. This
method corresponds to an approximation of functions $c_{n,F}(x)$ by step
functions instead of one suitably chosen number.

\begin{figure}
  \centering
  \includegraphics{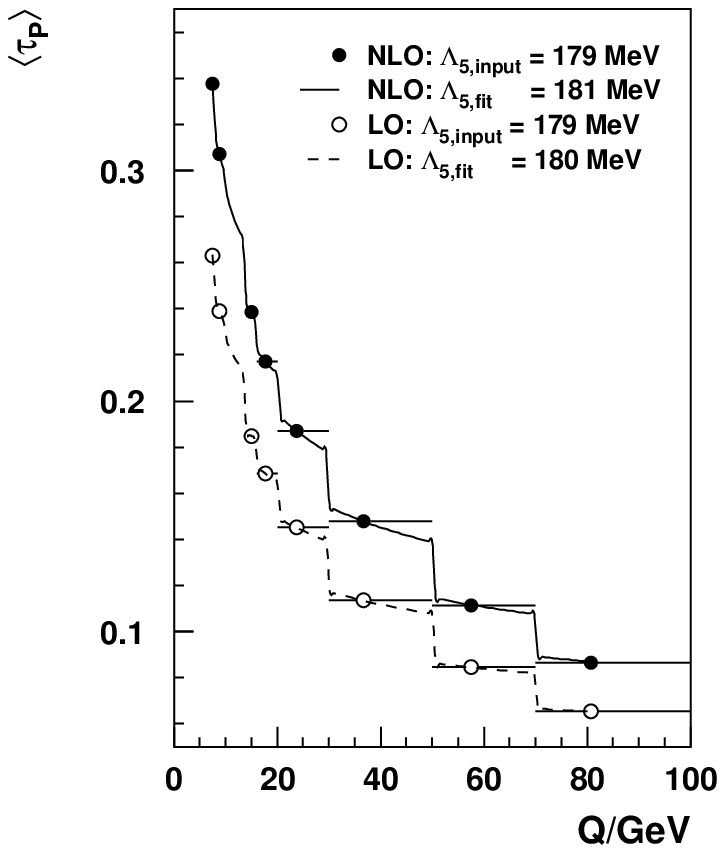}\hftwo%
  \includegraphics{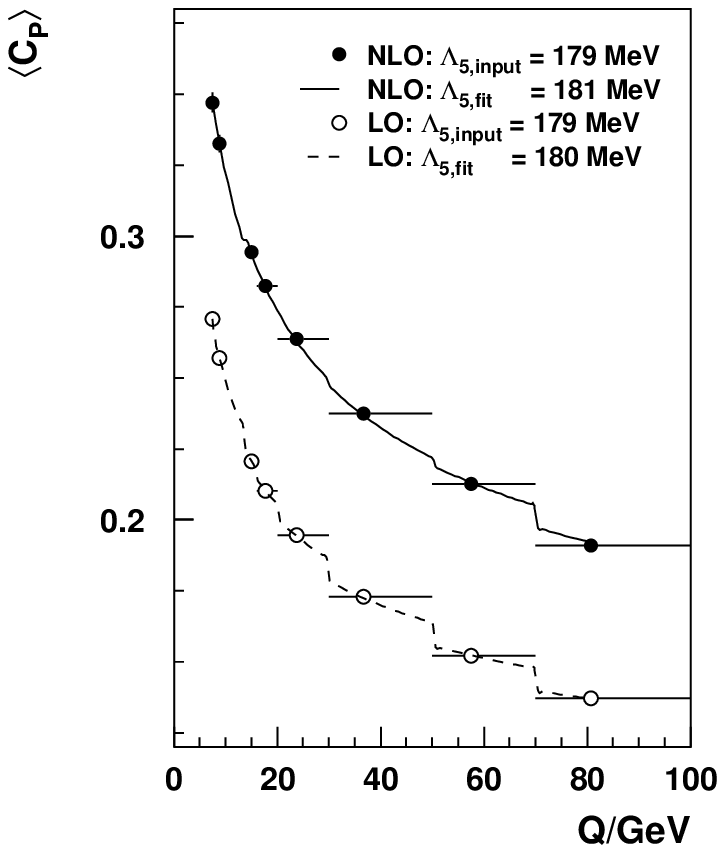}
  \caption[Fits of $\lmsb$ for the LO and NLO means of $\tau_P$ and $C_P$.]
  {Fits (dashed and full lines) of $\lmsb$ for the LO (hollow symbols) and NLO
    (full symbols) means of $\tau_P$ (left) and $C_P$ (right). The steps
    express the $x$-dependence of the perturbative coefficients $c_{1,F}$ and
    $c_{2,F}$.}
  \label{fig:refit}
\end{figure}

As a consistency check, the perturbative mean values at LO and NLO are fit
with the coefficients derived in the same computation.  The outcome for
$\lmsb$ should be identical to the initial choice, that is $\lmsb = 179\,\mev$
for MRSA'-115\@.  The achieved fits for $\mean{\tau_P}$ and $\mean{C_P}$ are
displayed in fig.~\ref{fig:refit}. Again, the much steeper steps for
$\mean{\tau_P}$ indicate its stronger $x$-dependence. The $\lmsb$ values
extracted turn out to be very well compatible with the expectation for all
event shapes except for the jet broadening $B_P$ in NLO\@! This may be a hint
that there is a problem in the DISENT~0.1 calculations at \lowq.

\section{Final Perturbative Coefficients}
\label{sec:fincoeff}

The final results obtained with DISENT~0.1 are compiled in the
tables~\ref{tab:disentsigma}--\ref{tab:disentmeanykt}. The first one contains
the mean $Q$- and $x$-values to NLO for the eight bins of
section~\ref{sec:fincut} as well as the LO and NLO cross sections.  The latter
are differentiated by indices \qi 1H\qo\ and \qi 2H\qo\ to indicate whether
the CH alone (1H, energy cut no.~\ref{cut:QBreit}) or both hemispheres (2H, no
energy cut) are allowed to contribute to the hadronic final state.  To LO
there is no difference because the only parton available has $E\Bf = Q/2$ and
runs into the CH by definition.

Tables~\ref{tab:disentmeantaup}--\ref{tab:disentmeanykt} comprise the mean
values of all event shapes for our standard pdf set MRSA'-115 and the
perturbative coefficients $c_{1,F}$ and $c_{2,F}$ needed for the
$\asmz$-dependent parameterization of the means.

\begin{table}
  \centering
  \begin{tabular}{|c||r|r|r|r|}
    \hline
    \multicolumn{1}{|c||}{\rbthm$\mean{Q}/\gev$} & 
    \multicolumn{1}{|c|}{$\mean{x}$} & 
    \multicolumn{1}{|c|}{$\sigma_{\rm LO}/pb$} &
    \multicolumn{1}{|c|}{$\sigma_{\rm NLO,1H}/pb$} &
    \multicolumn{1}{|c|}{$\sigma_{\rm NLO,2H}/pb$} \\\hline\hline
    $7.46$ & $4.58\cdot10^{-3}$ &
    $2519.9\pm 0.3$ & $2232.5\pm 0.4$ & $2402.6\pm 0.4$\rbtrr\\\hline
    $8.80$ & $7.21\cdot10^{-3}$ &
    $2357.5\pm 0.4$ & $2111.9\pm 0.5$ & $2245.8\pm 0.5$\rbtrr\\\hline
    $14.95$ & $1.60\cdot10^{-2}$ &
    $534.4\pm 0.1$ & $484.4\pm 0.1$ & $503.8\pm 0.1$\rbtrr\\\hline
    $17.73$ & $2.16\cdot10^{-2}$ &
    $584.7\pm 0.1$ & $533.4\pm 0.1$ & $551.1\pm 0.1$\rbtrr\\\hline
    $23.75$ & $3.16\cdot10^{-2}$ &
    $465.8\pm 0.1$ & $428.4\pm 0.1$ & $438.9\pm 0.1$\rbtrr\\\hline
    $36.69$ & $5.80\cdot10^{-2}$ &
    $156.2\pm 0.1$ & $145.9\pm 0.1$ & $147.9\pm 0.1$\rbtrr\\\hline
    $57.61$ & $1.09\cdot10^{-1}$ &
    $24.79\pm 0.02$ & $23.63\pm 0.02$ & $23.75\pm 0.02$\rbtrr\\\hline
    $80.76$ & $1.74\cdot10^{-1}$ &
    $7.502\pm 0.006$ & $7.291\pm 0.006$ & $7.305\pm 0.006$\rbtrr\\\hline
  \end{tabular}
  \caption[Mean $Q$- and $x$-values as well as the LO and NLO
  cross sections.]
  {Mean $Q$- and $x$-values as well as the LO and NLO
    cross sections as calculated by DISENT~0.1.
    The indices \qi 1H\qo\ and \qi 2H\qo\ differentiate between
    cross sections with and without the minimal energy cut in the
    CH\@.}
    \label{tab:disentsigma}
  \end{table}

\begin{table}
  \centering
  \begin{tabular}{|c||r|r|r|r|}
    \hline
    \multicolumn{1}{|c||}{\rbthm$\mean{Q}/\gev$} &
    \multicolumn{1}{|c|}{$\mean{\tau_P}_{\rm LO}$} & 
    \multicolumn{1}{|c|}{$\mean{\tau_P}_{\rm NLO}$} & 
    \multicolumn{1}{|c|}{$c_{1,\tau_P}$} &
    \multicolumn{1}{|c|}{$c_{2,\tau_P}$} \\\hline\hline
    $7.46$ &
    $0.2633\pm 0.0001$ & $0.3377\pm 0.0014$ &
    $1.4559\pm 0.0006$ & $2.250\pm 0.041$\rbtrr\\\hline 
    $8.80$ &
    $0.2391\pm 0.0001$ & $0.3070\pm 0.0012$ &
    $1.3735\pm 0.0008$ & $2.226\pm 0.038$\rbtrr\\\hline
    $14.95$ &
    $0.1847\pm 0.0001$ & $0.2387\pm 0.0009$ &
    $1.1895\pm 0.0008$ & $2.222\pm 0.038$\rbtrr\\\hline
    $17.73$ &
    $0.1684\pm 0.0001$ & $0.2173\pm 0.0008$ &
    $1.1217\pm 0.0007$ & $2.159\pm 0.037$\rbtrr\\\hline
    $23.75$ &
    $0.1453\pm 0.0002$ & $0.1871\pm 0.0013$ &
    $1.0209\pm 0.0011$ & $2.065\pm 0.062$\rbtrr\\\hline
    $36.69$ &
    $0.1138\pm 0.0001$ & $0.1480\pm 0.0007$ &
    $0.8622\pm 0.0011$ & $1.978\pm 0.041$\rbtrr\\\hline
    $57.61$ &
    $0.0846\pm 0.0002$ & $0.1113\pm 0.0008$ &
    $0.6909\pm 0.0014$ & $1.811\pm 0.055$\rbtrr\\\hline
    $80.76$ &
    $0.0654\pm 0.0002$ & $0.0864\pm 0.0006$ &
    $0.5622\pm 0.0022$ & $1.604\pm 0.043$\rbtrr\\\hline 
  \end{tabular}
  \caption[Mean values and perturbative coefficients for $\tau_P$.]
  {Mean values and perturbative coefficients for $\tau_P$ as derived
    from DISENT~0.1.}
  \label{tab:disentmeantaup}
\end{table}

\begin{table}
  \centering
  \begin{tabular}{|c||r|r|r|r|}
    \hline
    \multicolumn{1}{|c||}{\rbthm$\mean{Q}/\gev$} &
    \multicolumn{1}{|c|}{$\mean{B_P}_{\rm LO}$} & 
    \multicolumn{1}{|c|}{$\mean{B_P}_{\rm NLO}$} & 
    \multicolumn{1}{|c|}{$c_{1,B_P}$} &
    \multicolumn{1}{|c|}{$c_{2,B_P}$} \\\hline\hline
    $7.46$ &
    $0.3923\pm 0.0001$ & $0.2524\pm 0.0018$ &
    $2.1696\pm 0.0007$ & $-3.444\pm 0.049$\rbtrr\\\hline 
    $8.80$ &
    $0.3604\pm 0.0002$ & $0.2510\pm 0.0018$ &
    $2.0699\pm 0.0009$ & $-2.894\pm 0.052$\rbtrr\\\hline
    $14.95$ &
    $0.2879\pm 0.0001$ & $0.2309\pm 0.0014$ &
    $1.8545\pm 0.0009$ & $-1.847\pm 0.051$\rbtrr\\\hline
    $17.73$ &
    $0.2659\pm 0.0001$ & $0.2215\pm 0.0012$ &
    $1.7713\pm 0.0008$ & $-1.513\pm 0.047$\rbtrr\\\hline
    $23.75$ &
    $0.2350\pm 0.0002$ & $0.2063\pm 0.0013$ &
    $1.6515\pm 0.0013$ & $-1.034\pm 0.056$\rbtrr\\\hline
    $36.69$ &
    $0.1921\pm 0.0002$ & $0.1795\pm 0.0009$ &
    $1.4565\pm 0.0015$ & $-0.437\pm 0.047$\rbtrr\\\hline
    $57.61$ &
    $0.1521\pm 0.0002$ & $0.1528\pm 0.0007$ &
    $1.2431\pm 0.0021$ & $0.245\pm 0.042$\rbtrr\\\hline
    $80.76$ &
    $0.1254\pm 0.0002$ & $0.1309\pm 0.0005$ &
    $1.0798\pm 0.0023$ & $0.571\pm 0.034$\rbtrr\\\hline 
\end{tabular}
\caption[Mean values and perturbative coefficients for $B_P$.]
{Mean values and perturbative coefficients for $B_P$ as derived
  from DISENT~0.1.}
\label{tab:disentmeanbp}
\end{table}

\begin{table}
  \centering
  \begin{tabular}{|c||r|r|r|r|}
    \hline
    \multicolumn{1}{|c||}{\rbthm$\mean{Q}/\gev$} &
    \multicolumn{1}{|c|}{$\mean{\rho_E}_{\rm LO}$} & 
    \multicolumn{1}{|c|}{$\mean{\rho_E}_{\rm NLO}$} & 
    \multicolumn{1}{|c|}{$c_{1,\rho_E}$} &
    \multicolumn{1}{|c|}{$c_{2,\rho_E}$} \\\hline\hline
    $7.46$ &
    $0.0464\pm 0.0001$ & $0.0538\pm 0.0006$ &
    $0.2564\pm 0.0002$ & $0.243\pm 0.017$\rbtrr\\\hline 
    $8.80$ &
    $0.0437\pm 0.0001$ & $0.0522\pm 0.0005$ &
    $0.2509\pm 0.0002$ & $0.293\pm 0.014$\rbtrr\\\hline
    $14.95$ &
    $0.0367\pm 0.0001$ & $0.0464\pm 0.0004$ &
    $0.2363\pm 0.0002$ & $0.405\pm 0.015$\rbtrr\\\hline
    $17.73$ &
    $0.0347\pm 0.0001$ & $0.0445\pm 0.0003$ &
    $0.2309\pm 0.0003$ & $0.434\pm 0.014$\rbtrr\\\hline
    $23.75$ &
    $0.0316\pm 0.0001$ & $0.0406\pm 0.0009$ &
    $0.2222\pm 0.0004$ & $0.448\pm 0.040$\rbtrr\\\hline
    $36.69$ &
    $0.0274\pm 0.0001$ & $0.0367\pm 0.0003$ &
    $0.2076\pm 0.0004$ & $0.537\pm 0.019$\rbtrr\\\hline
    $57.61$ &
    $0.0233\pm 0.0001$ & $0.0321\pm 0.0004$ &
    $0.1907\pm 0.0007$ & $0.589\pm 0.028$\rbtrr\\\hline
    $80.76$ &
    $0.0204\pm 0.0001$ & $0.0279\pm 0.0002$ &
    $0.1761\pm 0.0009$ & $0.569\pm 0.017$\rbtrr\\\hline 
\end{tabular}
  \caption[Mean values and perturbative coefficients for $\rho_E$.]
{Mean values and perturbative coefficients for $\rho_E$ as derived
  from DISENT~0.1.}
\label{tab:disentmeanrhoe}
\end{table}

\begin{table}
  \centering
  \begin{tabular}{|c||r|r|r|r|}
    \hline
    \multicolumn{1}{|c||}{\rbthm$\mean{Q}/\gev$} &
    \multicolumn{1}{|c|}{$\mean{\tau_C}_{\rm LO}$} & 
    \multicolumn{1}{|c|}{$\mean{\tau_C}_{\rm NLO}$} & 
    \multicolumn{1}{|c|}{$c_{1,\tau_C}$} &
    \multicolumn{1}{|c|}{$c_{2,\tau_C}$} \\\hline\hline
    $7.46$ &
    $0.0701\pm 0.0001$ & $0.0907\pm 0.0010$ &
    $0.3876\pm 0.0003$ & $0.620\pm 0.028$\rbtrr\\\hline 
    $8.80$ &
    $0.0664\pm 0.0001$ & $0.0872\pm 0.0008$ &
    $0.3816\pm 0.0004$ & $0.675\pm 0.025$\rbtrr\\\hline
    $14.95$ &
    $0.0569\pm 0.0001$ & $0.0770\pm 0.0007$ &
    $0.3666\pm 0.0004$ & $0.814\pm 0.026$\rbtrr\\\hline
    $17.73$ &
    $0.0542\pm 0.0001$ & $0.0738\pm 0.0006$ &
    $0.3607\pm 0.0004$ & $0.854\pm 0.025$\rbtrr\\\hline
    $23.75$ &
    $0.0500\pm 0.0001$ & $0.0688\pm 0.0006$ &
    $0.3515\pm 0.0006$ & $0.911\pm 0.030$\rbtrr\\\hline
    $36.69$ &
    $0.0443\pm 0.0001$ & $0.0618\pm 0.0005$ &
    $0.3357\pm 0.0006$ & $0.999\pm 0.030$\rbtrr\\\hline
    $57.61$ &
    $0.0389\pm 0.0001$ & $0.0550\pm 0.0005$ &
    $0.3179\pm 0.0007$ & $1.077\pm 0.034$\rbtrr\\\hline
    $80.76$ &
    $0.0348\pm 0.0001$ & $0.0490\pm 0.0004$ &
    $0.2991\pm 0.0009$ & $1.073\pm 0.028$\rbtrr\\\hline 
\end{tabular}
  \caption[Mean values and perturbative coefficients for $\tau_C$.]
{Mean values and perturbative coefficients for $\tau_C$ as derived
  from DISENT~0.1.}
\label{tab:disentmeantauc}
\end{table}

\begin{table}
  \centering
  \begin{tabular}{|c||r|r|r|r|}
    \hline
    \multicolumn{1}{|c||}{\rbthm$\mean{Q}/\gev$} &
    \multicolumn{1}{|c|}{$\mean{C_P}_{\rm LO}$} & 
    \multicolumn{1}{|c|}{$\mean{C_P}_{\rm NLO}$} & 
    \multicolumn{1}{|c|}{$c_{1,C_P}$} &
    \multicolumn{1}{|c|}{$c_{2,C_P}$} \\\hline\hline
    $7.46$ &
    $0.2708\pm 0.0002$ & $0.3470\pm 0.0035$ &
    $1.4977\pm 0.0010$ & $2.304\pm 0.096$\rbtrr\\\hline 
    $8.80$ &
    $0.2570\pm 0.0002$ & $0.3325\pm 0.0031$ &
    $1.4765\pm 0.0013$ & $2.472\pm 0.091$\rbtrr\\\hline
    $14.95$ &
    $0.2207\pm 0.0002$ & $0.2945\pm 0.0024$ &
    $1.4220\pm 0.0015$ & $3.000\pm 0.094$\rbtrr\\\hline
    $17.73$ &
    $0.2103\pm 0.0002$ & $0.2826\pm 0.0021$ &
    $1.4008\pm 0.0015$ & $3.151\pm 0.088$\rbtrr\\\hline
    $23.75$ &
    $0.1945\pm 0.0003$ & $0.2638\pm 0.0024$ &
    $1.3679\pm 0.0022$ & $3.371\pm 0.112$\rbtrr\\\hline
    $36.69$ &
    $0.1728\pm 0.0003$ & $0.2374\pm 0.0019$ &
    $1.3109\pm 0.0024$ & $3.689\pm 0.107$\rbtrr\\\hline
    $57.61$ &
    $0.1522\pm 0.0004$ & $0.2126\pm 0.0019$ &
    $1.2451\pm 0.0033$ & $4.049\pm 0.124$\rbtrr\\\hline
    $80.76$ &
    $0.1371\pm 0.0004$ & $0.1909\pm 0.0013$ &
    $1.1796\pm 0.0044$ & $4.063\pm 0.095$\rbtrr\\\hline 
\end{tabular}
  \caption[Mean values and perturbative coefficients for $C_P$.]
{Mean values and perturbative coefficients for $C_P$ as derived
  from DISENT~0.1.}
\label{tab:disentmeancp}
\end{table}

\begin{table}
  \centering
  \begin{tabular}{|c||r|r|r|r|}
    \hline
    \multicolumn{1}{|c||}{\rbthm$\mean{Q}/\gev$} &
    \multicolumn{1}{|c|}{$\mean{y_{fJ}}_{\rm LO}$} & 
    \multicolumn{1}{|c|}{$\mean{y_{fJ}}_{\rm NLO}$} & 
    \multicolumn{1}{|c|}{$c_{1,y_{fJ}}$} &
    \multicolumn{1}{|c|}{$c_{2,y_{fJ}}$} \\\hline\hline
    $7.46$ &
    $0.1694\pm 0.0001$ & $0.1575\pm 0.0007$ &
    $0.9369\pm 0.0004$ & $-0.197\pm 0.020$\rbtrr\\\hline 
    $8.80$ &
    $0.1553\pm 0.0001$ & $0.1493\pm 0.0007$ &
    $0.8918\pm 0.0005$ & $-0.043\pm 0.021$\rbtrr\\\hline
    $14.95$ &
    $0.1219\pm 0.0001$ & $0.1253\pm 0.0006$ &
    $0.7852\pm 0.0005$ & $0.259\pm 0.021$\rbtrr\\\hline
    $17.73$ &
    $0.1119\pm 0.0001$ & $0.1172\pm 0.0005$ &
    $0.7452\pm 0.0005$ & $0.341\pm 0.020$\rbtrr\\\hline
    $23.75$ &
    $0.0976\pm 0.0001$ & $0.1054\pm 0.0005$ &
    $0.6859\pm 0.0007$ & $0.473\pm 0.024$\rbtrr\\\hline
    $36.69$ &
    $0.0779\pm 0.0001$ & $0.0875\pm 0.0004$ &
    $0.5905\pm 0.0007$ & $0.618\pm 0.019$\rbtrr\\\hline
    $57.61$ &
    $0.0594\pm 0.0001$ & $0.0697\pm 0.0003$ &
    $0.4854\pm 0.0007$ & $0.739\pm 0.018$\rbtrr\\\hline
    $80.76$ &
    $0.0470\pm 0.0001$ & $0.0564\pm 0.0002$ &
    $0.4052\pm 0.0009$ & $0.739\pm 0.014$\rbtrr\\\hline 
\end{tabular}
  \caption[Mean values and perturbative coefficients for $y_{fJ}$.]
{Mean values and perturbative coefficients for $y_{fJ}$ as derived
  from DISENT~0.1.}
\label{tab:disentmeanyfj}
\end{table}

\begin{table}
  \centering
  \begin{tabular}{|c||r|r|r|r|}
    \hline
    \multicolumn{1}{|c||}{\rbthm$\mean{Q}/\gev$} &
    \multicolumn{1}{|c|}{$\mean{y_{k_t}}_{\rm LO}$} & 
    \multicolumn{1}{|c|}{$\mean{y_{k_t}}_{\rm NLO}$} & 
    \multicolumn{1}{|c|}{$c_{1,y_{k_t}}$} &
    \multicolumn{1}{|c|}{$c_{2,y_{k_t}}$} \\\hline\hline
    $7.46$ &
    $0.2138\pm 0.0001$ & $0.2942\pm 0.0004$ &
    $1.1825\pm 0.0004$ & $2.529\pm 0.016$\rbtrr\\\hline 
    $8.80$ &
    $0.1803\pm 0.0001$ & $0.2366\pm 0.0003$ &
    $1.0348\pm 0.0005$ & $1.929\pm 0.016$\rbtrr\\\hline
    $14.95$ &
    $0.1214\pm 0.0001$ & $0.1514\pm 0.0002$ &
    $0.7818\pm 0.0004$ & $1.299\pm 0.013$\rbtrr\\\hline
    $17.73$ &
    $0.1045\pm 0.0001$ & $0.1288\pm 0.0002$ &
    $0.6960\pm 0.0004$ & $1.123\pm 0.011$\rbtrr\\\hline
    $23.75$ &
    $0.0830\pm 0.0001$ & $0.1008\pm 0.0002$ &
    $0.5830\pm 0.0005$ & $0.920\pm 0.013$\rbtrr\\\hline
    $36.69$ &
    $0.0562\pm 0.0001$ & $0.0676\pm 0.0001$ &
    $0.4256\pm 0.0004$ & $0.686\pm 0.009$\rbtrr\\\hline
    $57.61$ &
    $0.0349\pm 0.0001$ & $0.0423\pm 0.0001$ &
    $0.2851\pm 0.0007$ & $0.515\pm 0.008$\rbtrr\\\hline
    $80.76$ &
    $0.0235\pm 0.0001$ & $0.0287\pm 0.0001$ &
    $0.2015\pm 0.0009$ & $0.405\pm 0.005$\rbtrr\\\hline 
\end{tabular}
  \caption[Mean values and perturbative coefficients for $y_{k_t}$.]
{Mean values and perturbative coefficients for $y_{k_t}$ as derived
  from DISENT~0.1.}
\label{tab:disentmeanykt}
\end{table}


\chapter{Power Corrections}
\label{chap:powcorr}

\begin{figure} 
  \centering
  \includegraphics{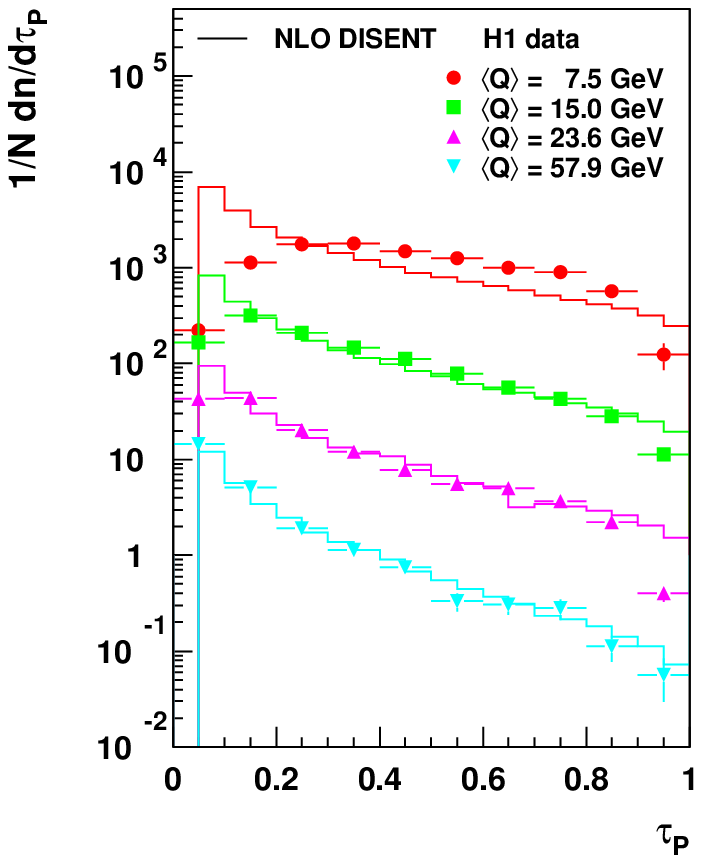}\hftwo%
  \includegraphics{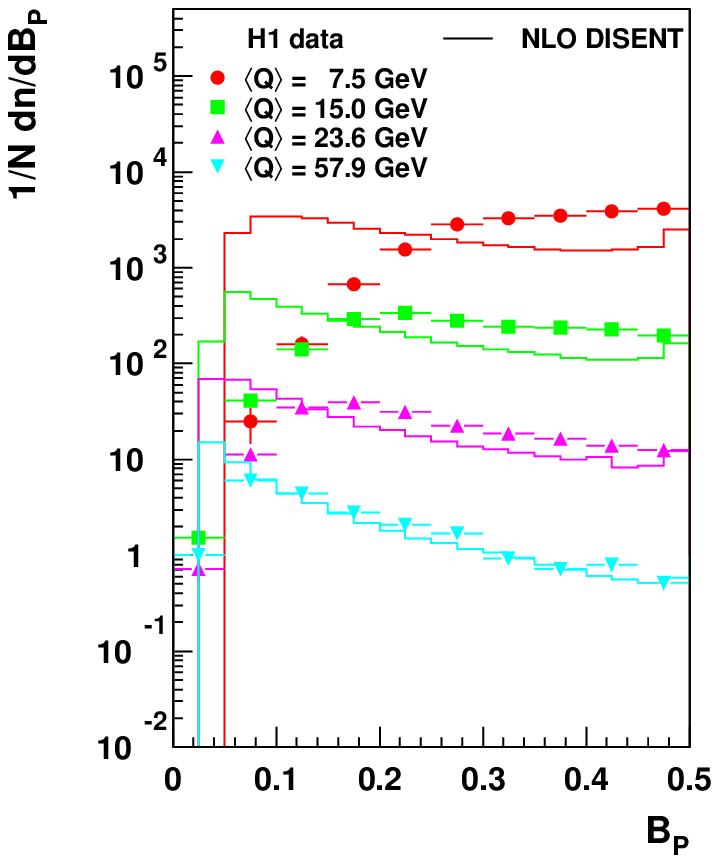}
  \includegraphics{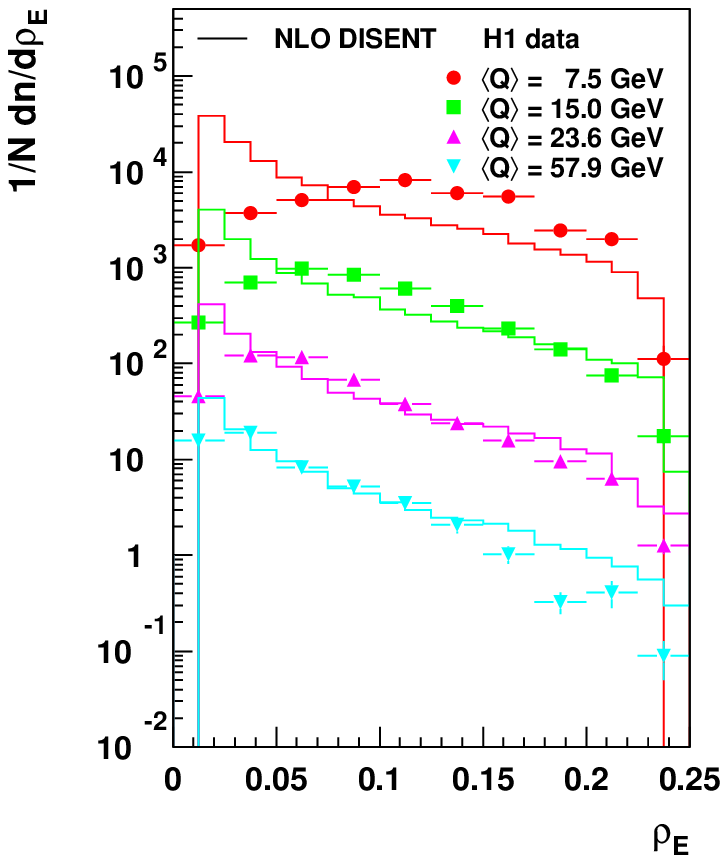}\hftwo%
  \includegraphics{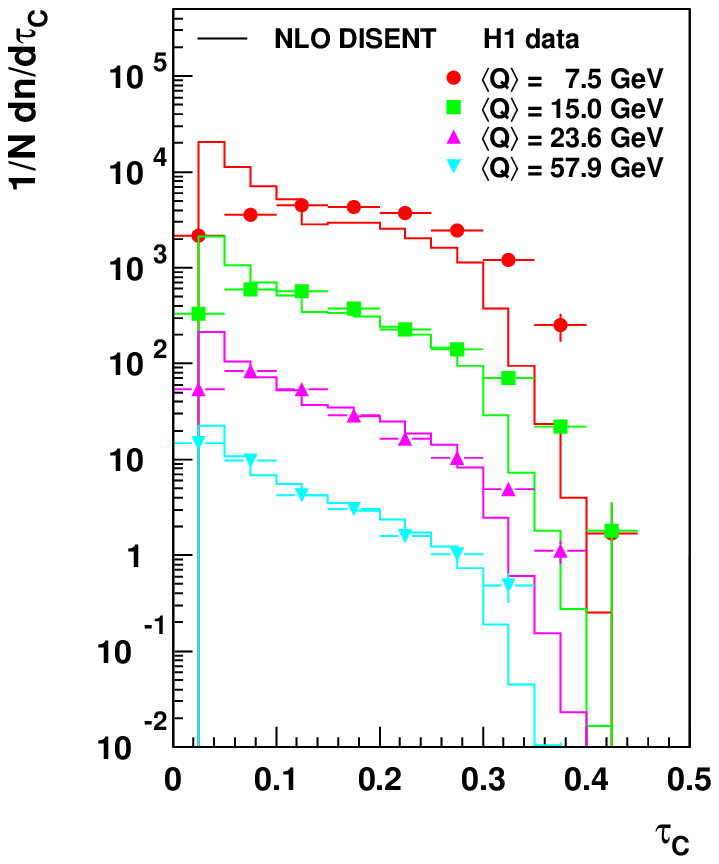}
  \caption[Normalized differential distributions of
  $\tau_P$, $B_P$, $\rho_E$ and $\tau_C$ corrected to hadron level in
  comparison with NLO predictions.]  {Normalized differential distributions of
    the event shapes $\tau_P$, $B_P$, $\rho_E$ and $\tau_C$ corrected to
    hadron level.  H1 data (full symbols) are compared with DISENT~0.1 NLO
    calculations (full lines) for four out of eight investigated bins in $Q$.
    The spectra for $\mean{Q} = 7.5$--$57.9\gev$ are multiplied by factors of
    $10^n$, $n=0,1,2,3$.  The error bars represent statistical uncertainties
    only.  Note that the first data bin was left out in both normalizations.}
  \label{fig:dndFhl1}
\end{figure}

\begin{figure} 
  \centering \includegraphics{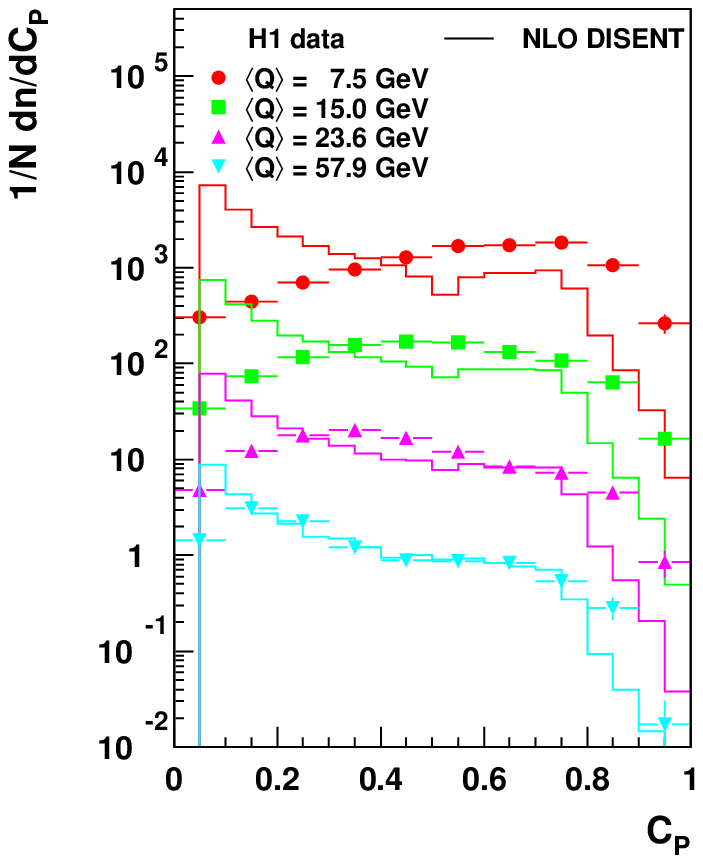}
  \includegraphics{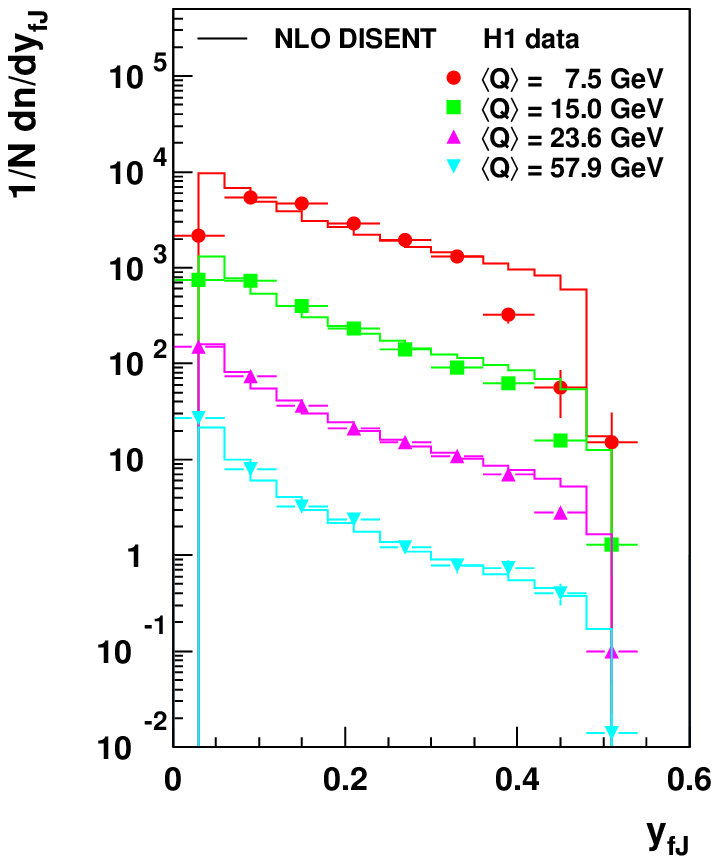}\hftwo%
  \includegraphics{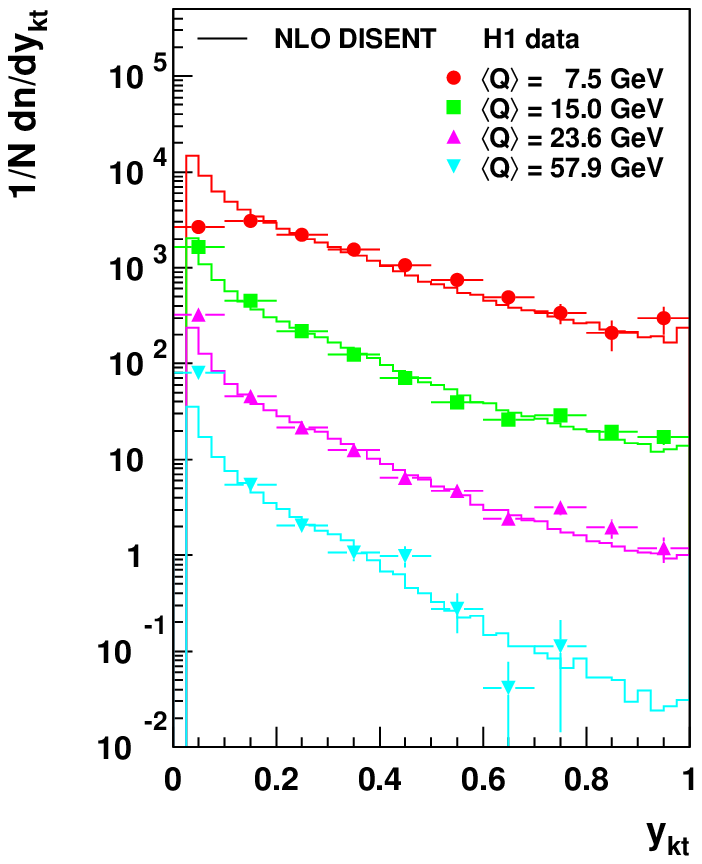}
  \caption[Normalized differential distributions of
  $C_P$, $y_{fJ}$ and $y_{k_t}$ corrected to hadron level in comparison with
  NLO predictions.]  {Normalized differential distributions of the event
    shapes $C_P$, $y_{fJ}$ and $y_{k_t}$ corrected to hadron level.  H1 data
    (full symbols) are compared with DISENT~0.1 NLO calculations (full lines)
    for four out of eight investigated bins in $Q$. The spectra for $\mean{Q}
    = 7.5$--$57.9\gev$ are multiplied by factors of $10^n$, $n=0,1,2,3$.  The
    error bars represent statistical uncertainties only.  Note that the first
    data bin was left out in both normalizations.}
  \label{fig:dndFhl2}
\end{figure}

To get a first impression of the extent to which pQCD is able to approximate
the data, figs.~\ref{fig:dndFhl1} and~\ref{fig:dndFhl2} present the normalized
differential distributions of the seven investigated event shapes, corrected
to hadron level, in comparison with DISENT~0.1 NLO calculations.  Note that
the first data bin was left out in both normalizations. This is necessary
since according to section~\ref{sec:NLOtech} it is not possible to go down to
$F=0$ in $\order(\as^2)$. Starting with bin one, the NLO histogram could
alternatively be normalized to $\sigma_{\rm NLO}$ yielding a slight mismatch
by a factor of $\sigma_{\rm NLO}/\sigma_{\rm NNLO}$.  Except for a small shift
in height complicating a comparison of shapes, the conclusions would not
change.

Obviously, there are large discrepancies between data and the NLO predictions.
Common to all event shapes, however, is the clear tendency that the
description of data improves considerably with rising $Q$; to what degree
depends on the variable under consideration.  $y_{fJ}$ and $y_{k_t}$ exhibit a
very fast convergence hinting at small hadronization corrections, whereas the
prediction for $C_P$ approaches the data slowly.

Turning back to the mean values $\fmean$, we can, except for the $y$ shapes,
expect them to be substantially larger than predicted by pQCD with decreasing
differences for increasing $Q$.  Indeed, physical quantities receive, in
addition to pQCD, non-perturbative contributions which basically behave
power-like, i.e.~$\propto 1/Q^p$, hence called {\em power corrections}.  The
cross section $\ee \rightarrow {\rm hadrons}$, $\sigma_{\rm had}$, for example
exhibits a $1/Q^4$-term which is also understandable on theoretical
grounds~\cite{QCD:Mueller}.  Less inclusive variables like event shapes are
not as well-behaved as $\sigma_{\rm had}$ and reveal $1/Q$-corrections that
are sizable even at $Q = M_Z$~\cite{QCD:WICHEP94}.  They can be made plausible
in the framework of a longitudinal phase space or {\em tube
  model}~\cite{Feynman}, which essentially is the simplest version of the
string fragmentation implemented in JETSET\@. One possibility to compare with
pQCD is based on hadronization models like the one implemented in JETSET by
unfolding the data in a third correction step to some kind of ({\bf
  ambiguous}) parton level.

We do not follow this recipe. Instead, we will call upon explicit formulae
trying to parameterize these effects.  They are presented in
sections~\ref{sec:tubemod} and~\ref{sec:DWmod} for the tube model and a new
approach respectively.  The main goal of this work is to test these models.

\section{The Tube Model}
\label{sec:tubemod}

Measurements of the momentum spectra of secondary particles produced by a
primary parton of energy $E_0 = Q/2$ approximately obey rather simple scaling
laws~\cite{Perkins}.  Characterizing a particular secondary by $E$, $p_t$,
$p_l$ and $m$ with respect to the direction of the original parton, the
distribution in $p_t$, $\rho(p_t)$, decreases exponentially almost
independently of $E_0$ and $p_l$, whereas the distribution in $p_l$ rises
proportional to $E_0$.  Making use of the rapidity variable $y_R$ (s.\ 
eq.~(\ref{eqn:rapidity})),
\begin{equation}
  y_R := \frac{1}{2} \ln\frac{E+p_l}{E-p_l} =
  \ln\frac{E+p_l}{\sqrt{p_t^2 + m^2}}\,,
\end{equation}
with the property $dy_R = dp_l/E$, this can be expressed in the form of
\begin{equation}
  \Phi(y_R,p_t) = B\cdot\rho(p_t)\,.
\end{equation}
$B$ is the (constant) number of hadrons per unity in rapidity and $\Phi$ the
distribution function of light hadrons in the produced jet corresponding to a
tube in $(y_R,p_t)$-space.  Following refs.~\cite{ESW,Kuhlen}, the energy
$E_j$ and longitudinal momentum $P_{l,j}$ of a jet of length $y_{R,{\rm max}}$
in rapidity can be estimated by:
\begin{eqnarray}
  E_j = & \int\limits_0^{y_{R,{\rm max}}}\int\limits_0^\infty E \cdot
  B\rho(p_t)dp_tdy_R & \approx
  \lambda\sinh y_{R,{\rm max}} \approx p_{l,{\rm max}} \approx E_0\,,\\
  P_{l,j} = & \int\limits_0^{y_{R,{\rm max}}}\int\limits_0^\infty p_{l,j} \cdot
  B\rho(p_t)dp_tdy_R & \approx
  \lambda(\cosh y_{R,{\rm max}} -1) \approx E_0 - \lambda\,.
\end{eqnarray}
Recall that $E=m_t\cosh y_R$ and $p_l=m_t\sinh y_R$ with the transverse mass
\begin{equation}
  m_t := \sqrt{p_t^2 + m^2} \approx p_t
\end{equation}
for light hadrons.  $\lambda$ is proportional to the average transverse
momentum:
\begin{equation}
  \lambda := B \int\limits_0^\infty p_t\rho(p_t) dp_t\,.
\end{equation}
The power corrections to mean $\tau = 1-$thrust and jet mass now become
\begin{eqnarray}
  \mean{\tau}^{\rm pow} &
  = 1-\frac{P_{l,j}}{E_j} \approx 1-\frac{E_0-\lambda}{E_0} = &
  \frac{2\lambda}{Q}\,,\\
  \mean{\rho}^{\rm pow} &
  \approx \frac{E_j^2 - P_{l,j}^2}{Q^2} \approx
  \frac{E_0^2 - (E_0-\lambda)^2}{Q^2} \approx & \frac{\lambda}{Q}\,.
\end{eqnarray}
With $B\approx 2$ and $\mean{p_t}\approx 0.3\gev$~\cite{Kuhlen}, we can
estimate $\lambda\approx 0.6\gev$.  Testing this simple ansatz with event
shapes in $ep$ DIS is the topic of the next section.

\section[$1/Q^\symbol{112}$-Fits]
{\boldmath$1/Q^\symbol{112}$\unboldmath\ -Fits}
\label{sec:tubefits}

The input for all fits performed here and in section~\ref{sec:DWfits} consists
of the corrected data means of table~\ref{tab:finalmeans} and the perturbative
coefficients given in tables~\ref{tab:disentmeantaup}--\ref{tab:disentmeanykt}
together with their statistical uncertainties.  Systematic uncertainties are
not taken into account; they will be analyzed in section~\ref{sec:syserr}.

The standard fit procedure employs the method of least squares (s.\ 
e.g.~\cite{Frodesen}) to determine the best estimates for the parameters of
the supplied model. Technically, this is accomplished by using the program
MINUIT~\cite{MINUIT} which additionally provides the $\chi^2$-value and, where
appropriate, the maximally encountered correlation $\kappa$. In the tables
collecting the results, the reduced $\chi^2$-values, i.e.\ $\chi^2$ divided by
the number of {\bf d}egrees {\bf o}f {\bf f}reedom (dof), are entered. For
one- and two-parameter fits we have seven respectively six dofs yielding a
maximal $\chin\approx 2$ at a confidence level of $5\%$.

The first test carried out is to allow a variation of $\asmz$ in the
perturbative expression
\begin{equation}
  \label{eqn:Fpert}
  \fpert = c_{1,F}(x)\,\as(\mr) + \left(c_{2,F}(x) +
    \frac{\beta_0}{2\pi}\ln\frac{\mr}{Q}c_{1,F}(x)\right)\as^2(\mr)
\end{equation}
only, where, for the time being, $\mr$ is identified with $Q$.  As can be seen
from table~\ref{tab:1parmufit}, this works surprisingly well except for $B_P$.
Seemingly, the decrease of $\fmean$ with $Q$ can be accounted for by the
logarithmic $Q$-dependence of $\as$ as long as $\asmz$ may be adapted to the
corresponding hadronization correction.  Compared to the world average
\mbox{$\asmz = 0.119$}~\cite{PDG}, we obtain high results for $\tau_P$ and
$B_P$, very high ones for $\rho_E$, $\tau_C$, $C_P$ and lower values for
$y_{fJ}$, $y_{k_t}$.  Taking this as an indication for the impact of
hadronization, it can be concluded that our event shapes are affected by
medium, large and small ({\bf negative}) corrections respectively.

Sometimes, the ratio of NLO to LO results, the $K$-factor,
is considered to signal big effects of higher orders. 
However, although evaluating them for $\tau_P$ and $\rho_E$ in $Q$-bin one to
be $K_{\tau_P} = 1.28$ and $K_{\rho_E} = 1.16$, we get ratios of the data
means with respect to NLO of $1.30$ and $2.07$ calling this argument into
question.

\begin{table}
  \centering
  \begin{tabular}{|c||r|r||r|r||r|r|}
    \hline
    \multicolumn{1}{|c||}{\rbthm$\fmean$} &
    \multicolumn{1}{|c|}{$\asmz$} & 
    \multicolumn{1}{|c||}{$\chin$} & 
    \multicolumn{1}{|c|}{$\lambda/\gev$} & 
    \multicolumn{1}{|c||}{$\chin$} &
    \multicolumn{1}{|c|}{$\mu/\gevq$} & 
    \multicolumn{1}{|c|}{$\chin$} \\\hline\hline
    $\mean{\tau_P}$ &
    $0.1311\pm 0.0005$ & $  0.5$ &
    $0.705 \pm 0.028$  & $  1.4$ &
    $7.17  \pm 0.32$   & $ 22.8$ \rbtrr\\\hline
    $\mean{B_P}$ &
    $0.1341\pm 0.0005$ & $ 24.0$ &
    $0.549 \pm 0.016$  & $  0.7$ &
    $5.22  \pm 0.16$   & $ 28.0$ \rbtrr\\\hline
    $\mean{\rho_E}$ &
    $0.1601\pm 0.0006$ & $  0.6$ &
    $0.539 \pm 0.007$  & $ 38.4$ &
    $5.17  \pm 0.08$   & $282.5$ \rbtrr\\\hline
    $\mean{\tau_C}$ &
    $0.1520\pm 0.0005$ & $  1.5$ &
    $0.734 \pm 0.011$  & $ 29.2$ &
    $7.12  \pm 0.13$   & $212.2$ \rbtrr\\\hline
    $\mean{C_P}$ &
    $0.1481\pm 0.0004$ & $  5.4$ &
    $2.325 \pm 0.032$  & $ 66.2$ &
    $20.82 \pm 0.36$   & $326.6$ \rbtrr\\\hline
    $\mean{y_{fJ}}$ &
    $0.1110\pm 0.0005$ & $  1.7$ &
    $-0.185\pm 0.012$  & $  2.0$ &
    $-1.74 \pm 0.13$   & $ 10.5$ \rbtrr\\\hline
    $\mean{y_{k_t}}$ &
    $0.1082\pm 0.0007$ & $  1.8$ &
    $-0.360\pm 0.021$  & $  1.4$ &
    $-4.21 \pm 0.27$   & $  9.3$ \rbtrr\\\hline
  \end{tabular}
  \caption[Results of one-parameter fits without power corrections
  and according to the tube model.]
  {Results of one-parameter fits according to
    eqs.~(\ref{eqn:Fpert})--(\ref{eqn:m/Q2}) for the event shape means.
    Uncertainties are statistical only.}
  \label{tab:1parmufit}
\end{table}

Writing
\begin{equation}
  \label{eqn:Fsum}
  \fmean = \fpert + \fpow
\end{equation}
according to~\cite{pc:DWform1}, we extend our fit to include terms of the form
\begin{equation}
  \label{eqn:l/Q}
  \fpow = \frac{\lambda}{Q}
\end{equation}
and
\begin{equation}
  \label{eqn:m/Q2}
  \fpow = \frac{\mu}{Q^2}\,.
\end{equation}
At first, $\asmz$ will be kept fixed to $0.119$ yielding estimates for
$\lambda$ and $\mu$ as given in table~\ref{tab:1parmufit}.  With the exception
of $B_P$, the $\chi^2$-values worsen for both formulae, even dramatically in
the case of $\mu/Q^2$.  Simple $1/Q^2$-corrections are therefore ruled out,
but notably $y_{k_t}$ exhibits the best behaviour of all.  Also, pure
$1/Q$-terms do not work very well, yet, at least the order of magnitude for
$\lambda \approx 0.5$--$1.0\gev$ is confirmed for $\tau_P$ and $\rho_E$,
although a relative factor of two is not found between them.  According to our
previous expectations, $C_P$ receives a larger $\lambda$, whereas for $y_{fJ}$
and $y_{k_t}$ negative values are assumed.

In summary, this suggests that logarithmic as well as power-like
$Q$-dependences play a role.  The idea to fit both, $\asmz$ and $\lambda$,
however, does not lead to the desired outcome as shown in
table~\ref{tab:2parmufit}.  Due to the extreme anti-correlation found between
them, there is a tendency to minimize the power contribution at the cost of a
very high $\asmz$. Merely $B_P$ and the $y$ event shapes produce correlated
but reasonable numbers.  Triggered by the $x$ dependence of $\fpert$, one
could assume that $\fpow$ should be a function of $x$ as well. Yet, for
corrections with power $p=1$ this is not expected~\cite{prc:Dasgupta}.  For an
improvement a more sophisticated model is needed.

At last, an explicit check on the power $p$ was performed by fitting
\begin{equation}
  \fpow = \frac{\nu}{Q^p}
\end{equation}
keeping again $\asmz$ fixed.  The results for $p$ lie between $0.5$ and $1.0$,
but once more $\nu$ and $p$ are strongly correlated.

\begin{table}
  \centering
  \begin{tabular}{|c||r|r|r|r||r|r|r|r|}
    \hline
    \multicolumn{1}{|c||}{\rbthm$\fmean$} &
    \multicolumn{1}{|c|}{$\asmz$} & 
    \multicolumn{1}{|c|}{$\lambda/\gev$} & 
    \multicolumn{1}{|c|}{$\chin$} & 
    \multicolumn{1}{|c||}{$\kappa/\%$} &
    \multicolumn{1}{|c|}{$\asmz$} & 
    \multicolumn{1}{|c|}{$\mu/\gevq$} & 
    \multicolumn{1}{|c|}{$\chin$} & 
    \multicolumn{1}{|c|}{$\kappa/\%$} \\\hline\hline
    $\mean{\tau_P}$ &
    $0.1318$ & $-0.044$ & $0.6$ & $-99$ &
    $0.1314$ & $-0.23 $ & $0.6$ & $-89$ \rbtrr\\\hline
    $\mean{B_P}$ &
    $0.1195$ & $ 0.532$ & $0.8$ & $-92$ &
    $0.1279$ & $ 3.07 $ & $1.7$ & $-68$ \rbtrr\\\hline
    $\mean{\rho_E}$ &
    $0.1652$ & $-0.089$ & $0.5$ & $-99$ &
    $0.1614$ & $-0.27 $ & $0.4$ & $-87$ \rbtrr\\\hline
    $\mean{\tau_C}$ &
    $0.1573$ & $-0.153$ & $1.3$ & $-99$ &
    $0.1535$ & $-0.57 $ & $1.1$ & $-86$ \rbtrr\\\hline
    $\mean{C_P}$ &
    $0.1609$ & $-1.393$ & $2.0$ & $-99$ &
    $0.1519$ & $-4.84 $ & $1.0$ & $-85$ \rbtrr\\\hline
    $\mean{y_{fJ}}$ &
    $0.1142$ & $-0.076$ & $1.6$ & $-97$ &
    $0.1119$ & $-0.28 $ & $1.7$ & $-82$ \rbtrr\\\hline
    $\mean{y_{k_t}}$ &
    $0.1216$ & $-0.453$ & $1.6$ & $-99$ &
    $0.1050$ & $ 1.33 $ & $1.6$ & $-93$ \rbtrr\\\hline
  \end{tabular}
  \caption[Results of two-parameter fits according to the tube model.]
  {Results of two-parameter fits according to
    eqs.~(\ref{eqn:Fpert})--(\ref{eqn:m/Q2}) for the event shape means.}
  \label{tab:2parmufit}
\end{table}

\section{The Model of Dokshitzer, Webber et al.}
\label{sec:DWmod}

For shorthand notation the model introduced in the following is labelled after
Dokshitzer and Webber, the authors of ref.~\cite{pc:DWform1}, where the
essential formula~(\ref{eqn:Fpow0}) needed here is first presented.  Note that
besides these initiating authors many others contributed to the currently very
active field of power corrections in QCD, s.~e.g.~\cite{pc:BenekeRenormalons}
and refs.\ therein.

As demonstrated above, pQCD applied in the form of NLO integration programs is
not sufficient to describe our data. Ignoring the fact that the calculation of
still higher orders will become forbidding due to the tremendous amount of
Feynman diagrams to compute, it may not even be a good idea, since the QCD
perturbation series is not necessarily expected to converge!  One known source
of singularities are diagrams with a chain of fermion loops inserted into a
gluon line.  These {\em renormalon}\/ chains lead to a factorial divergence of
the coefficients in the series expansion~(\ref{eqn:expansion}).  At first
sight, we are at a loss here. But on the contrary, one can even gain more
insight into non-perturbative effects of QCD by studying these
divergences~\cite{QCD:Mueller,pc:BenekeRenormalons}.  In a manner of speaking,
one approaches the non-perturbative regime from the perturbative side.
Employing this renormalon approach~\cite{pc:W1}, one ends up with ambiguities
of the series expansion of the general form $(\Lambda/\mr)^p$ modulo
logarithms $\ln^q(\mr/\Lambda)$, yet the exact result for an observable has to
be single-valued, of course.  Hence, power-like terms have to be added to
regain definiteness.

Practically, this scheme can be exploited to give the leading corrections in
terms of powers $p$ and $q$ for an observable, but the absolute normalization
is unknown.  Introducing the notion of a universal infrared-finite effective
coupling $\aeff(\mr)$ responsible for soft gluon emissions, the coefficients
for different observables to the same power $p$ can be related at the expense
of a new but universal non-perturbative parameter
$\overline{\alpha}_{p-1}(\mi)$.  Above an {\em infrared matching}\/ scale
$\mi$ with
\begin{equation}
  \label{eqn:mi}
  \Lambda \ll \mi \approx 2\gev \ll Q\,,
\end{equation}
this effective coupling has to coincide with the normal perturbative one
$\as(\mr)$.  For event shapes in \ee annihilation, this concept was first
applied in~\cite{pc:DWform1} and extended in~\cite{pc:WDIS95}:
\begin{equation}
  \label{eqn:Fpow0}
  \fpow = \frac{\hat{a}_F}{p} \left(\frac{\mi}{\mr}\right)^p
  \ln^q\left(\frac{\mr}{\mi}\right)
  \left[ \overline{\alpha}_{p-1}(\mi) - \as(\mr) - \frac{\beta_0}{2\pi}
    \left(\ln\frac{\mr}{\mi} + \frac{K}{\beta_0} + \frac{1}{p}\right)
    \as^2(\mr)\right]\,.
\end{equation}
Here, $\hat{a}_F$ is an $F$-dependent but calculable coefficient and
\begin{equation}
  \label{eqn:K}
  K = \frac{67}{6} - \frac{\pi^2}{2} - \frac{5}{9}N_f 
\end{equation}
accounts for differences between the $\overline{\rm MS}$ and the CMW
renormalization scheme~\cite{QCD:CMW,pc:Brevisited}:
\begin{equation}
  \label{eqn:MStoCMW}
  \alpha_{s,{\rm CMW}} = \alpha_{s,{\rm \overline{MS}}} +
  \frac{K}{2\pi} \alpha_{s,{\rm \overline{MS}}}^2\,.
\end{equation}
The subtractions proportional to $\as$ and $\as^2$ should be evaluated for the
same number of flavours, here $N_f = 5$, as in the perturbative part
eq.~(\ref{eqn:Fpert}) and serve to avoid double counting.

In terms of the effective coupling, $\overline{\alpha}_{p-1}(\mi)$ can be
written as the moment
\begin{equation}
  \overline{\alpha}_{p-1}(\mi) :=
  \frac{p}{\mi^p} \int\limits_0^{\mi} k^{p-1}\aeff(k)dk\,,
\end{equation}
i.e.~$\an(\mi)$ essentially corresponds to an average effective coupling at
low scales and is deduced from \ee data on thrust to be $\approx 0.5$ for
$\mi=2\gev$~\cite{pc:DWform1}.

For our event shapes, $p=1$ except for $y_{k_t}$ where $p=2$ is
expected~\cite{pc:WDIS95,pc:WDDIS}.  In that case, other mechanisms can give
contributions to the same order $1/Q^2$, preventing a straightforward
calculation of $\hat{a}_{y_{k_t}}$.  The other coefficients can be gathered
from~\cite{pc:WDIS95,pc:WDDIS} when performing the transition
\begin{equation}
  \hat{a}_F \rightarrow \frac{16}{3\pi} a_F
\end{equation}
to normalize to the $\hat{a}_F$-value for $\tau$ in \ee
physics~\cite{pc:DWform1}.

In the form of the {\em dispersive}\/ approach~\cite{pc:DWdisp}, this
technique was applied to event shape distributions~\cite{pc:DWdistr} and other
subjects as well~\cite{pc:WDstrf,pc:WDfragf1}.  In the meantime, however, it
was discovered that some ambiguities unfortunately remain in deriving
$a_F$-values.  Refining the investigations to two-loop
level~\cite{pc:Milan,pc:Salam}, the problem could be resolved and universality
was reconstituted provided an additional factor, the {\em Milan factor}
\begin{equation}
  {\cal M} = 1 + \frac{7.311-0.052 N_f}{\beta_0}
\end{equation}
is applied. Recently, it was confirmed to hold also for DIS event
shapes~\cite{pc:WDMilanDIS}. Due to a mismatch in definitions compared
to~(\ref{eqn:Fpow0}), however, we have to multiply by $2/\pi\cdot{\cal
  M}\approx 1.143$ where, in general, $N_f=3$ should be used~\cite{prc:Salam}.

Putting everything together, we get for $\fpow$
\begin{equation}
  \label{eqn:Fpow}
  \fpow = a_F\frac{32}{3\pi^2}\frac{\cal M}{p} \left(\frac{\mi}{\mr}\right)^p
  \left[ \overline{\alpha}_{p-1}(\mi) - \as(\mr) - \frac{\beta_0}{2\pi}
    \left(\ln\frac{\mr}{\mi} + \frac{K}{\beta_0} + \frac{1}{p}\right)
    \as^2(\mr)\right]
\end{equation}
with the coefficients given in table~\ref{tab:2parDWfit}.

Note that there is no logarithm with power $q$ any more.  Such a term with
$q=1$ was predicted for $B_P$ only.  In the previous
publications~\cite{KR:DIS97,H1:Shapes}, this could not be supported and the
theoretical calculations for it have been reexamined~\cite{pc:Brevisited},
leading to the much more complicated factor
\begin{equation}
  \label{eqn:Brevisited}
  a'_{B_P} = \frac{\pi}{2\sqrt{\frac{8}{3}\alpha_{s,{\rm CMW}}(e^{-3/4}Q)}}
  + \frac{3}{4} - \frac{\beta_0}{16} - 0.6137 + \order(1)\,.
\end{equation}

\section{Fits \alDW\ et al.}
\label{sec:DWfits}
\subsection{Two-Parameter Fits}
\label{sec:DW2parfits}

Two-parameter fits according to eqs.~(\ref{eqn:Fpert}) and~(\ref{eqn:Fpow})
are performed for each event shape separately.  The results are compiled in
table~\ref{tab:2parDWfit}, where for comparison the fit of $\mean{B_P}$
without the new factor $a'_{B_P}$ is given as well.  For a discussion of the
quoted total systematic uncertainties s.\ section~\ref{sec:syserr}.

Omitting at first the $y$ variables, one immediately recognizes reduced
correlations as compared to the tube model ---~with the exception of
$\tau_P$~--- and reasonable $\chin$-values.  The five obtained $\anmi$'s
scatter around the expectation of $\approx 0.5$ by about $20\%$ and within
errors are compatible with it.  The uncertainties are dominated by systematics
of $5\%$--$10\%$ being at least twice as large as the statistical ones.

Turning to $\asmz$, one observes a two-fold ambiguity. The event shapes
employing $z\Bf$ as event axis prefer low couplings, whereas the other three
lead to $\asmz\approx0.130$.  Considering the errors of $1\%$--$3\%$
statistically and $5\%$--$9\%$ systematically, they are consistent with each
other. Yet, it does evoke some suspicion.

\begin{table}
  \centering
  \begin{tabular}{|c|c|c||c|c|r|r|}
    \hline
    \multicolumn{1}{|c|}{\rbthm$\fmean$} &
    \multicolumn{1}{|c|}{$a_F$} & 
    \multicolumn{1}{|c||}{$p$} & 
    \multicolumn{1}{|c|}{$\overline{\alpha}_{p-1}(\mi=2\gev)$} & 
    \multicolumn{1}{|c|}{$\asmz$} & 
    \multicolumn{1}{|c|}{$\chin$} &
    \multicolumn{1}{|c|}{$\kappa/\%$} \\\hline\hline
    $\mean{\tau_P}$ & $1$ & $1$ &
    $0.480 \pm 0.028~^{+0.048}_{-0.062}$ &
    $0.1174 \pm 0.0030~^{+0.0097}_{-0.0081}$ &
    $0.5$ & $-97$ \rbtrr\\\hline
    $\mean{B_P}$ & $1/2\cdot a'_{B_P}$ & $1$ &
    $0.491 \pm 0.005~^{+0.032}_{-0.036}$ &
    $0.1106 \pm 0.0012~^{+0.0060}_{-0.0057}$ &
    $0.7$ & $-58$ \rbtrr\\\hline
    $\mean{\rho_E}$ & $1/2$ & $1$ &
    $0.561 \pm 0.004~^{+0.051}_{-0.058}$ &
    $0.1347 \pm 0.0015~^{+0.0111}_{-0.0100}$ &
    $1.2$ & $+7$ \rbtrr\\\hline
    $\mean{\tau_C}$ & $1$ & $1$ &
    $0.475 \pm 0.003~^{+0.044}_{-0.048}$ &
    $0.1284 \pm 0.0014~^{+0.0100}_{-0.0092}$ &
    $1.3$ & $+19$ \rbtrr\\\hline
    $\mean{C_P}$ & $3\pi/2$ & $1$ &
    $0.425 \pm 0.002~^{+0.033}_{-0.039}$ &
    $0.1273 \pm 0.0009~^{+0.0104}_{-0.0093}$ &
    $0.9$ & $+63$ \rbtrr\\\hline\hline
    $\mean{B_P}$ & $1/2$ & $1$ &
    $0.602 \pm 0.014~^{+0.029}_{-0.042}$ &
    $0.1164 \pm 0.0011~^{+0.0055}_{-0.0049}$ &
    $0.7$ & $-84$ \rbtrr\\\hline\hline
    $\mean{y_{fJ}}$ & $1$ & $1$ &
    $0.258 \pm 0.004$ &
    $0.1044 \pm 0.0017$ &
    $1.9$ & $-61$ \rbtrr\\\hline
    $\mean{y_{fJ}}$ & $-1/4^*$ & $1$ &
    $0.460 \pm 0.079$ &
    $0.1183 \pm 0.0027$ &
    $1.5$ & $+99$ \rbtrr\\\hline\hline
    $\mean{y_{k_t}}$ & $1^\dagger$ & $1$ &
    $0.431 \pm 0.034$ &
    $0.0462 \pm 0.0129$ &
    $2.6$ & $-99$ \rbtrr\\\hline
    $\mean{y_{k_t}}$ & $1^\dagger$ & $2$ &
    $0.586 \pm 0.182$ &
    $0.1047 \pm 0.0019$ &
    $1.6$ & $+93$ \rbtrr\\\hline
    $\mean{y_{k_t}}$ & $-30^*$ & $2$ &
    $0.326 \pm 0.034$ &
    $0.1206 \pm 0.0040$ &
    $0.3$ & $+99$ \rbtrr\\\hline
  \end{tabular}
  \caption[Results of two-parameter fits \alDW\ et al.]
  {Results of two-parameter fits \alDW\ et al.,
    eqs.~(\ref{eqn:Fpert}) and~(\ref{eqn:Fpow}),
    for the event shape means.
    In case of $B_P$, the outcome without the additional factor
    $a'_{B_P}$, eq.~\ref{eqn:Brevisited}, is given as well.
    The starred coefficients of the $y$ variables are derived
    from a {\bf circular} fit procedure described in
    section~\ref{sec:DWaffits}, whereas the coefficients marked
    with $^\dagger$'s are guesstimated.
    Where reasonable, statistical {\bf and} total systematic
    uncertainties are presented.}
  \label{tab:2parDWfit}
\end{table}

\begin{figure} 
  \centering
  \includegraphics{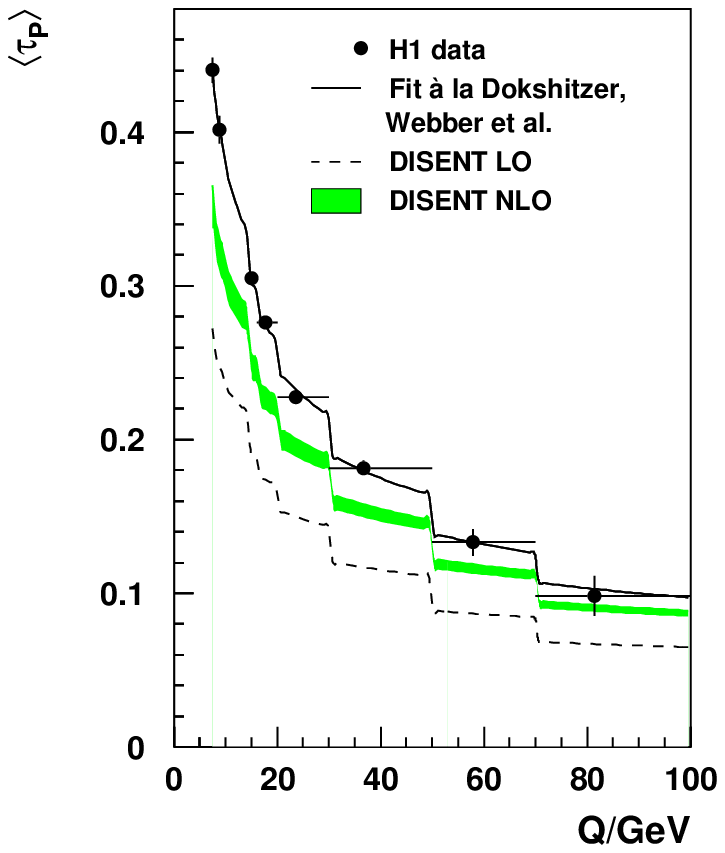}\hftwo%
  \includegraphics{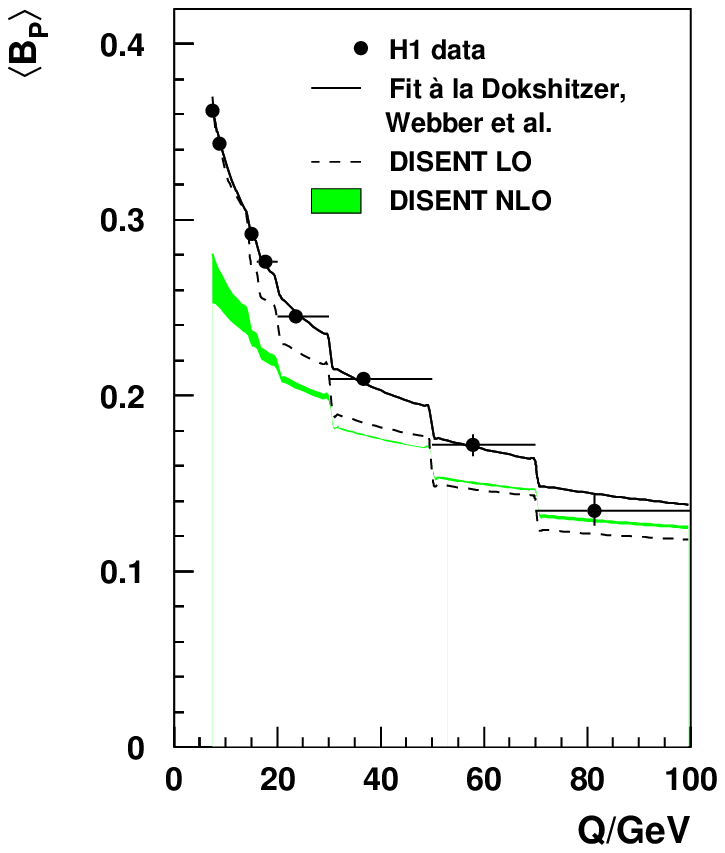}
  \includegraphics{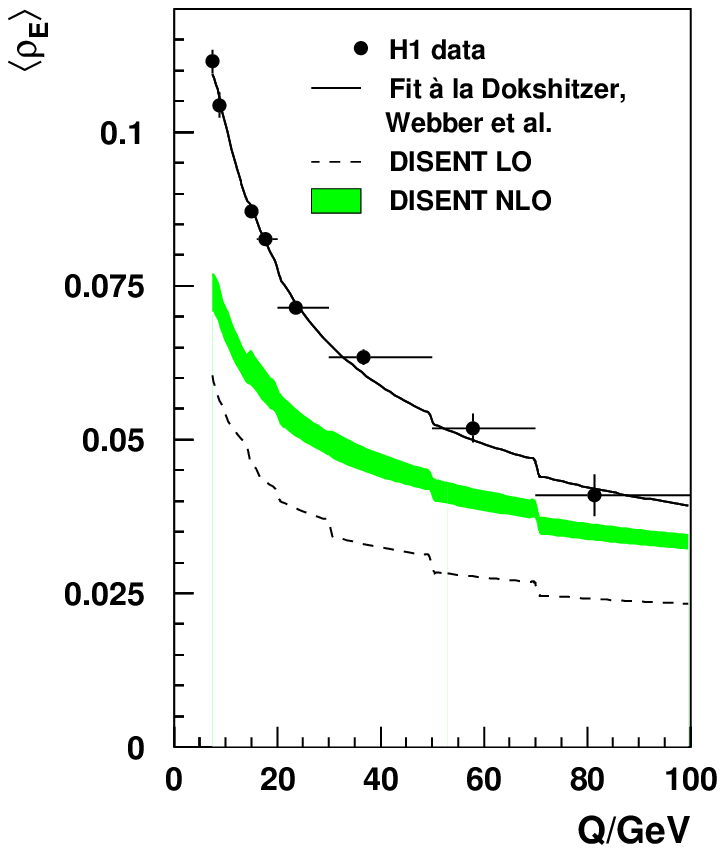}\hftwo%
  \includegraphics{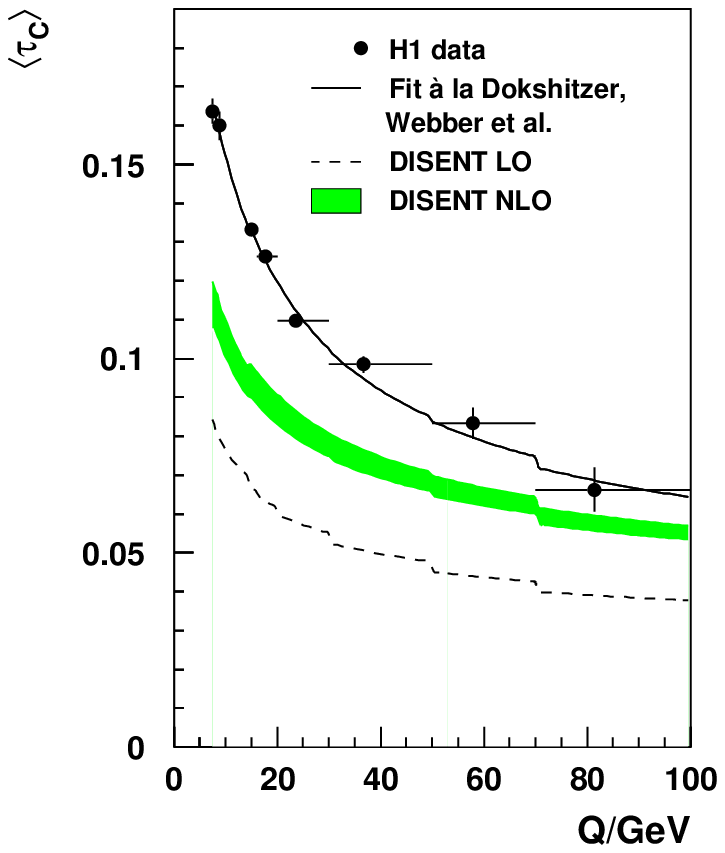}
  \caption[Power correction fits to
  corrected means of $\tau_P$, $B_P$, $\rho_E$ and $\tau_C$ together with the
  LO and NLO predictions.]  {Corrected means (full symbols) of $\tau_P$,
    $B_P$, $\rho_E$ and $\tau_C$ as a function of $Q$. The error bars
    represent statistical uncertainties only.  The full line corresponds to a
    power correction fit according to eqs.~(\ref{eqn:Fpert})
    and~(\ref{eqn:Fpow}).  For comparison the LO (dashed) and NLO (band)
    predictions of DISENT~0.1 are shown. The band reflects a variation of the
    renormalization scale of $1/2 \leq \mr^2/\gevq \leq 2$.}
  \label{fig:DWfit1}
\end{figure}

\begin{figure} 
  \centering \includegraphics{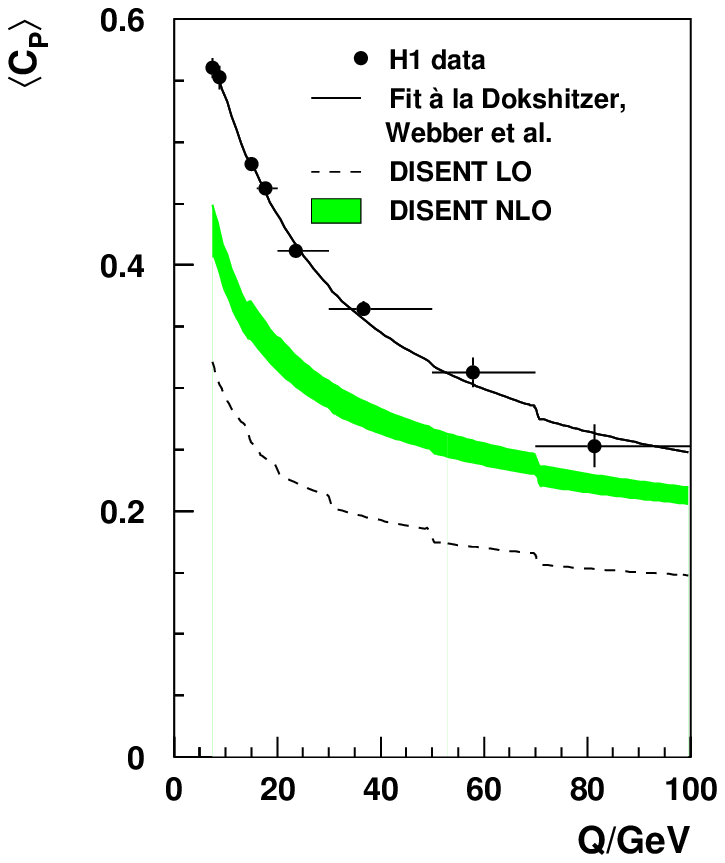}
  \includegraphics{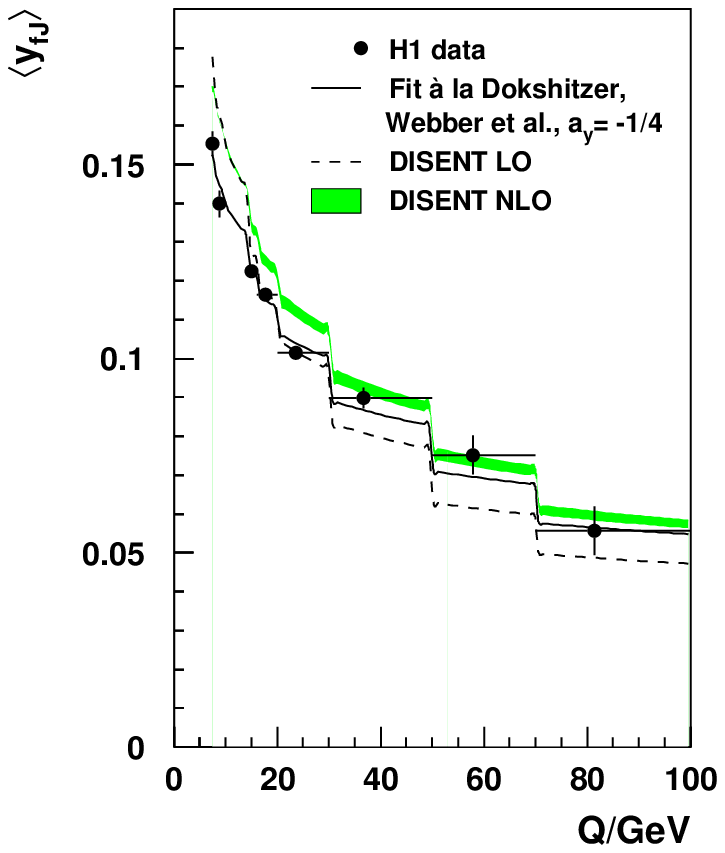}\hftwo%
  \includegraphics{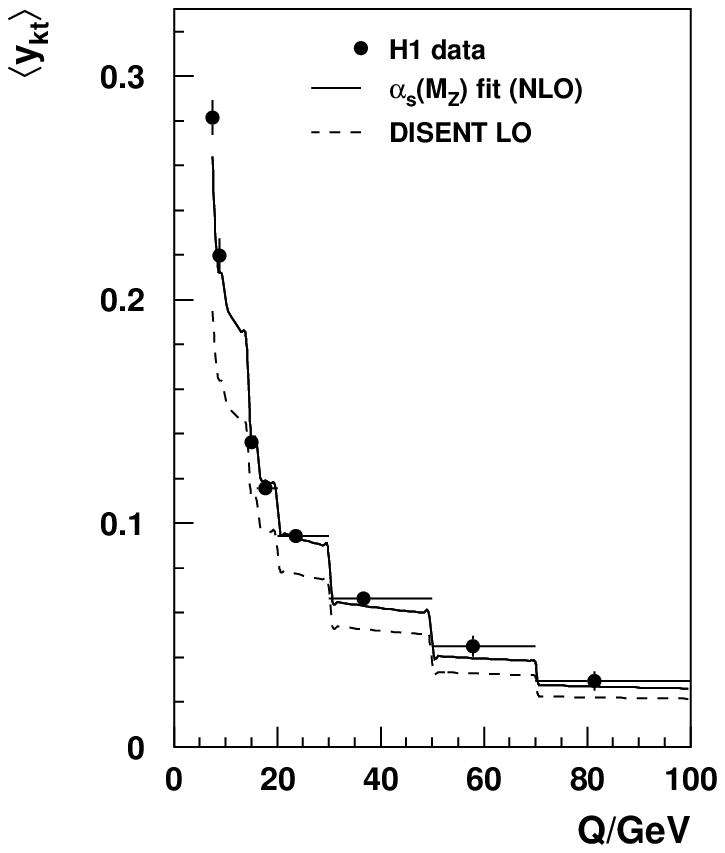}
  \caption[Power correction fits to
  corrected means of $C_P$ and $y_{fJ}$ and a fit without power correction to
  the corrected means of $y_{k_t}$, together with the LO and NLO predictions.]
  {Corrected means (full symbols) of $C_P$, $y_{fJ}$ and $y_{k_t}$ as a
    function of $Q$. The error bars represent statistical uncertainties only.
    The full line corresponds to a power correction fit according to
    eqs.~(\ref{eqn:Fpert}) and~(\ref{eqn:Fpow}) except for $y_{k_t}$ where no
    power contribution was assumed.  For comparison the LO (dashed) and, if
    appropriate, NLO (band) predictions of DISENT~0.1 are shown. The band
    reflects a variation of the renormalization scale of $1/2 \leq \mr^2/\gevq
    \leq 2$.}
  \label{fig:DWfit2}
\end{figure}

\begin{figure} 
  \centering \includegraphics{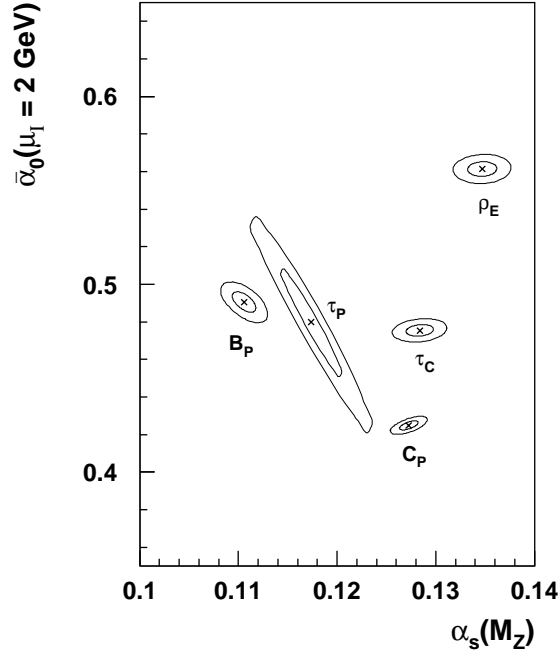}
  \caption[Results of the two-parameter fits \alDW\ et al.\ for the means
  of $\tau_P$, $B_P$, $\rho_E$, $\tau_C$ and $C_P$ in the $(\as,\an)$-plane.]
  {Results of the two-parameter fits \alDW\ et al.\ for the means of $\tau_P$,
    $B_P$, $\rho_E$, $\tau_C$ and $C_P$ in the form of $1\sigma$- and
    $2\sigma$-contours in the $(\as,\an)$-plane for statistical uncertainties
    only.}
  \label{fig:ellipses}
\end{figure}

Figs.~\ref{fig:DWfit1} and~\ref{fig:DWfit2} show the fit results in comparison
with the corrected data as well as the LO and NLO predictions of DISENT~0.1
with respect to each corresponding $\asmz$. The NLO band represents a
variation of the renormalization scale of \mbox{$1/2 \leq \mr^2/\gevq \leq
  2$}, s.\ section~\ref{sec:syserr} for details. Each fit taken separately,
the data are nicely described.  Putting all five in terms of $(\as,\an)$-pairs
into the one fig.~\ref{fig:ellipses}, the discussed discrepancies become
obvious. But recall again that systematic uncertainties are treated
individually and therefore only statistical uncertainties are included!

With $y_{fJ}$, one obtains a good fit but rather low numbers for $\anmi$ and
$\asmz$, leading to the conclusion that the coefficient $a_{y_{fJ}}$ given in
the proceedings~\cite{pc:WDIS95} may not be appropriate.  Referring back to
section~\ref{sec:tubefits}, $y_{fJ}$ does exhibit smaller hadronization
corrections than thrust.  The curve in fig.~\ref{fig:DWfit2} was produced with
$a_{y_{fJ}}=-1/4$ derived in section~\ref{sec:DWaffits}.

Concerning $y_{k_t}$, the coefficient $a_{y_{k_t}}$ is not known and the
$1^\dagger$ in table~\ref{tab:2parDWfit} is pure guesswork.  The fit for $p=1$
works not very well supporting the conjecture of a $1/Q^2$-term instead of
$1/Q$.  In fact, both results obtained with $p=2$ look more reasonable.
Nevertheless, fig.~\ref{fig:DWfit2} contains the outcome according to
eq.~(\ref{eqn:Fpert}) without any power correction.

\subsection{Combined Fits}
\label{sec:DWcombfits}

Based on QCD, the universality of $\as$ is on firm theoretical grounds. Not
so, however, $\an$ since here additional assumptions are made. It was
therefore checked if fits allowing for different $\an$'s are possible while
enforcing one $\asmz$.  Correlations between the observables were ignored for
that purpose.  According to the systematic behaviour found previously
concerning the $\asmz$-results, several combinations are tried.  The outcome
is collected in table~\ref{tab:combDWfits}.

The first two arrangements reflect the separation in $\tau_P$, $B_P$ on the
one hand, and $\rho_E$, $\tau_C$, $C_P$ on the other.  As expected, the
numbers show resemblance to what we have obtained before.  Yet, including step
by step first $\tau_P$ and then $B_P$ into the $\rho_E$, $\tau_C$, $C_P$
triple, the $\chin$-values worsen considerably.\footnote{Note that with
  increasing dofs from $6$ for the two-parameter fits to $34$ in the last
  combination, the limit for the $5\%$ confidence level falls below $1.5$.}
Both times the fit results are dominated by the latter, which in addition to
be more numerous, also have the smaller statistical uncertainties.  $\asmz$ is
forced to remain high, and thereby $\anmi$ for $\tau_P$ and $B_P$ is shifted.

One concludes that combined fits with one $\asmz$ are possible, but
notwithstanding the freedom to choose an individual $\an$ for each event
shape, they suffer from systematic effects.  In order to determine $\asmz$, it
would be interesting to know what their origin is and what biases are implied.

\begin{table}
  \centering
  \begin{tabular}{|c||c|c|c|c|}
    \hline
    \multicolumn{1}{|c||}{\rbthm Comb.\ fits:} &
    \multicolumn{1}{|c|}{$\tau_P$, $B_P$} & 
    \multicolumn{1}{|c|}{$\rho_E$, $\tau_C$, $C_P$} & 
    \multicolumn{1}{|c|}{$\tau_P$, $\rho_E$, $\tau_C$, $C_P$} & 
    \multicolumn{1}{|c|}{$\tau_P$, $B_P$, $\rho_E$, $\tau_C$, $C_P$}\\
    \hline\hline
    $\an(\tau_P)$ &
    $0.532 \pm 0.012$ & --- &
    $0.374 \pm 0.010$ & $0.414 \pm 0.009$ \rbtrr\\\hline
    $\an(B_P)$ &
    $0.488 \pm 0.005$ & --- &
    --- & $0.443 \pm 0.006$ \rbtrr\\\hline
    $\an(\rho_E)$ &
    --- & $0.561 \pm 0.004$ &
    $0.561 \pm 0.004$ & $0.562 \pm 0.004$ \rbtrr\\\hline
    $\an(\tau_C)$ &
    --- & $0.476 \pm 0.003$ &
    $0.476 \pm 0.003$ & $0.474 \pm 0.003$ \rbtrr\\\hline
    $\an(C_P)$ &
    --- & $0.428 \pm 0.002$ &
    $0.427 \pm 0.002$ & $0.421 \pm 0.002$ \rbtrr\\\hline\hline
    $\asmz$ &
    $0.1115 \pm 0.0011$ & $0.1291 \pm 0.0007$ &
    $0.1285 \pm 0.0007$ & $0.1245 \pm 0.0006$ \rbtrr\\\hline\hline
    $\chin$ &
    $11.6/13$ & $38.4/20$ & $55.5/27$ & $222/34$ \rbtrr\\\hline
  \end{tabular}
  \caption[Results of combined fits \alDW\ et al.]
  {Results of combined fits \alDW\ et al.\ allowing for
    a separate $\anmi$ for each event shape, but enforcing
    one $\asmz$. Several combinations are tried;
    uncertainties are statistical only.}
  \label{tab:combDWfits}
\end{table}

\subsection[$\symbol{97}_F$-Fits]
{\boldmath$\symbol{97}_F$\unboldmath{}-Fits}
\label{sec:DWaffits}

Motivated by the problems encountered with the $a_F$-coefficients for the $y$
event shapes, one can try to do a three-parameter fit of $\an$, $\asmz$ and
$a_F$. However, this set of parameters is strongly correlated and with eight
data points only it does not properly converge for $\tau_P$, $B_P$, $\rho_E$,
$\tau_C$ or $C_P$.  The $y$ variables are somewhat of an exception here,
giving $\anmi = 0.58\pm 0.20$, $\asmz = 0.134\pm 0.015$ and
$a_{y_{fJ}}=-0.71\pm 0.18$ at $\chin = 1.6$ for $y_{fJ}$, and $\ao(\mi=2\gev)
= 0.35\pm 0.05$, $\asmz = 0.123\pm 0.006$ and $a_{y_{k_t}}=-33.2\pm 3.0$ at
$\chin = 0.23$ for $y_{k_t}$ respectively.  Despite looking quite reasonable,
one nevertheless has to exercise caution because of the very strong
correlations.

Being constrained that way to less than three parameters, we start from the
other side by deriving $a_F$ while presupposing the validity of
eq.~\ref{eqn:Fpow} with $\anmi=0.5$ and $\asmz=0.119$.
Table~\ref{tab:afDWfits} compiles the determined coefficients, which, although
some rather large $\chin$'s occur, are of a similar magnitude as the
theoretical ones.  Notably, $y_{fJ}$ gives $a_F\approx -1/4$ instead of $1$!

As a consistency check, this new value was reinserted to produce the
two-parameter fit in fig.~\ref{fig:DWfit2} with $\anmi=0.46$ and
$\asmz=0.118$, s.~again table~\ref{tab:2parDWfit}.  Performing the same, i.e.\ 
$p=1$, circular procedure with $y_{k_t}$, one gets $\anmi=0.34$ and
$\asmz=0.112$ which looks much better than the entry in
table~\ref{tab:2parDWfit}, but still is not satisfactory. Since $\ao$ is
basically unknown, we must determine it in addition to $a_F$ in case of $p=2$.
Refitting then with $a_F=-30$, one obtains $\ao(\mi=2\gev)=0.33$ and
$\asmz=0.121$, s.~table~\ref{tab:2parDWfit}.  Both refits, however, exhibited
strong correlations between $\an$ respectively $\ao$ and $\asmz$.

In a last test, all coefficients were multiplied by factors of $2$
respectively $1/2$ to study the effect on the $\anmi$, $\asmz$ results.
Having fig.~\ref{fig:ellipses} in mind, the outcome corresponds approximately
to shifts along the main diagonal with large $a_F$'s inducing smaller
$\anmi$'s and $\asmz$'s.

\begin{table}
  \centering
  \begin{tabular}{|c||c|c||c||c|c||c||c|c|}
    \hline
    \multicolumn{1}{|c||}{\rbthm $F$} &
    \multicolumn{1}{|c|}{$a_F$} & 
    \multicolumn{1}{|c||}{$\chin$} & 
    \multicolumn{1}{|c||}{$F$} & 
    \multicolumn{1}{|c|}{$a_F$} & 
    \multicolumn{1}{|c||}{$\chin$} & 
    \multicolumn{1}{|c||}{$F$} & 
    \multicolumn{1}{|c|}{$a_F$} &
    \multicolumn{1}{|c|}{$\chin$} \\\hline\hline
    $\tau_P$  & $0.84 \pm 0.03$  & $0.4$ &
    $\rho_E$  & $0.65 \pm 0.01$  & $11.2$ &
    $y_{fJ}$  & $-0.23 \pm 0.01$ & $1.3$ \rbtrr\\\hline
    $B_P$     & $0.41 \pm 0.01$  & $6.0$ &
    $\tau_C$  & $0.88 \pm 0.01$  & $9.0$ &
    $y_{k_t}$ & $-0.41 \pm 0.02$ & $1.8$ \rbtrr\\\hline
    &&& $C_P$ & $2.85 \pm 0.04$  & $27.7$ &&&\rbtrr\\\hline
  \end{tabular}
  \caption[Results for fits of the coefficients $a_F$ of the
  model of Dokshitzer, Webber et al.]
  {Results for fits of the coefficients $a_F$ while
    presupposing the validity of eq.~\ref{eqn:Fpow} with
    $\anmi=0.5$ and $\asmz=0.119$.
    Uncertainties are statistical only.}
  \label{tab:afDWfits}
\end{table}

\section{Evaluation of Systematic Uncertainties}
\label{sec:syserr}

The general procedure followed here is to repeat the fits for a variation in
every prominent origin of systematic effects. The obtained discrepancy
compared to the standard result is attributed to a corresponding uncertainty.
In case of two deviations in one direction for the same primary source,
e.g.~an up- and downwards modification of an energy scale, only the larger one
is considered for the evaluation of the total uncertainty.  The latter is
derived from all contributions by adding them up quadratically.  An exception
is the unfolding, whose influence is estimated in the same way as explained in
section~\ref{sec:finmeans} on the final results of the data means.

\pagebreak Altogether, the following studies have been performed to estimate
systematic effects:
\begin{itemize}
\item Experimental uncertainties:
  \begin{enumerate}
  \item Usage of four unfolding procedures
  \item Variation of the electromagnetic energy scale of the calorimeters by
    $\pm 1\%$, $2\%$, $3\%$ depending on $z_{\rm imp}$
  \item Variation of the hadronic energy scale of the LAr calorimeter by $\pm
    4\%$
  \end{enumerate}
\item Theoretical uncertainties:
  \begin{enumerate}
  \item Variation of the renormalization scale $\mr^2 = Q^2$ by factors of $2$
    and $1/2$
  \item Variation of the factorization scale $\mf^2 = Q^2$ by factors of $4$
    and $1/4$
  \item Variation of the infrared matching scale $\mi$ by $\pm 1/2\gev$
  \item Selection of the MRSA'-105 and MRSA'-130 pdfs~\cite{QCD:MRSAp} with
    lower respectively higher intrinsic $\asmz$ than in the standard pdfs
  \item Usage of the completely different pdf set CTEQ4A2~\cite{QCD:CTEQ4}
    with similar $\asmz$
  \end{enumerate}
\end{itemize}

The experimental sources have already been described in
section~\ref{sec:finmeans}, but concerning the theoretical ones, some remarks
are in order.  The renormalization scale $\mr$ as well as the factorization
scale $\mf$ (s.~\cite{ESW} for details) are arbitrary in the sense that in a
complete theory the calculations are not allowed to vary with any specific
choice. Yet, in reality we have only an approximative theory at our disposal
yielding residual dependences due to the neglected higher orders. To avoid the
appearance of large logarithms in the computations, it is recommended to
identify them with a process relevant scale which for our purposes is always
$Q$.  To estimate the effect of omitted higher orders, it is {\bf
  conventional} to vary them by an {\bf arbitrary} factor of $4$ concerning
the squares $\mr^2$, $\mf^2$.  In the case of $\mr^2$, we had to reduce this
factor to $2$ because of the condition~(\ref{eqn:mi}): $\Lambda \ll \mi
\approx 2\gev \ll Q$.  Taking this into account, varied results can be
achieved quickly via eq.~(\ref{eqn:c2mrdep}). With respect to $\mf$, however,
the complete calculations have to be redone.

The change of the infrared matching scale $\mi$ by $\pm 1/2$ follows from the
values used in the original proposal~\cite{pc:DWform1}.  Note that this
contributes only to the uncertainty for $\asmz$!  $\an$ explicitly depends on
$\mi$.

The next point accounts for the fact that implicitly $\asmz$ has already been
used in deriving the pdfs from data which in turn may bias our computations.
The same is true for the choice of parameterization of the pdfs. For those
reasons, three alternative sets, two with different assumptions on $\asmz$ and
one with approximately the same $\asmz$ but another parameterization, have
been selected for a reevaluation with DISENT~0.1.

\begin{figure} 
  \centering
  \includegraphics{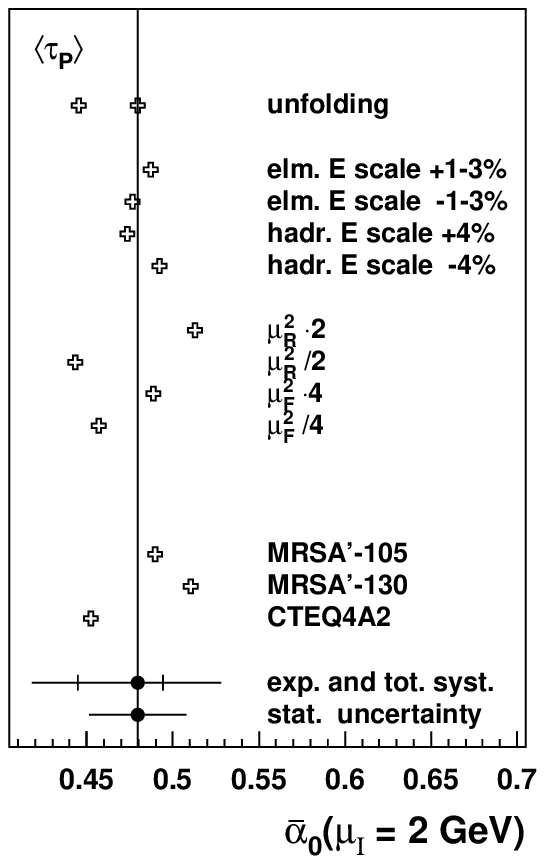}\hftwo%
  \includegraphics{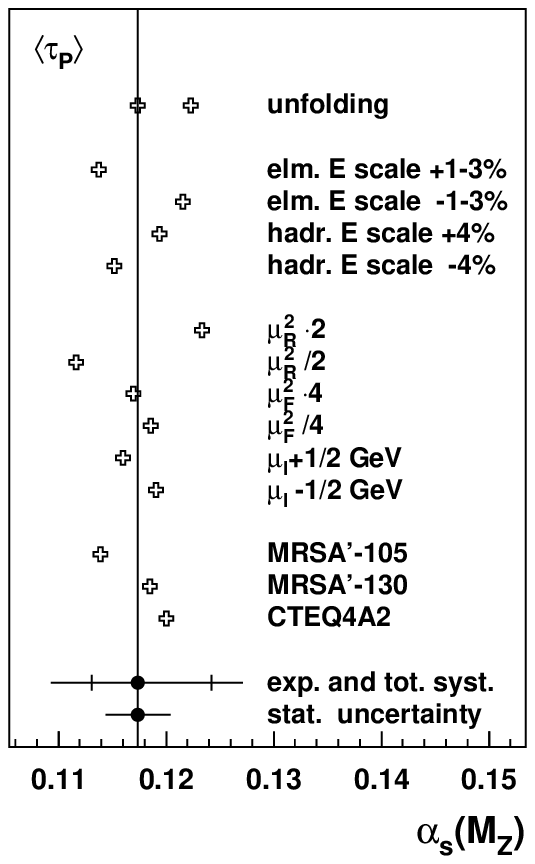}
  \includegraphics{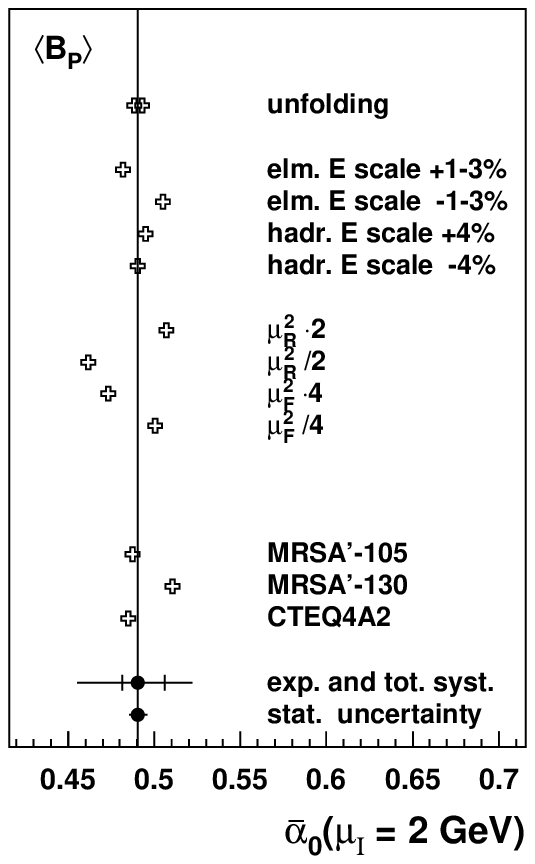}\hftwo%
  \includegraphics{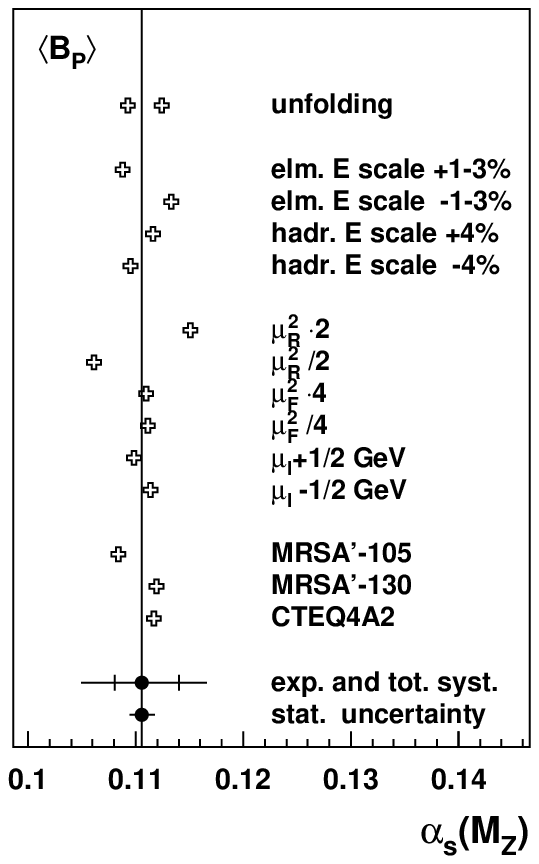}
  \caption[Systematic uncertainties of $\anmi$ and $\asmz$
  for $\tau_P$ and $B_P$.]  {Systematic uncertainties of $\anmi$ (left) and
    $\asmz$ (right) for $\tau_P$ (top) and $B_P$ (bottom).}
  \label{fig:syserr1}
\end{figure}

\begin{figure} 
  \centering
  \includegraphics{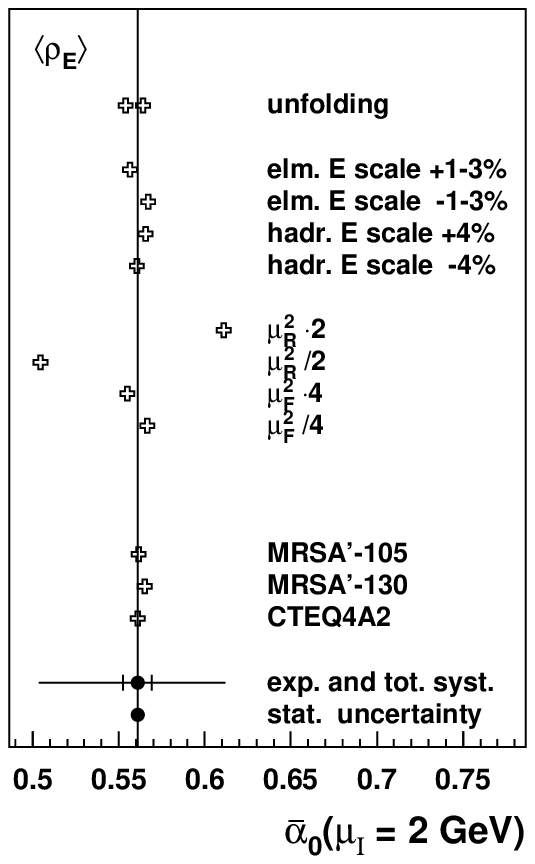}\hftwo%
  \includegraphics{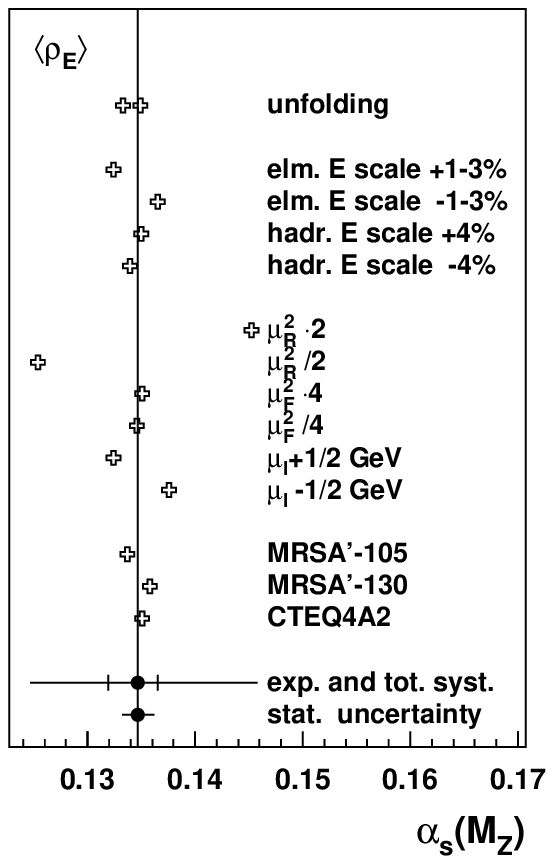}
  \includegraphics{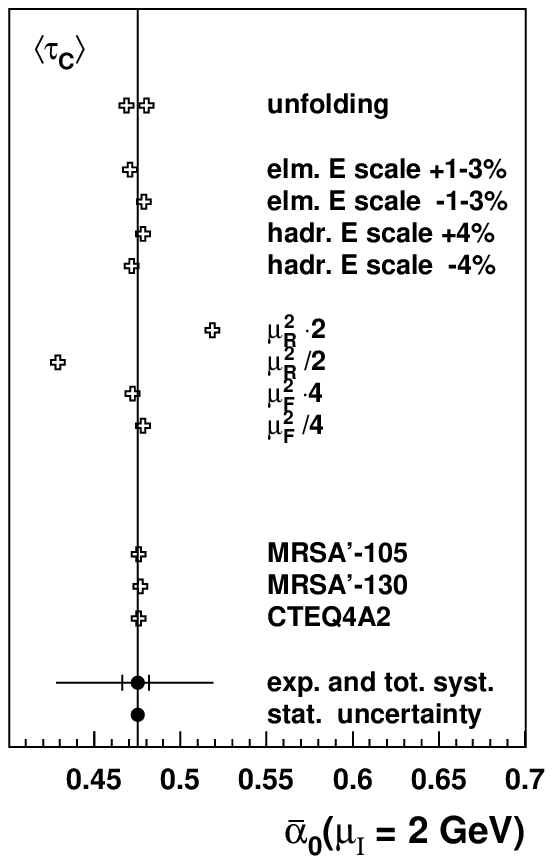}\hftwo%
  \includegraphics{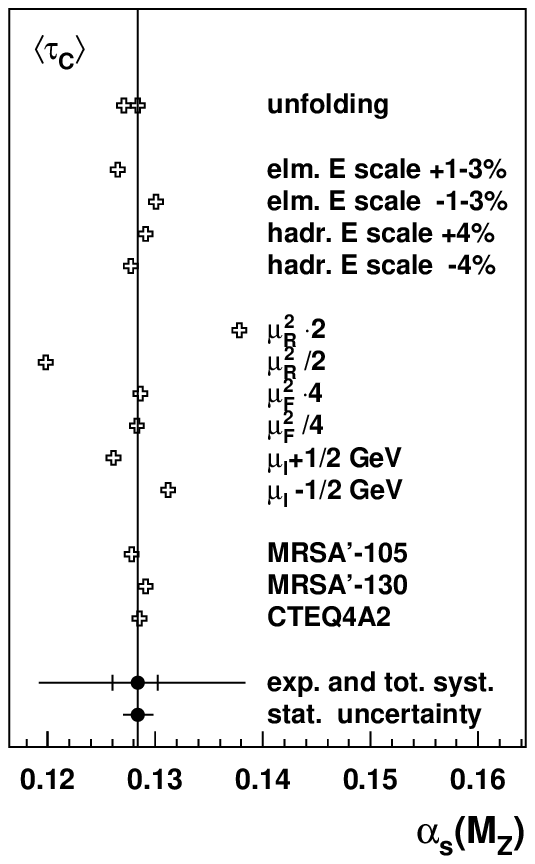}
  \caption[Systematic uncertainties of $\anmi$ and $\asmz$
  for $\rho_E$ and $\tau_C$.]  {Systematic uncertainties of $\anmi$ (left) and
    $\asmz$ (right) for $\rho_E$ (top) and $\tau_C$ (bottom).}
  \label{fig:syserr2}
\end{figure}

\begin{figure} 
  \centering
  \includegraphics{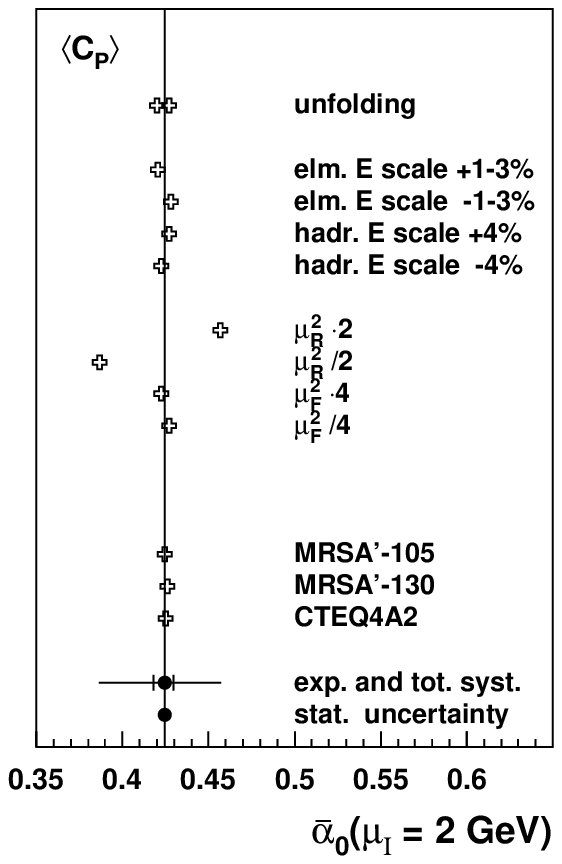}\hftwo%
  \includegraphics{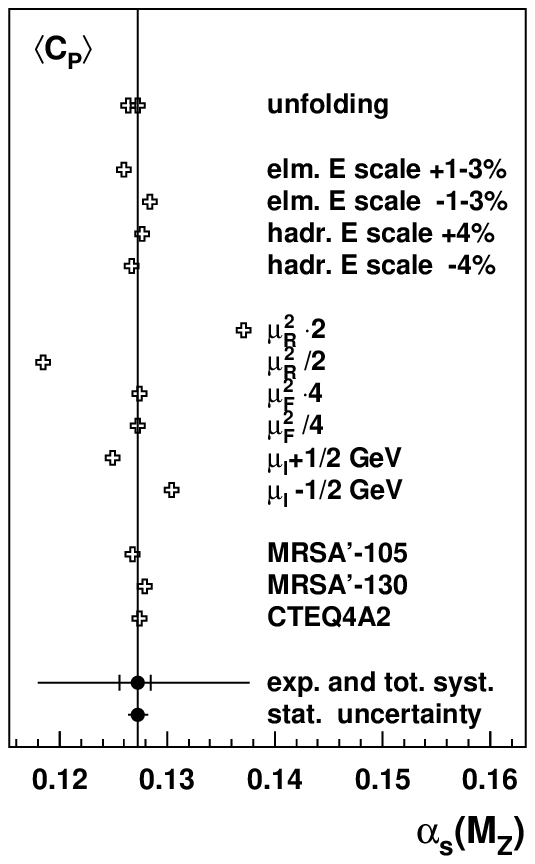}
  \caption[Systematic uncertainties of $\anmi$ and $\asmz$
  for $C_P$.]  {Systematic uncertainties of $\anmi$ (left) and $\asmz$ (right)
    for $C_P$.}
  \label{fig:syserr3}
\end{figure}

All uncertainties described are presented graphically in
figs.~\ref{fig:syserr1}--\ref{fig:syserr3}, separately for each of the five
event shapes where the coefficients $a_F$ are known.  Without such a
prediction for $y_{fJ}$ and $y_{k_t}$, a study of systematic effects is
omitted.

Inspecting the figs.~\ref{fig:syserr1}--\ref{fig:syserr3} again, it stands out
that $\tau_P$ and $B_P$ behave differently compared to $\rho_E$, $\tau_C$ and
$C_P$. For the latter, the systematic uncertainties are clearly dominated by
the variation of the renormalization scale.  In case of $\tau_P$ and $B_P$,
this is much less pronounced.  The larger influence of experimental
uncertainties probably can be related to the explicit reference to the boson
axis implied in the definitions of $\tau_P$ and~$B_P$.

\section{Cross-Checks}

Concerning systematic effects some additional cross-checks were performed:

\begin{enumerate}
\item In order to ensure a reliable measurement of the scattered electron,
  partially inefficient regions between calorimeter modules ($\phi$-cracks) or
  wheels ($z$-cracks) are avoided. This selectively diminishes the
  contribution to certain phase space regions and is corrected for by the
  procedure described in section~\ref{sec:corrdet}. Nevertheless, it was
  checked that the actual influence is negligible, even without unfolding.
  
\item The derivation of the power correction coefficients does not account for
  phase space constraints such as the cuts nos.~\ref{cut:Ee}
  and~\ref{cut:thetae} imposed on the scattered electron. Experimentally, they
  are absolutely necessary, but for testing purposes the NLO calculations were
  repeated without these cuts. Except for $\mean{y_{k_t}}$, which increased by
  about $8\%$ in the \lowq\ region, all other mean values changed by less than
  $2\%$.

\end{enumerate}


\chapter{Summary and Outlook}
\label{chap:summ}

The first measurement of event shapes in $ep$ DIS~\cite{KR:DIS97,H1:Shapes},
covering a range in $Q$ from $7\gev$ up to $100\gev$ in a single experiment,
has been substantially improved and extended to include the additional
variables $\rho_E$, $C_P$, $y_{fJ}$ and $y_{k_t}$ as well as new data taken in
$1997$. The available statistics in the \highq\ sample was more than doubled,
facilitating a bisection of the highest $Q$-bin.  The improvements encompass
i.a.\ a more precise calibration of the energy of the scattered electron, an
unfolding of the data distributions applying a Bayesian approach and a full
account of radiative QED corrections. Taking into consideration the
predominant sources of systematic influences, i.e.\ the accuracy of the
electromagnetic and hadronic energy scale of the calorimeters and the
unfolding procedure, corrected mean values of the event shape spectra together
with estimates on their statistical and systematic uncertainty could be
derived.

All event shape means exhibit a strong $Q$-dependence. They decrease with
rising $Q$, i.e.\ the energy flow becomes more collimated.  For the event
shapes defined in the current hemisphere alone, it is interesting to compare
with corresponding ones in \ee collisions.  This was already done
in~\cite{H1:Shapes} for $\mean{\tau_C}$ and $\mean{\tau_P}$ as well as
$\mean{\rho_Q}$. Despite the fact that the QCD dynamics in $ep$ DIS is
different,\footnote{There are neither BGF diagrams in \ee physics, nor is
  there a predefined event axis. In addition, it is impossible to have an
  empty hemisphere.}  a large similarity not only in the $Q$-dependence, but
also in absolute value to their \ee counterparts could be established.  When
consulting reanalyzed data of the JADE Collaboration~\cite{JADE:ICHEP98}, this
still holds for the $C$ parameter, but not for the jet broadening. $B_P$ is
about a factor two larger than the wide as well as the total jet broadening.

Employing calculations in pQCD to NLO accuracy, the influence of parton
fragmentation and hadronization could be studied.  As an improvement, the
approximative treatment of the $x$-dependence of the perturbative coefficients
applied in~\cite{H1:Shapes} has been replaced by a more precise bin-wise
evaluation.  A direct comparison with data revealed that the investigated
event shapes can roughly be categorized into three classes:
\pagebreak
\begin{enumerate}
\item The event shapes $\tau_P$ and $B_P$ employing the $z\Bf$-axis as event
  axis are sensitive to QED radiation. They show hadronization corrections of
  medium size.
\item $\rho_E$, $\tau_C$ and $C_P$ are strongly affected by the
  non-perturbative transition to hadrons. Their by far largest systematic
  uncertainty is due to the variation of the renormalization scale $\mr$.
\item Both event shapes based on jet algorithms, $y_{fJ}$ and $y_{k_t}$,
  exhibit small (even negative) hadronization corrections.
\end{enumerate}

Fixing $\asmz$ to $0.119$, the current world average~\cite{PDG}, simple power
corrections $\lambda/Q$ and $\mu/Q^2$ to parameterize the discrepancies
between data and theory were tested.  For $\lambda$, reasonable results of
$0.7\gev$ and $0.5\gev$ could be obtained for $\tau_P$ and $B_P$, whereas
$y_{fJ}$ and $y_{k_t}$ yielded negative values of $-0.2\gev$ and $-0.4\gev$
respectively.  For the other event shapes the fit delivered poor $\chin$'s.
This is generally the case for $\mu/Q^2$-terms; except for maybe $y_{fJ}$ and
$y_{k_t}$, they are ruled out.  Fitting $\asmz$ in addition to $\lambda$ or
$\mu$ did not help either due to very large correlations.

Invoking the approach for power corrections initiated by Dokshitzer and
Webber, the situation improves significantly.  For $\tau_P$, $B_P$, $\rho_E$,
$\tau_C$ and $C_P$, where predictions for the coefficients $a_F$ are
available, correlations were reduced ---~although not as far as desirable~---
and the fits converged nicely.  Note that for $B_P$ the new coefficient
$a'_{B_P}$ replacing the formerly predicted logarithmic term had to be
included.  Within $20\%$, all determined $\an$'s are compatible with a
universal value of $\anmi \approx 0.5$.  Very similar results are achieved
with this model when applied to \ee
data~\cite{JADE:ICHEP98,Wicke:QCD97,DELPHI:pc} except for the wide jet
broadening.

Concerning $\asmz$, an inconvenient spread ranging from $0.111$--$0.135$ is
observed. Although all five values turned out to be consistent with each
other, a simple averaging seems to be inappropriate since the outcomes for
$\asmz$ again reflect two of the categories defined above. $\tau_P$ and $B_P$
mark the lower bound of the range in $\asmz$ as opposed to $\rho_E$, $\tau_C$
and $C_P$ at the upper end.  Combining several of these event shapes, the fits
did work, but their quality severely declined, even allowing an individual
$\an$ for each.\footnote{Note that correlations were ignored.}  As long as
this systematic behaviour is not better understood, a precise determination of
$\asmz$ is not feasible that way.  Derived from the spread above, it can be
stated at best that $\asmz = 0.123\pm0.012$.

Returning to the $y$ event shapes, the conjectured coefficient of
$a_{y_{fJ}}=1$ has been excluded. Derived from data keeping $\anmi = 0.5$ and
$\asmz = 0.119$ fixed, a value of about $-1/4$ can be predicted.  Without any
coefficient given for $y_{k_t}$, it can only be checked for the power $p$ of
the $1/Q^p$ term. Due to the smallness of the observed hadronization
corrections in comparison with experimental uncertainties, however, it is very
difficult to extract them from data. In fact, the performed investigations
hint at powers $p>1$, but they are not conclusive enough to claim more.

\pagebreak\noindent The achieved measurements can be complimented in several
ways:
\begin{itemize}
\item Instead of looking for the $Q$-dependence alone, the data could
  additionally be divided into bins in $x$. This is done by the ZEUS
  Collaboration~\cite{ZEUS:Shapes}.
\item In \ee annihilation the model of Dokshitzer and Webber et al.\ has been
  extended to differential distributions~\cite{pc:DWdistr,pc:Brevisited}.
  Although no predictions for $ep$ DIS exist, it is straightforward to just
  look what comes out, since the unfolded event shape spectra are at hand.
  Actually, a first check was already performed~\cite{HUM:DIS98} and revealed
  inconsistencies even between fits of the mean values and distributions of
  one shape variable. Resummed calculations which are not available for $ep$
  DIS may be helpful.
\item By further subdividing a distribution into intervals as done by the
  DELPHI Collaboration, it is possible to examine where specific power-like
  contributions occur~\cite{DELPHI:pc,Wicke:Moriond98}.
\item Alternatively, higher moments of already investigated event shapes can
  be studied. First steps into that direction were taken by the OPAL
  Collaboration~\cite{OPAL:EPSHEP97}.
\item Another obvious extension is the usage of new variables, possibly with
  small hadronization corrections.
\end{itemize}  

Especially concerning the last point, however, the question arises of what to
do with them. For $y_{k_t}$ it was already noticed that with the given
experimental uncertainties power contributions can not be reliably extracted.
Yet, just in this case it could be interesting to perform direct fits of
$\asmz$ with the distributions at \highq. Remaining hadronization corrections
must then be estimated by the usual procedure of correcting to the parton
level of MC models. At the same time, this could be done for the event shapes
of this analysis, such that a comparison of the fit results in $\asmz$ could
provide insight into systematic effects in one (or both) of these methods.

At last, it should be mentioned that the {\em per se}\/ sharp angular cut-off
of $90\grad$ in the Breit frame is neither theoretically nor experimentally
unproblematic, (s.~sections~\ref{sec:shapesdef} and~\ref{sec:fincut}).  It has
been suggested to replace it by a kind of weighting procedure to ensure a
smooth transition.  On the other hand, one could try to separate the remnant
by employing e.g.\ the $k_t$-algorithm and treat then all remaining objects as
input for the derivation of event shapes.


\cleardoublepage
\addcontentsline{toc}{chapter}{List of Figures}
\listoffigures
\cleardoublepage
\addcontentsline{toc}{chapter}{List of Tables}
\listoftables
\cleardoublepage


\cleardoublepage
\pagestyle{empty}
\addcontentsline{toc}{chapter}{Curriculum Vitae}
\chapter*{Curriculum Vitae} \thispagestyle{empty} 

\begin{tabular}{ll}
  Geburtsdatum:         & 13. Januar 1967\\
  Geburtsort:           & M"onchengladbach\\
  Familienstand:        & ledig\\
  Staatsangeh"origkeit: & deutsch\\
  Eltern:               & Matthias Rabbertz, Weber\\
  & Gertrud Rabbertz, geb. Hermanns, Bankangestellte\\
  Geschwister:          & Beate Reitler\\
  & Gerda Boms
\end{tabular}\\[0.5cm]
\begin{tabular}{lll}
  \parbox[t]{3.5cm}{{\bf Schulbildung}:} & 01.08.73 - 06.07.77 &
  \parbox[t]{8.2cm}{Gemeinschaftsgrundschule Wegberg}\\
  & 01.08.77 - 12.06.86 &
  \parbox[t]{8.2cm}{St"adtisches Maximilian-Kolbe-Gymnasium\\ Wegberg\\
    Abschlu"s: Allgemeine Hochschulreife}\\
  &&\\
  {\bf Wehrdienst}:
  & 01.10.86 - 31.12.87 &
  \parbox[t]{8.2cm}{Grundausbildung in Pinneberg\\
    T"atigkeit als Sprechfunker in Geilenkirchen}\\
  &&\\
  {\bf Studium}:
  & 01.10.87 - 08.26.94 &
  \parbox[t]{8.2cm}{Rhei\-nisch-West\-f"a\-lische
    Tech\-nische Hoch\-schule Aachen,
    Mathe\-matisch-Natur\-wissen\-schaft\-liche Fa\-kul\-t"at\\
    Studienfach: Physik\\
    Fachrichtung: Elementarteilchenphysik\\
    Abschlu"s: Diplom-Physiker}\\
  &&\\
  & 01.11.89 - 15.10.92 &
  \parbox[t]{8.2cm}{T"atigkeit als studentische Hilfskraft am Institut f"ur
    Mathematik, Prof.~Dr.\ G.~Hellwig}\\
  &&\\
  & seit 01.07.94 &
  \parbox[t]{8.2cm}{Wissen\-schaft\-licher An\-gestell\-ter am I.\
    Phy\-sika\-lischen Institut der Rhei\-nisch-West\-f"a\-lischen
    Tech\-nischen Hoch\-schule Aachen, Beginn des Promotionsvorhabens bei
    Prof.~Dr.\ \mbox{Ch.~Berger}}\\
  &&\\
  & seit 01.01.95 &
  \parbox[t]{8.2cm}{Mitglied der H1-Kollaboration am DESY, Hamburg}
\end{tabular}\\[0.5cm]
%


\cleardoublepage
\addcontentsline{toc}{chapter}{Acknowledgements}
\chapter*{Acknowledgements}
\thispagestyle{empty}

First of all, many thanks to my advisor, Prof.~Dr.\ Christoph Berger,
who put me on track of this exciting new topic, and to my collaborator
Dr.~Hans-Ulrich Martyn. Their constant and patient support were
invaluable to me. Both had, at some point or another, to suffer from
my stubbornness.

I am thankful to Prof.~Dr.\ Siegfried Bethke for accepting the task
of being my second referee.

An experiment as complicated as H1 can not be successfully operated
without the dedicated and knowledgeable work of many people. I wish
to thank the members of the H1 Collaboration, who contributed to
this analysis in manifold ways, for their hospitality and the
fascinating opportunity of researching in an international community.

For enlightening discussions and answering many questions, I am
indebted to M.~Dasgupta, Yu.L.~Dokshitzer, D.~Graudenz, E.~Mirkes,
G.P.~Salam, M.H.~Seymour and B.R.~Webber.

On the important field of social interactions, I enjoyed, and still do so,
very much the friendly and inspiring work climate of the Aachen groups,
which sometimes led to less work and more climate \ldots,
thank you:\\
Simone Baer-Lang, Sascha Caron, Carlo Duprel, Thomas Hadig,
Martin Hampel, Heiko Itterbeck, Claus Keuker, Torsten K"ohler,
Carsten Krauss, Boris Leissner, Peer-Oliver Meyer, Christian Niedzballa,
Konrad Rosenbauer, Wilhelm Rottkirchen, J"urgen Scheins,
Lars Sonnenschein, Markus Wobisch and \ldots\ all the people
I might have forgotten.

Special thanks to Thomas Hadig, Martin Hampel, Thorsten Wengler and
Markus Wobisch for proofreading and lots of helpful suggestions.

Last but not least, I am heartily grateful to my family and especially
my parents for their generous support and their confidence in me.
It is very easy to take comforts for granted.


\cleardoublepage
\end{document}